\documentclass[%
 aip,
 amsmath,amssymb,
 preprint,%
]{revtex4-2}

\usepackage{graphicx}
\usepackage{bm}

\usepackage[utf8]{inputenc}
\usepackage[T1]{fontenc}
\usepackage{mathptmx}
\usepackage{etoolbox}

\renewcommand{\vec}[1]{\boldsymbol{#1}}
\newcommand{\vy}{\vec{y}}
\newcommand{\N}{\mathcal{N}}
\newcommand{\vzero}{\vec{0}}
\newcommand{\vI}{\vec{I}}

\newcommand{\transpose}{\intercal}
\DeclareMathOperator*{\E}{\mathbb{E}}
\DeclareMathOperator{\diag}{diag}

\newcommand{\SizeTime}{m}
\newcommand{\SizeDynSys}{n}
\newcommand{\SizeAlphabet}{c}
\newcommand{\IndexTime}{i}
\newcommand{\IndexDynSys}{j}
\newcommand{\IndexAlphabet}{\ell}

\newcommand{\g}{\dot{\vec{X}}_\IndexDynSys} 
\renewcommand{\H}{\vec{\Theta}}
\newcommand{\f}{\vec{\Xi}_\IndexDynSys}
\newcommand{\hatf}{\hat{\vec{\Xi}}_\IndexDynSys}
\newcommand{\noise}{\vec{\epsilon}_{\IndexDynSys}}
\newcommand{\Vepsilon}{\vec{V}_{\vec{\epsilon}_\IndexDynSys}}
\newcommand{\Vf}{\vec{V}_{\vec{\Xi}_\IndexDynSys}}
\newcommand{\model}{{f}}
\newcommand{\statec}{x}
\newcommand{\dotstatec}{\dot{x}}
\newcommand{\state}{\vec{x}}
\newcommand{\dotstate}{\dot{\vec{x}}}
\newcommand{\statemat}{\vec{X}}
\newcommand{\dotstatemat}{\dot{\vec{X}}}
\newcommand{\matparam}{\vec{\Xi}}
\newcommand{\matparamc}{\xi}
\newcommand{\meanpriorf}{\vec{\mu}_\IndexDynSys}
\newcommand{\regparam}{\psi}
\newcommand{\BPostMean}{\hat{\matparam}_\IndexDynSys}
\newcommand{\BPostCovMat}{\hat{\vec{\Delta}}_\IndexDynSys}
\newcommand{\POR}{\text{POR}_\IndexDynSys}
\newcommand{\BEvidMean}{\hat{\dotstatemat}_\IndexDynSys}
\newcommand{\BEvidCovMat}{\hat{\vec{\Omega}}_\IndexDynSys}
\newcommand{\Vepsilondiag}{\vec{v}_{\vec{\epsilon}_\IndexDynSys}}
\newcommand{\meanVepsilondiag}{\overline{\vec{v}}_{\vec{\epsilon}_\IndexDynSys}}
\newcommand{\Vfdiag}{\vec{v_{\vec{\Xi_\IndexDynSys}}}}

\newcommand{\VepsilondiagInd}[1]{v_{\epsilon_{#1}}}
\newcommand{\VfdiagInd}[1]{v_{\Xi_{#1}}}
\newcommand{\eizero}{\epsilon_{\IndexTime \IndexDynSys}}
\newcommand{\fizero}{\Xi_{\IndexAlphabet \IndexDynSys}}

\newcommand{\AIC}{\text{AIC}_\IndexDynSys}
\newcommand{\AICc}{{\text{AIC}}_\IndexDynSys^\text{c}}
\newcommand{\BIC}{\text{BIC}_\IndexDynSys}

\newcommand{\EB}{\text{EB}_\IndexDynSys}

\begin{document}


\title[Dynamical System Identification by Bayesian Inference]{Dynamical System Identification, Model Selection and Model Uncertainty Quantification by Bayesian Inference}

\author{Robert K. Niven}
 \email{r.niven@unsw.edu.au}
 \affiliation{School of Engineering and Technology, The University of New South Wales, Canberra, ACT, 2600, Australia}
 
\author{Laurent Cordier}
\affiliation{Institut Pprime, CNRS – Université de Poitiers – ISAE-ENSMA, 86360 Chasseneuil-du-Poitou, France}%

\author{Ali Mohammad-Djafari}%
\affiliation{Laboratoire des signaux et syst\`emes (L2S), CentraleSup\'elec, 91190 Gif-sur-Yvette, France}%

\author{Markus Abel}
\affiliation{Ambrosys GmbH, 14482 Potsdam, Germany}%

\author{Markus Quade}
\affiliation{Ambrosys GmbH, 14482 Potsdam, Germany}%

\date{\today}

\begin{abstract}

This study presents a Bayesian maximum \textit{a~posteriori} (MAP) framework for dynamical system identification from time-series data. This is shown to be equivalent to a generalized Tikhonov regularization, providing a rational justification for the choice of the residual and regularization terms, respectively, from the negative logarithms of the likelihood and prior distributions. In addition to the estimation of model coefficients, the Bayesian interpretation gives access to the full apparatus for Bayesian inference, including the ranking of models, the quantification of model uncertainties and the estimation of unknown (nuisance) hyperparameters. Two Bayesian algorithms, joint maximum \textit{a~posteriori} (JMAP) and variational Bayesian approximation (VBA), are compared to the {LASSO, ridge regression and SINDy algorithms for sparse} regression, by application to several dynamical systems with added {Gaussian or Laplace} noise. For multivariate Gaussian likelihood and prior distributions, the Bayesian formulation gives Gaussian posterior and evidence distributions, in which the numerator terms can be expressed in terms of the Mahalanobis distance or ``Gaussian norm'' $||\vy-\hat{\vy}||^2_{M^{-1}} = (\vy-\hat{\vy})^\top {M^{-1}} (\vy-\hat{\vy})$, where $\vy$ is a vector variable, $\hat{\vy}$ is its estimator and $M$ is the covariance matrix. The posterior Gaussian norm is shown to provide a robust metric for quantitative model selection {for the different systems and noise models examined.}

\end{abstract}

\maketitle

\begin{quotation}
A large number of natural and human phenomena are controlled by dynamical systems, commonly expressed in the form of differential or difference equations for the evolution of state variables in time.  Examples include 
conservation laws for the flow of fluids, chemical species, momentum and energy; 
the laws of mechanics for harmonic oscillators, 
machines and planetary motion; 
biological population models for species growth, competition and predator-prey relations; 
financial demand-supply models for prices, incomes and economic growth; 
and combat models in military operations research.
One of the great challenges of inference is to identify a dynamical system from its discrete time-series data, including its governing differential equation system and parameter values. 
While many methods have been used, few researchers have applied Bayesian inference, the fundamental method for the discovery of models from data.
This study provides a Bayesian perspective on dynamical system identification, and illustrates several advantages over existing regularization methods. 
\end{quotation}

\section{Introduction}

Many physical phenomena can be described by a continuous dynamical system, commonly represented by:
\begin{equation}
\dot{\state}(t) = {\model}(\state(t)),
\label{eq:dynsys}
\end{equation}
where $\state \in \mathbb{R}^\SizeDynSys$ is the observable state vector, generally a function of time $t$ (and/or some other parameters), $\dot{\state} \in \mathbb{R}^\SizeDynSys$ is its time derivative and $\model$ is the system function or model, in general a nonlinear function of $\state$. Given a sample of discrete time-series data $[\state(t_1), \state(t_2), \state(t_3), ...]$,  
how should we identify the model $\model$?  In its most general form, this can be described as \emph{system identification}, while for $\model$ selected from a known class of functions, the problem reduces to one of \emph{parameter identification}. These functions may be chosen from a set of basis functions applicable to the system, or constructed from an alphabet of the data variables. 

In recent years, there has been considerable interest in the use of sparse regularization methods for dynamical system identification \citep[e.g.][]{Brunton_etal_2016, Schaeffer_2016, Mangan_etal_2017, Rudy_etal_2017, Kaiser_etal_2018, Quade_etal_2018, Mangan_etal_2019, Champion_etal_2019, Champion_etal_2020, Kaheman etal_2020, Lai_2021, Fasel_etal_2022}. 
These are used to infer a matrix of coefficients which can reproduce the dynamical system very precisely.
Such methods usually incorporate a sparsification component to reduce the number of coefficients, such as a regularization term in the objective function. However, both the regularization term and its multiplying parameter (penalization coefficient) tend to be chosen in a heuristic or \textit{ad hoc} manner, without much theoretical insight. 

It must be noted that the fundamental method for the solution of inverse problems such as \eqref{eq:dynsys} is {\it Bayesian inference}. This involves the application of Bayes' rule to calculate the ``probability of an hypothesis or model, given the data'', or in other words, the degree of belief in the hypothesis or model, given what is known \cite{Bayes_1763, Laplace_1774, Jaynes_2003, vonToussaint_2011}. 
The Bayesian interpretation of probabilities as ``degrees of belief'' or ``plausibilities'' has been shown to be universal and mathematically rigorous \cite{Polya_1954, Cox_1961, Jaynes_2003}, overcoming the late-19th century restriction of probabilities to measurable frequencies \cite{Venn_1888, Fisher_1925, Neyman_Pearson_1933, Feller_1966}.
Bayesian inference provides a common platform for the analysis of all uncertainties, including 
inverse model identification, model ranking, hypothesis testing, estimation, interpolation, extrapolation, classification, experimental design or any other analysis of data, and can be used in subsequent decision frameworks such as utility or loss analysis and risk assessment.  

The aim of this study is to present a Bayesian framework for dynamical system identification from time-series data, based on the Bayesian maximum \textit{a posteriori} (MAP) estimate. This was originally developed for image analysis \cite{MD_2015, MD_Dimutru_2015, MD_2016, Teckentrup_MaxEnt2018, Calvetti_Somersalo_2018}, but  there have been some applications to dynamical systems \citep[e.g.][]{Pan_etal_2016, Zhang_Lin_2018, Niven_etal_MaxEnt_2019, Chiuso_Pillonetto_2019, Chen_Lin_2021, Hirsch_etal_2022, Yuan_etal_2023, Lin_etal_2023, Zhang_etal_2023, Taghavi_2024, Fung_etal_2024, Klishin_etal_2024}.
The method bears specific similarities to Bayesian dynamical system reconstruction by Gaussian processes \cite{Roberts_etal_2013, Svensson_Schon_2017, Raissi_Karniadakis_2018, Teckentrup_2020, Zhang_etal_2023}, 
posterior exploratory methods \cite{Hirsh_etal_2022, Rinkens_etal_2023} 
or constrained computational methods \cite{Kontogiannis_etal_2022}. 
In \S\ref{sect:reg} we first examine the formulation of regularization methods, followed in \S\ref{sect:Bayes_reg} by a Bayesian reinterpretation. 
The MAP estimate is shown to be equivalent to a generalized Tikhonov regularization, {providing} a rational justification for the choices of the residual and regularization terms, respectively, as the negative logarithms of the likelihood and prior distributions.
We then explore further features of the Bayesian inverse method, including 
the Bayesian posterior and model uncertainty quantification (\S\ref{sect:UQ}), 
the posterior Bayes factor and Bayesian model selection (\S\ref{sect:model_sel}), 
the Bayesian evidence (\S\ref{sect:evidence}), 
and 
two Bayesian algorithms for the estimation of unknown hyperparameters: joint maximum \textit{a posteriori} (JMAP) and variational Bayesian approximation (VBA) (\S\ref{sect:joint_Bayes}). 
In \S\ref{sect:apps}, we then apply JMAP and VBA to the analysis of three dynamical systems with added {Gaussian or Laplace} noise.  {These are compared to the LASSO, ridge regression and SINDy algorithms for sparse} regression. 
The findings are summarized in the conclusions in \S\ref{sect:concl}.

\section{\label{sect:theory} Theoretical Foundations}
\subsection{\label{sect:reg} Sparse Regression (Regularization) Methods}

The application of sparse regression for dynamical system identification \citep[e.g.][]{Brunton_etal_2016, Schaeffer_2016, Mangan_etal_2017, Rudy_etal_2017, Lai_2021} generally proceeds by the following steps.  First, a set of time-series data, consisting of $\SizeTime$ time steps of an $\SizeDynSys$-dimensional variable $\state(t)$ with elements $\statec_\IndexDynSys(t_\IndexTime)$, and of its derivative $\dot{\state}(t)$ with elements $\dot{\statec}_\IndexDynSys(t_\IndexTime)$, are assembled into the $\SizeTime \times \SizeDynSys$ matrices:
\begin{align}
\statemat &= 
\begin{bmatrix}
\state^{\transpose}(t_1) \\ \vdots \\ \state^{\transpose}(t_\SizeTime)
\end{bmatrix}
=
\begin{bmatrix}
\statec_1(t_1) & \dots & \statec_\SizeDynSys(t_1) \\ 
\vdots & & \vdots \\ 
\statec_1(t_\SizeTime) & \dots & \statec_\SizeDynSys(t_\SizeTime) 
\end{bmatrix} \text{ and}
\label{eq:X}
\\
\dotstatemat &= 
\begin{bmatrix}
\dotstate^{\transpose}(t_1) \\ \vdots \\ \dotstate^{\transpose}(t_\SizeTime)
\end{bmatrix}
=
\begin{bmatrix}
\dotstatec_1(t_1) & \dots & \dotstatec_\SizeDynSys(t_1) \\ 
\vdots & & \vdots \\ 
\dotstatec_1(t_\SizeTime) & \dots & \dotstatec_\SizeDynSys(t_\SizeTime) 
\end{bmatrix}.
\label{eq:dotX}
\end{align}
Second, an alphabet of $\SizeAlphabet$ functions with index $\IndexAlphabet$ -- chosen by a classification method or from theoretical insights into the problem -- is applied to $\statemat$ (and also if desired to $\dotstatemat$), to generate an $\SizeTime \times \SizeAlphabet$ library matrix, for example:
\begin{equation}
\H(\statemat) = 
\begin{bmatrix}
1 & \statemat & \statemat^2 & \statemat^3 & \dots & \sin(\statemat) & \cos(\statemat) & \dots 
\end{bmatrix},
\label{eq:Theta}
\end{equation}
here expressed in terms of functions of the columns of \eqref{eq:X}. 
The dynamical system \eqref{eq:dynsys} is thereby converted to the discrete linear map:
\begin{equation}
\dotstatemat = \H(\statemat) \matparam,
\label{eq:sys}
\end{equation}
in which $\matparam$ is a $\SizeAlphabet \times \SizeDynSys$ matrix of coefficients $\matparamc_{\IndexAlphabet \IndexDynSys} \in \mathbb{R}$, representing the model $\model$. 
For multivariate data, \eqref{eq:sys} can be resolved columnwise, i.e., by considering a succession of problems in which $\g$ and $\f$ correspond to the $j$th columns of $\dotstatemat$ and $\matparam$, respectively. 

Third, \eqref{eq:sys} is solved to determine the matrix $\matparam$. 
The simplest inversion method is least squares regression, involving minimization of an objective function based on the residual sum of squares \citep[e.g.,][]{Aster_etal_2013, Bardsley_2018}:  
\begin{align}
\begin{split}
{\hatf}^\text{LS} &= \arg \min\limits_{\f} J^\text{LS}(\f),
\\
J^\text{LS}(\f) &= ||\g - \H(\statemat) \f||^2_2 
\end{split}
\label{eq:least_sq}
\end{align}
where 
$\hatf^\text{LS}$ is the inferred solution and $||\vy||_2=\sqrt{\vy^\transpose \vy}$ is the $L_2$ or Euclidean norm for the vector quantity $\vy$. 
However, in practice \eqref{eq:sys} will contain noise, causing computational difficulties and the generation of spurious coefficients in 
$\hatf^\text{LS}$.  This is generally handled by a sparse regression method, in which an additional regularization term is added to the objective function \cite{Tikhonov_1963, Hoerl_Kennard_1970, Santosa_Symes_1986, Kaipio_Somersalo_2005, Aster_etal_2013, Bardsley_2018}:
\begin{align}
\begin{split}
\hatf^\text{reg} &= \arg \min\limits_{\f} {J}^\text{reg}(\f),
\\
{J}^\text{reg}(\f) &= ||\g - \H(\statemat) \f||^\alpha_\beta + \regparam ||\f||^\gamma_\delta,
\end{split}
\label{eq:modJ}
\end{align}
\noindent where in general 
${\| \vy \|_{p}=(\sum\nolimits_{i} |y_{i}|^{p})^{1/p}}$  
is the $L_p$ norm with $p \in \mathbb{R}$ for the vector quantity $\vy$, 
$\regparam \in \mathbb{R}$ is the regularization coefficient and $\alpha, \beta, \gamma, \delta \in \mathbb{R}$ are constants. 
For dynamical system identification, 
\eqref{eq:modJ} has
been implemented with $\alpha, \gamma \in \{1,2\}$, $\beta=2$ and $\delta \in \{0,1,2 \}$, 
including ridge regression \cite{Hoerl_Kennard_1970, Bardsley_2018} {with $\regparam=\theta_j$} for $\alpha, \beta, \gamma,\delta=2$, 
the least absolute shrinkage and selection operator (LASSO) \cite{Santosa_Symes_1986, Tibshirani_1996} 
or  
sparse optimization \cite{Schaeffer_2016}  {with $\regparam=\kappa_j$} for $\alpha, \beta=2$ and $\gamma, \delta=1$, 
and strong sparsity \cite{Rudy_etal_2017} {with $\regparam=\phi_j$} for $\alpha, \beta=2$, $\gamma=1$ and $\delta=0$.
{The convergence properties of these algorithms, and their connection to basis functions and reconstruction theorems for nonlinear dynamical systems, have been reported \cite{Stark_1999, Sauer_2004, Yao_Bollt_2006, Wang_etal_2011}.
Regularization} methods have close connections to other {order reduction methods} such as principal orthogonal decomposition (POD), singular value decomposition (SVD), dynamic mode decomposition (DMD) and Koopman analysis \citep[e.g.][]{Aster_etal_2013, Brunton_etal_2016_PLOS, Brunton_etal_2017_NatureComm, Taira_etal_2017}.  

Alternatively, instead of \eqref{eq:modJ}, some authors have enforced a sparse model by iterative thresholding, termed the sparse identification of nonlinear dynamics (SINDy) method \cite{Brunton_etal_2016, Mangan_etal_2017, Kaiser_etal_2018, Quade_etal_2018, Mangan_etal_2019, Champion_etal_2019, Champion_etal_2020, Kaheman etal_2020, Fasel_etal_2022}:
\begin{equation}
{J}^\text{IT}(\f) = ||\g - \H(\statemat) \f||^2_2 \hspace{10pt} \text{with} \hspace{10pt} |\matparamc_{\IndexAlphabet \IndexDynSys}| \ge \lambda, \hspace{2pt} 
\forall\IndexAlphabet\,,\forall\IndexDynSys 
\label{eq:SINDy_reg}
\end{equation}
where $\lambda$ is a threshold parameter. 
{While the original SINDy method is described as a variant of LASSO \cite{Brunton_etal_2016}, its code is based on a sequential thresholded least-squares (STLSQ) algorithm \eqref{eq:SINDy_reg}, designed to be more computationally efficient than LASSO for large data sets. This was later} shown \cite{Zhang_Schaeffer_2018}  to converge to \eqref{eq:modJ} with $\alpha=\beta=2$ and $\gamma=0$, {or strong sparsity \cite{Rudy_etal_2017}}. 

We note that in most previous applications, the regularization term in \eqref{eq:modJ} and parameter $\regparam$ -- or the threshold $\lambda$ in \eqref{eq:SINDy_reg} -- are selected by heuristic methods, to prefer solutions with smaller norms and/or based on user judgments of the numerical performance. Some authors have selected the SINDy $\lambda$ based on metrics such as the Akaike information criterion (AIC), to prefer models with fewer coefficients \cite{Mangan_etal_2017}. Other methods for selection of $\regparam$ include cross-validation and generalized cross-validation, unbiased predictive risk estimators, the discrepancy principle and the L-curve principle \cite{Aster_etal_2013, Bardsley_2018}.
The selection of the appropriate model class, and comparisons between individual models, are also largely based on qualitative criteria and user judgment.
Some authors 
 \citep[e.g.][]{Pan_etal_2016, Zhang_Lin_2018, Niven_etal_MaxEnt_2019, Chiuso_Pillonetto_2019, Chen_Lin_2021, Hirsch_etal_2022, Yuan_etal_2023, Lin_etal_2023, Zhang_etal_2023, Taghavi_2024, Fung_etal_2024, Klishin_etal_2024} 
have applied Bayesian regularization methods to dynamical system identification, the subject of this study. 

\subsection{\label{sect:Bayes_reg} Bayesian Regularization}

To construct a Bayesian solution to the above inverse problem \citep[e.g.][]{Aster_etal_2013, MD_2015, MD_Dimutru_2015, MD_2016, Teckentrup_MaxEnt2018, Calvetti_Somersalo_2018}, we first recognize that each component of the discrete linear map \eqref{eq:sys} should be written as:
\begin{equation}
\g= \H(\statemat) \f + \noise 
\label{eq:sys2}
\end{equation}
where $\noise$ is an 
$\SizeTime$-dimensional error (noise) vector,
representing the uncertainty in the measurement data. The variables $\g$, $\statemat$, $\f$ and $\noise$ are considered to be probabilistic, each represented by a probability density function (pdf) defined over their domain. Instead of inverting \eqref{eq:sys2}, the Bayesian inversion proceeds by calculating the {posterior} probability density function (pdf) of $\f$ conditioned on the observed data $\g$, given by Bayes' rule \cite{Bayes_1763, Laplace_1774, Jaynes_2003, vonToussaint_2011}:
\begin{equation}
p(\f|\g) = \frac{p(\g|\f)p(\f)}{p(\g)} \propto  p(\g|\f)p(\f)
\label{eq:Bayes}
\end{equation}
where 
$p(\f)$ is the {prior pdf}, 
$p(\g|\f)$ is the {likelihood} 
and $p(\g)$ is the {evidence}. 
The simplest Bayesian method is the maximum {\it a posteriori} (MAP) point estimate of $\f$, given by the maximum of \eqref{eq:Bayes}. For higher fidelity, it is usual to calculate the logarithmic maximum, from \eqref{eq:Bayes} giving the estimator:
\begin{align}
\begin{split}
\hatf^\text{MAP}  
\cong \arg \max\limits_{\f} \bigl[ \ln p(\g|\f) + \ln p(\f) \bigr].
\end{split}
\label{eq:MAP_log}
\end{align}
in which the evidence $p(\g)$ is assumed {constant}.

To compute the solution to \eqref{eq:MAP_log}, we must assign expressions to the probability distributions $p(\g|\f)$ and $p(\f)$. First, we assume that $\noise$ consists of unbiased multivariate Gaussian noise with the $\SizeTime \times \SizeTime$ covariance matrix $\Vepsilon$ \cite{Teckentrup_MaxEnt2018}:
\begin{equation}
p(\noise|\f) = \N(\noise;\vzero, \Vepsilon) 
= \frac{ \exp \bigl( - \tfrac{1}{2} ||\noise||^2_{\Vepsilon^{-1}} \bigr) }{ \sqrt{ (2\pi)^\SizeTime \det{\bigl(\Vepsilon\bigr)}}},
\label{eq:noise}
\end{equation}

\noindent where $\vzero$ is a zero vector, $\det(\cdot)$ is the determinant and $|| \vy ||^2_{A} := \vy^\transpose A \vy$ is the Mahalanobis distance of the vector $\vy$ with respect to a symmetric positive semi-definite matrix $A$, referred to here as a ``Gaussian norm''. 
This gives:
\begin{equation}
p(\noise |\f) \propto   \exp \bigl( - \tfrac{1}{2} ||\noise||^2_{\Vepsilon^{-1}} \bigr).
\label{eq:noise2}
\end{equation}
{Substitution of \eqref{eq:sys2} into \eqref{eq:noise2}} gives the likelihood:
\begin{align}
\begin{split}
p(\g | \f) 
&= \N(\g;\H(\statemat) \f, \Vepsilon) 
\\
&\propto   \exp \bigl( - \tfrac{1}{2} ||\g - \H(\statemat) \f ||^2_{\Vepsilon^{-1}} \bigr ).
\end{split}
\label{eq:likelihood}
\end{align}
Second, we adopt a multivariate Gaussian prior for $\f$, with zero mean 
$\meanpriorf=\vzero$ 
and $\SizeAlphabet \times \SizeAlphabet$ covariance matrix $\Vf$:
\begin{equation}
\begin{split}
p(\f) &= 
\N(\f;\meanpriorf, \Vf)  = \N(\f;\vzero, \Vf) \\ 
&= \frac{ \exp \bigl( - \tfrac{1}{2} ||\f ||^2_{\Vf^{-1}} \bigr ) }{ \sqrt{ (2\pi)^\SizeAlphabet \det\bigl({\Vf}\bigr)}} 
\propto   \exp \bigl( - \tfrac{1}{2} ||\f ||^2_{\Vf^{-1}} \bigr ). 
\end{split}
\label{eq:prior}
\raisetag{50pt}
\end{equation}
This represents the \textit{a priori} belief that the model coefficients  
$\matparamc_{\IndexAlphabet \IndexDynSys}$ 
will be positive or negative, without bias, but will not approach $\pm \infty$. The Gaussian prior gives a simple distribution of this form, which will preferentially select coefficients close to zero, yet provides support over the entire parameter space $\matparamc_{\IndexAlphabet \IndexDynSys} \in \mathbb{R}$. 
For constant 
values $\SizeTime$ and $\SizeAlphabet$, covariance matrices and evidence,  
\eqref{eq:MAP_log} becomes \cite{Aster_etal_2013, Teckentrup_MaxEnt2018}:
\begin{align}
\begin{split}
\hatf^\text{MAP}
& \cong \arg \max\limits_{\f} \Bigl[ \ln \exp \bigl( - \tfrac{1}{2} ||\g - \H(\statemat) \f ||^2_{\Vepsilon^{-1}} \bigr ) 
\\& \qquad + \ln \exp \bigl( - \tfrac{1}{2} ||\f ||^2_{\Vf^{-1}} \bigr )  \Bigr]
\\
&= \arg \max\limits_{\f} -\tfrac{1}{2}\Bigl[||\g - \H(\statemat) \f ||^2_{\Vepsilon^{-1}}  + ||\f ||^2_{\Vf^{-1}}   \Bigr]
\\
&= \arg \min\limits_{\f} \Bigl[   ||\g - \H(\statemat) \f ||^2_{\Vepsilon^{-1}}  + ||\f ||^2_{\Vf^{-1}}   \Bigr].
\end{split}
\label{eq:MAP_simp2}
\end{align}

\noindent A more complete formulation of \eqref{eq:MAP_simp2} is examined in Appendix \ref{sect:ApxA}.
The Bayesian MAP estimator thus provides a generalized form of the objective function used in {traditional regularization methods \eqref{eq:modJ}.} 
Furthermore, as shown in Appendix \ref{sect:ApxA}, for isotropic variances of the noise $\Vepsilon = \sigma^2_{\noise} \vI_\SizeTime$ and prior $\Vf = \sigma^2_{\f} \vI_\SizeAlphabet$, where $\vI_s$ is the identity matrix of size $s$, \eqref{eq:MAP_simp2} reduces to {ridge regression} \eqref{eq:modJ} with $\alpha, \beta, \gamma, \delta=2$ and $\regparam=\sigma^2_{\noise}/\sigma^2_{\f}$. 

{We emphasize that the choice of multivariate Gaussian noise \eqref{eq:noise} -- hence likelihood \eqref{eq:likelihood} -- and prior \eqref{eq:prior} distributions is not essential for the Bayesian MAP estimator \eqref{eq:MAP_log}, but is a convenient choice, exploiting the conjugacy property of Gaussian distributions to give a closed-form solution for the posterior (see \S\ref{sect:UQ}). 
Alternative closed-form solutions could be obtained by choosing a different conjugate prior with its associated likelihood function. 
For example, as shown in Appendix \ref{sect:ApxB}, a gamma prior with an independent and identically distributed ({\it iid}) Laplace likelihood distribution gives a gamma posterior. 
For non-conjugate choices of the prior and likelihood functions, numerical optimization of \eqref{eq:MAP_log} may be necessary.
The combination of Gaussian likelihood and Laplace prior distributions, for example, does not have a closed-form solution, but simplifies for the {\it iid} case to LASSO optimization. This is demonstrated in Appendix \ref{sect:ApxC}.}

{We see from \eqref{eq:MAP_log}, \eqref{eq:MAP_simp2} and Appendices \ref{sect:ApxA}-\ref{sect:ApxC}} that regularization by the Bayesian MAP method is equivalent to a generalized Tikhonov regularization \cite{Tikhonov_1963, Aster_etal_2013, Calvetti_Somersalo_2018}, {which reduces to ridge regression for isotropic Gaussian likelihood and prior distributions, and to the LASSO function for an isotropic Gaussian likelihood} {with an {\it iid} Laplace prior.} This insight provides a Bayesian rationale for the selection of residual and regularisation terms, respectively, as the negative logarithms of the likelihood and prior distributions. It also provides a rule for selecting the regularization parameter $\regparam$, which can be extended to anisotropic variances. 
More generally, \eqref{eq:MAP_log} provides a general Bayesian regularization framework for the handling of probability distributions of any form.
The Bayesian interpretation also provides access to the full Bayesian inversion apparatus, outlined in the following sections, offering many advantages over traditional regularization methods. 

\subsection{\label{sect:UQ} Bayesian Posterior and Model Uncertainty Quantification} 

It can be shown \cite{Tarantola_2005, Bishop_2006, Hoff_2009, MD_2016, Tenorio_2017, Calvetti_Somersalo_2018, Bontekoe_2023} that the posterior \eqref{eq:Bayes} obtained from multivariate Gaussian likelihood and prior distributions \eqref{eq:likelihood}-\eqref{eq:prior} is also multivariate Gaussian: 
\begin{equation}
p(\f | \g) = \N(\f;\BPostMean, \BPostCovMat)
= \frac{ \exp \bigl( - \tfrac{1}{2} || \f - \BPostMean ||^2_{{\BPostCovMat^{-1}}}  \bigr ) }{ \sqrt{ (2\pi)^\SizeAlphabet \det\bigl({\BPostCovMat}\bigr)}},
\label{eq:posterior}
\end{equation}
based on the $\SizeAlphabet$-dimensional mean vector and $\SizeAlphabet \times \SizeAlphabet$ covariance matrix, respectively: 
\begin{align} 
\begin{split}
\BPostMean &= \BPostCovMat 
                  \, \bigl( \H^{\transpose} \Vepsilon^{-1} \g   + \Vf^{-1} \meanpriorf \bigr)
\\
\BPostCovMat &=\bigl( \H^{\transpose} \Vepsilon^{-1} \H + \Vf^{-1} \bigr)^{-1}.
\end{split}
\label{eq:posterior_estimators}
\end{align}

\noindent {This arises from the conjugate property of Gaussian prior and posterior distributions with respect to a Gaussian likelihood. Furthermore, for a Gaussian posterior, the MAP estimate corresponds to the expected value, $\BPostMean = \E[p(\f | \g)]$.}

{Equation \eqref{eq:posterior_estimators} gives} an analytical solution for the model coefficients $\hat{\matparamc}_{\IndexAlphabet \IndexDynSys}$ 
in \eqref{eq:MAP_simp2}. Furthermore, the standard deviations $\hat{\sigma}_{\matparamc_{\IndexAlphabet \IndexDynSys}}$ of each coefficient $\hat{\matparamc}_{\IndexAlphabet \IndexDynSys}$, calculated from the square roots of diagonal terms $\bigl(\BPostCovMat\bigr)_{\IndexAlphabet \IndexAlphabet}$ in the covariance matrices  
$\BPostCovMat$, provide quantitative estimates of the error in each coefficient -- thereby quantifying the model uncertainty \citep[e.g.][]{Niven_etal_MaxEnt_2019, Hirsh_etal_2022}.

While the matrix inversions in \eqref{eq:posterior_estimators} are computationally expensive, they are feasible for low-dimensional problems such as those in this study. For high-dimensional problems, the posterior mean and covariance can be computed efficiently using the optimization in \eqref{eq:MAP_simp2} or \eqref{eq:J_Gaussian}.

\subsection{\label{sect:model_sel} Bayesian Model Selection}

A strong advantage of the Bayesian framework is that it enables the quantitative ranking of models. Consider two models \eqref{eq:sys2} for the same data, respectively:
\begin{equation}
\g = \H^{(1)}(\statemat) \f^{(1)} + \noise, \hspace{10pt} \g = \H^{(2)} (\statemat) \f^{(2)}  + \noise
\label{eq:sys_2model}
\end{equation}
where $(1)$ or $(2)$ is the model index. 
The models can be ranked by the ratio of their posteriors, known as the 
posterior odds ratio (POR) \cite{Jaynes_2003, Hoff_2009}:
\begin{equation}
\POR^{(1)/(2)}
= \frac{p(\f^{(1)} |\g )}{p(\f^{(2)} |\g )} 
\label{eq:POR}
\end{equation}
If the pdfs are expressed as point values using the Bayesian MAP estimates \eqref{eq:MAP_simp2}, then \eqref{eq:POR} provides a useful metric for model selection ({\it ``naive Bayes''}), with model $(1)$ preferred for $\POR >1$.
For Gaussian posteriors \eqref{eq:posterior} within the same model class (constant $\SizeAlphabet$) and {constant} covariance matrices, \eqref{eq:POR} reduces to the logarithmic form:
\begin{equation}
\begin{split}
&\ln \POR^{(1)/(2)} 
= \ln p(\f^{(1)} |\g ) - \ln p(\f^{(2)} |\g )\\
& \approx   - \tfrac{1}{2} \Bigl( || \f - \BPostMean^{(1)} ||^2_{\bigl(\BPostCovMat^{(1)}\bigr)^{-1}} -  || \f - \BPostMean^{(2)} ||^2_{\bigl(\BPostCovMat^{(2)}\bigr)^{-1}} \Bigr)
\end{split}
\label{eq:log_POR}
\end{equation}
with model $(1)$ preferred for $\ln \POR >0$. For a fixed reference model, \eqref{eq:log_POR} can be applied in an absolute sense: 
\begin{equation}
\ln \POR^{(\ast)}
\approx  - \tfrac{1}{2} || \f - \BPostMean^{(\ast)} ||^2_{\bigl(\BPostCovMat^{(\ast)}\bigr)^{-1}}
\label{eq:log_POR2}
\end{equation}
The maximization of \eqref{eq:log_POR2} therefore provides a quantitative rule for model selection, equivalent -- for the assumptions made here -- to minimization of the posterior Gaussian norm. 

\subsection{\label{sect:metrics} Other Model Selection Metrics}

Many other metrics have also been proposed for model selection by traditional or Bayesian methods. These include the Akaike information criterion (AIC) and Bayes information criterion (BIC), respectively \cite{Burnham_Anderson_2002}:
\begin{align}
\begin{split}
\AIC &=-2 \ln p(\g|\hatf) + 2 \SizeAlphabet \\
\BIC &=-2 \ln p(\g|\hatf) + \SizeAlphabet \ln \SizeTime
\end{split}
\label{eq:AIC_BIC}
\end{align}
where $\ln p(\g|\hatf)$ is the maximized log-likelihood based on the estimated coefficients $\BPostMean$.
For finite sample sizes, $\AIC$ requires a correction given by \cite{Mangan_etal_2017}:
\begin{align}
\begin{split}
\AICc &= \AIC + \dfrac{2(\SizeAlphabet+1)(\SizeAlphabet+2)}{\SizeTime-\SizeAlphabet-2}
\end{split}
\label{eq:AICc}
\end{align}
Some authors \cite{Burnham_Anderson_2002, Mangan_etal_2017} have approximated 2-norm forms of the AIC and BIC based on $\ln p(\g|\hatf) \approx - \frac{\SizeTime}{2} \ln (||\g - \H(\statemat) \f||^2_2 /{\SizeTime})$. 
\noindent An alternative metric is the ``error bar'', defined as the sum of squared coefficients of variation of the posterior \cite{Zhang_Lin_2018, Chen_Lin_2021}:
\begin{align}
\EB 
= \sum\limits_{\substack{\IndexAlphabet = 1 \\ \hat{\matparamc}_{\IndexAlphabet \IndexDynSys} \ne 0}}^\SizeAlphabet \dfrac{\bigl(\BPostCovMat\bigr)_{\IndexAlphabet \IndexAlphabet}}{\hat{\matparamc}_{\IndexAlphabet \IndexDynSys}^2}
\label{eq:EB}
 \end{align}
It is claimed \cite{Chen_Lin_2021} that this should be minimized to choose models with lower variance.

\subsection{\label{sect:evidence} Bayesian Evidence} 

A further advantage of the Bayesian apparatus is that the evidence $p(\g)$ in Bayes' rule \eqref{eq:Bayes} -- sometimes termed the prior predictive distribution or marginal distribution -- provides a measure of the goodness of fit of the model to the data \cite{Bishop_2006, vonToussaint_2011}. For the analysis here, since the likelihood, prior and posterior are multivariate Gaussian, the evidence is also multivariate Gaussian \cite{Bishop_2006, Tenorio_2017, Bontekoe_2023}:
\begin{align} 
p(\g) = \N(\g; \BEvidMean, \BEvidCovMat)
= \frac{ \exp \bigl( - \tfrac{1}{2} || \g - \BEvidMean ||^2_{{\BEvidCovMat^{-1}}}  \bigr ) }{ \sqrt{ (2\pi)^\SizeTime \det\bigl(\BEvidCovMat\bigr)}},
\label{eq:evidence}
\end{align}
with the $\SizeTime$-dimensional mean vector and $\SizeTime \times \SizeTime$ covariance matrix for each dimension: 
\begin{align} 
\begin{split}
\BEvidMean &=\H \meanpriorf = \vzero
\\
\BEvidCovMat &= \H \Vf \H^\transpose + \Vepsilon
\end{split}
\label{eq:evidence_estimators}
\end{align}
Thus for the Gaussian assumptions made in this study, the evidence can be calculated analytically. However, this requires the inversion of potentially large $\SizeTime \times \SizeTime$ covariance matrices $\BEvidCovMat$, which introduce computational difficulties. For this reason, Bayesian practitioners usually use exploratory numerical methods such as Markov chain Monte Carlo (MCMC) \cite{Tarantola_2005, Bishop_2006, vonToussaint_2011, Bardsley_2018} or nested sampling \cite{Skilling_2004} to explore the shape of the posterior and compute the evidence. 

\subsection{\label{sect:joint_Bayes} Extended Bayesian Regularization with Nuisance Parameters}

The Bayesian regularization method in \S \ref{sect:Bayes_reg} assumes multivariate Gaussian distributions for the error \eqref{eq:noise}-\eqref{eq:likelihood} and prior \eqref{eq:prior}, with specified covariance matrices $\Vepsilon$ and $\Vf$, respectively. In most situations these matrices are \textit{a priori} not known. Their estimation as nuisance parameters can in fact be incorporated into the Bayesian framework.  Consider the joint posterior pdf $p(\f, \Vepsilon, \Vf |\g)$ of the model and unknown matrices, subject to the data. Applying Bayes' rule:
\begin{align}
\begin{split}
p(\f, \Vepsilon, \Vf |\g) 
= \frac{p(\g|\f, \Vepsilon, \Vf) p(\f, \Vepsilon, \Vf)} {p(\g)}  
\end{split}
\label{eq:Bayes_joint}
\end{align}
By separation of the prior $p(\f, \Vepsilon, \Vf) = p(\f| \Vepsilon, \Vf)$ $p(\Vepsilon) p(\Vf)$ for independent covariance matrices, and using the simplifications $p(\f| \Vepsilon, \Vf)  =p(\f|\Vf) $ (independence of the model from the noise) and $p(\g|\f, \Vepsilon, \Vf)=p(\g|\f, \Vepsilon)$ (independence of the data from the prior), \eqref{eq:Bayes_joint} simplifies to:
\begin{align}
\begin{split}
p(\f, \Vepsilon, \Vf |\g) 
= \frac{p(\g|\f, \Vepsilon) p(\f | \Vf) p(\Vepsilon) p(\Vf)} {p(\g)}  
\end{split}
\label{eq:Bayes_joint2}
\end{align}
To simplify \eqref{eq:Bayes_joint2}, we again choose Gaussian likelihood and prior distributions \eqref{eq:likelihood}-\eqref{eq:prior}, based on anisotropic diagonal error and model covariance matrices: 
\begin{align}
\begin{split}
\Vepsilon &= \diag[\Vepsilondiag] \quad\text{with}\quad
\Vepsilondiag = [\VepsilondiagInd{1 \IndexDynSys}, \cdots, \VepsilondiagInd{\IndexTime \IndexDynSys}, \cdots, \VepsilondiagInd{\SizeTime \IndexDynSys}]^\transpose
\\
\Vf &= \diag[\Vfdiag]\quad\text{with}\quad 
\Vfdiag = [\VfdiagInd{1 \IndexDynSys}, \cdots, \VfdiagInd{\IndexAlphabet \IndexDynSys}, \cdots, \VfdiagInd{\SizeAlphabet \IndexDynSys}]^\transpose
\end{split}
\end{align}
where $\diag$ maps a vector to the diagonal of a matrix. The {covariance priors} $p(\Vepsilon)$ and $p(\Vf)$ are assumed separable, given by the product of individual Inverse Gamma (IG) distributions:
\begin{align}
\begin{split}
p(\Vepsilon) 
&= p(\Vepsilondiag) 
= \prod\limits_{\IndexTime=1}^\SizeTime p(\VepsilondiagInd{\IndexTime \IndexDynSys}) 
= \prod\limits_{\IndexTime=1}^\SizeTime \text{IG}(\VepsilondiagInd{\IndexTime \IndexDynSys}; \alpha_{\eizero}, \beta_{\eizero})
\\
p(\Vf) 
&= p(\Vfdiag) 
= \prod\limits_{\IndexAlphabet=1}^\SizeAlphabet p(\VfdiagInd{\IndexAlphabet \IndexDynSys}),
= \prod\limits_{\IndexAlphabet=1}^\SizeAlphabet \text{IG}(\VfdiagInd{\IndexAlphabet \IndexDynSys}; \alpha_{\fizero}, \beta_{\fizero})
\end{split}
\label{eq:priors_sep}
\end{align}
in which $\{\alpha_\iota,\beta_\iota\}$ for  
$\iota \in \{ \eizero, \fizero \}$ 
are the shape and scale hyperparameters, respectively. 
The IG distribution can be derived as the marginal posterior distribution for an unknown variance of a Gaussian distribution with an uninformative prior \cite{Marin_Robert_2007, Clyde_etal_2015}, and is commonly used for variance estimation in Bayesian inference. It can be interpreted as a two-parameter positive generalization of the Student's $t$-distribution with a small positive peak and heavy tail, with the mean and variance \cite{MD_Dimutru_2015}:
\begin{align}
\mathsf{E}_\iota & = \E[v_\iota] = \frac{\beta_\iota}{\alpha_\iota-1},\quad\text{for}\quad \alpha_\iota > 1\\ 
\mathsf{V}_\iota & = Var[v_\iota] =\frac{\beta_\iota^2}{(\alpha_\iota-1)^2(\alpha_\iota-2)}\quad\text{for}\quad \alpha_\iota > 2.
\label{eq:exp_var_IG}
\end{align}
\noindent The IG hyperparameters $\{\alpha_\iota, \beta_\iota \}$ or alternatively $\{\mathsf{E}_\iota, \mathsf{V}_\iota \}$ can be chosen based on prior belief or experience, or computed iteratively until a convergence criterion is achieved.

To compute the joint posterior \eqref{eq:Bayes_joint2} with the {covariance priors} \eqref{eq:priors_sep}, two algorithms are commonly used \cite{MD_Dimutru_2015}:
\begin{itemize}

\item The \emph{joint maximum a posteriori} (JMAP) method, defined by the logarithmic optimization  \cite{MD_Dimutru_2015}:
\begin{align}
&(\BPostMean^\text{JMAP}, \hat{\Vepsilon}^\text{JMAP}, \hat{\Vf}^\text{JMAP}) \\
&= \arg \max\limits_{\f, \Vepsilon, \Vf} \bigl[ \ln p(\f, \Vepsilon, \Vf |\g) \bigr]
\label{eq:JMAP}
\end{align}
This can be implemented iteratively using the analytical solution \eqref{eq:posterior_estimators}, for which the pseudocode of an algorithm has been reported \cite{MD_Dimutru_2015}. 

\item  The \emph{variational Bayes approximation} (VBA) method, which approximates the posterior by the separable distribution $q(\f, \Vepsilon, \Vf) =q_1(\f) q_2(\Vepsilon) q_3(\Vf)$ using the Gaussian prior \eqref{eq:prior} for $q_1$ and IG {covariance} priors \eqref{eq:priors_sep} for $q_2$ and $q_3$. 
These can be computed iteratively by minimizing the Kullback-Leibler divergence between $p(\f, \Vepsilon, \Vf |\g)$ and $q$, for which the pseudocode of an algorithm has been reported \cite{MD_Dimutru_2015}. 

\end{itemize}

\section{\label{sect:apps} Applications to Dynamical Systems}
\subsection{\label{sect:method} Dynamical Systems and Methodology}

To compare the traditional and Bayesian methods for dynamical system identification, three dynamical systems were analyzed by the {SINDy, LASSO, ridge regression, JMAP and VBA algorithms}:

\newcounter{Lcount1}
\begin{list}{(\alph{Lcount1})}{\usecounter{Lcount1} \topsep 4pt \itemsep 0pt \parsep 4pt \leftmargin 15pt \rightmargin 0pt \listparindent 0pt \itemindent 0pt}

\item The Lorenz system, a simplified two-dimensional atmospheric convection system given by \cite{Lorenz_1963, Lynch_2004}:
\begin{equation}
\dot{\state} = [\sigma(y-x), \hspace{5pt} x(\rho-z)-y, \hspace{5pt} xy-\beta z]^\transpose
\end{equation}
The parameter values $[\sigma, \rho, \beta] = [10, \frac{8}{3}, 28]$ give chaotic behavior.

\item The Vance system, a one-predator two-prey ecological system given by \cite{Vance_1978, Gilpin_1979}:
\begin{equation}
\dot{\state} = (\vec{r} - \vec{\alpha} \vec{x}) \odot \vec{x}
\end{equation}
where $\odot$ is the element-wise (Hadamard) vector product. The parameter values $\vec{r} =  \biggl[ \begin{smallmatrix} 1 \\ 1 \\ -1 \end{smallmatrix}\biggr]$ and $\vec{\alpha}= \frac{1}{1000} \biggl[ \begin{smallmatrix} 1 & 1 & 10 \\ 1.5 & 1 & 1 \\ -5 & -0.5 & 0 \end{smallmatrix}\biggr]$ give chaotic behavior.

\item The Shil'nikov system, a cubic modification of the Lorenz system given by \cite{Shilnikov_etal_1993, Lynch_2004}:
\begin{equation}
\dot{\state} = [y, \hspace{5pt} x(1-z)-Bx^3-\lambda y, \hspace{5pt} -\alpha (z-x^2)]^\transpose
\end{equation}
The parameter values $[\alpha, \lambda, B] = \left[\frac{4 \sqrt{30}}{135},\frac{11 \sqrt{30}}{90},\frac{2}{13}\right]$ give chaotic behavior.

\end{list}

\noindent Phase plots of these systems with added {Gaussian} noise (see below) are shown in Figures \ref{fig:dyn_sys}(a)-(c).

The analyses were conducted {using a new Bayesian Dynamical System Identification (BDSI) code written in MATLAB for this study. This was executed} in MATLAB 2021b on a MacBook Pro with 2.3 GHz Intel Core i9, with numerical integration by the ode45 function. 
{For each system, the data $\statemat$ were generated and then augmented by additive noise to give $\statemat_{noisy} = \statemat + \varepsilon \statemat_{distrib}$, where $\statemat_{distrib}$ was sampled from a univariate noise distribution and $\varepsilon$ is a scale parameter.}
{For most analyses, the noise was sampled from a normal distribution:
\begin{align}
\text{Gaussian}(x| \mu, \sigma) = \frac{1}{\sigma \, \sqrt{2 \pi} } \exp \biggl( -\frac{(x-\mu)^2}{2 \sigma^2} \biggr)
\label{eq:univar_Gaussian}
\end{align}
with random variable $x$, mean $\mu=0$ and standard deviation $\sigma$. To test the robustness of the Bayesian algorithms, some analyses were instead augmented with unbiased Laplace noise, drawn from:
\begin{align}
\text{Laplace}(x | \mu, b) = \frac{1}{2b} \exp \biggl( -\frac {| x-\mu|}{b} \biggr)
\label{eq:univar_Laplace}
\end{align}
with mean $\mu=0$ and scale parameter $b=\sigma/\sqrt{2}$.}
The noise-moderated derivative data $\dotstatemat$ were then calculated from the noisy data by a dynamical system call. 

For the {system identification} process, an alphabet was selected to construct $\H(\statemat)$, with the possibility of linear, polynomial (up to third order) and other functions.  
The main code computes the dynamical system, and then {calls the original SINDy code based on the STLSQ algorithm \cite{Brunton_etal_2016, Mangan_etal_2017}, the {LASSO\cite{Santosa_Symes_1986, Tibshirani_1996} and ridge regression\cite{Hoerl_Kennard_1970, Bardsley_2018} algorithms using inbuilt MATLAB functions, and} modified forms of the JMAP and VBA algorithms \cite{MD_Dimutru_2015}. 
Each method was implemented using an inner iteration for fixed value(s) of its hyperparameter(s) -- in the Bayesian algorithms using a convergence criterion based on $\ln \det \BPostCovMat^{-1}$ -- 
and an outer iteration to explore changes in the hyperparameter(s).}
{LASSO and ridge regression were iterated over each regularization parameter $\regparam$ in \eqref{eq:modJ}, while SINDy was iterated over the threshold $\lambda$ in \eqref{eq:SINDy_reg}.}
For the Bayesian algorithms, the prior \eqref{eq:prior} hyperparameters $\{ \alpha_{\fizero}, \beta_{\fizero} \}$ were {held constant} to give a broad Gaussian distribution --  expressing a consistent degree of belief in the expected model coefficients -- while the outer iterations were conducted over the likelihood \eqref{eq:likelihood} hyperparameters $\{ \alpha_{\eizero}, \beta_{\eizero} \}$ as expressed by the IG mean $\mathsf{E}_{\eizero}$ \eqref{eq:exp_var_IG}, to discover the error in the data. 
This iteration sequence is shown in Figure \ref{fig:dyn_sys}(d), {proceeding in our code from high to low values (right to left)}.
The code also makes subroutine or function calls to a number of utility files \cite{MD_Dimutru_2015, Brunton_etal_2016, Mangan_etal_2017, Taghavi_2024} for the purpose of analyzing, plotting and exporting the data.  

For the Bayesian algorithms, the code calculates the mean and covariance estimators for the posterior \eqref{eq:posterior_estimators} and evidence \eqref{eq:evidence_estimators}, and the Gaussian norms within the likelihood \eqref{eq:likelihood}, prior \eqref{eq:prior}, posterior \eqref{eq:posterior} and evidence \eqref{eq:evidence}. Attempts were made to calculate the full Gaussian distributions for the posterior \eqref{eq:posterior} and evidence \eqref{eq:evidence}, using the logdet function \cite{Lin_2024} for the denominator determinant terms, but these were affected by numerical underflow in many instances. 
Some numerical difficulties were also encountered in the evidence Gaussian norms. 
The code also computes the metrics defined in \eqref{eq:AIC_BIC}-\eqref{eq:EB} and their 2-norm approximations \cite{Taghavi_2024}.


\begin{figure*}[t]
\begin{center}
\setlength{\unitlength}{0.55pt}
  \begin{picture}(800,540)
   \put(0,270){\includegraphics[height=50mm]{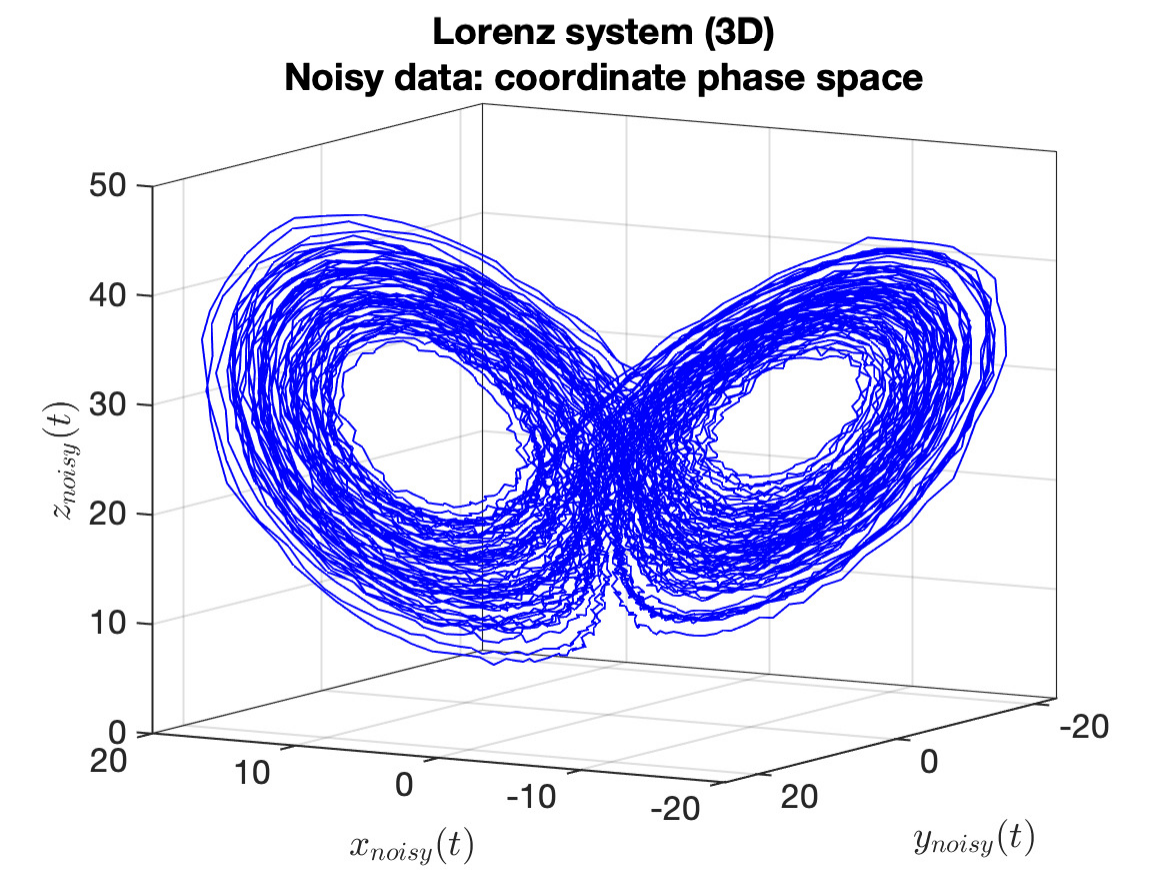} }
   \put(0,270){\small (a)}
   \put(400,270){\includegraphics[height=50mm]{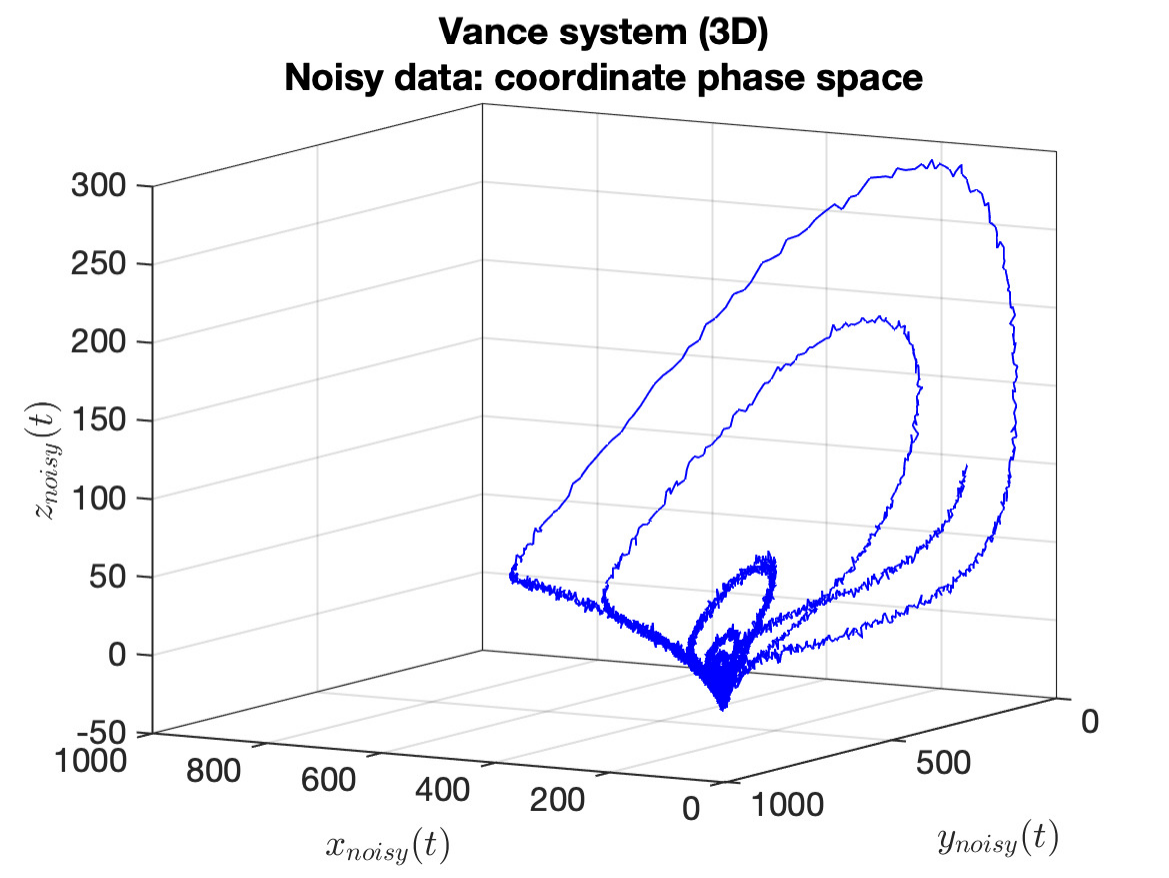} }
   \put(400,270){\small (b)}
   \put(0,0){\includegraphics[height=50mm]{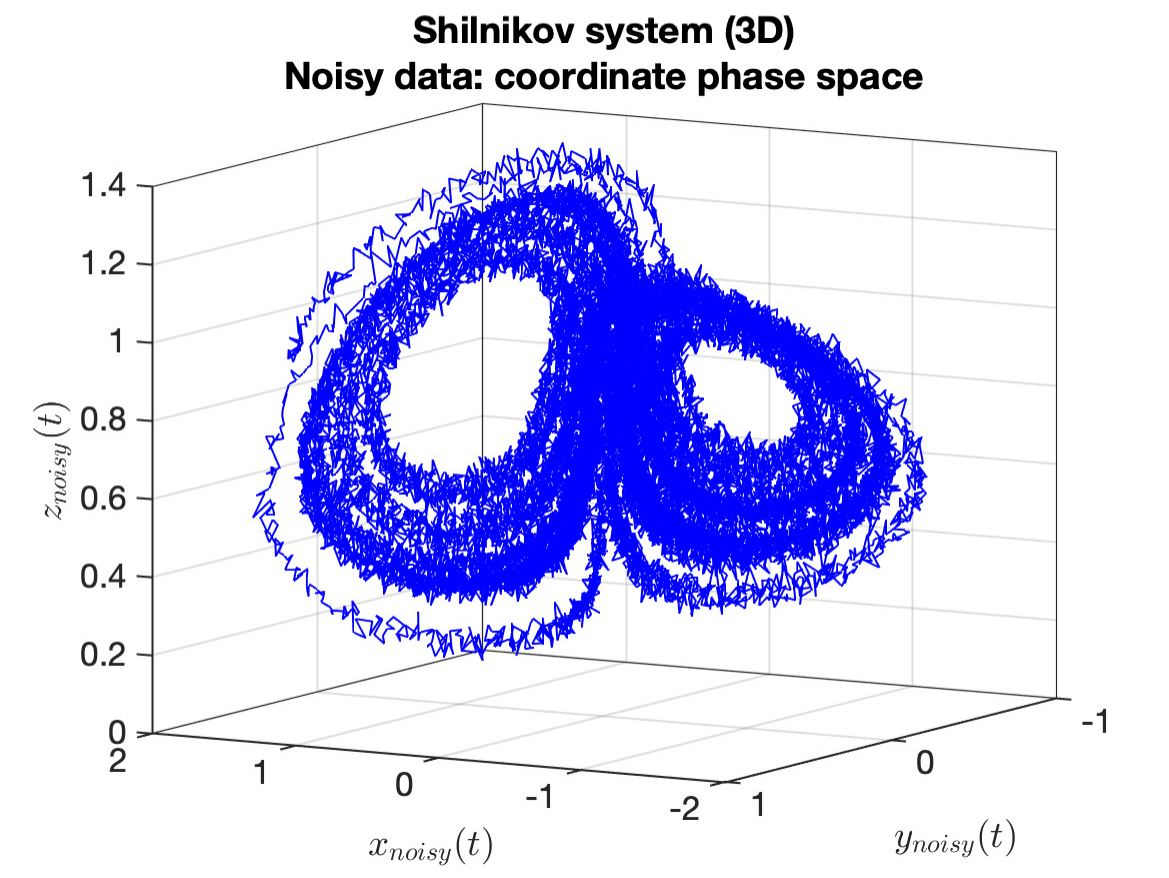} }
   \put(0,0){\small (c)}
   \put(400,0){\includegraphics[height=50mm]{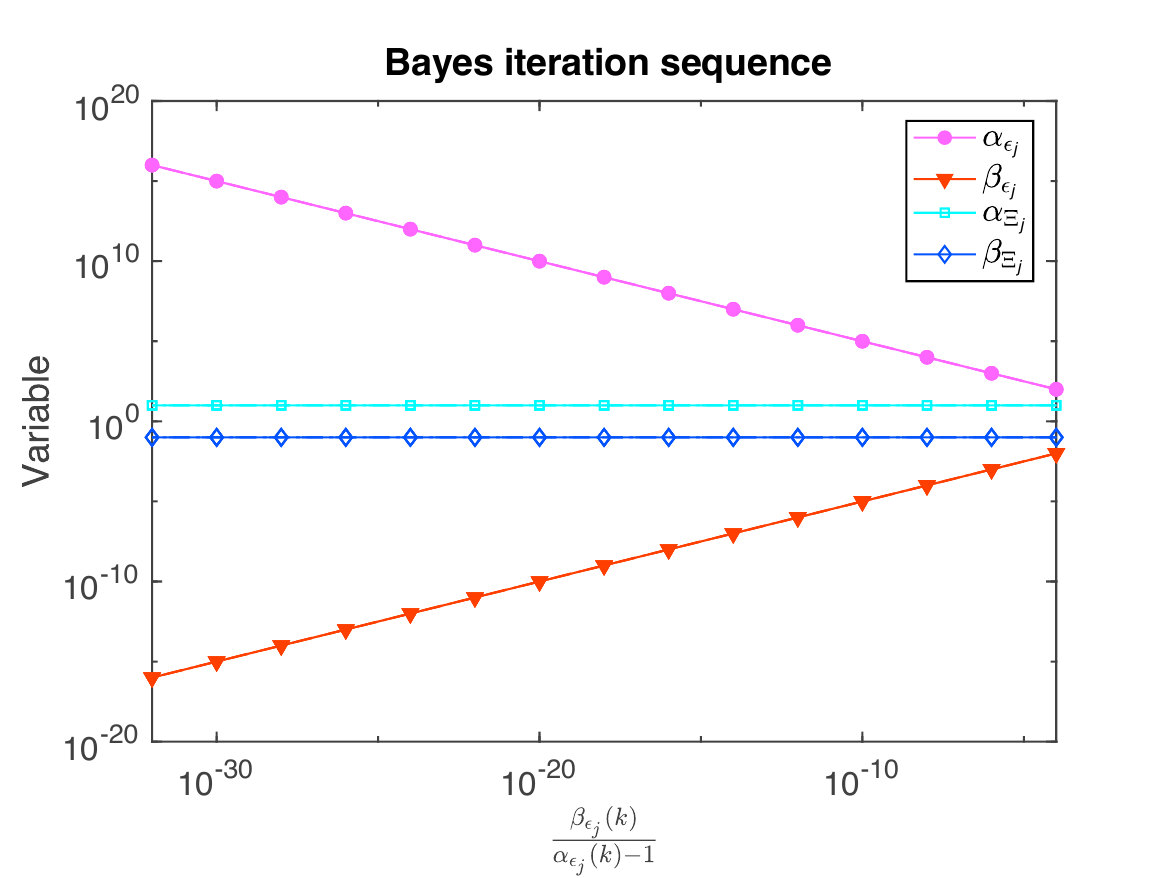} }
   \put(400,0){\small (d)}
  \end{picture}
\end{center}
\caption{Phase plots {with added Gaussian noise} for dynamical systems in this study: 
(a) Lorenz system with $\varepsilon=0.2$;
(b) Vance system with $\varepsilon=2.0$; 
(c) Shil'nikov system with $\varepsilon=0.02$; and
(d) outer iteration sequence used for the JMAP and VBA algorithms {(in our code, from right to left)}.}
\label{fig:dyn_sys}
\end{figure*}
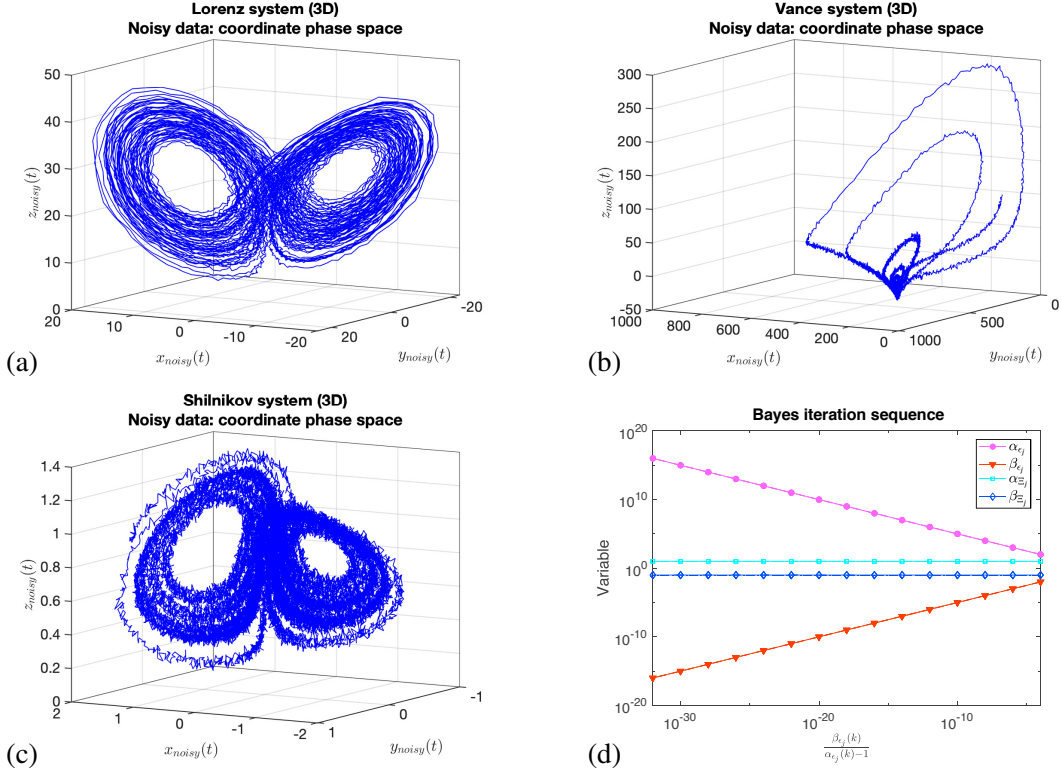


\begin{figure*}[ht]
\begin{center}
\setlength{\unitlength}{0.55pt}
  \begin{picture}(800,800)
   \put(0,540){\includegraphics[height=51mm]{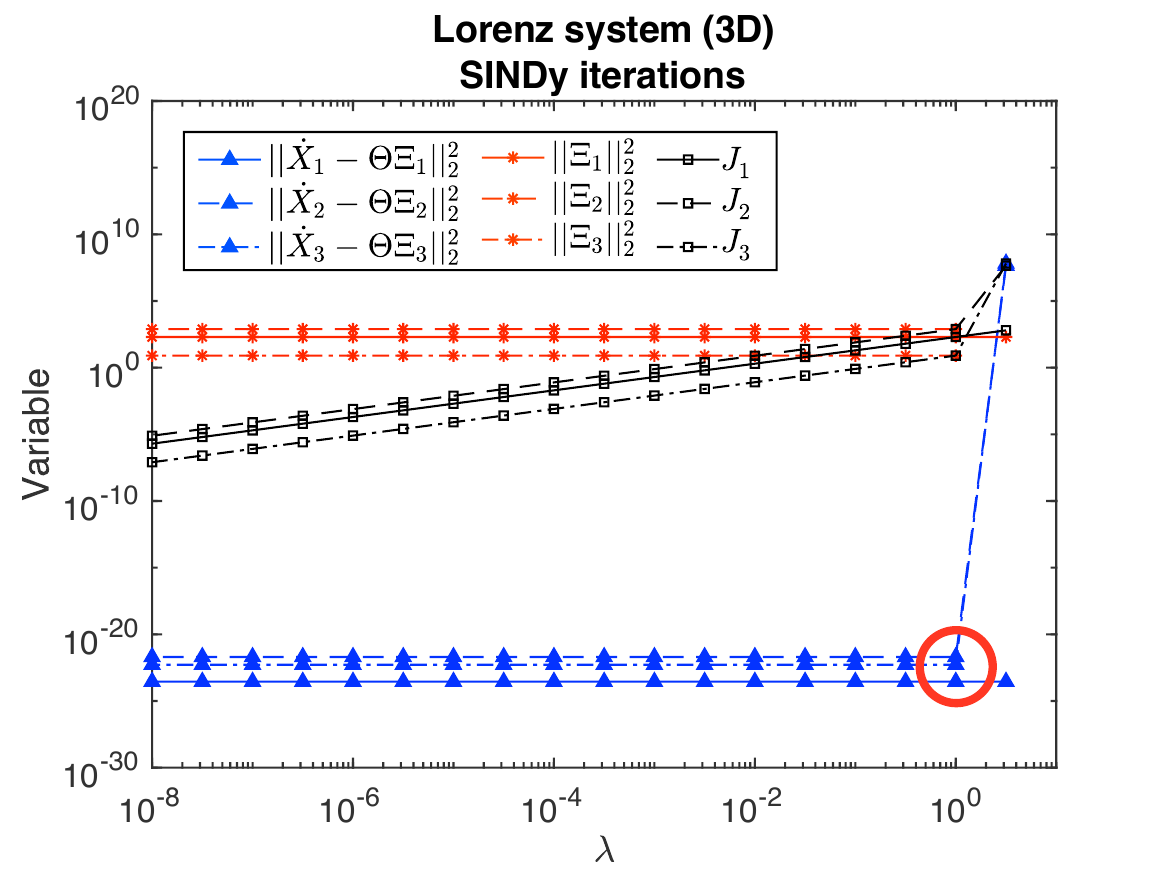} }
   \put(0,540){\small (a)}
   \put(417,540){\includegraphics[height=50mm]{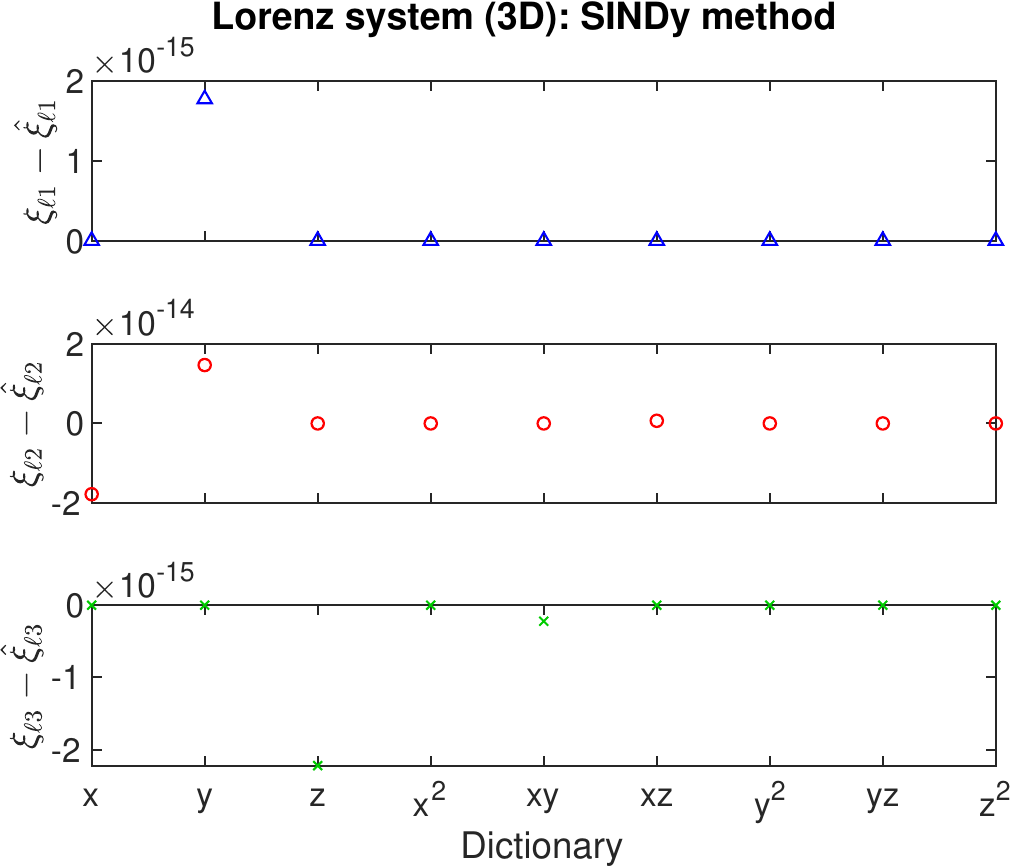} }
   \put(417,540){\small (b)}
   \put(0,270){\includegraphics[height=51mm]{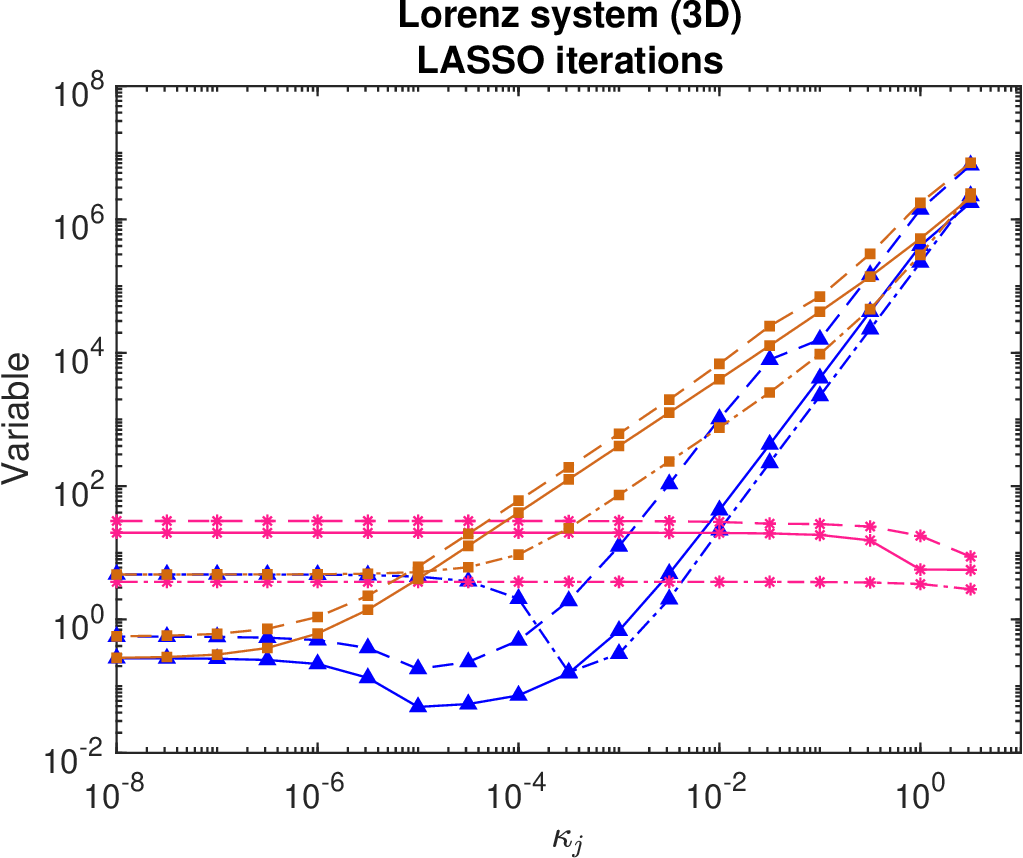} }
   \put(0,270){\small (c)}
   \put(417,270){\includegraphics[height=50mm]{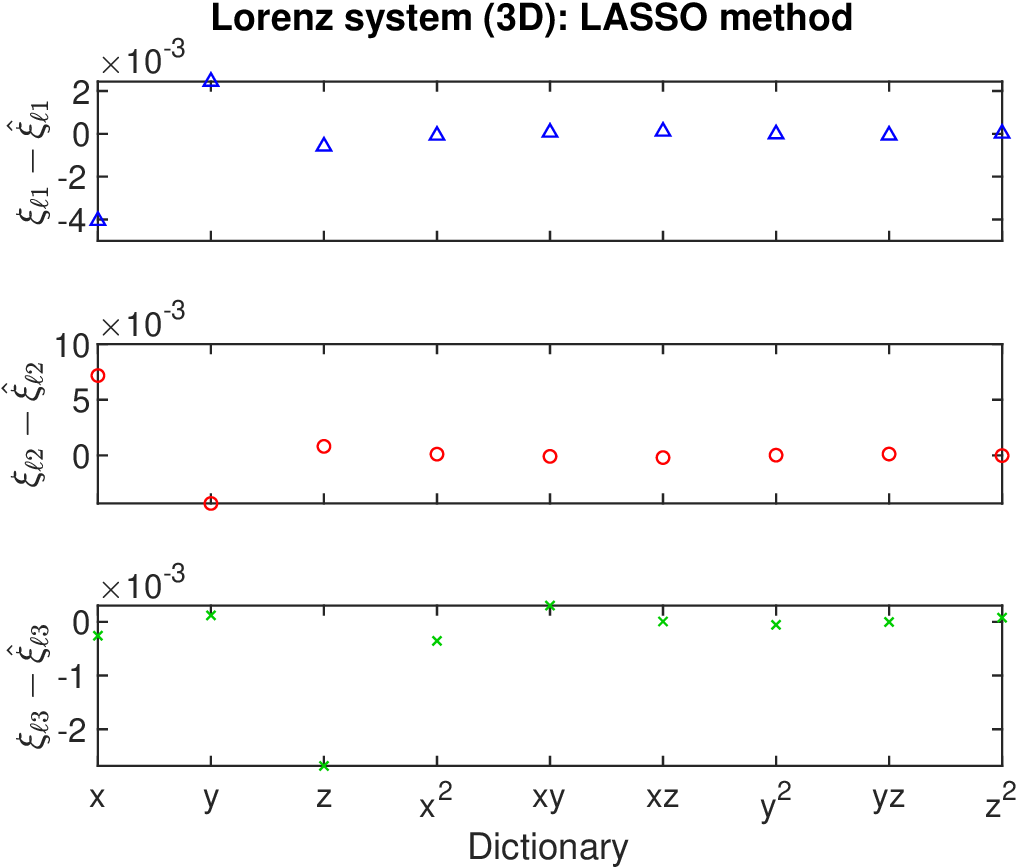} }
   \put(417,270){\small (d)}
   \put(0,0){\includegraphics[height=51mm]{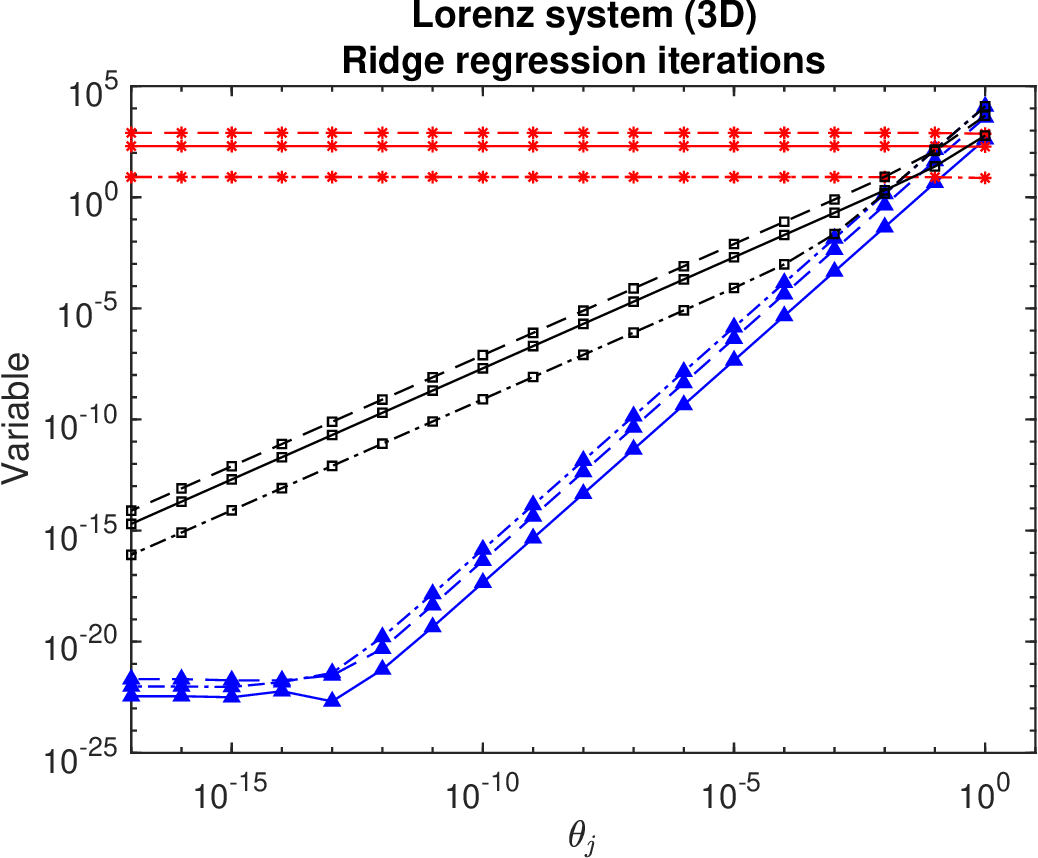} }
   \put(0,0){\small (e)}
   \put(417,0){\includegraphics[height=50mm]{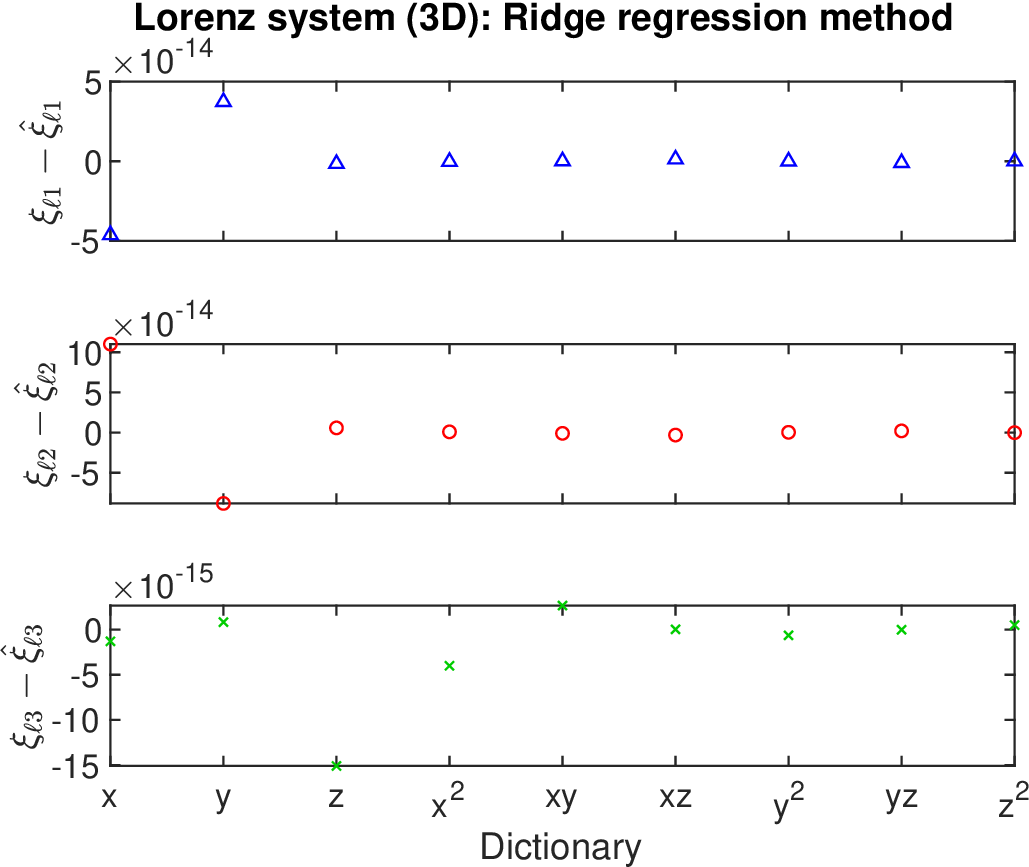} }
   \put(417,0){\small (f)}
  \end{picture}
\end{center}
\caption{{Lorenz system with added Gaussian noise for} {period $T=100$, time step $t_{step}=0.01$ and scaling parameter $\varepsilon=0.2$},  
{for (a)-(b) SINDy, (c)-(d) LASSO and (e)-(f) ridge regression, showing
(a),(c),(e) plots of residual, regularization and objective functions \eqref{eq:modJ} 
with decreasing hyperparameter(s), showing the optimal iteration; and
(b),(d),(f) optimal error in predicted coefficients $\matparamc_{\ell j}-\hat{\matparamc}_{\ell j}$.}}
\label{fig:Lorenz_reg}
\end{figure*}

\subsection{\label{sect:result} Results}

\begin{figure*}[t]
\begin{center}
\setlength{\unitlength}{0.55pt}
  \begin{picture}(800,510)
   \put(0,270){\includegraphics[height=50mm]{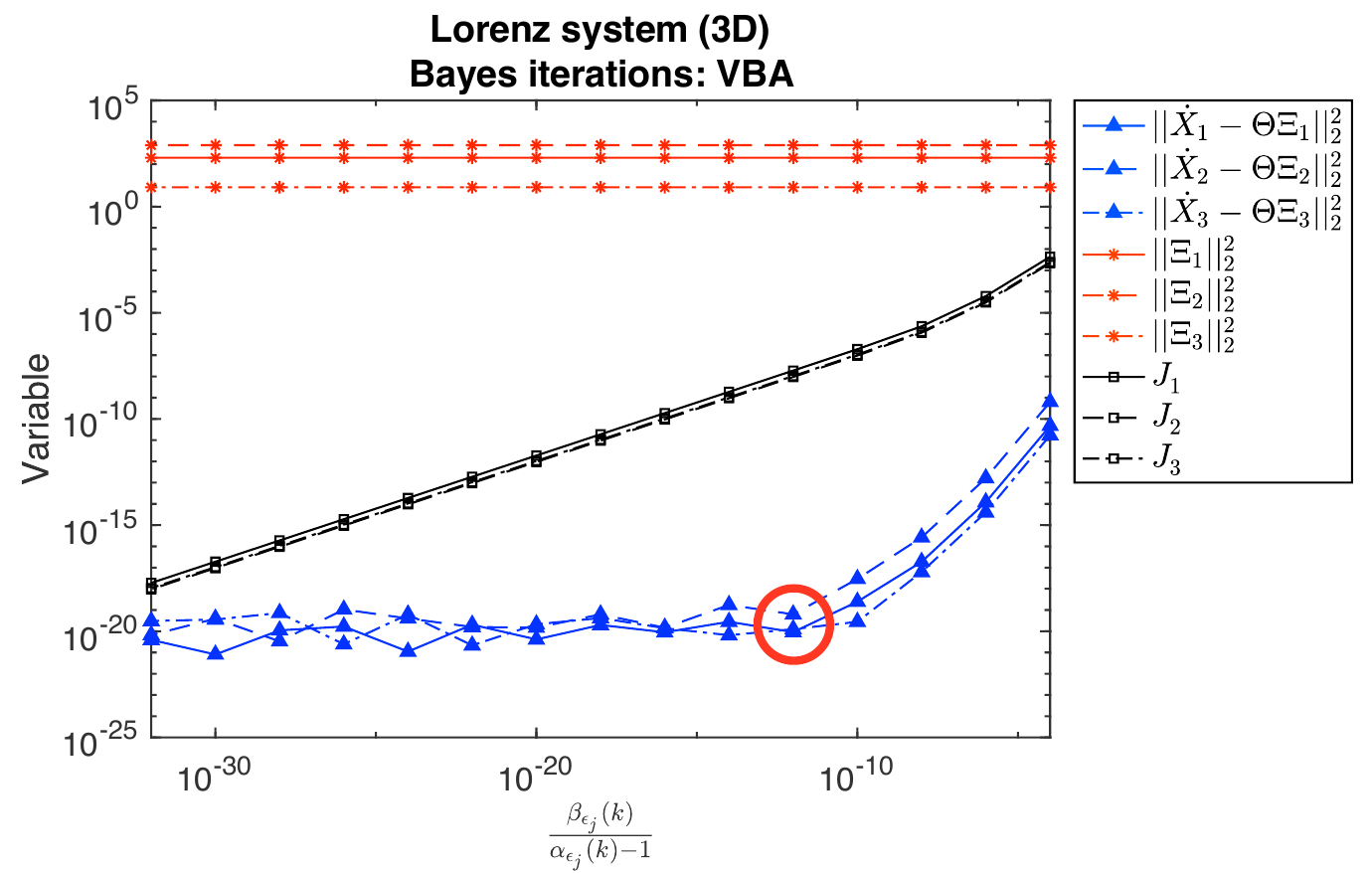} }
   \put(0,270){\small (e)}
   \put(430,270){\includegraphics[height=50mm]{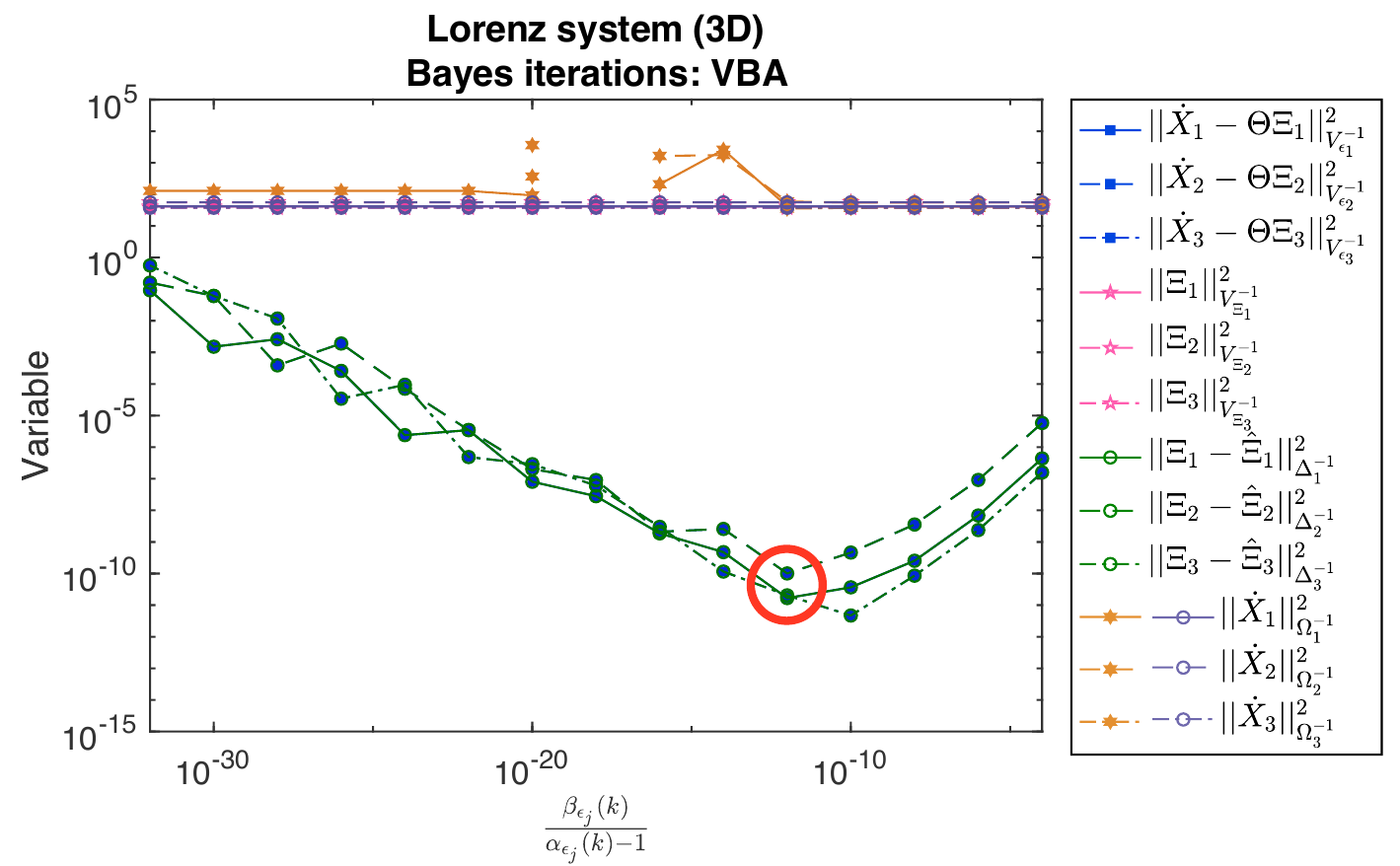} }
   \put(430,270){\small (f)}
   \put(17,0){\includegraphics[height=50mm]{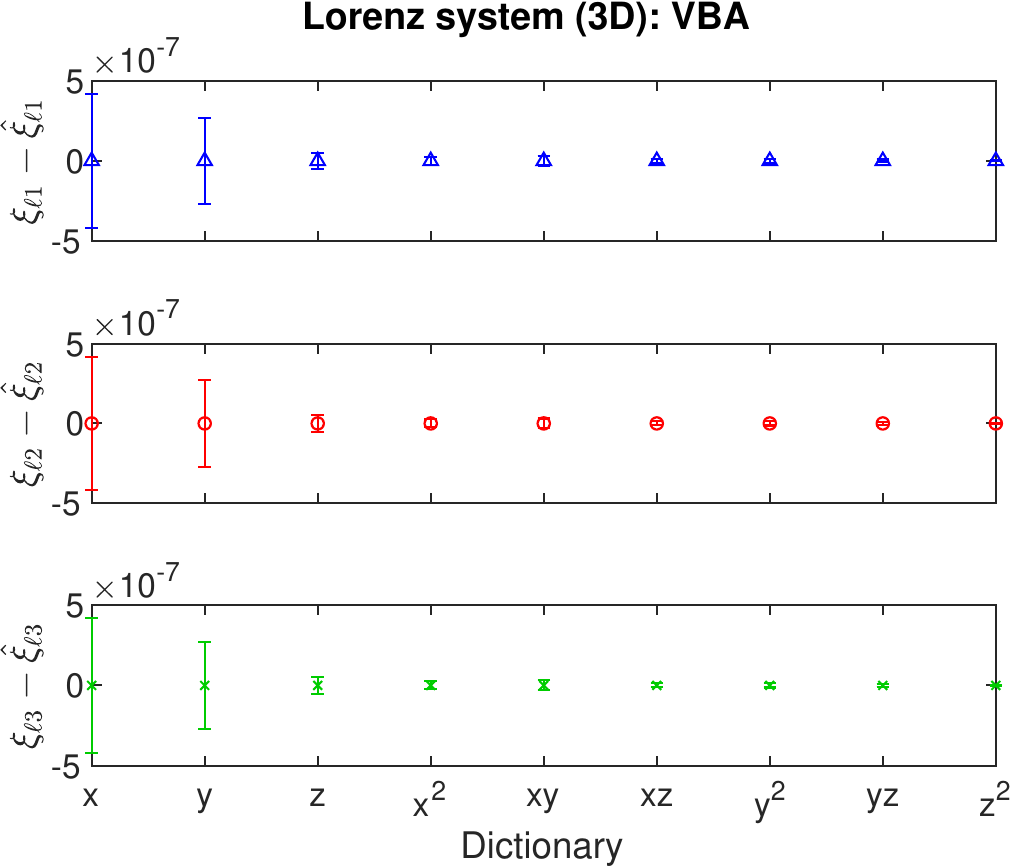} }
   \put(0,0){\small (g)}
   \put(447,0){\includegraphics[height=48mm]{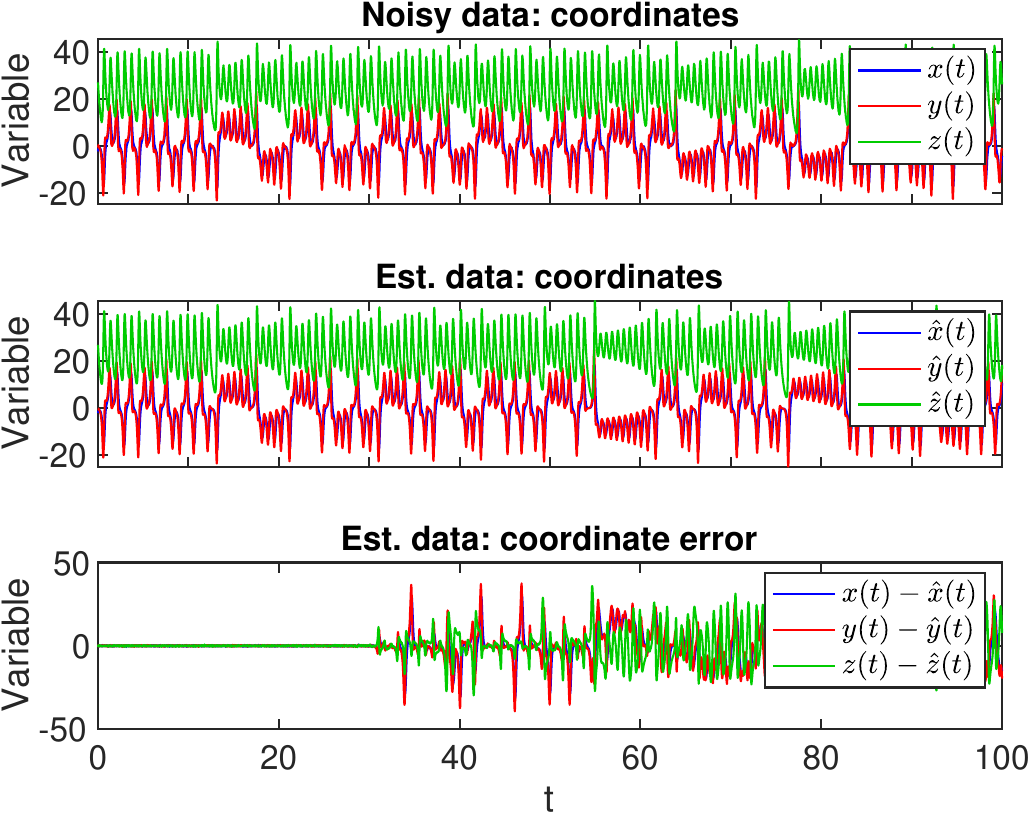} }
   \put(430,0){\small (h)}
  \end{picture}
\end{center}
\caption{Lorenz system {with added Gaussian noise for} {$T=100$, $t_{step}=0.01$, $\varepsilon=0.2$}, VBA regularization: 
(a) iteration sequence with decreasing $\mathsf{E}_{\eizero}$, showing the 2-norm residual, regularization and objective functions \eqref{eq:modJ} and the optimal iteration; 
(b) Gaussian norms for the prior, likelihood, posterior and evidence, showing the optimal iteration; 
(c) optimal error in predicted coefficients $\matparamc_{\ell j}-\hat{\matparamc}_{\ell j}$, with error bars calculated from the posterior covariance \eqref{eq:posterior_estimators}; and
(d) original and optimal predicted data and their differences.}
\label{fig:Lorenz_VBA_Gauss}
\end{figure*}

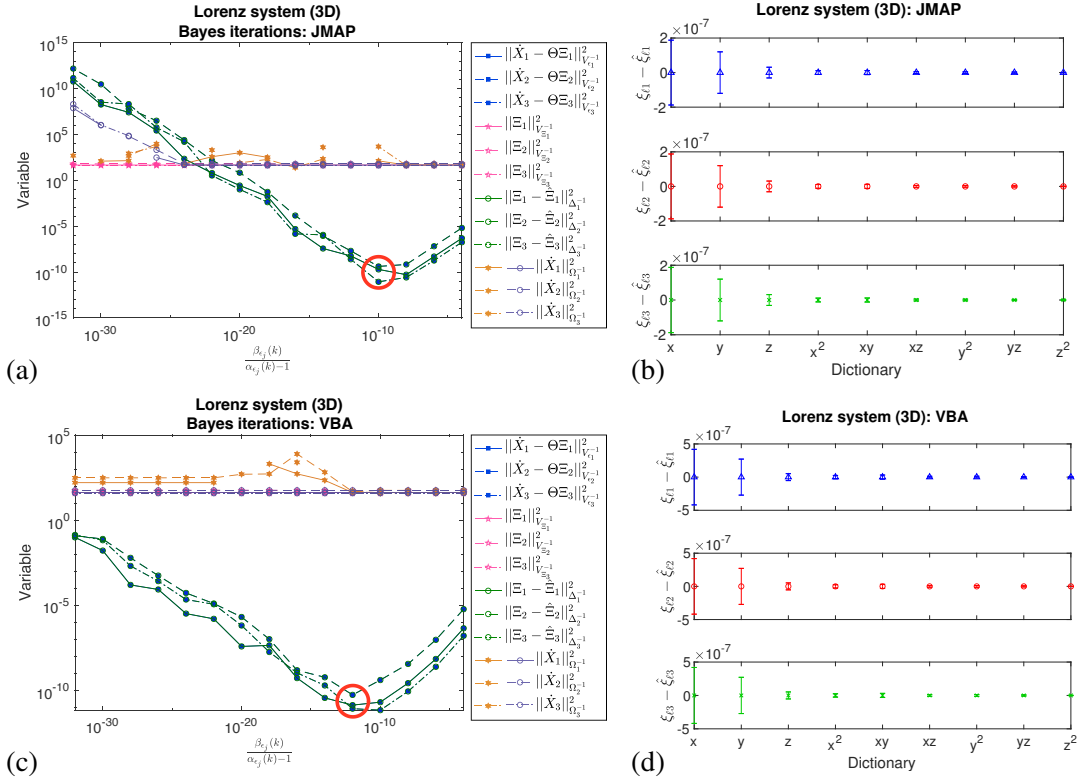
\begin{figure*}[ht]
\begin{center}
\setlength{\unitlength}{0.55pt}
  \begin{picture}(800,510)
   \put(0,270){\includegraphics[height=50mm]{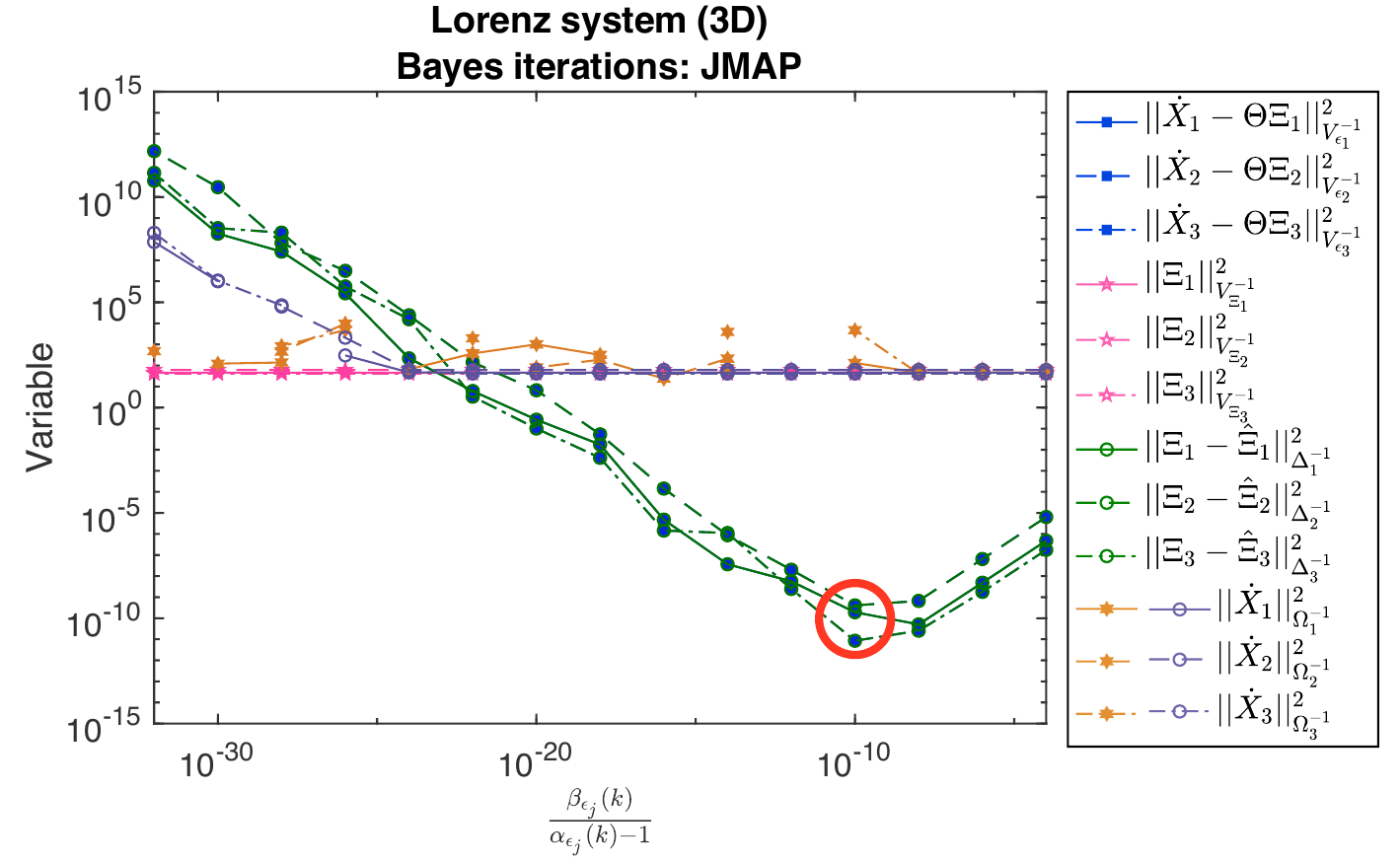} }
   \put(0,270){\small (a)}
   \put(430,270){\includegraphics[height=50mm]{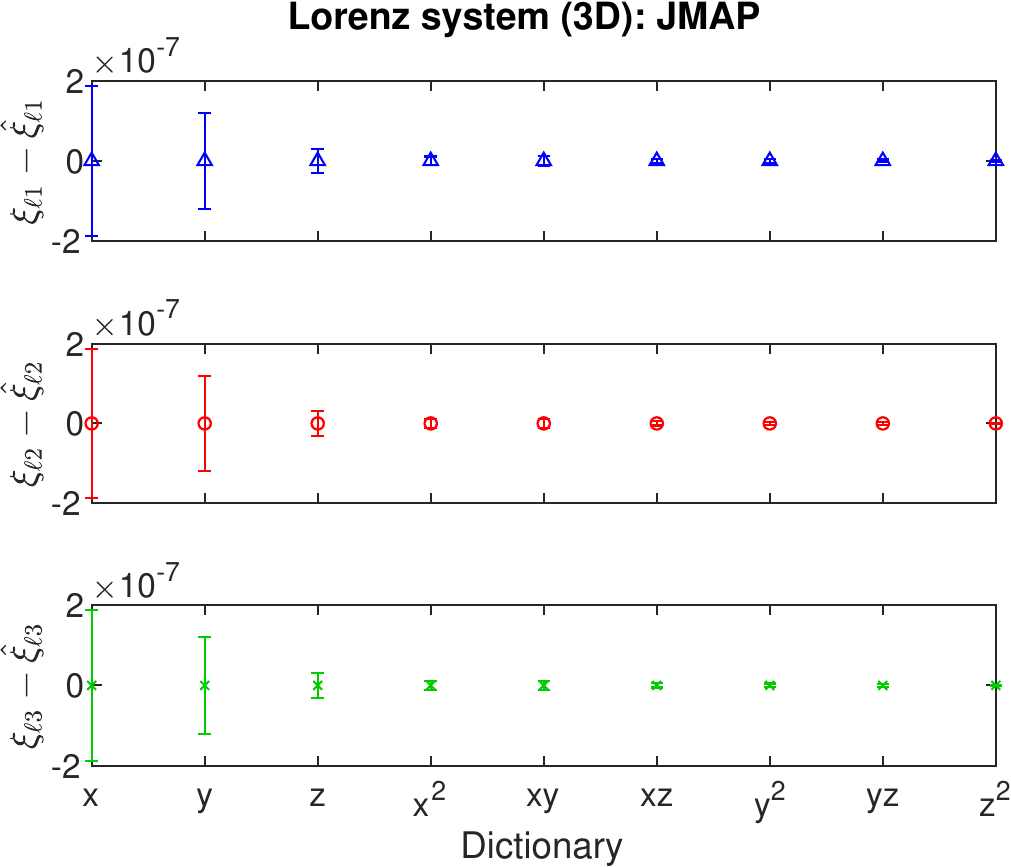} }
   \put(430,270){\small (b)}
   \put(0,0){\includegraphics[height=50mm]{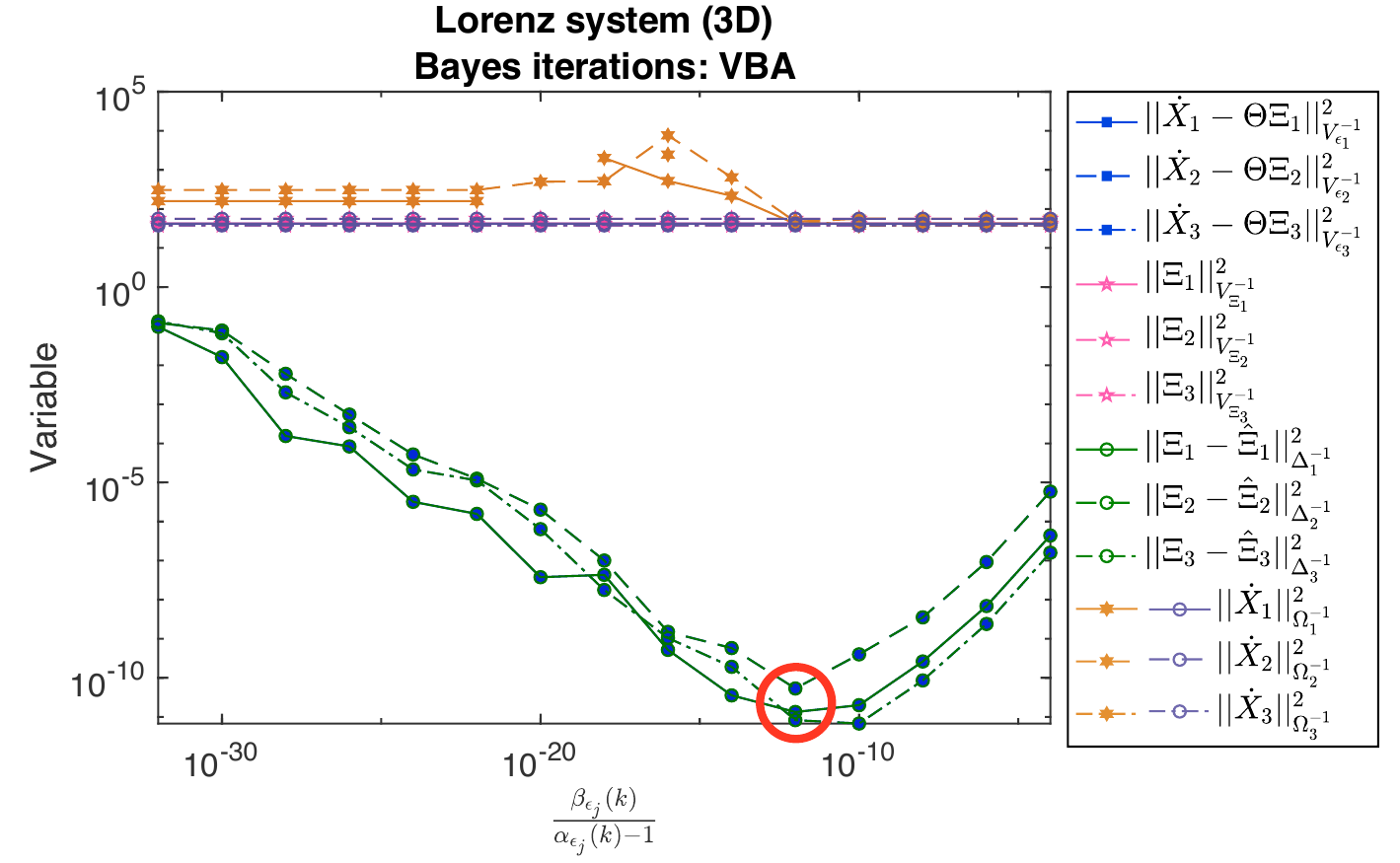} }
   \put(0,0){\small (c)}
   \put(447,0){\includegraphics[height=48mm]{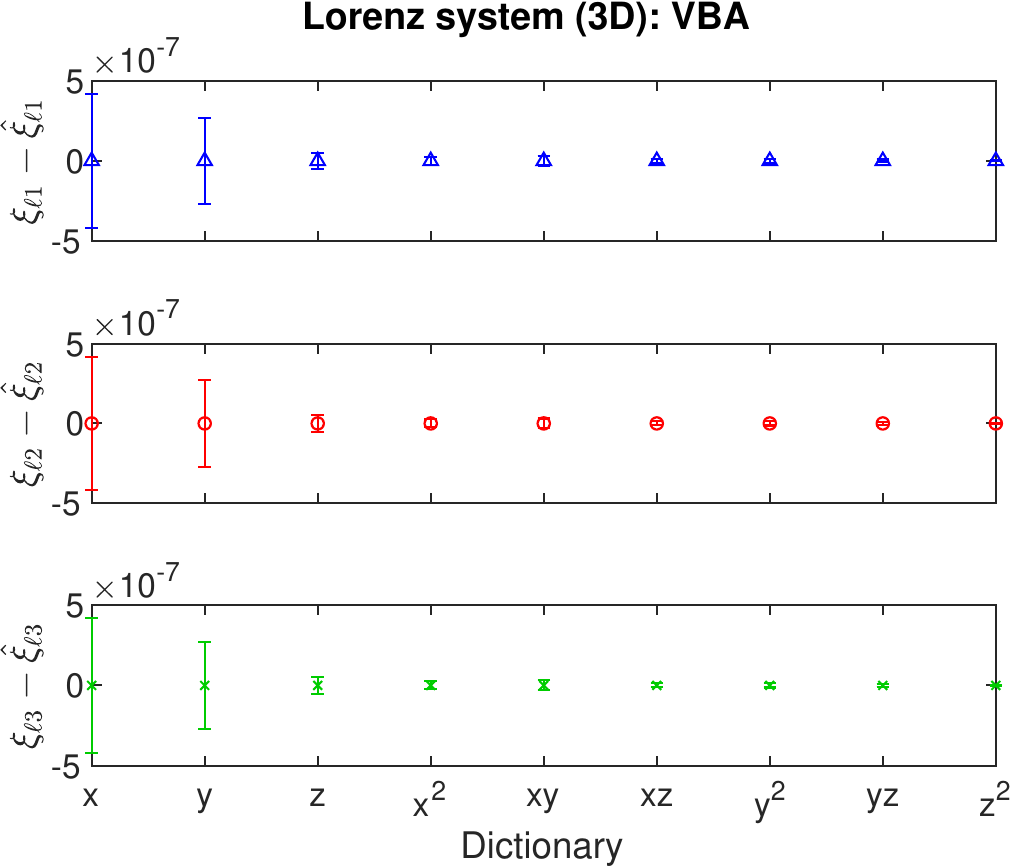} }
   \put(430,0){\small (d)}
  \end{picture}
\end{center}
\caption{{Lorenz system with added Laplace noise for $T=100$, $t_{step}=0.01$, $\varepsilon=0.2$, (a)-(b) JMAP regularization and (c)-(d) VBA regularization: 
(a),(c) Gaussian norms for the prior, likelihood, posterior and evidence, showing the optimal iteration; and
(b),(d) optimal error in predicted coefficients $\matparamc_{\ell j}-\hat{\matparamc}_{\ell j}$, with error bars calculated from the posterior covariance \eqref{eq:posterior_estimators}.}}
\label{fig:Lorenz_JMAP_VBA_Laplace}
\end{figure*}

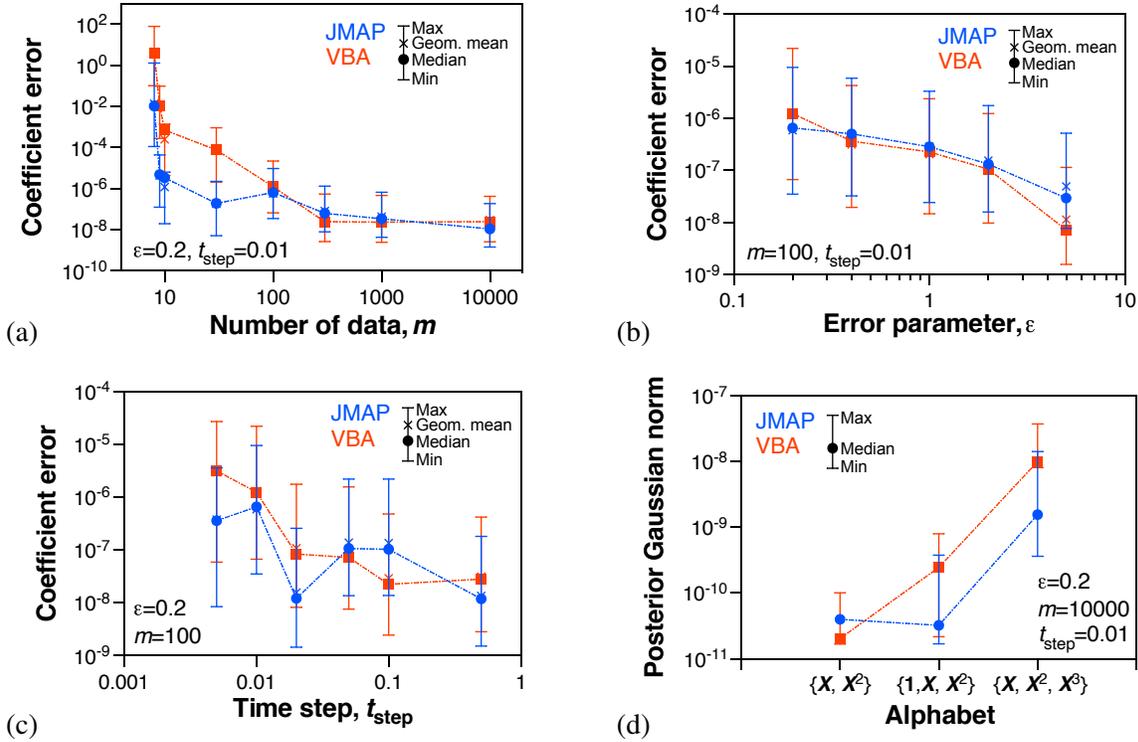
\begin{figure*}[t]
\begin{center}
\setlength{\unitlength}{0.55pt}
  \begin{picture}(800,500)
   \put(10,270){\includegraphics[height=45mm]{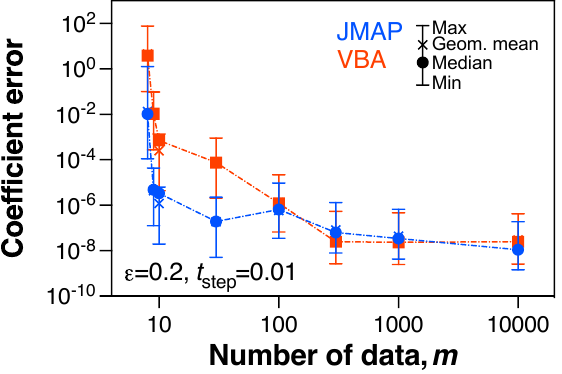} }
   \put(0,270){\small (a)}
  \put(440,270){\includegraphics[height=45mm]{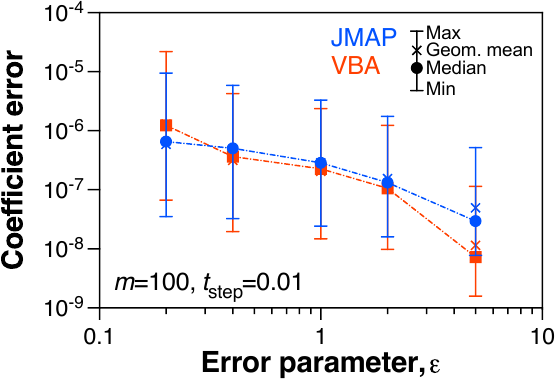} }
   \put(420,270){\small (b)}
   \put(20,0){\includegraphics[height=47mm]{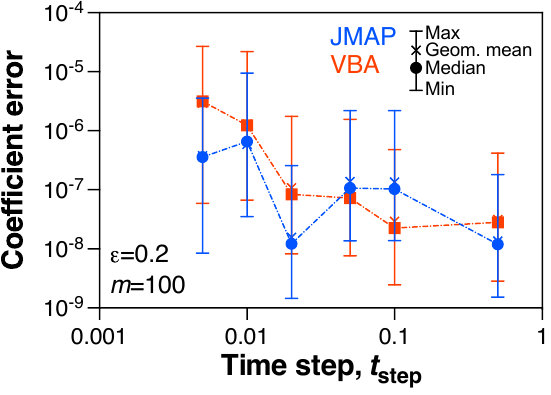} }
   \put(0,0){\small (c)}
   \put(438,0){\includegraphics[height=49mm]{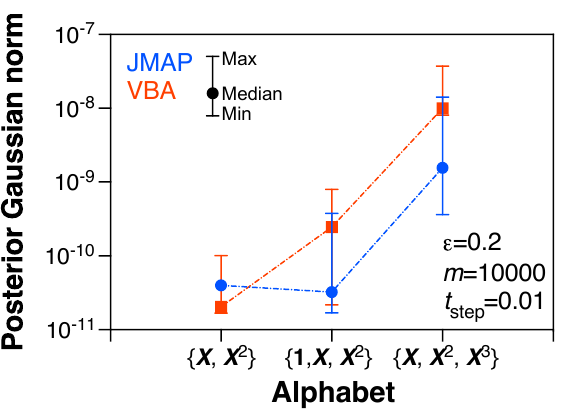} }
   \put(420,0){\small (d)}
  \end{picture}
\end{center}
\caption{Lorenz system {with added Gaussian noise}, effect of parameters on JMAP and VBA algorithms:
(a) training data length;
(b) added noise; 
(c) time step interval; and 
(d) choice of alphabet.}
\label{fig:Lorenz_params}
\end{figure*}

\begin{figure*}[ht]
\begin{center}
\setlength{\unitlength}{0.55pt}
  \begin{picture}(800,520)
   \put(10,270){\includegraphics[height=50mm]{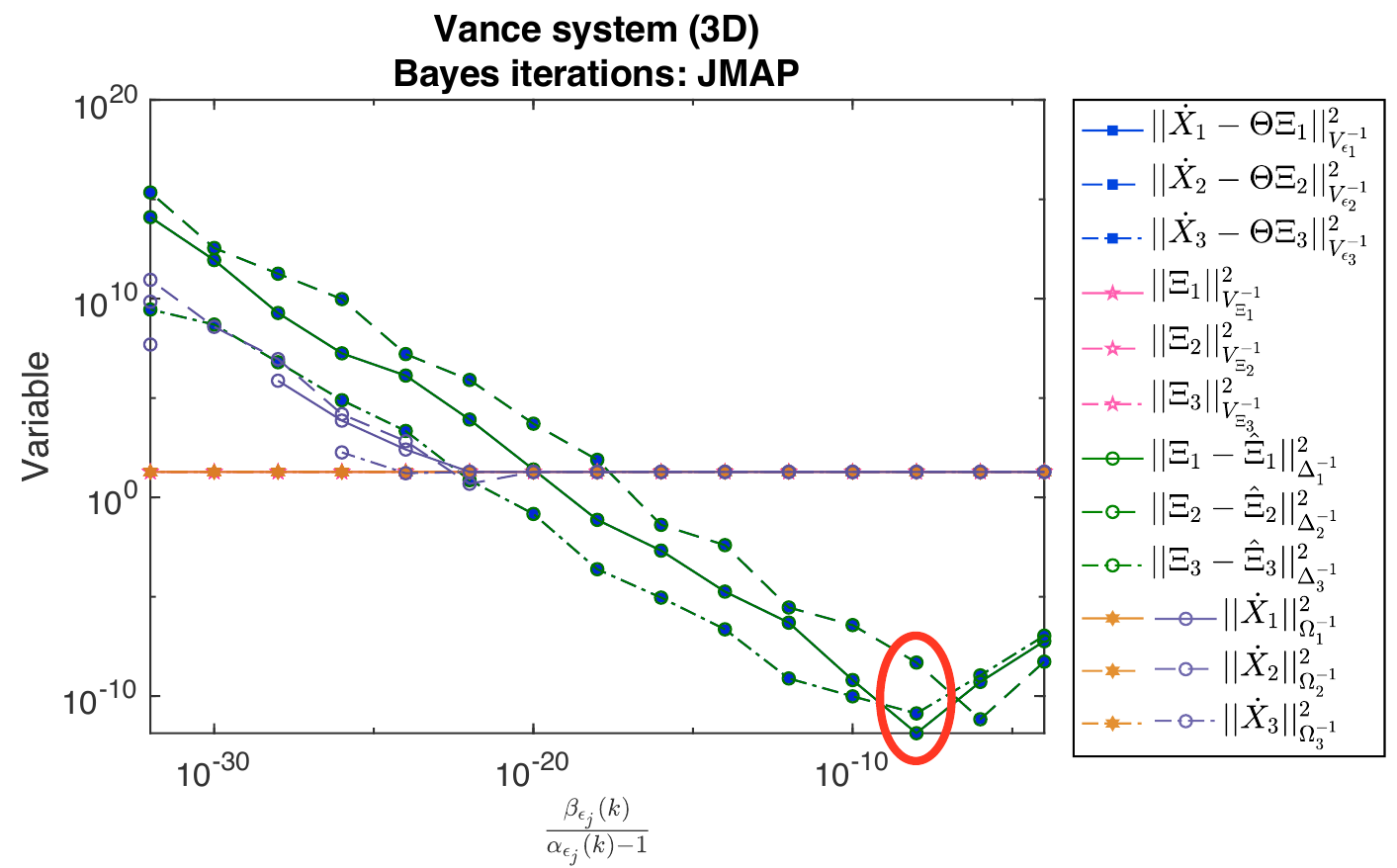}}
   \put(0,270){\small (a)}
   \put(467,270){\includegraphics[height=48mm]{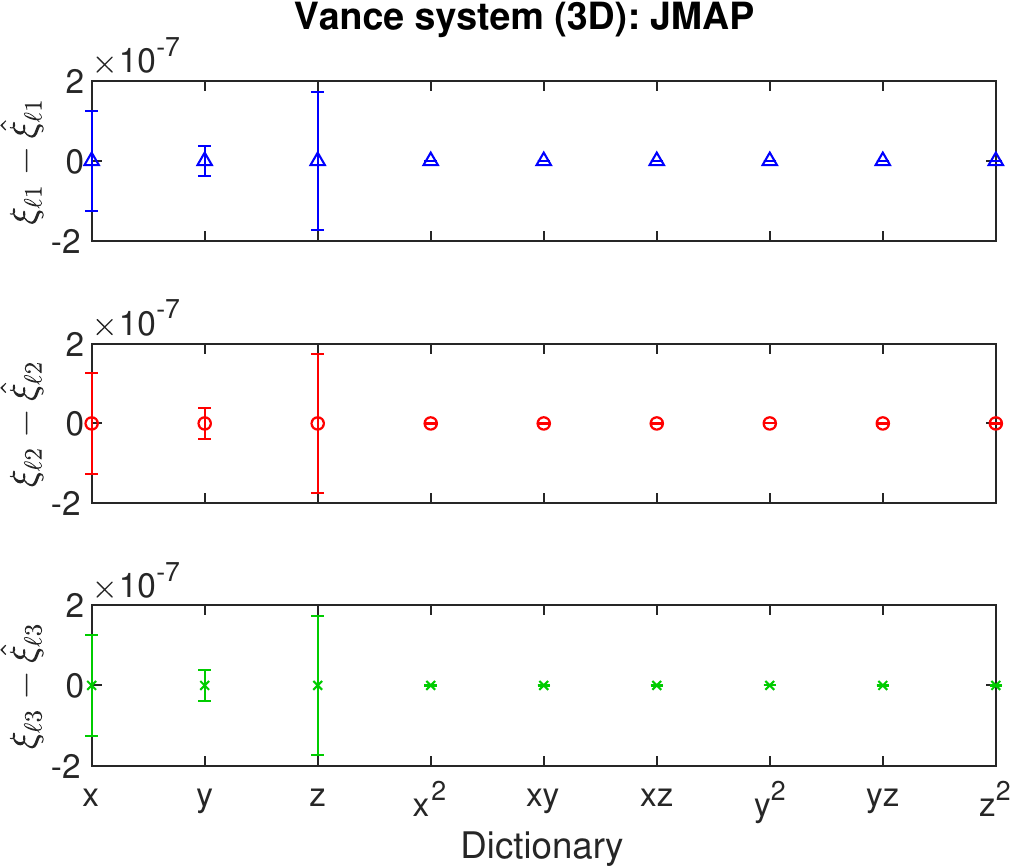} }
   \put(460,270){\small (b)}
   \put(10,0){\includegraphics[height=50mm]{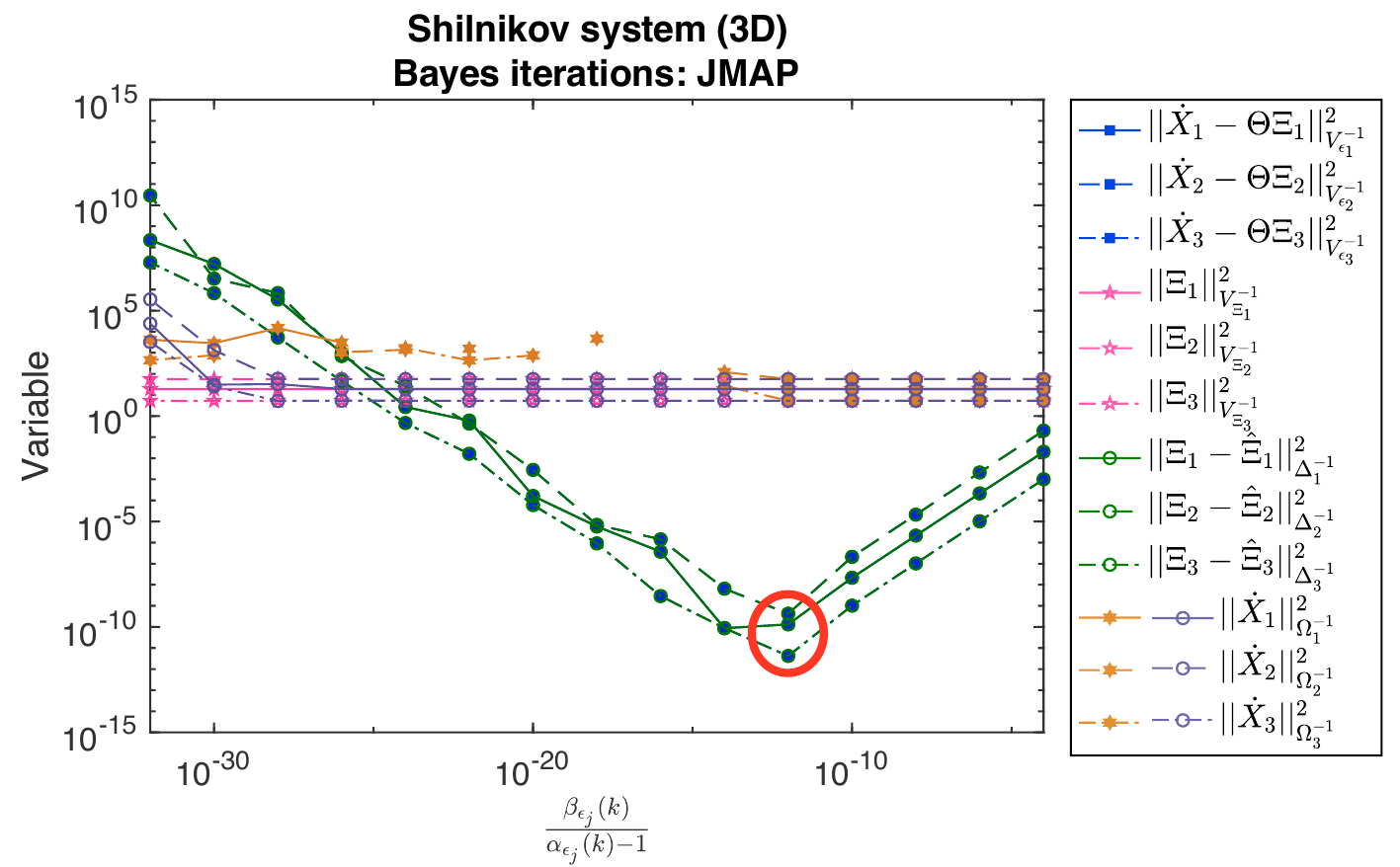} }
   \put(0,0){\small (c)}
   \put(467,0){\includegraphics[height=48mm]{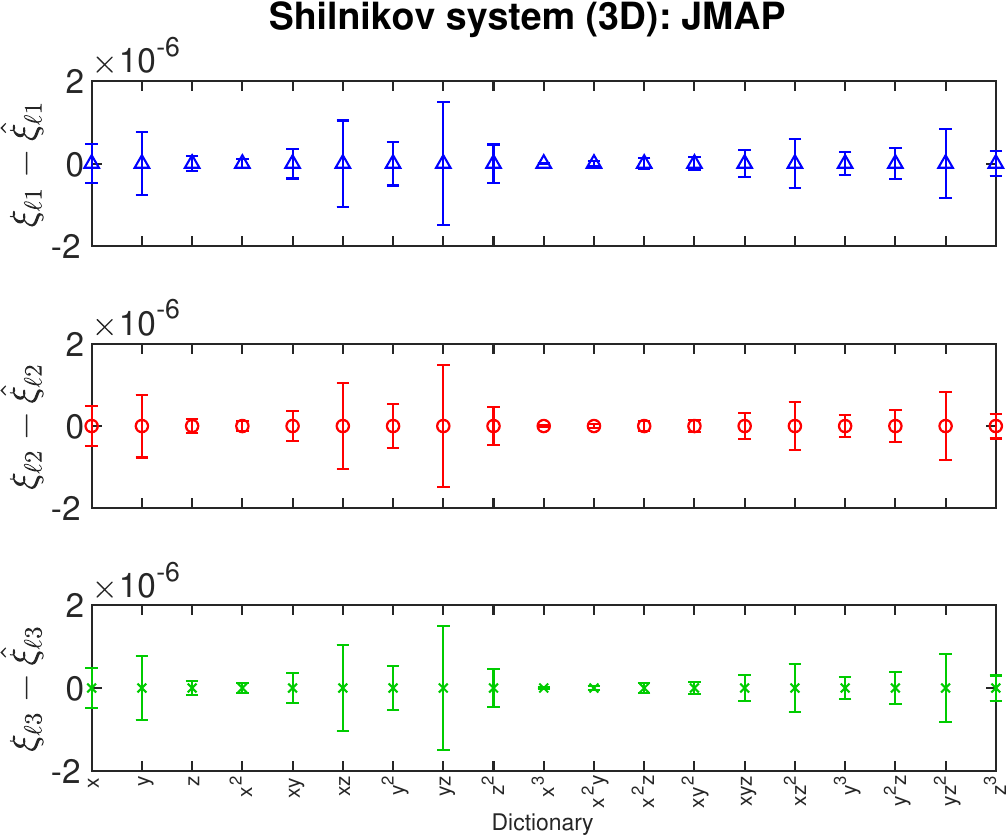} }
   \put(460,0){\small (d)}
  \end{picture}
\end{center}
\caption{Vance (a)-(b) and Shil'nikov (c)-(d) systems with {with added Gaussian noise for $T=500$}, $t_{step}=0.02$, showing JMAP regularization: 
(a),(c) Gaussian norms for the prior, likelihood, posterior and evidence, with the optimal iteration; 
(b),(d) optimal error in predicted coefficients $\matparamc_{\ell j}-\hat{\matparamc}_{\ell j}$, with error bars calculated from the posterior covariance \eqref{eq:posterior_estimators}.}
\label{fig:Vance_Shilnikov_summary}
\end{figure*}

The analyses are illustrated in Figures \ref{fig:Lorenz_reg}-\ref{fig:Vance_Shilnikov_summary}, with more complete sets of {figures in the {supplementary material} (Figs \ref{fig:SI_noise}-\ref{fig:SI_Shilnikov_sys_VBA_metrics}).}  A number of graphs are presented for each system, as follows:


\begin{figure*}[t]
\begin{center}
\setlength{\unitlength}{0.55pt}
  \begin{picture}(800,510)
    \put(0,270){\includegraphics[height=50mm]{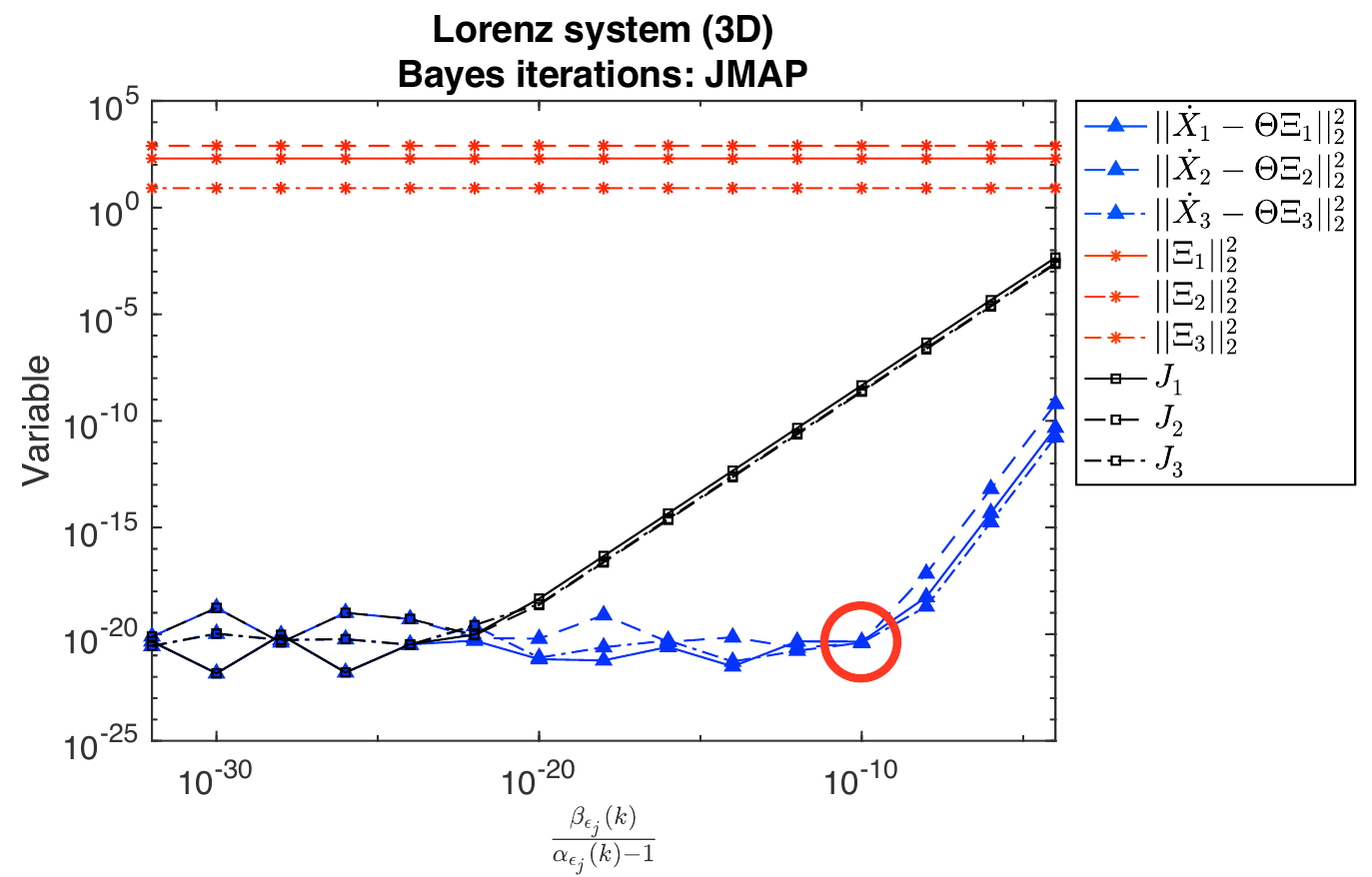} }
   \put(0,270){\small (a)}
   \put(430,270){\includegraphics[height=50mm]{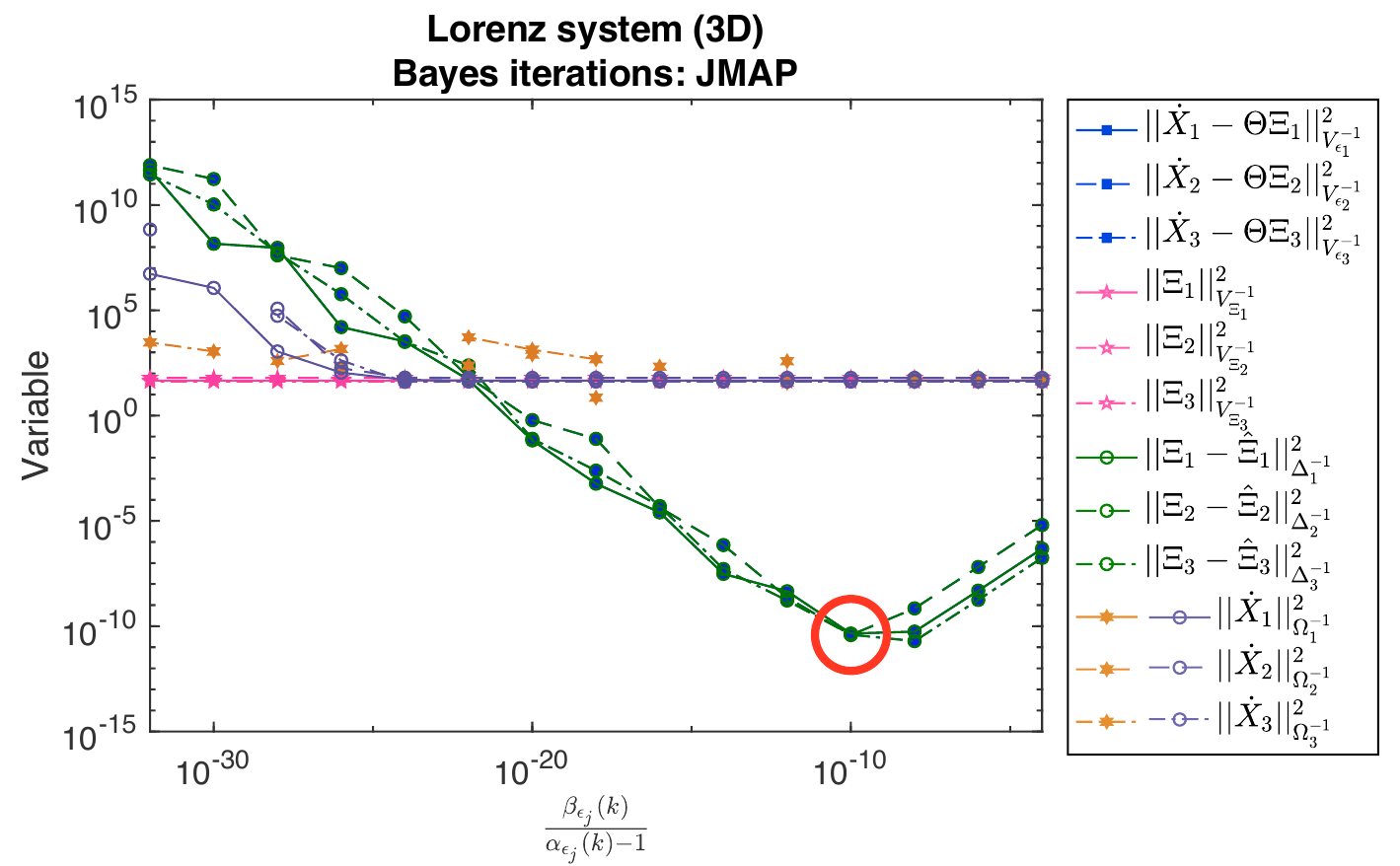} }
   \put(430,270){\small (b)}
   \put(17,0){\includegraphics[height=50mm]{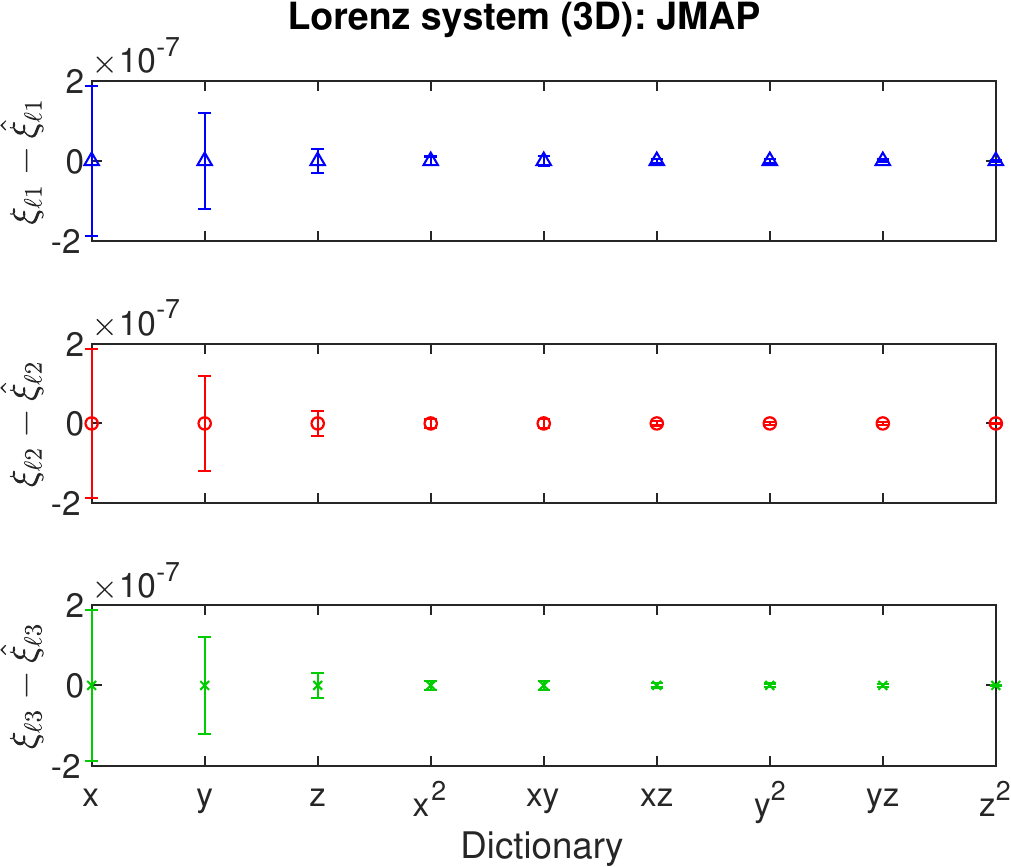} }
   \put(0,0){\small (c)}
   \put(447,0){\includegraphics[height=48mm]{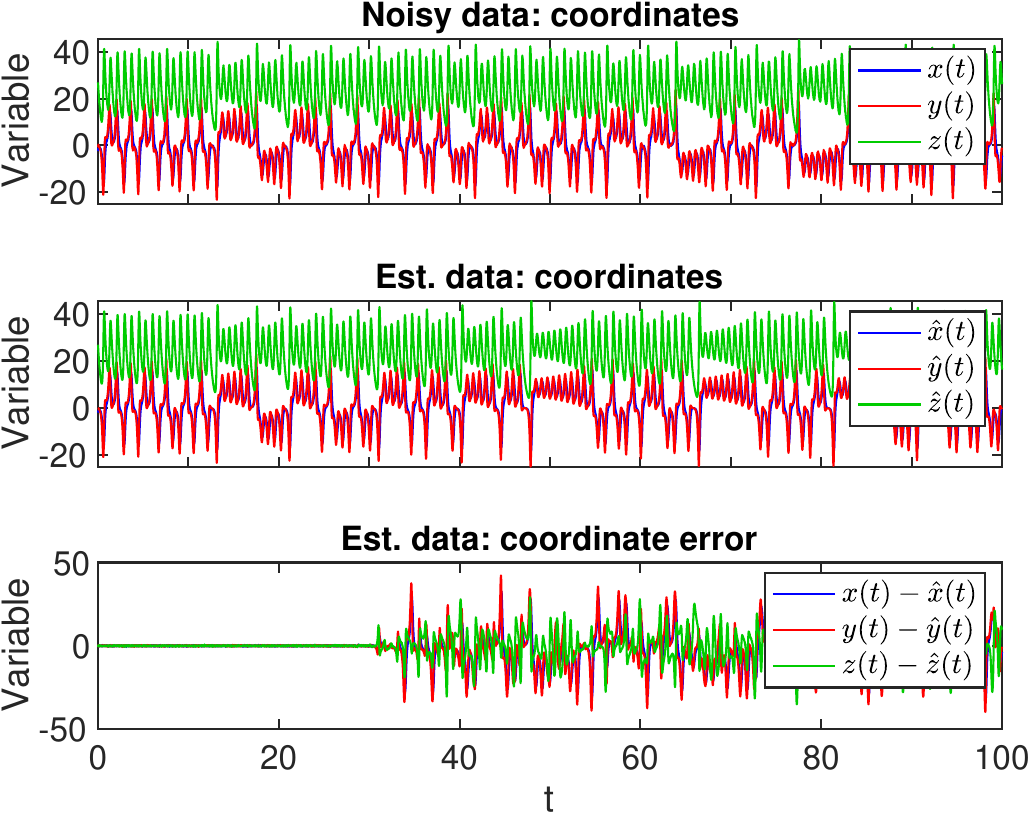} }
   \put(430,0){\small (d)}
  \end{picture}
\end{center}
\caption{Lorenz system {with added Gaussian noise for} {$T=100$, $t_{step}=0.01$, $\varepsilon=0.2$}, JMAP regularization: 
(a) iteration sequence with decreasing $\mathsf{E}_{\eizero}$, showing the 2-norm residual, regularization and objective functions \eqref{eq:modJ} and the optimal iteration; 
(b) Gaussian norms for the prior, likelihood, posterior and evidence, showing the optimal iteration; 
(c) optimal error in predicted coefficients $\matparamc_{\ell j}-\hat{\matparamc}_{\ell j}$, with error bars calculated from the posterior covariance \eqref{eq:posterior_estimators}; and
(d) original and optimal predicted data and their differences.}
\label{fig:Lorenz_JMAP_Gauss}
\end{figure*}

\begin{list}{$\bullet$}{\topsep 4pt \itemsep 0pt \parsep 4pt \leftmargin 15pt \rightmargin 0pt \listparindent 0pt \itemindent 0pt}

\item For all algorithms {except LASSO}, plots are provided of 2-norms including the residual terms $||\g - \H \f||^2_2$ (dark blue), regularization terms $||\f||^2_2$ (red) and objective functions ${J}_j = ||\g - \H \f||^2_2 + \psi ||\f||^2_2$ \eqref{eq:modJ} (black), using $\psi=\lambda$ for SINDy, {$\psi=\theta_j$ for ridge regression} and $\psi =  \min(\meanVepsilondiag \vec{1} \oslash \Vfdiag)$ for JMAP and VBA, where $\meanVepsilondiag$ is the mean $\Vepsilondiag$, $\vec{1}$ is the 1 vector and {$\oslash$ is the element-wise Hadamard division operator}. These are plotted to show the outer iteration sequence, for SINDy against $\lambda$, {for ridge regression against $\theta_j$} and for JMAP and VBA against $\mathsf{E}_{\eizero} = {\beta_{\eizero}}/({\alpha_{\eizero}-1})$ \eqref{eq:exp_var_IG}.

\item {For LASSO, plots are provided of the residual terms $||\g - \H \f||^2_2$ (dark blue), regularization terms $||\f||_1$ (red) and objective functions ${J}_j^{LASSO} = ||\g - \H \f||^2_2 + \kappa_j ||\f||_1$ \eqref{eq:modJ} (brown). These are plotted against each $\kappa_j$.}

\item For JMAP and VBA, plots are {also} provided of the Gaussian norms for the likelihood $||\g - \H \f ||^2_{\Vepsilon^{-1}}$ (medium blue), prior $||\f ||^2_{\Vf^{-1}}$ (pink), posterior $||\f - \BPostMean ||^2_{{\BPostCovMat^{-1}}}$ (green) and evidence $||\g - \BEvidMean ||^2_{{\BEvidCovMat^{-1}}} = ||\g ||^2_{{\BEvidCovMat^{-1}}}$, {the latter} by direct calculation (orange) or from Bayes' rule (slate blue). These are plotted against $\mathsf{E}_{\eizero}$.

\item For all algorithms, plots are provided of the differences between the true and optimal predicted model coefficients $\matparamc_{\IndexAlphabet \IndexDynSys} -\hat{\matparamc}_{\IndexAlphabet \IndexDynSys}$, over the alphabet of functions used. These show the precision of the predicted coefficients. For the Bayesian algorithms, they allow the error bars on each coefficient to be illustrated clearly.

\item For all algorithms, a triple plot is provided of the noisy time-series data, the predicted data (using the optimal coefficients) and their differences. These illustrate the degree of success of the inference method used.

\end{list}

The Lorenz system with {Gaussian noise is illustrated in Figures \ref{fig:dyn_sys}(a) and \ref{fig:SI_noise}(a)-(b). The SINDy, LASSO, ridge regression, JMAP and VBA analyses are shown in Figures \ref{fig:Lorenz_reg}-\ref{fig:Lorenz_VBA_Gauss}, \ref{fig:SI_Lorenz_SINDy}-\ref{fig:SI_Lorenz_Gauss_JMAP} and \ref{fig:SI_Lorenz_Gauss_VBA}, with the metrics in Figures \ref{fig:SI_Lorenz_Gauss_JMAP_metrics} and \ref{fig:SI_Lorenz_Gauss_VBA_metrics}.}
As shown:

\begin{list}{$\bullet$}{\topsep 4pt \itemsep 0pt \parsep 4pt \leftmargin 15pt \rightmargin 0pt \listparindent 0pt \itemindent 0pt}

\item The SINDy algorithm rapidly finds an optimal solution with decreasing $\lambda$; this is somewhat ambiguous but can be assigned to the turning point in the 2-norm residual (Figure \ref{fig:Lorenz_reg}(a)). The calculated coefficients have a precision of $\sim 10^{-14}$ to $10^{-15}$ (Figure \ref{fig:Lorenz_reg}(b)), with excellent recovery of the time series to $t \sim 35$ (Figure \ref{fig:SI_Lorenz_SINDy}(c)).

\item In contrast, the JMAP and VBA algorithms converge more slowly with decreasing $\mathsf{E}_{\eizero}$, but both find optimal solutions {within 5-10 iterations}, identified by the minimum posterior Gaussian norm in \eqref{eq:posterior} (Figures \ref{fig:Lorenz_JMAP_Gauss}(b) and \ref{fig:Lorenz_VBA_Gauss}(b)). This is equivalent to the maximum absolute log POR \eqref{eq:log_POR2}, and clearly illustrates the operation of the Bayesian MAP method. For the {fixed priors used}, each optimum also corresponds to the minimum in the likelihood Gaussian norm. 
They also correspond to (or are close to) the turning points in the 2-norm residuals (Figures \ref{fig:Lorenz_JMAP_Gauss}(a) and \ref{fig:Lorenz_VBA_Gauss}(a)).  

\item {Ridge regression (Figure \ref{fig:SI_Lorenz_ridge}) gives similar model coefficients to the SINDy and Bayesian algorithms, with slow convergence. The calculated precision ($\sim10^{-14}$ to $10^{-15}$) is comparable to SINDy.} 

\item {The LASSO algorithm (Figure \ref{fig:SI_Lorenz_LASSO}) did not converge well, giving poor precision ($\sim10^{-3}$) in the estimated coefficients and poor reproduction of the time-series for all cases examined. This weakness of LASSO for parameter identification has been reported by other researchers \cite{Pan_etal_2016, Mangan_etal_2019, Champion_etal_2020}.}

\item The JMAP and VBA algorithms {were less efficient than SINDy, but viable for all systems examined.} For the analyses in Figures \ref{fig:Lorenz_reg}-\ref{fig:Lorenz_VBA_Gauss}, {the average times for a single  outer iteration were SINDy (0.167 min), LASSO (0.22 min), ridge regression (0.44 min), JMAP (1.33 min) and VBA (2.2 min).}

\item For the JMAP and VBA algorithms, the evidence norms were approximately constant, with some numerical dispersion evident at low values of $\mathsf{E}_{\eizero}$. This supports the assumption made in the derivation of the MAP estimator in \eqref{eq:MAP_log}. 

\item {An important advantage of the JMAP and VBA algorithms is that they provide} error bars on the predicted coefficients (Figures \ref{fig:Lorenz_JMAP_Gauss}(c) and \ref{fig:Lorenz_VBA_Gauss}(c)), extracted from the posterior covariance matrix (Section \ref{sect:UQ}). The calculated errors are $\sim10^{-7}$, much larger than {the precision of SINDy or ridge regression, with comparable time windows for recovery of the time series.} Together, these suggest overfitting of the data by {SINDy or ridge regression}, to an unjustifiably high precision.

\item The model selection metrics AIC and BIC \eqref{eq:AIC_BIC} do not provide any discernible optima for either JMAP or VBA (Figures \ref{fig:SI_Lorenz_Gauss_JMAP_metrics}(a) and \ref{fig:SI_Lorenz_Gauss_VBA_metrics}(a)). The EB \eqref{eq:EB} does exhibit an optimum solution, but contrary to previous arguments \cite{Chen_Lin_2021}, this corresponds to the point of maximum rather than minimum variance. The 2-norm variants AIC$_2$, AIC$^c_2$ and BIC$_2$ (Figures \ref{fig:SI_Lorenz_Gauss_JMAP_metrics}(b) and \ref{fig:SI_Lorenz_Gauss_VBA_metrics}(b)) reveal an optimum, but this corresponds to the turning point of the 2-norm residual, obviating any need for these metrics.
 
\end{list}

{The Lorenz system with Laplace noise is illustrated in Figures \ref{fig:SI_noise}(c)-(d), and the results of JMAP and VBA are shown in Figures \ref{fig:Lorenz_JMAP_VBA_Laplace} and \ref{fig:SI_Lorenz_Laplace_JMAP}-\ref{fig:SI_Lorenz_Laplace_VBA}. Laplace noise does not match the assumption of a Gaussian likelihood function \eqref{eq:likelihood}, so this provides a test of the Bayesian algorithms to a different noise model. 
The results are remarkably similar to the analyses of Gaussian noise, with optimal prediction of the error hyperparameters within 5-10 iterations, and calculated error bars of $\sim10^{-7}$. These and other analyses indicate the JMAP and VBA algorithms presented here to be surprisingly robust to the choice of noise model.} 


{Examining the effect of different parameters on the JMAP and VBA analyses of the Lorenz system {with Gaussian noise}:}

\begin{list}{$\bullet$}{\topsep 4pt \itemsep 0pt \parsep 4pt \leftmargin 15pt \rightmargin 0pt \listparindent 0pt \itemindent 0pt}

\item {The effect of length of the time-series training data -- expressed by the number of data points $m$ -- on the absolute coefficient errors $|\matparamc_{\ell j}-\hat{\matparamc}_{\ell j}|$ is shown in Figure \ref{fig:Lorenz_params}(a), with supporting data for JMAP shown in Figure \ref{fig:SI_Lorenz_Gauss_JMAP_time}. As shown, the errors increase with fewer data points, but remain quite acceptable for time series from 10000 points down to 10 points, sharply rising only for time series with less than 10 points. The JMAP algorithm was found to be more robust to smaller data sets.  If applicable generally, the  results support the viability of real-time applications of Bayesian inference to incoming dynamical system data.}

\item {For the above examples at low $m$, breakdown of the JMAP and VBA algorithms is revealed by (i) very large error bars on some predicted coefficients (e.g., Figure \ref{fig:SI_Lorenz_Gauss_JMAP_time}(e)); (ii) separation of the likelihood and posterior Gaussian norms (e.g., Figure \ref{fig:SI_Lorenz_Gauss_JMAP_time}(f)), which should follow similar trends for a constant prior; (iii) unexpected changes in the evidence Gaussian norm; and/or (iv) negative diagonal terms in the posterior covariance matrix $\BPostCovMat$, giving complex standard deviations $\hat{\sigma}_{\matparamc_{\IndexAlphabet \IndexDynSys}}$. In contrast, the SINDy method exhibits a turning point in its 2-norm residual, implying the existence of a solution, without giving any indication that something is amiss.} 

\item {The effect of {the magnitude of} added Gaussian noise -- expressed by the parameter $\varepsilon$ -- is shown in Figure \ref{fig:Lorenz_params}(b). 
Unexpectedly, the error decreases with larger added noise. Some differences were evident between the JMAP and VBA algorithms.}

\item {The effect of the time step interval, for data series of the same length $m$, is shown in Figure \ref{fig:Lorenz_params}(c). 
These data broadly indicate that the coefficient errors decrease with larger time steps, but with some fluctuations, presumably due to interactions between the step interval and intrinsic timescales of the Lorenz system.}

\item The effect of different choices of alphabet, including second-order polynomials with a column of 1s, or third-order polynomials, is shown in Figure \ref{fig:Lorenz_params}(d), with supporting data for JMAP shown in Figure \ref{fig:SI_Lorenz_Gauss_JMAP_poly}. As shown, the inclusion of unnecessary functions in the alphabet produces larger error bars on the predicted coefficients. It is also revealed by higher posterior Gaussian norms. 

\item Different processors (Mac Intel 8-core or 4-core) were found to give small differences in the 2-norm residuals and posterior Gaussian norms, especially for large data sets or at low values of $\mathsf{E}_{\eizero}$, but these did not affect the choice of optimal iterations, the predicted coefficients or the main conclusions. 


\end{list}

The Vance system with added {Gaussian} noise is illustrated in Figure \ref{fig:dyn_sys}(b), and the analyses in Figures \ref{fig:Vance_Shilnikov_summary}(a)-(b) and \ref{fig:SI_Vance_SINDy}-\ref{fig:SI_Vance_sys_VBA_metrics}. Here the SINDy algorithm requires more iterations to find the optimum, due to the presence of small coefficients. Otherwise, the three algorithms are broadly similar to the Lorenz system. The JMAP and VBA optima are again identified by the minimum posterior norms, corresponding to a turning points in the 2-norm residuals. The Bayesian methods give coefficient errors of $\sim 10^{-6}$ to $10^{-7}$, much higher than the SINDy precision ($\sim10^{-15}$ to $10^{-16}$). 

The Shil'nikov system with added {Gaussian} noise is illustrated in Figure \ref{fig:dyn_sys}(c), and the analyses in Figures \ref{fig:Vance_Shilnikov_summary}(c)-(d) and \ref{fig:SI_Shilnikov_SINDy}-\ref{fig:SI_Shilnikov_sys_VBA_metrics}.  Despite requiring a third-order polynomial alphabet, the three algorithms behave similarly to the Lorenz system, with optimal coefficient errors of $\sim10^{-6}$. For this system, the Bayesian algorithms infer significant error bars for many coefficients.


\section{\label{sect:concl} Conclusions}

This study presents a maximum \textit{a~posteriori} (MAP) Bayesian framework for the identification of a dynamical system \eqref{eq:dynsys} from its discrete time series data. The MAP estimate of the posterior distribution is shown to be equivalent to a generalized Tikhonov regularization (\S\ref{sect:Bayes_reg}), providing a rational justification for the choice of the residual and regularization terms, respectively, from the negative logarithms of the likelihood and prior distributions. 
The Bayesian regularization can be used to infer a matrix of model coefficients associated with an alphabet of functions of the data, to reproduce the dynamical system very precisely.
The Bayesian interpretation also gives access to the full apparatus for Bayesian inference, including 
the Bayesian posterior and model uncertainty quantification (\S\ref{sect:UQ}), 
the posterior Bayes factor and Bayesian model selection (\S\ref{sect:model_sel}), 
the Bayesian evidence (\S\ref{sect:evidence}), 
and 
the JMAP and VBA algorithms for the estimation of unknown hyperparameters (\S\ref{sect:joint_Bayes}). 
In \S\ref{sect:apps}, the JMAP and VBA algorithms are applied to the analysis of three dynamical systems with added {Gaussian or Laplace} noise, {using a new Bayesian Dynamical System Identification (BDSI) code written in MATLAB. The results are compared to those of the LASSO, ridge regression and SINDy algorithms for sparse regression.}

From the theoretical analyses and the three case studies, the following conclusions can be drawn:

\begin{list}{$\bullet$}{\topsep 4pt \itemsep 0pt \parsep 4pt \leftmargin 15pt \rightmargin 0pt \listparindent 0pt \itemindent 0pt}

\item The assumption of a multivariate Gaussian likelihood \eqref{eq:likelihood} and {conjugate} prior \eqref{eq:prior}
distributions -- while not required for Bayesian inference -- enables a dramatic simplification of Bayes' rule \eqref{eq:Bayes}, to give {closed-form solutions for the multivariate Gaussian posterior \eqref{eq:posterior_estimators} and evidence \eqref{eq:evidence_estimators} distributions. These simplify the construction of MAP algorithms and computations for low-dimensional data. Furthermore, for Gaussian functions the MAP estimate corresponds to the posterior mean. Despite their Gaussian foundations, the JMAP and VBA algorithms were found to be robust to all choices of dynamical system and noise model examined.}

\item The Bayesian framework also enables uncertainty quantification in the predicted model coefficients, expressed as standard deviations extracted from the posterior covariance matrix \eqref{eq:posterior_estimators}. This is an important advantage of the Bayesian framework over traditional regularization methods \cite{Niven_etal_MaxEnt_2019, Hirsh_etal_2022}. Analyses of three dynamical systems indicates coefficient errors of $\sim10^{-6}$ to $10^{-7}$ for time-series of length $m=10000$, much larger than the precision {($\sim10^{-15}$ to $10^{-14}$) reported by SINDy or ridge regression, for comparable time windows of predictability. This suggests overfitting of the data by {the latter two methods}}.

\item The Mahalanobis distance $|| \f - \BPostMean ||^2_{{\BPostCovMat^{-1}}}$, extracted from \eqref{eq:posterior} and referred to as the ``posterior Gaussian norm'', provides a point estimate of the posterior distribution \eqref{eq:posterior}. Its minimization corresponds to maximization of the absolute log posterior odds ratio \eqref{eq:log_POR2} based on the MAP estimate, for systems with a fixed alphabet, fixed data length and {constant} prior and likelihood covariances. As demonstrated by the three case studies, the posterior Gaussian norm provides a robust metric for Bayesian model selection, giving a clearly defined minimum for iterations over the error hyperparameters. 

\item For JMAP and VBA iterations using constant prior hyperparameters, the minimum posterior Gaussian norm corresponds to the minimum likelihood Gaussian norm $||\g - \H \f ||^2_{\Vepsilon^{-1}}$. It also corresponds to (or is close to) a turning point in the 2-norm residual $||\g - \H \f||^2_2$. 
{For SINDy and ridge regression, the optimal value of the hyperparameter(s) $\lambda$ or $\theta_j$ can be assigned to this residual turning point, while for LASSO, it is necessary to consider $\kappa_j$ at the turning point in the objective function.}

\item The model selection metrics AIC and BIC \eqref{eq:AIC_BIC} and their 2-norm counterparts do not incorporate the Bayesian prior, so must be interpreted as maximum likelihood rather than Bayesian metrics. For the case studies examined, no benefit was gained by their calculation. In contrast, the MAP solution corresponds to a maximum in the error bar metric \eqref{eq:EB}, at the position of maximum variance. 

\item Contrary to the suggestions of many Bayesian practitioners \cite{vonToussaint_2011, Tenorio_2017, Bontekoe_2023}, the evidence distribution \eqref{eq:evidence} -- represented by the evidence Gaussian norm $|| \g - \BEvidMean ||^2_{{\BEvidCovMat^{-1}}}$ point estimate -- was relatively constant (or subject to numerical problems) over the hyperparameter choices examined. It was therefore {unusable} for Bayesian model selection. 

\item The Bayesian algorithms require longer computational times than {SINDy, LASSO or ridge regression, but remain viable on a laptop computer for all systems examined. Importantly, the Gaussian MAP formulation avoids the need for more computationally expensive exploratory methods such as MCMC or nested sampling, to establish the shape of the posterior.} 

\item Sensitivity analysis of the Bayesian computations indicate that the coefficient errors decrease with increasing training data length $m$, added Gaussian noise scale $\varepsilon$ and time step interval $t_{step}$, and increase with unnecessarily longer alphabets (Figure \ref{fig:Lorenz_params}). These trends can be used to optimize these parameters to reduce the computational cost, yet achieve a specified uncertainty. Furthermore, the coefficient errors remain acceptable ($\sim 10^{-5}$) using JMAP for Lorenz time-series down to only 10 points, suggesting the possibility of real-time applications of Bayesian inference to incoming streams of dynamical system data. 

\item For the analyses of the Lorenz system, the JMAP and VBA algorithms exhibit visible signs of breakdown for short time series, including (i) large coefficient errors; (ii) separation of the likelihood and posterior Gaussian norms; (iii) inconsistencies in the evidence Gaussian norm; and/or (iv) complex standard deviations. In contrast, the SINDy algorithm gives solutions for the model coefficients for short time series, without any indication of computational problems. 

\end{list}

Further research is required on the Bayesian regularization framework developed herein, as part of a broader program for the implementation of Bayesian inference for all data analysis problems throughout science and engineering.  
This includes deeper theoretical insights into the MAP solution and the Bayesian posterior and evidence norms. 
{Further analysis of other dynamical systems with non-Gaussian error or model distributions is also warranted.} 
The role of the evidence norm in Bayesian model selection under the MAP framework requires further investigation. 
The outer iterative scheme given here can also be extended into more elaborate optimisation {schemes}, to allow for more complicated hyperparametric representations or posterior distributions with complicated landscapes. 
The connections between the present MAP framework and related Bayesian methods should be examined in more detail, including 
functional fitting using Gaussian processes \cite{Roberts_etal_2013, Svensson_Schon_2017, Raissi_Karniadakis_2018, Teckentrup_2020}, 
latent variable or structure discovery \cite{Pearl_1988, Psorakis_etal_2011}, 
and exploratory analyses of the posterior and evidence distributions {by MCMC or nested sampling} \cite{Hirsh_etal_2022, Rinkens_etal_2023, vonToussaint_2011, Skilling_2004}.

\begin{acknowledgments}
The first author benefited from funding and travel support from UNSW, Canberra, Australia; Institute Pprime / CNRS / Université de Poitiers, Chasseneuil-du-Poitou, France; CentraleSup\'elec, Gif-sur-Yvette, France and Ambrosys GmbH, Potsdam, Germany. 

\end{acknowledgments}










\appendix
\section{\label{sect:ApxA} Equivalence of Bayesian MAP Estimation {based on Gaussian Functions with Ridge Regression}}

The negative logarithm of Bayes' rule \eqref{eq:Bayes} provides the general objective function for minimization in Bayesian inference:
\begin{align}
\begin{split}
 J^\text{B}(\f)
 &= - \ln p(\f|\g) 
\\&= - \ln p(\g|\f) - \ln p(\f) + \ln p(\g)
\end{split}
\label{eq:MAP_est}
\end{align}
For the assumptions of multivariate Gaussian noise \eqref{eq:noise}-\eqref{eq:likelihood}, prior \eqref{eq:prior} and evidence \eqref{eq:evidence} made herein, with the analytical solution for the evidence \eqref{eq:evidence_estimators}, this reduces to:
\begin{align}
\begin{split}
&J^\text{B}(\f) 
= 
-\ln \frac{ \exp \Bigl( - \tfrac{1}{2} || \noise ||^2_{\Vepsilon^{-1}} \Bigr ) }{ \sqrt{ (2\pi)^\SizeTime \det\bigl({\Vepsilon}\bigr)}} 
- \ln \frac{ \exp \Bigl( - \tfrac{1}{2} || \f ||^2_{\Vf^{-1}} \Bigr ) }{ \sqrt{ (2\pi)^\SizeAlphabet \det\bigl({\Vf}\bigr)}}  
\\& \hspace{50pt} 
+ \ln \frac{ \exp \bigl( - \tfrac{1}{2} || \g - \BEvidMean ||^2_{{\BEvidCovMat^{-1}}}  \bigr ) }{ \sqrt{ (2\pi)^\SizeTime \det\bigl(\BEvidCovMat\bigr)}},
\\
&=  \tfrac{1}{2} \Bigl[ ||\g - \H(\statemat) \f ||^2_{\Vepsilon^{-1}}   +  ||\f ||^2_{\Vf^{-1}}   -  || \g ||^2_{{\BEvidCovMat^{-1}}}  \Bigr]
\\& \hspace{10pt}
+ \tfrac{1}{2}\ln \det\bigl({\Vepsilon}\bigr)   
+ \tfrac{1}{2}\ln  \det\bigl({\Vf}\bigr) 
- \tfrac{1}{2}\ln  \det\bigl({\H \Vf \H^\transpose + \Vepsilon}\bigr) 
\\& \hspace{10pt} +  (\SizeTime+\tfrac{\SizeAlphabet}{2} ) \ln  (2\pi) 
\end{split}
\label{eq:J_Gaussian}
\raisetag{10pt}
\end{align}
For constant $\SizeTime$, $\SizeAlphabet$, $\Vepsilon$ and $\Vf$, dependent variable data $\g$ and alphabet $\H$, \eqref{eq:J_Gaussian} reduces to the argument of \eqref{eq:MAP_simp2}. 

Now consider isotropic variances for the noise 
$\Vepsilon = \sigma^2_{\noise} \vI_\SizeTime$ 
and prior 
$\Vf = \sigma^2_{\f} \vI_\SizeAlphabet$, 
where $\vI_s$ is the identity matrix of size $s$, then 
$\Vepsilon^{-1} = {\vI_\SizeTime}/{\sigma^2_{\noise}} $, 
$\Vf^{-1} = {\vI_\SizeAlphabet}/{\sigma^2_{\f}} $, 
$\det\bigl(\Vepsilon\bigr)=\sigma_{\noise}^{2m}$, 
$\det\bigl(\Vf\bigr)=\sigma_{\f}^{2c}$ and 
$\BEvidCovMat = \sigma^2_{\f} \H \H^\transpose + \sigma^2_{\noise} \vI_\SizeTime$. 
The objective function \eqref{eq:J_Gaussian} becomes:
\begin{align}
\begin{split}
 &J^\text{B}(\f)
=  \frac{1}{2} \biggl[ \frac{1}{\sigma^2_{\noise}}  ||\g - \H(\statemat) \f ||^2_2  +  \frac{1}{\sigma^2_{\f}} ||\f ||^2_2   -  || \g ||^2_{{\BEvidCovMat^{-1}}}  \biggr]
\\& \hspace{10pt}
+ \SizeTime \ln \sigma_{\noise} 
+ \SizeAlphabet \ln  \sigma_{\f} 
- \tfrac{1}{2}\ln  \det\bigl({\sigma^2_{\f} \H \H^\transpose + \sigma^2_{\noise} \vI_\SizeTime}\bigr) 
\\& \hspace{10pt} +  (\SizeTime+\tfrac{\SizeAlphabet}{2} ) \ln  (2\pi) 
\label{eq:J_Gaussian_isotrop}
\raisetag{30pt}
\end{split}
\end{align}
For constant $\SizeTime$, $\SizeAlphabet$, $\sigma_{\noise}$ and $\sigma_{\f}$, dependent variable data $\g$ and alphabet $\H$, the minimization problem reduces to \cite{Bishop_2006, Aster_etal_2013, Bardsley_2018, Calvetti_Somersalo_2018}:
\begin{align} 
\begin{split}
\hatf^{\text{L}_2}
&=  \arg \min\limits_{\f} \, J^{\text{L}_2}(\f)\\
J^{\text{L}_2}(\f) &=   ||\g - \H(\statemat) \f ||^2_2  +  \kappa_j ||\f ||^2_2    
\end{split}
\label{eq:MAP_isotrop3}
\end{align}
which we {identify as ridge regression with residual term $||\g - \H(\statemat) \f ||^2_2$, regularization term $||\f ||^2_2$ and regularization coefficient $\kappa_j =  {\sigma^2_{\noise}}/{\sigma^2_{\f}}$.}  

The AIC and BIC metrics \eqref{eq:AIC_BIC} can also be interpreted\cite{Bishop_2006} as crude approximations of \eqref{eq:J_Gaussian} or \eqref{eq:J_Gaussian_isotrop}. However, since the regularization term (hence the prior) is absent, these provide metrics for a modified maximum likelihood method rather than for Bayesian inference. 

\section{\label{sect:ApxB} {Conjugate Prior for the Laplace Distribution}}

{Consider an {\it iid} sample $X_{1:n} = (X_1, X_2, ..., X_n)$ from univariate Laplace distributions \eqref{eq:univar_Laplace}. 
Here, we estimate only the scale parameter $b$.}

{For convenience, we replace ${b}^{-1}$ by $\zeta$, giving the local likelihood distributions:
\begin{align}
L_i(x_i \mid \mu,\zeta) = \frac{\zeta}{2}\exp(-\zeta\lvert x_i-\mu\rvert)\quad\text{with}\quad \zeta, x_i > 0.
\end{align}
which combine to give the total likelihood:
\begin{align}
L(\vec{x} \mid \mu,\zeta) 
= \prod\limits_{i=1}^n L_i(x_i\mid\mu,\zeta)
=  \frac{\zeta^n}{2^n}  \exp(-\zeta \sum\limits_{i=1}^n \lvert x_i-\mu\rvert) 
\label{eq:Laplace_iid}
\end{align}
We consider the prior for $\zeta$ is given by the gamma distribution with shape parameter $\alpha > 0$ and rate parameter $\beta > 0$:
\begin{align}
q(\zeta \mid \alpha,\beta) 
= \frac{\beta^{\alpha}}{\Gamma(\alpha)} \zeta^{\alpha - 1}\exp(-\zeta \beta)
\end{align}
where $\Gamma$ is the Gamma function given by:
\begin{align}
\displaystyle \Gamma(z) = \int_{0}^{\infty} t^{z-1} \exp(-t)\text{d}t.
\end{align}
}

{The posterior density function of $\zeta$ for the random sample $X$ is given by:
\begin{align}
p(\zeta\mid \vec{x}) 
= \dfrac{L(\vec{x} \mid \mu,\zeta) \, q(\zeta\mid\alpha,\beta)} 
{\int_0^{+\infty} L(\vec{x} \mid \mu,\zeta) \, q(\zeta\mid\alpha,\beta) \, d \zeta}
= \frac{N}{E}
\end{align}
in which the numerator $N$ is:
\begin{align}
N=   \frac{\zeta^{\alpha +n-1} \beta^{\alpha}}{2^n \Gamma(\alpha)} 
\exp \Bigl(-\zeta \Bigl[ \sum\limits_{i=1}^n \lvert x_i-\mu\rvert + \beta \Bigr] \Bigr)
\end{align}
The evidence is $E=\int_0^{+\infty} N d\zeta=({\beta^{\alpha} }/{2^n \Gamma(\alpha)}) \,I $, with
\begin{align}
\begin{split}
I&=   {\int_0^{+\infty}
\zeta^{\alpha+n-1}\,\exp \bigl(-\zeta \bigl[ \sum_{i=1}^n\lvert  x_i-\mu\rvert+\beta \bigr] \bigr) \, \text{d}\zeta} 
\end{split}
\end{align}
To find this integral, we set:
\begin{align}
A = \alpha+n \qquad\text{and}\qquad B=\sum_{i=1}^n\lvert  x_i-\mu\rvert+\beta
\end{align}
hence
\begin{align}
\begin{split}
I & = \int_0^{+\infty}\,\zeta^{A-1}\exp(-\zeta B) \, \text{d}\zeta
\\
 & = \frac{1}{B^A}\int_0^{+\infty}\,t^{A-1}\exp(-t) \, \text{d}t = \frac{\Gamma(A)}{B^A} \quad \text{where}\quad t = \zeta B
\end{split}
\end{align}
The evidence becomes
\begin{align}
E=\frac{\Gamma(A) \beta^{\alpha}}{2^n \Gamma(\alpha) B^A} 
=\frac{\Gamma(\alpha+n) \beta^{\alpha}}{2^n \Gamma(\alpha) \bigl[\sum_{i=1}^n\lvert  x_i-\mu\rvert+\beta \bigr]^{\alpha+n}} 
\end{align}
We thus obtain the posterior:
\begin{align}
\begin{split}
p(\zeta\mid \vec{x}) 
&= \frac{ B^A}
{\Gamma(A)} \,
\zeta^{A-1}\,\exp \bigl( -  \zeta B  \bigr)
\\
&= \frac{ \bigl[ \sum_{i=1}^n\lvert  x_i-\mu\rvert+\beta \bigr]^{\alpha+n}}
{\Gamma(\alpha+n)}\,\zeta^{\alpha+n-1} 
\\& \times 
\exp \Bigl( - \zeta \Bigl[ \sum_{i=1}^n\lvert  x_i-\mu\rvert+\beta \Bigr] \Bigr)
\end{split}
\end{align}
which implies $p(\zeta\mid \vec{x})\sim q(\zeta\mid\alpha+n,\sum_{i=1}^n\lvert  x_i-\mu\rvert+\beta)$. The prior and posterior distributions therefore belong to the same family, so the gamma distribution is a conjugate prior for the {\it iid} Laplace distribution \citep{Hasan_Baizid_2016}.
}

{
If however we replace the {\it iid} Laplace distribution \eqref{eq:Laplace_iid} with a multivariate Laplace distribution, which contains the product of a Gaussian norm (raised to a power) and a modified Bessel function of the second kind in place of the exponential function, the foregoing analysis does not apply.}

\section{\label{sect:ApxC} {Laplace Prior and the LASSO Algorithm}}

{The aim of this Appendix is to show that an {\it iid} Laplace prior distribution for the coefficients $\matparamc_{\IndexAlphabet \IndexDynSys}$ ($\IndexAlphabet=1,\cdots,\SizeAlphabet$) of the Bayesian MAP estimator \eqref{eq:MAP_log}, combined with a Gaussian noise distribution \eqref{eq:noise}, reduces to LASSO regularization.
}

{The univariate Laplace distribution is given by \eqref{eq:univar_Laplace}. Consider a zero-mean Laplace prior on all coefficients $\matparamc_{\IndexAlphabet \IndexDynSys}$. Assuming an {\it iid} set of coefficients $\matparamc_{\IndexAlphabet \IndexDynSys}$, the negative logarithm of the prior is given by:}
{\begin{align}
\begin{split}
- \ln p(\f) & = 
- \ln 
\left[
\prod_{\IndexAlphabet=1}^{\SizeAlphabet}
\frac{1}{2b}
\exp\left(-\frac{\lvert\matparamc_{\IndexAlphabet \IndexDynSys}\rvert}{b}\right)
\right]\\
& = 
\SizeAlphabet\ln\left(2b\right) + 
\frac{1}{b}
\sum_{\IndexAlphabet=1}^\SizeAlphabet\lvert\matparamc_{\IndexAlphabet \IndexDynSys}\rvert\\
& = 
\SizeAlphabet\ln\left(2b\right) + 
\frac{1}{b}
\lVert \f \rVert_1
\end{split}
\label{eq:Laplace_prior}
\end{align}
Replacing the contribution of the prior  in \eqref{eq:MAP_isotrop3} by \eqref{eq:Laplace_prior}, assuming isotropic Gaussian noise and constant parameter values, the minimization problem simplifies to:}
{
\begin{align} 
\begin{split}
\hatf^{\text{L}_1}
&=  \arg \min\limits_{\f} \, J^{\text{L}_1}(\f)\\
J^{\text{L}_1}(\f) &=   ||\g - \H(\statemat) \f ||^2_2  +  \theta_j ||\f ||_1    
\end{split}
\label{eq:MAP_LASSO}
\end{align}
which we identify as LASSO regression with residual term $||\g - \H(\statemat) \f ||^2_2$, regularization term $||\f ||_1$ and regularization coefficient $\theta_j =  {\sigma^2_{\noise}}/{b}$.}


\begin{thebibliography}{999}



\bibitem{Brunton_etal_2016} S.L. Brunton, J.L. Proctor \& J.N. Kutz, ``Discovering governing equations from data by sparse identification of nonlinear dynamical systems'', {\em PNAS} {\bf 113}(15), 3932-3937 (2016).

\bibitem{Schaeffer_2016} H. Schaeffer, ``Learning partial differential equations via data discovery and sparse optimization'', {\em Proc. R. Soc. A} {\bf 473}, 20160446 (2016).  

\bibitem{Mangan_etal_2017} N.M. Mangan, J.N. Kutz, S.L. Brunton \& J.L. Proctor, ``Model selection for dynamical systems via sparse regression and information criteria'', {\em Roy Soc Proc A} {\bf 473}, 20170009  (2017).

\bibitem{Rudy_etal_2017} S.H. Rudy, S.L. Brunton, J.L. Proctor \& J.N. Kutz, ``Data-driven discovery of partial differential equations'', {\em Sci. Adv.} {\bf 3}, e1602614 (2017). 

\bibitem{Kaiser_etal_2018} E. Kaiser, J.N. Kutz \& S.L. Brunton, ``Sparse identification of nonlinear dynamics for model predictive control in the low-data limit'', {\em Proc. R. Soc. A} {\bf 474}, 20180335  (2018). 

\bibitem{Quade_etal_2018} M. Quade, M. Abel, J.N. Kutz \& S.L. Brunton, ``Sparse identification of nonlinear dynamics for rapid model recovery'', {\em Chaos} {\bf 28}, 063116  (2018).

\bibitem{Mangan_etal_2019} N.M. Mangan, T. Askham, S.L. Brunton, J.N. Kutz \& J.L. Proctor, ``Model selection for hybrid dynamical systems via sparse regression'', {\em Proc. R. Soc. A} {\bf 475}, 20180534  (2019).

\bibitem{Champion_etal_2019} K. Champion, B. Lusch, J.N. Kutz \& S.L. Brunton, ``Data-driven discovery of coordinates and governing equations'', {\em PNAS} {\bf 116}(45), 22445-22451 (2019). 

\bibitem{Champion_etal_2020} K. Champion, P. Zheng, A.Y. Aravkin, S.L. Brunton \& J.N. Kutz,  ``A unified sparse optimization framework to learn parsimonious physics-informed models from data'', {\em IEEE Access} {\bf 8}, 169259-169271 (2020).

\bibitem{Kaheman etal_2020} K. Kaheman, J.N. Kutz \& S.L. Brunton, ``SINDy-PI: a robust algorithm for parallel implicit sparse identification of nonlinear dynamics'',  {\em Proc. R. Soc. A} {\bf 476}, 20200279  (2020).

\bibitem{Fasel_etal_2022} U. Fasel, J.N. Kutz, B.W. Brunton \& S.L. Brunton ``Ensemble-SINDy: Robust sparse model discovery in the low-data, high-noise limit, with active learning and control'',  {\em Proc. R. Soc. A} {\bf 478}, 20210904  (2022).

\bibitem{Lai_2021} {Y.-C. Lai,``Finding nonlinear system equations and complex network structures from data: A sparse optimization approach'', {\em Chaos: An Interdisciplinary Journal of Nonlinear Science} {\bf 31}, 082101 (2021).} 


\bibitem{Bayes_1763}
{T. Bayes (presented by R. Price)}, ``An essay towards solving a problem in the doctrine of chance'', {\em Phil. Trans. Royal Soc. London} {\bf 53}, 370-418 (1763). 

\bibitem{Laplace_1774}
{P. Laplace}, ``M\'emoire sur la probabilit\'e des causes par les \'ev\`enements'', {\em l'Acad\'emie Royale des Sciences}, {\bf 6}, 621-656 (1774). 

\bibitem{Jaynes_2003}
{E.T. Jaynes (G.L. Bretthorst, ed.),} {\em Probability Theory: The Logic of Science} (Cambridge Univ. Press, Cambridge, UK, 2003).

\bibitem{vonToussaint_2011} U. von Toussaint, ``Bayesian inference in physics'', {\em Rev. Mod. Phys.} {\bf 83}(3),  943-999 (2011).

\bibitem{Polya_1954} {G. Polya}, {\em Mathematics and Plausible Reasoning, Vol II, Patterns of Plausible Inference}, (Princeton Univ. Press, Princeton, NJ, USA, 1954).

\bibitem{Cox_1961} {R.T. Cox}, {\em The Algebra of Probable Inference} (John Hopkins Press, Baltimore, MD, USA, 1961).


\bibitem{Venn_1888} J. Venn, {\em The Logic of Chance} (Macmillan \& Co, London, UK, 1888).

\bibitem{Fisher_1925} R.A. Fisher, {\em Statistical Methods for Research Workers} (Oliver and Boyd, Edinburgh, UK, 1925).

\bibitem{Neyman_Pearson_1933}  J. Neyman \& E.S. Pearson, ``On the problem of the most efficient tests of statistical hypotheses'', {\em Phil. Trans. R. Soc. Lond. A} {\bf 231}(694-706), 289-337 (1933). 

\bibitem{Feller_1966} {W. Feller}, {\em An Introduction to Probability Theory and Its Applications}, Volume II (John Wiley and Sons, Inc., NY, USA, 1966).


\bibitem{MD_2015} A. Mohammad-Djafari, ``Inverse problems in signal and image processing and
Bayesian inference framework: from basic to advanced Bayesian computation'', Scube seminar, L2S, CentraleSup\'elec, Gif-sur-Yvette, France (2015).

\bibitem{MD_Dimutru_2015} A. Mohammad-Djafari \& M. Dumitru, ``Bayesian sparse solutions to linear inverse problems with non-stationary noise with Student-t priors'', {\em Digital Signal Proc.} {\bf 47}, 128-156 (2015).

\bibitem{MD_2016} A. Mohammad-Djafari, ``Approximate Bayesian computation for big data'', Tutorial, MaxEnt 2016, July 10-15, 2016, Ghent, Belgium (2016).

\bibitem{Teckentrup_MaxEnt2018} A. Teckentrup, ``Introduction to the Bayesian approach to inverse problems'', presentation, MaxEnt 2018, July 6, 2018, London, UK (2018).

\bibitem{Calvetti_Somersalo_2018} D. Calvetti \& E. Somersalo, ``Inverse problems: From regularization to Bayesian inference'', {\em Wiley Interdisc. Rev.: Comp. Stat.} {\bf 10}(2), e1427 (2018). 


\bibitem{Pan_etal_2016} W. Pan, Y. Yuan, J. Goncalves \& G-.B. Stan, ``A sparse Bayesian approach to the identification of nonlinear state- space systems'', {\em IEEE Trans. Automatic Control}, {\bf 61}(1), 182-187 (2016).

\bibitem{Zhang_Lin_2018} S. Zhang \& G. Lin, ``Robust data-driven discovery of governing physical laws with error bars'', {\em Proc. Royal Society A: Math. Phys. Eng. Sci.}, {\bf 474}(2217), 20180305 (2018).

\bibitem{Niven_etal_MaxEnt_2019} R.K. Niven, A. Mohammad-Djafari, L. Cordier, M. Abel \& M. Quade, ``Bayesian identification of dynamical systems'', {\em MDPI Proc.} {\bf 33}(1), 33 (2019).

\bibitem{Chiuso_Pillonetto_2019} A. Chiuso \& G. Pillonetto, ``System identification: a machine learning perspective'', {\em Annual Review of Control, Robotics, and Autonomous Systems} {\bf 2}, 281-304  (2019). 

\bibitem{Chen_Lin_2021} A. Chen \& G. Lin, ``Robust data-driven discovery of partial differential equations with time-dependent coefficients'', {\em arXiv}:2102.01432v1 (2021).  

\bibitem{Hirsch_etal_2022} Hirsh, S.M., Barajas-Solano, D.A. \& Kutz, J.N., ``Sparsifying priors for Bayesian uncertainty quantification in model discovery'', {\em R. Soc. Open Sci.} {\bf 9}, 211823 (2022). 

\bibitem{Yuan_etal_2023} Y. Yuan, X. Li, L. Li, F.J. Jiang, X. Tang, F. Zhang, J. Goncalves, H.U. Voss, H. Ding \& J. Kurths, ``Machine discovery of partial differential equations from spatiotemporal data: A sparse Bayesian learning framework'', {\em Chaos} {\bf 33}, 113122 (2023). 

\bibitem{Lin_etal_2023} S. Lin, G. Mengaldo, R. Maulik, ``Online data-driven changepoint detection for high-dimensional dynamical systems'',  {\em Chaos} {\bf 33}, 103112 (2023).

\bibitem{Zhang_etal_2023} Z. Zhang, Q. Shen, X. Wang, ``Parameter identification framework of nonlinear dynamical systems with Markovian switching'', {\em Chaos} {\bf 33}, 123117 (2023).

\bibitem{Taghavi_2024} N. Taghavi, {\em Developing a Geospatial Bayesian Probabilistic Method for Groundwater Vulnerability Assessment}, PhD Thesis, The University of New South Wales, Canberra, ACT, Australia (unpub.) (2024). 

\bibitem{Fung_etal_2024} {L. Fung, U. Fasel \& M. Juniper, Rapid Bayesian identification of sparse nonlinear dynamics from scarce and noisy data, {\em arXiv}:2402.15357v1 (2024).}

\bibitem{Klishin_etal_2024} {A.A. Klishin, J. Bakarji, J.N. Kutz \& K. Manohar, Statistical mechanics of dynamical dystem identification, {\em arXiv}:2403.01723v1 (2024)}


\bibitem{Roberts_etal_2013} S. Roberts, M. Osborne, M. Ebden, S. Reece, N. Gibson \& S. Aigrain, ``Gaussian processes for time-series modelling'', {\em Phil. Trans. R. Soc. A} {\bf 371}, 20110550  (2013).

\bibitem{Svensson_Schon_2017} A. Svensson \& T.B. Sch\"on, ``A flexible state-space model for learning nonlinear dynamical systems'', {\em Automatica} {\bf 80}, 189-199 (2017). 

\bibitem{Raissi_Karniadakis_2018} M. Raissi \& G.E. Karniadakis, ``Hidden physics models: Machine learning of nonlinear partial differential equations'', {\em J. Comp. Phys.} {\bf 357},125-141  (2018). 

\bibitem{Teckentrup_2020} A.L. Teckentrup, ``Convergence of Gaussian process regression with estimated hyper-parameters and applications in Bayesian inverse problems'', {\em SIAM/ASA J. Unc. Quant.} {\bf 8}(4), 1310-1337 (2020). 


\bibitem{Hirsh_etal_2022} S.M. Hirsh, D.A. Barajas-Solano \& J.N. Kutz, ``Sparsifying priors for Bayesian uncertainty quantification in model discovery'', {\em R. Soc. Open Sci.} {\bf 9}, 211823  (2022).

\bibitem{Rinkens_etal_2023} A. Rinkens, C.V. Verhoosel \& N.O. Jaensson, ``Uncertainty quantification for the squeeze flow of generalized Newtonian fluids'', {\em J. Non-Newtonian Fluid Mech.}, {\bf 322}, 105154  (2023). 


\bibitem{Kontogiannis_etal_2022} A. Kontogiannis, S.V. Elgersma, A.J. Sederman \& M.P. Juniper, ``Joint reconstruction and segmentation of noisy velocity images as an inverse Navier–Stokes problem'', {\em J. Fluid Mech.} {\bf 944}, A40  (2022).



\bibitem{Aster_etal_2013} R.C. Aster, B. Borchers \& C.H. Thurner, {\em Parameter Estimation and Inverse Problems}, 2nd ed. (Elsevier, Amsterdam, Netherlands, 2013).  

\bibitem{Bardsley_2018} J.M. Bardsley, {\em Computational Uncertainty Quantification for Inverse Problems} (SIAM,  Philadelphia, PA, USA, 2018).


\bibitem{Tikhonov_1963} A.N. Tikhonov, ``Solution of incorrectly formulated problems and the regularization method'', {\em Doklady Akademii Nauk SSSR} {\bf 151}, 501-504 (1963).

\bibitem{Hoerl_Kennard_1970} A.E. Hoerl \& R.W. Kennard, ``Ridge regression: Biased estimation for nonorthogonal problems'', {\em Technometrics} {\bf 12}(1), 55–67 (1970).

\bibitem{Kaipio_Somersalo_2005} {K. Kaipio \& E. Somersalo, {\em Statistical and Computational Inverse Problems}, (Springer Science \& Business Media, Springer-Verlag, New York, NY, USA), 2005.} 


\bibitem{Santosa_Symes_1986} F. Santosa \& W.W. Symes, ``Linear inversion of band-limited reflection seismograms'', {\em SIAM J. Sci. Stat. Comp.} {\bf 7}(4), 1307-1330  (1986). 

\bibitem{Tibshirani_1996} R. Tibshirani, ``Regression shrinkage and selection via the Lasso'', {\em J Royal Stat. Soc. B} {\bf 58}(1), 267-288 (1996).


\bibitem{Stark_1999} {J. Stark, ``Delay embeddings for forced systems. I. Deterministic forcing'', {\em J. Nonlinear Sci.},  {\bf 9}, 255-332 (1999).}

\bibitem{Sauer_2004} {T.D. Sauer, ``Reconstruction of shared nonlinear dynamics in a network'', {\em Phys. Rev. Lett.} {\bf 93}, 198701 (2004). }

\bibitem{Wang_etal_2011} {W.-X. Wang, R. Yang, Y.-C. Lai, V. Kovanis \& C. Grebogi, ``Predicting catastrophes in nonlinear dynamical systems by compressive sensing," {\em Phys. Rev. Lett.}, {\bf 106}, 154101 (2011). }

\bibitem{Yao_Bollt_2006} {C. Yao \& E. M. Bollt, ``Modeling and nonlinear parameter estimation with Kronecker product representation for coupled oscillators and spatiotemporal systems," {\em Physica D}, {\bf 227}, 78 (2007). }


\bibitem{Brunton_etal_2016_PLOS} S.L. Brunton, B.W. Brunton, J.L. Proctor, E. Kaiser \& J.N. Kutz, ``Koopman invariant subspaces and finite linear representations of nonlinear dynamical systems for control'', {\em PLOS One}, {\bf 11}(2), e0150171 (2016).

\bibitem{Brunton_etal_2017_NatureComm} S.L. Brunton, B.W. Brunton, J.L. Proctor, E. Kaiser \& J.N. Kutz, ``Chaos as an intermittently forced linear system'', {\em Nature Comm} {\bf 8}, 1  (2017).

\bibitem{Taira_etal_2017} K. Taira, S.L. Brunton, S.T.M. Dawson, C.W. Rowley, T. Colonius, B.J. McKeon, O.T. Schmidt, S. Gordeyev, V. Theofilis \& L.S. Ukeiley, ``Modal analysis of fluid flows: an overview'', {\em AIAA Journal} {\bf 55}(12), 4013-4041  (2017).


\bibitem{Zhang_Schaeffer_2018} L. Zhang \& H. Schaeffer, ``On the convergence of the SINDy algorithm'', {\em 
Multiscale Model. Simul.} {\bf 17}(3), 948-972 (2019).  

\bibitem{Tarantola_2005} A. Tarantola, {\em Inverse Problem Theory and Methods for Model Parameter Estimation} (SIAM, Philadelphia, PA, USA, 2005).

\bibitem{Bishop_2006} C.M. Bishop, {\em Pattern Recognition and Machine Learning} (Springer, MY, USA, 2006).


\bibitem{Hoff_2009} P.D. Hoff, {\em A First Course in Bayesian Statistical Methods} (Springer, Dordrecht, Germany, 2009). 

\bibitem{Tenorio_2017} L. Tenorio, {\em An Introduction to Data Analysis and Uncertainty Quantification for Inverse Problems} (SIAM, Philadelphia, PA, USA, 2017).

\bibitem{Bontekoe_2023} R. Bontekoe, {\em What is Your Model?: A Bayesian Tutorial} (Bontekoe Research, Amsterdam, Netherlands, 2023). 


\bibitem{Burnham_Anderson_2002} K. Burnham \& D. Anderson, {\em Model selection and multi-model inference}, 2nd ed. (Springer, Berlin, Germany, 2002).


\bibitem{Skilling_2004} J. Skilling, ``Nested Sampling'', {\em AIP Conf. Proc.} {\bf 735}, 395-405 (2004). 




\bibitem{Marin_Robert_2007} J.-M. Marin \& C. Robert, {\em Bayesian Core: A Practical Approach to Computational Bayesian Statistics} (Springer, New York, NY, USA, 2007). 

\bibitem{Clyde_etal_2015} M. Clyde, M. \c{C}etinkaya-Rundel, C. Rundel, D. Banks, C. Chai \& L. Huang, {\em An Introduction to Bayesian Thinking. A Companion to the Statistics with R Course}, https://statswithr.github.io/book/\_main.pdf, retrieved Dec 2023. 


\bibitem{Lorenz_1963} E.N. Lorenz, ``Deterministic nonperiodic flow'', {\em J. Atmos. Sci.} {\bf 20}(2), 130-141  (1963).

\bibitem{Lynch_2004} S. Lynch, {\em Dynamical Systems with Applications using MATLAB} (Birkh\"auserBoston, Springer, New York, NY, USA, 2004).

\bibitem{Vance_1978} R.R. Vance, ``Predation and resource partitioning in one predator-two prey model community'', {\em The American Naturalist}, {\bf 112}, 797-813 (1978).

\bibitem{Gilpin_1979} M.E. Gilpin, ``Spiral chaos in a predator-prey model'', {\em The American Naturalist} {\bf 113}(2), 306-308  (1979). 

\bibitem{Shilnikov_etal_1993} A.L. Shil'nikov, L.P. Shil'nikov \& D.V. Turaev, ``Normal forms and Lorenz attractors'', {\em Int. J. Bifurc. Chaos} {\bf 3}(5), 1123-1139 (1993).




\bibitem{Lin_2024} D. Lin, ``Safe computation of logarithm-determinant of large matrix'', MATLAB Central File Exchange, https://www.mathworks.com/matlabcentral/fileexchange/22026-safe-computation-of-logarithm-determinat-of-large-matrix, retrieved January 19, 2024.



\bibitem{Pearl_1988} J. Pearl, {\em Probabilistic Reasoning in Intelligent Systems: Networks of Plausible Inference} (Morgan Kaufmann, San Francisco, CA, USA, 1988).

\bibitem{Psorakis_etal_2011} I. Psorakis, S. Roberts, M. Ebden \& B. Sheldon, ``Overlapping community detection using Bayesian non-negative matrix factorization'', {\em Phys. Rev. E} {\bf 83}(6-2), 066114  (2011). 






\bibitem{Hasan_Baizid_2016} Hasan, M.R. \& Baizid, A.R., ``Bayesian estimation under different loss functions using gamma prior for the case of exponential distribution'', {\em J. Sci. Res.} {\bf 9}(1), 67-78 (2016).


\end{thebibliography}


\section*{Supplementary Material}

\setcounter{figure}{0}    
\renewcommand\thefigure{S\arabic{figure}}    

\begin{figure*}[h]
\begin{center}
\setlength{\unitlength}{0.6pt}
 \begin{picture}(800,600)
  \put(0,300){\includegraphics[height=60mm]{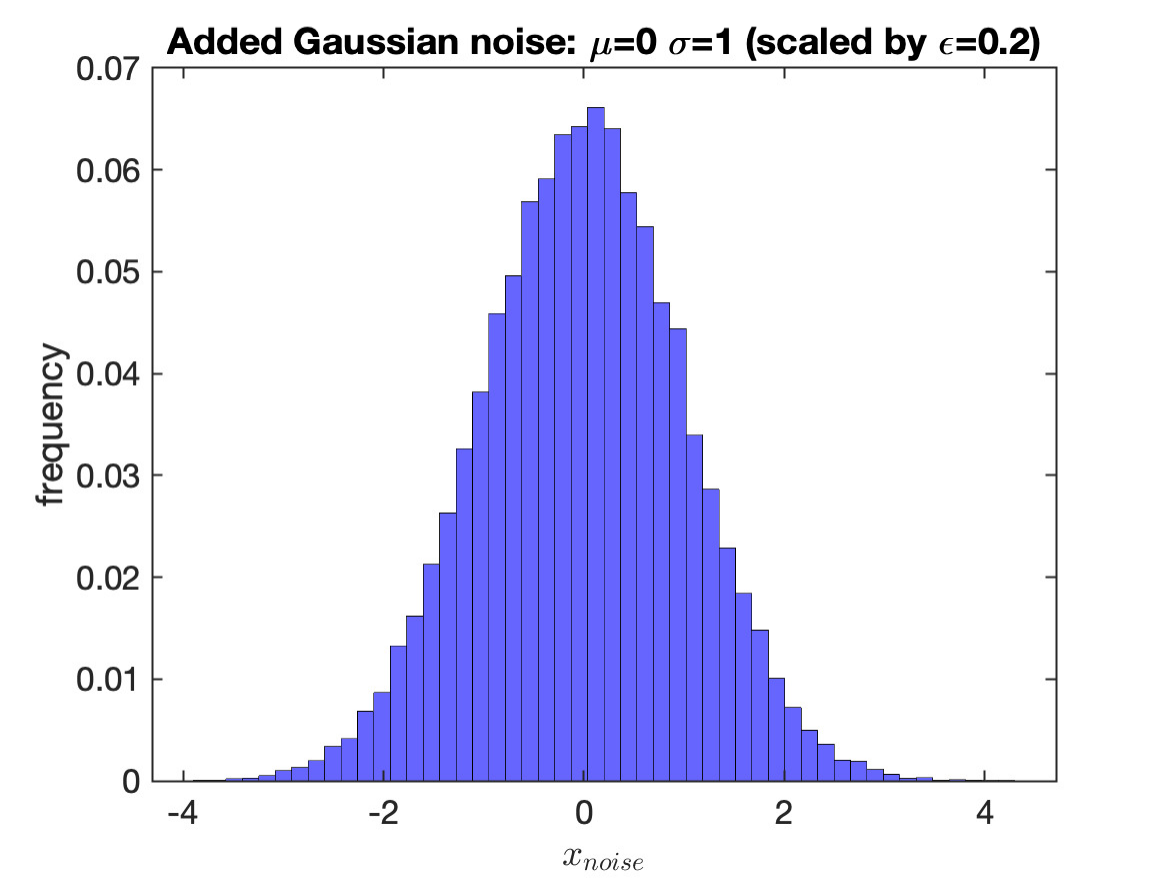} }
  \put(0,300){\small (a)}
  \put(378,300){\includegraphics[height=63mm]{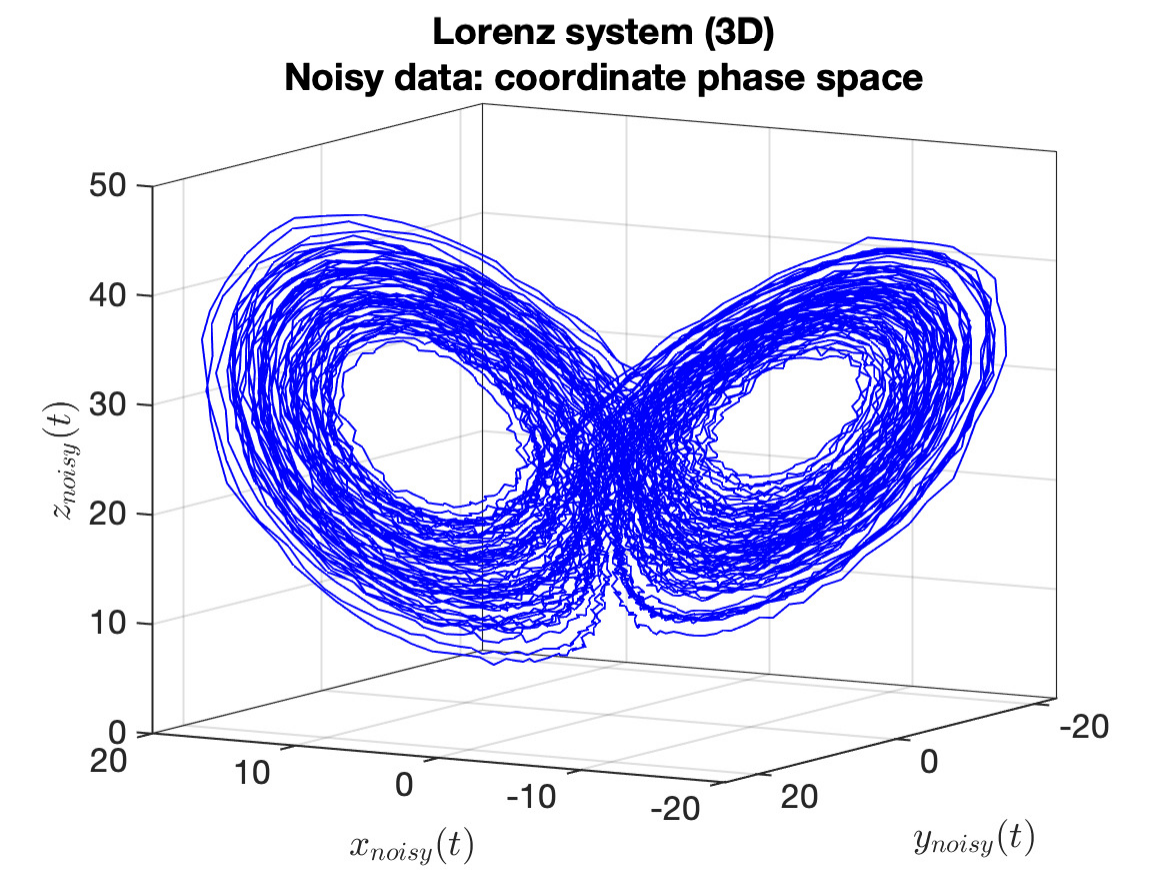} }
  \put(390,300){\small (b)}
  \put(0,0){\includegraphics[height=60mm]{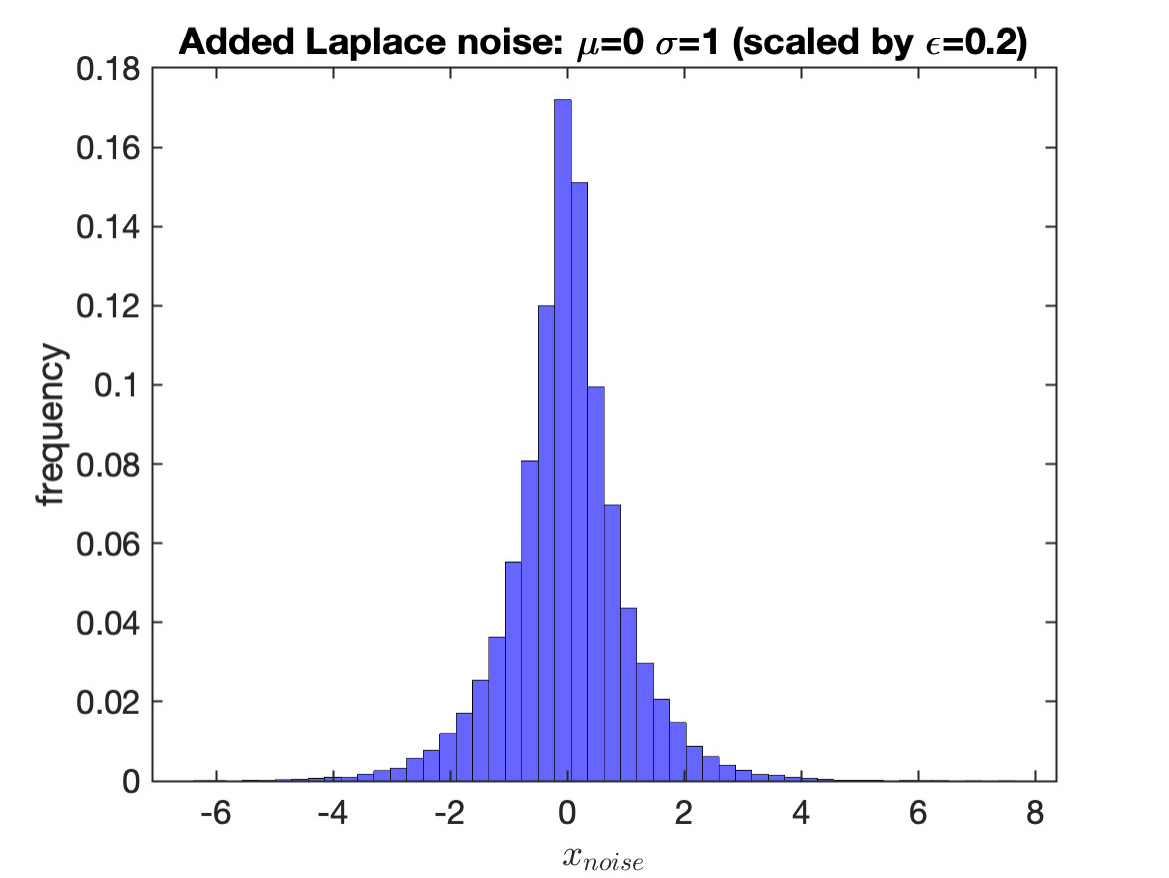} }
  \put(0,0){\small (c)}
  \put(390,0){\includegraphics[height=60mm]{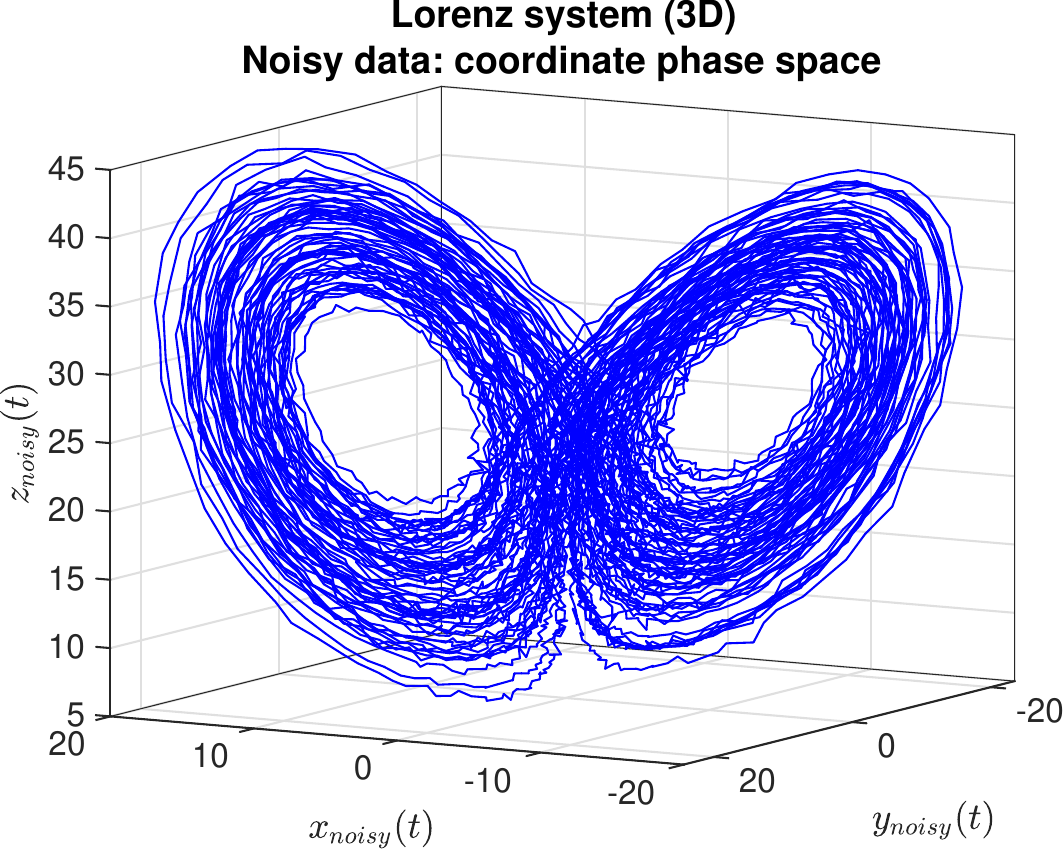} }
  \put(390,0){\small (d)}
 \end{picture}
\end{center}
\caption{{Noise models: (a) sample of univariate Gaussian noise with $\mu=0$, $\sigma=1$, (b) Lorenz system with added Gaussian noise, $\varepsilon = 0.2$; (c) sample of univariate Laplace noise with $\mu=0$, $\sigma=1$; (d) Lorenz system with added Laplace noise, $\varepsilon = 0.2$.}}
\label{fig:SI_noise}
\end{figure*}

\begin{figure*}[h]
\begin{center}
\setlength{\unitlength}{0.6pt}
 \begin{picture}(800,580)
  \put(0,300){\includegraphics[height=60mm]{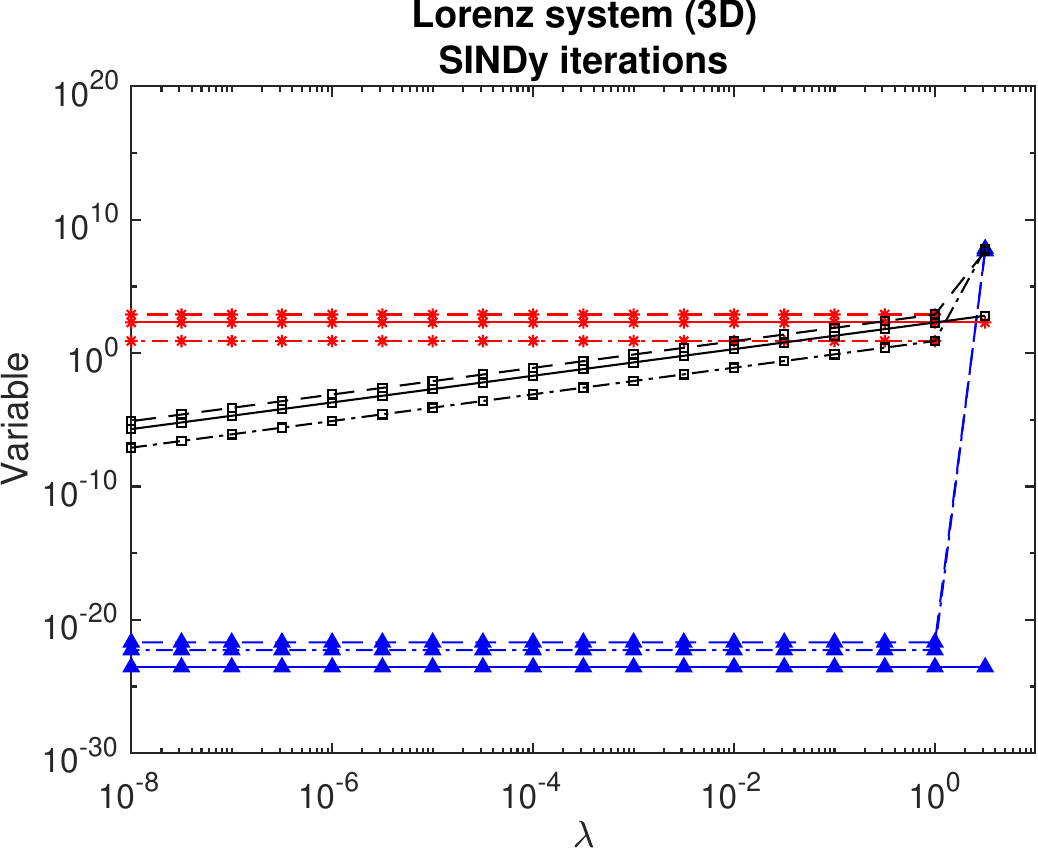} }
  \put(55,505){\includegraphics[height=9mm]{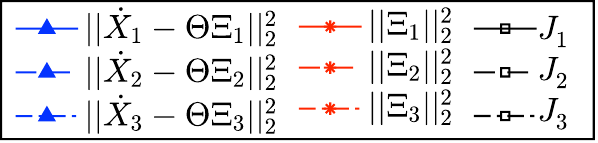} }
  \put(300,350){\includegraphics[height=6mm]{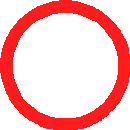} }
  \put(0,300){\small (a)}
  \put(370,300){\includegraphics[height=60mm]{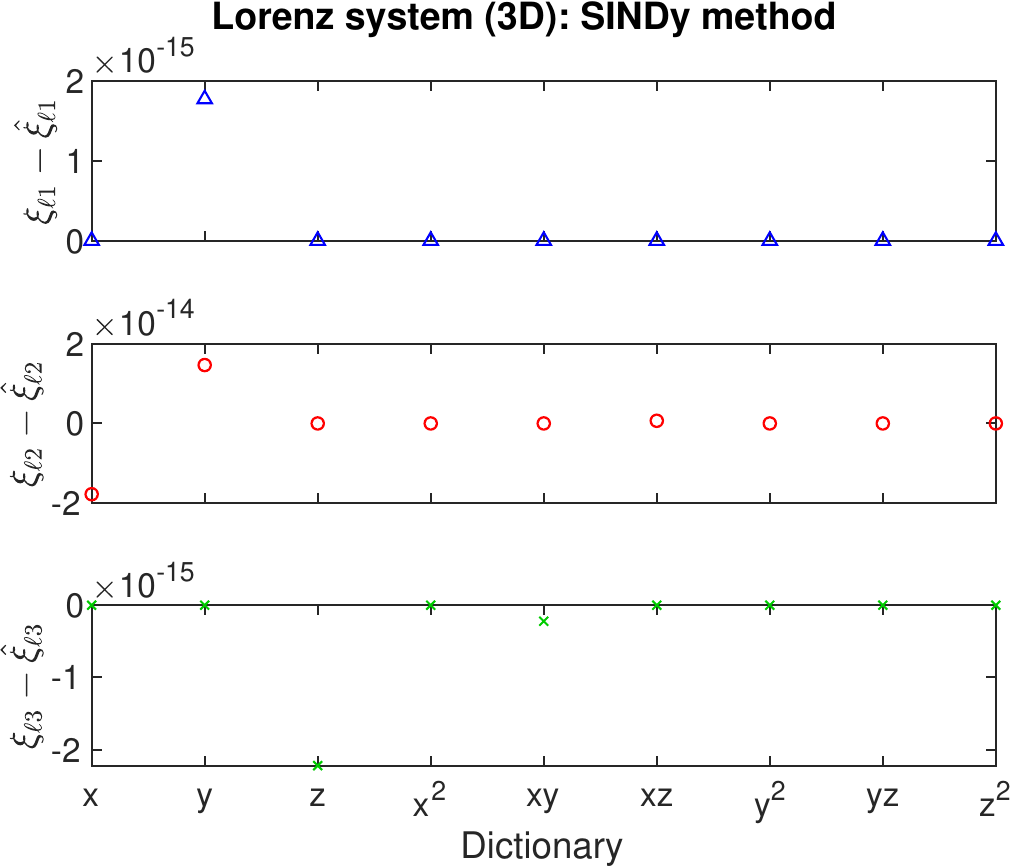} }
  \put(370,300){\small (b)}
  \put(0,0){\includegraphics[height=60mm]{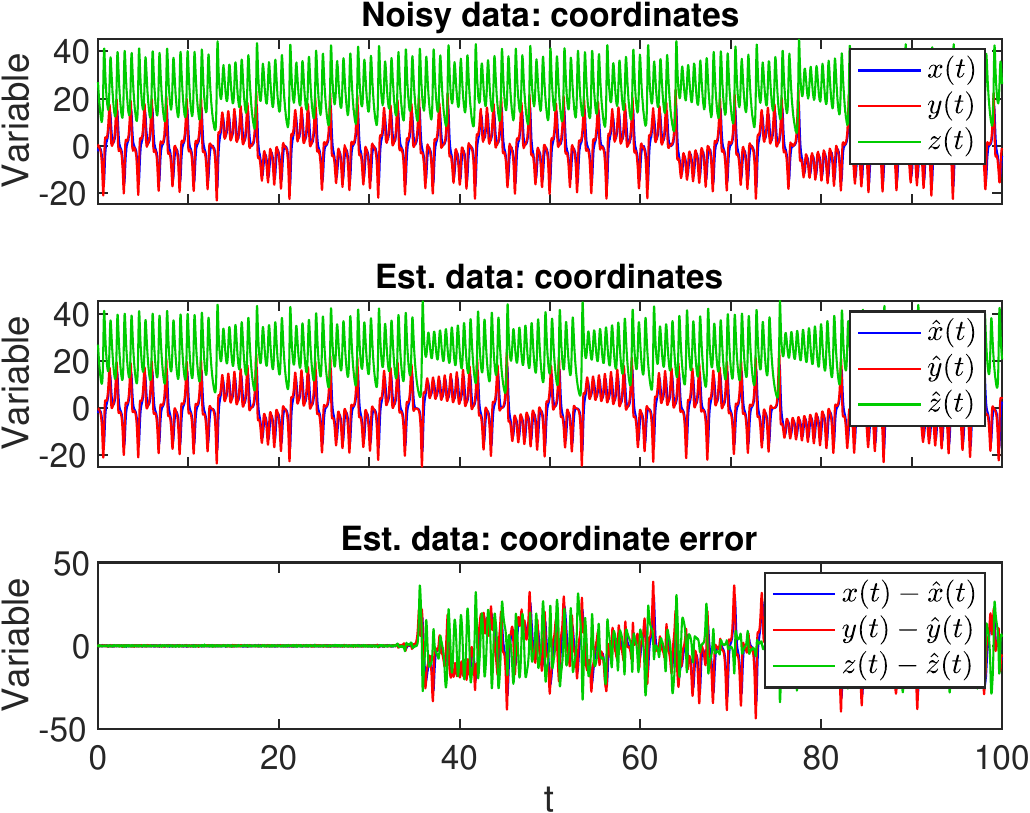} }
  \put(0,0){\small (c)}
 \end{picture}
\end{center}
\caption{Lorenz {system with added Gaussian noise}: second-order polynomial regularization using SINDy, $T=100$, $t_{step}=0.01$, $\varepsilon=0.2$: 
(a) plots of 2-norm residual, regularization and objective functions \eqref{eq:modJ} 
with decreasing $\lambda$, showing the optimal iteration ($k=3)$; 
(b) optimal error in predicted coefficients $\matparamc_{ij}-\hat{\matparamc}_{ij}$; and
(c) original and optimal predicted data and their differences.}
\label{fig:SI_Lorenz_SINDy}
\end{figure*}

\begin{figure*}[h]
\begin{center}
\setlength{\unitlength}{0.6pt}
 \begin{picture}(800,580)
 %
  \put(0,300){\includegraphics[height=60mm]{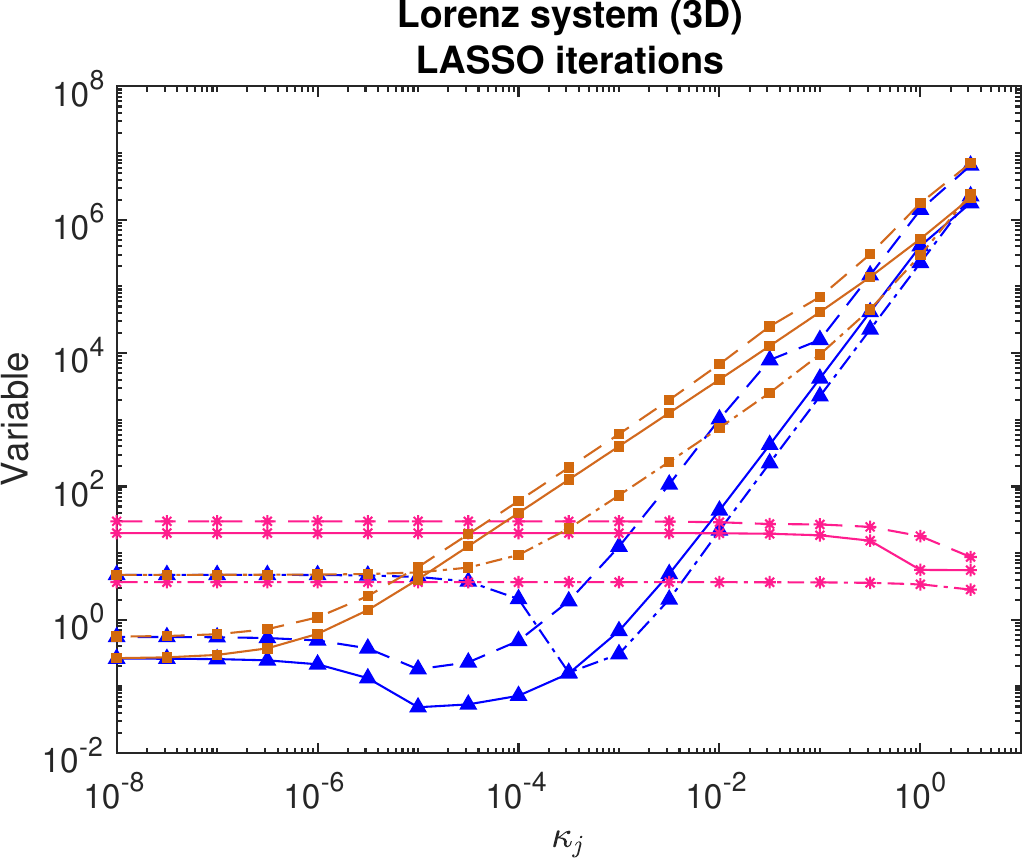} }
  \put(55,505){\includegraphics[height=9mm]{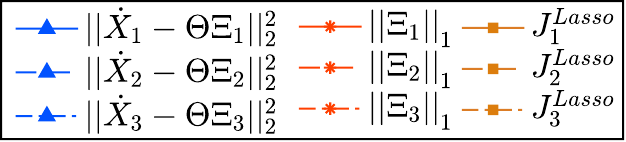} }
  \put(58,355){\includegraphics[height=6mm]{figs_suppl/figleg/circle} }
  \put(0,300){\small (a)}
  \put(370,300){\includegraphics[height=60mm]{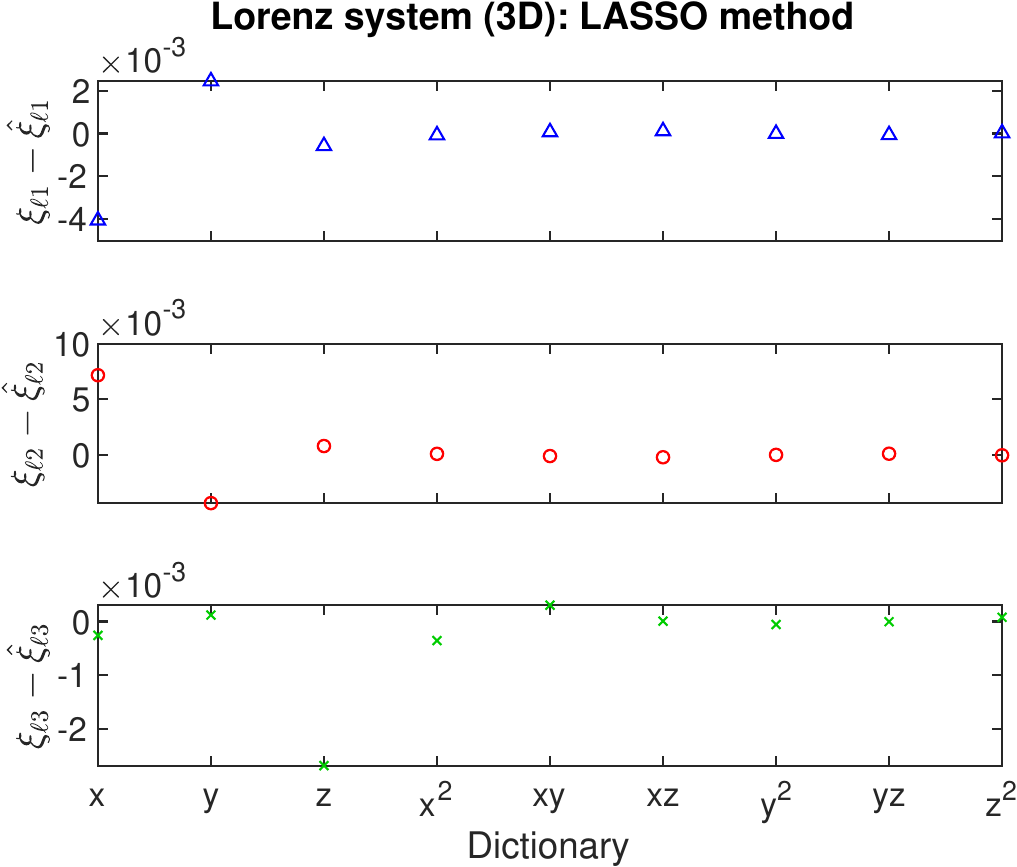} } 
  \put(370,300){\small (b)}
  \put(0,0){\includegraphics[height=60mm]{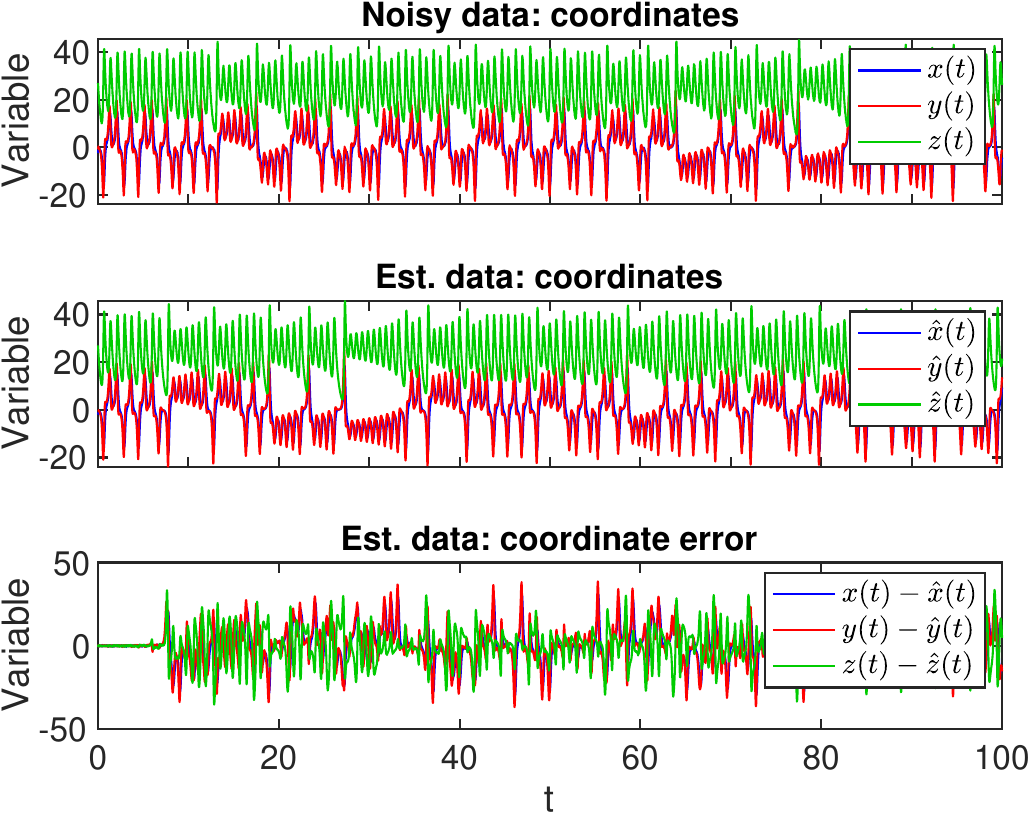} } 
  \put(0,0){\small (c)}
 \end{picture}
\end{center}
\caption{Lorenz {system with added Gaussian noise}: second-order polynomial regularization using LASSO, $T=100$, $t_{step}=0.01$, $\varepsilon=0.2$: 
(a) plots of 2-norm residual, 1-norm regularization terms and total objective functions with decreasing $\kappa_j$ in \eqref{eq:modJ}, showing the optimal iteration in the objective function ($k=17)$; 
(b) optimal error in predicted coefficients $\matparamc_{ij}-\hat{\matparamc}_{ij}$; and
(c) original and optimal predicted data and their differences.}
\label{fig:SI_Lorenz_LASSO}
\end{figure*}

\begin{figure*}[h]
\begin{center}
\setlength{\unitlength}{0.6pt}
 \begin{picture}(800,580)
  \put(0,300){\includegraphics[height=60mm]{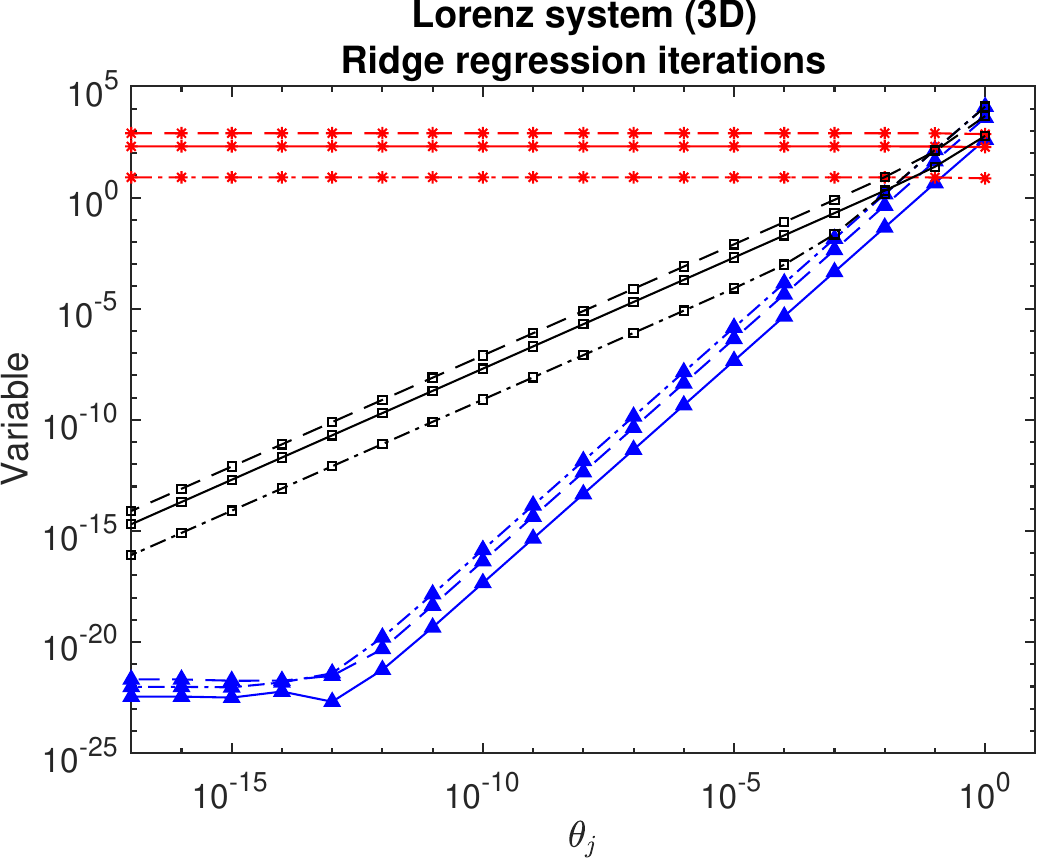} }
  \put(250,340){\includegraphics[height=25mm]{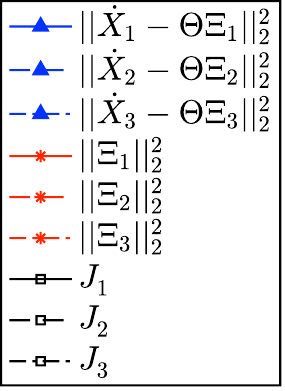} }
  \put(79,343){\includegraphics[height=6mm]{figs_suppl/figleg/circle} }
  \put(0,300){\small (a)}
  \put(370,300){\includegraphics[height=60mm]{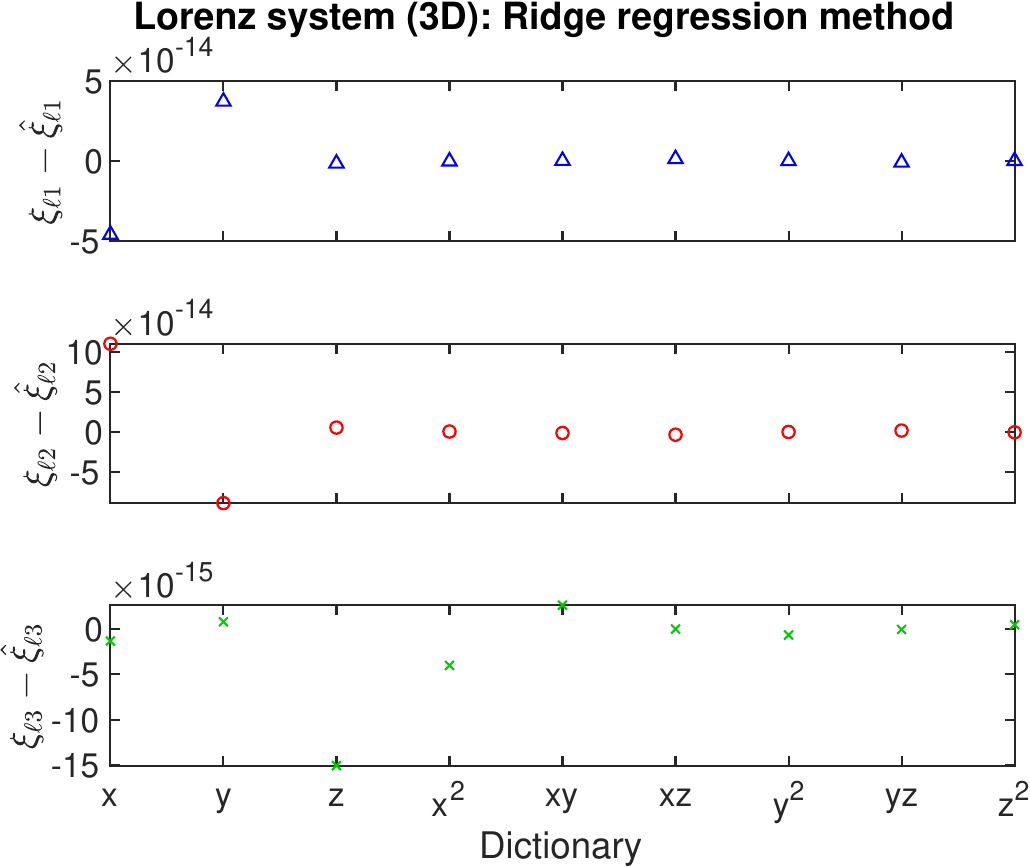} } 
  \put(370,300){\small (b)}
  \put(0,0){\includegraphics[height=60mm]{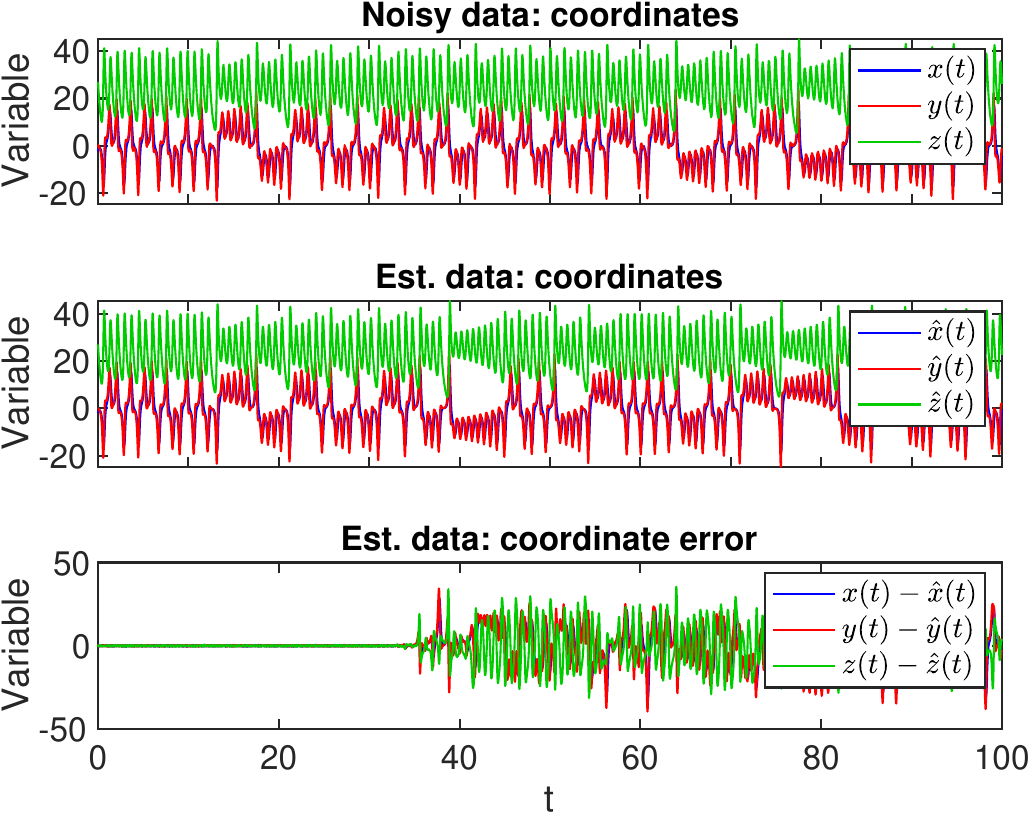} } 
  \put(0,0){\small (c)}
 \end{picture}
\end{center}
\caption{Lorenz {system with added Gaussian noise}: second-order polynomial regularization using ridge regression, $T=100$, $t_{step}=0.01$, $\varepsilon=0.2$: 
(a) plots of 2-norm residual, regularization and objective functions \eqref{eq:modJ} 
with decreasing regularization parameter $\theta_j$ in \eqref{eq:modJ}, showing the optimal iteration ($k=16)$; 
(b) optimal error in predicted coefficients $\matparamc_{ij}-\hat{\matparamc}_{ij}$; and
(c) original and optimal predicted data and their differences.}
\label{fig:SI_Lorenz_ridge}
\end{figure*}

\begin{figure*}[h]
\begin{center}
\setlength{\unitlength}{0.6pt}
 \begin{picture}(800,580)
  \put(0,300){\includegraphics[height=60mm]{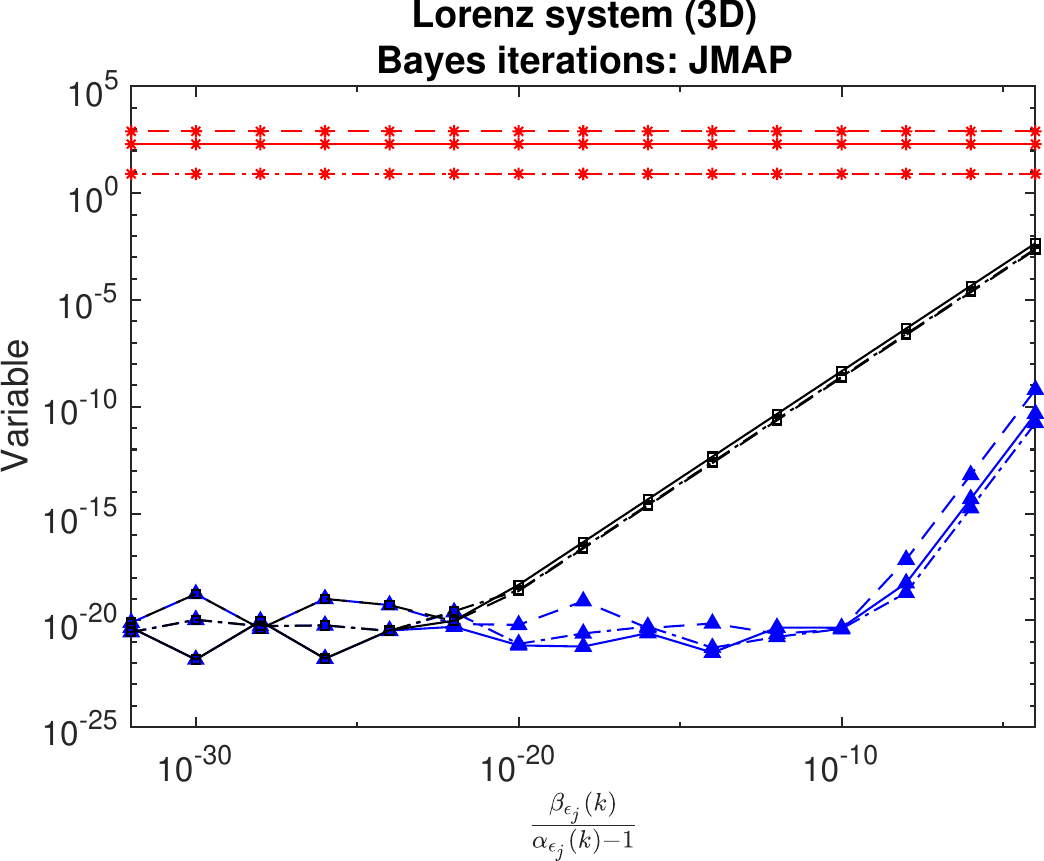} }
  \put(55,400){\includegraphics[height=25mm]{figs_suppl/figleg/Legend_2norms} }
  \put(265,362){\includegraphics[height=6mm]{figs_suppl/figleg/circle} }
  \put(0,300){\small (a)}
  \put(370,300){\includegraphics[height=60mm]{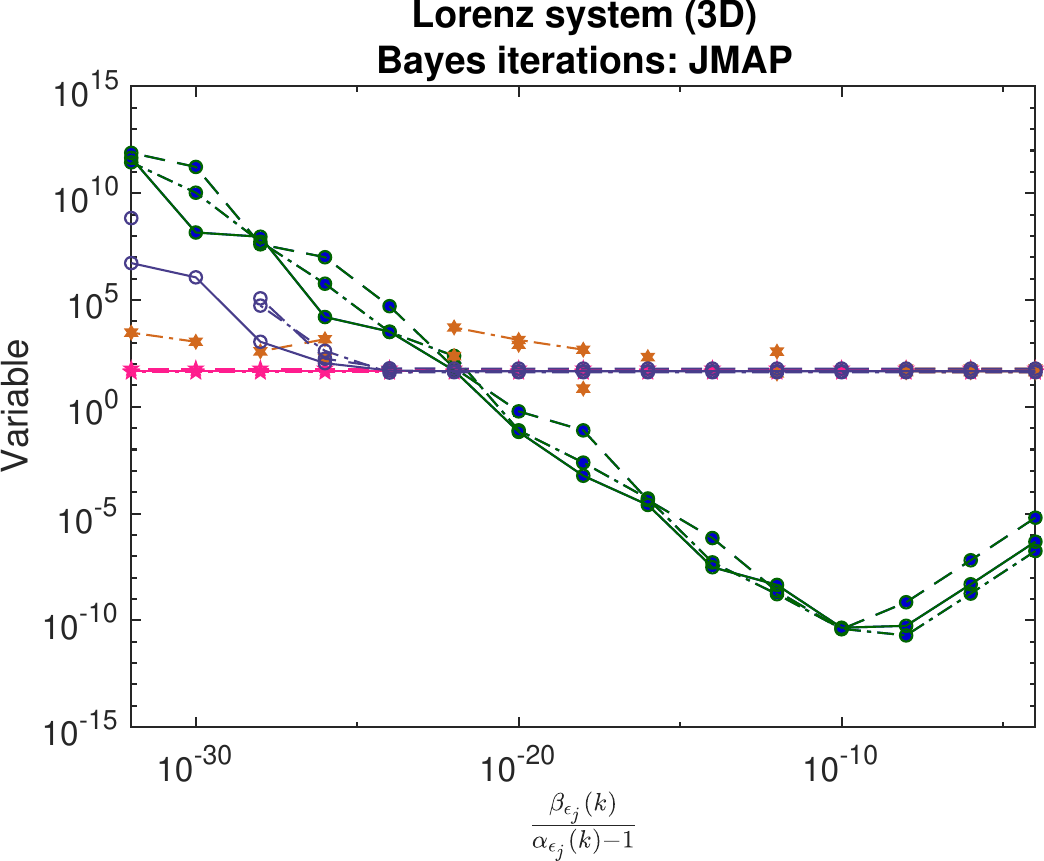} }
  \put(720,362){\includegraphics[height=41mm]{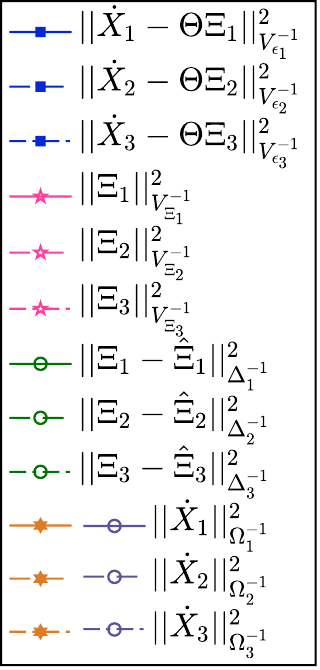} }
  \put(635,362){\includegraphics[height=6mm]{figs_suppl/figleg/circle} }
  \put(370,300){\small (b)}
  \put(0,0){\includegraphics[height=60mm]{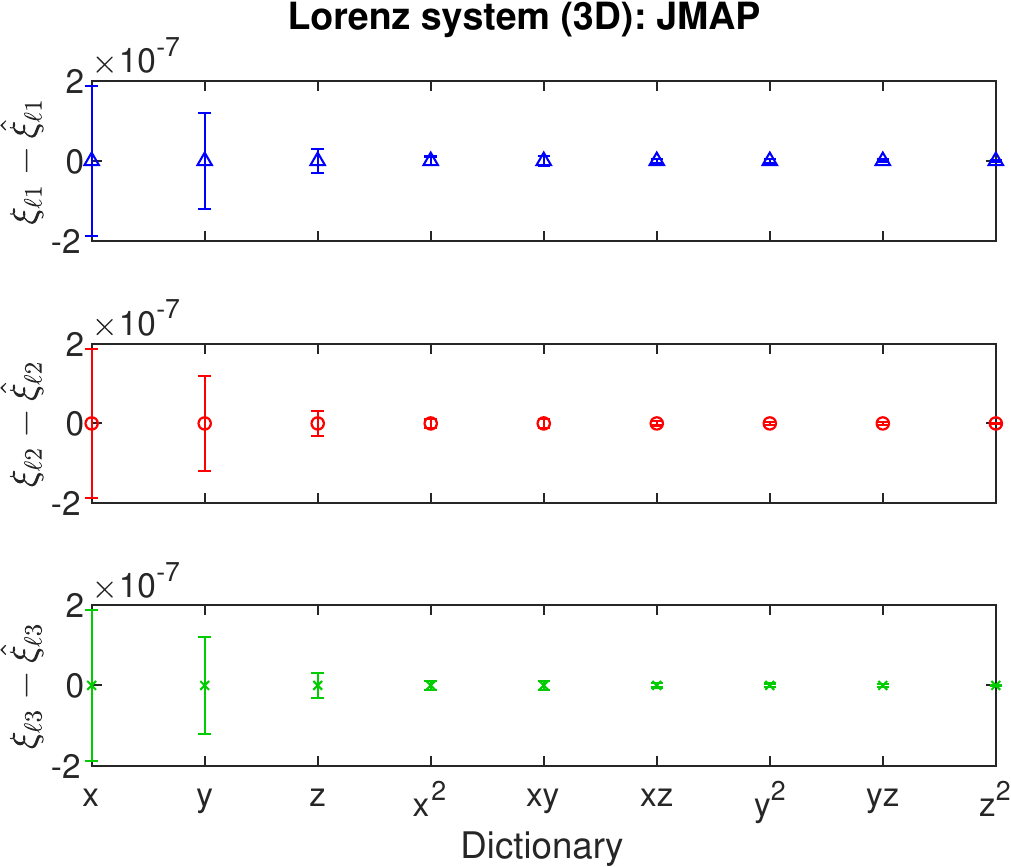} } 
  \put(0,0){\small (c)}
  \put(370,0){\includegraphics[height=60mm]{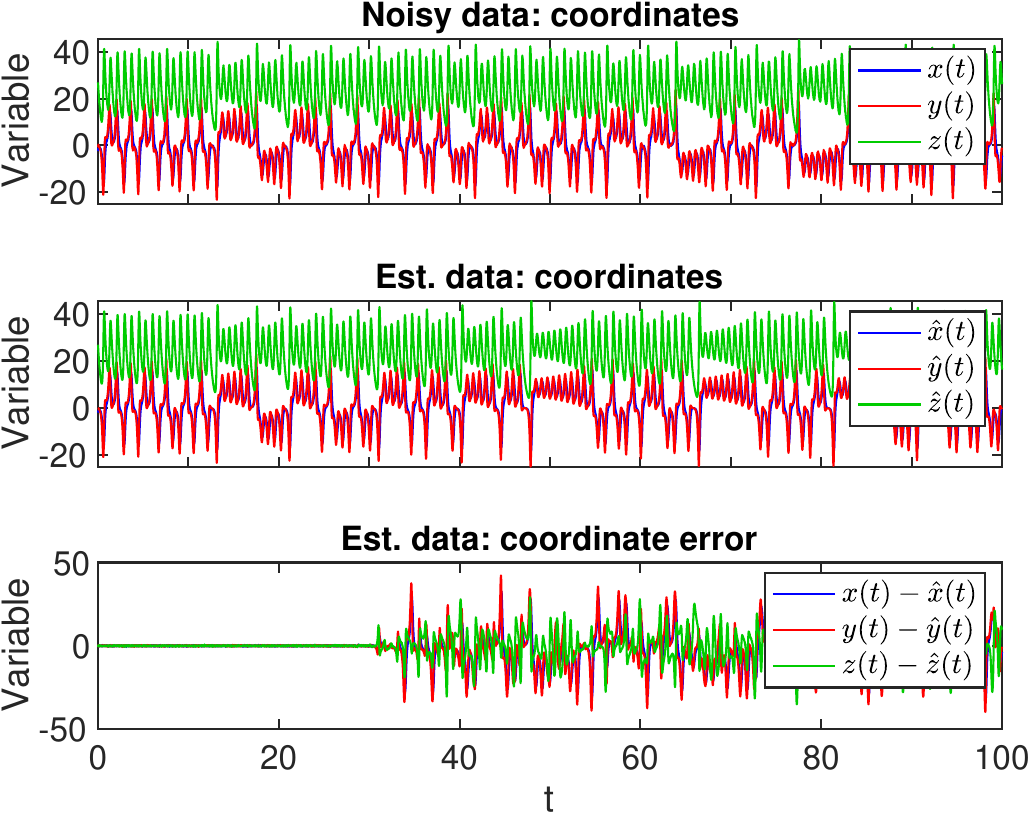} } 
  \put(370,0){\small (d)}
 \end{picture}
\end{center}
\caption{Lorenz {system with added Gaussian noise}: second-order polynomial regularization using JMAP, $T=100$, $t_{step}=0.01$, $\varepsilon=0.2$, showing the iteration sequence with decreasing $\mathsf{E}_{\eizero}$, including 
(a) 2-norm residual, regularization and objective functions \eqref{eq:modJ}, showing the optimal iteration ($k=5$); 
(b) Gaussian norms for the prior, likelihood, posterior and evidence, showing the optimal iteration ($k=5$); 
(c) optimal error in predicted coefficients $\matparamc_{ij}-\hat{\matparamc}_{ij}$, with error bars from the posterior covariance \eqref{eq:posterior_estimators}; and
(d) original and optimal predicted data and their differences.}
\label{fig:SI_Lorenz_Gauss_JMAP}
\end{figure*}

\begin{figure*}[h]
\begin{center}
\setlength{\unitlength}{0.6pt}
 \begin{picture}(800,300)
  \put(0,0){\includegraphics[height=65mm]{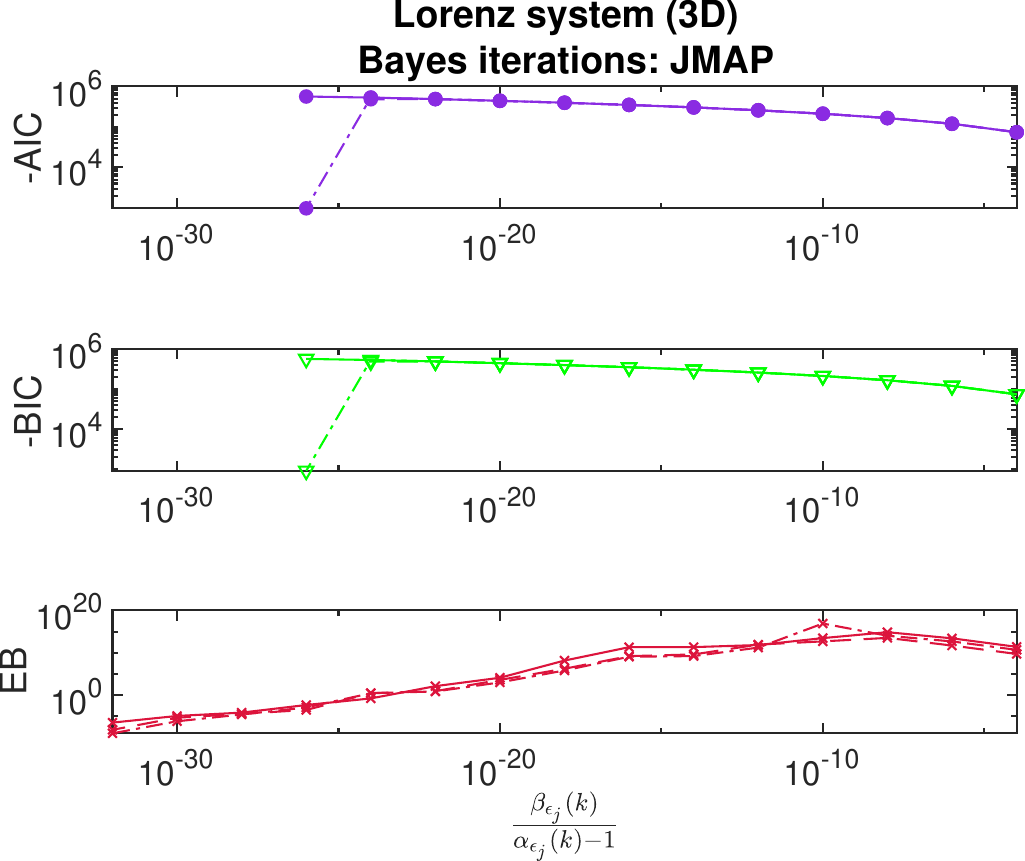} }
  \put(0,0){\small (a)}
  \put(400,0){\includegraphics[height=65mm]{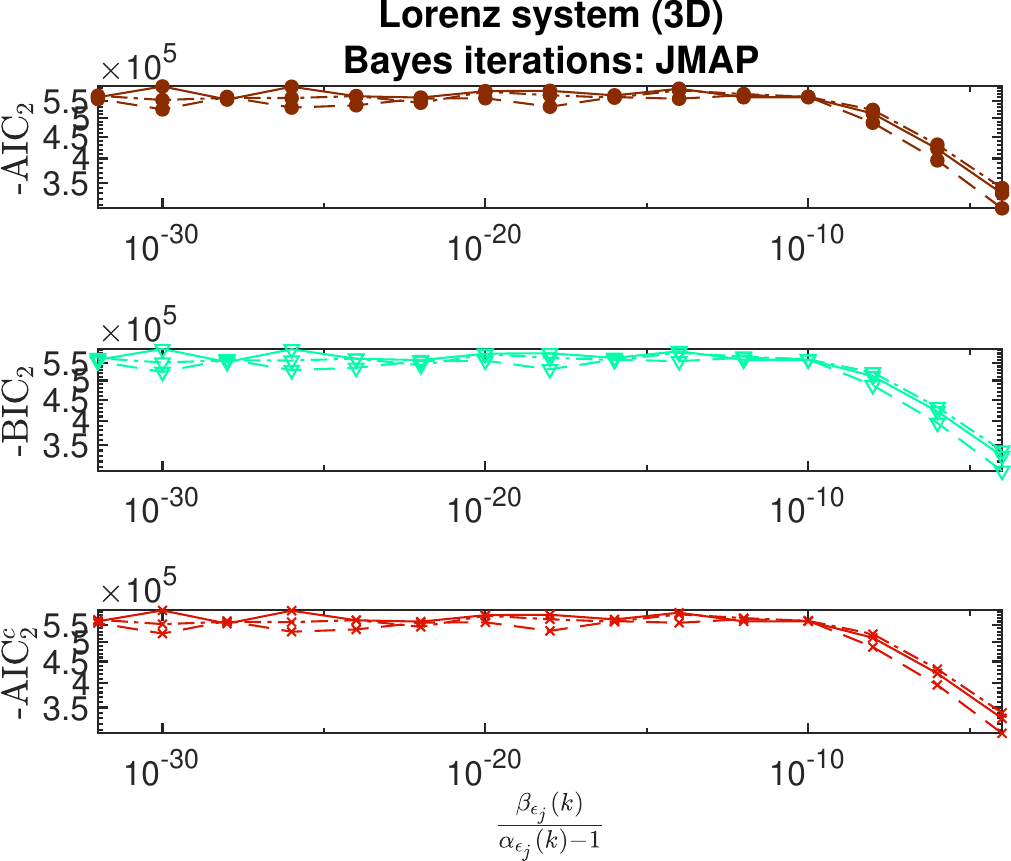} }
  \put(400,0){\small (b)}
 \end{picture}
\end{center}
\caption{Lorenz {system with added Gaussian noise}: second-order polynomial regularization using JMAP, $T=100$, $t_{step}=0.01$, $\varepsilon=0.2$, showing the alternative metrics by iteration sequence: (a) AIC, BIC and EB \eqref{eq:AIC_BIC}-\eqref{eq:EB}, and (b) 2-norm approximations to AIC, BIC and AIC$_c$.}
\label{fig:SI_Lorenz_Gauss_JMAP_metrics}
\end{figure*}

\begin{figure*}[h]
\begin{center}
\setlength{\unitlength}{0.6pt}
 \begin{picture}(800,580)
  \put(0,300){\includegraphics[height=60mm]{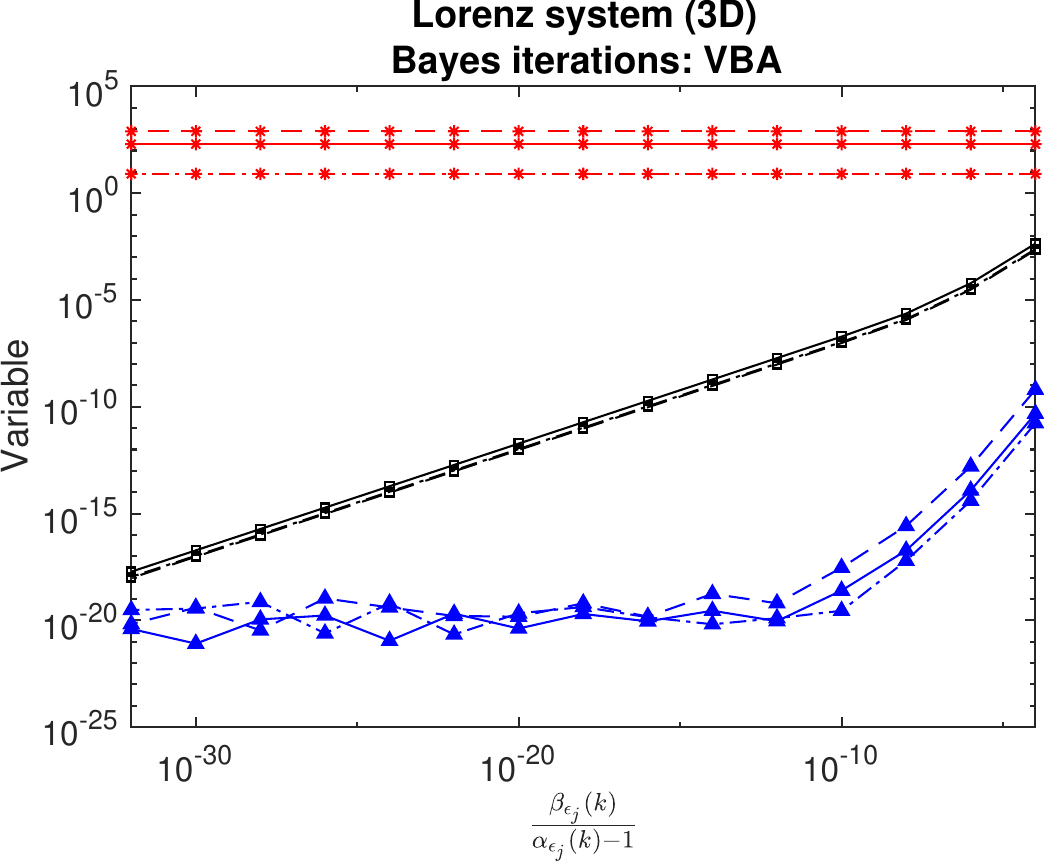} }
  \put(55,400){\includegraphics[height=25mm]{figs_suppl/figleg/Legend_2norms} }
  \put(243,367){\includegraphics[height=6mm]{figs_suppl/figleg/circle} }
  \put(0,300){\small (a)}
  \put(370,300){\includegraphics[height=60mm]{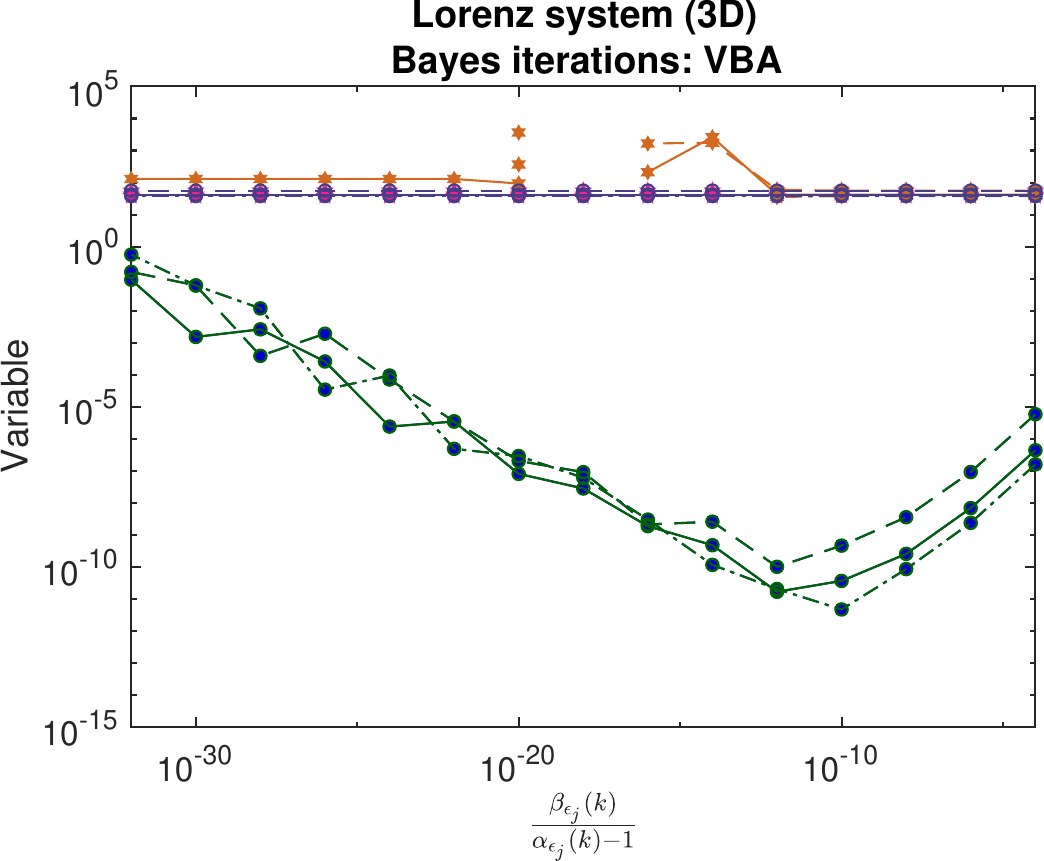} }
  \put(720,362){\includegraphics[height=41mm]{figs_suppl/figleg/Legend_Gnorms1} }
  \put(613,378){\includegraphics[height=6mm]{figs_suppl/figleg/circle} }
  \put(370,300){\small (b)}
  \put(0,0){\includegraphics[height=60mm]{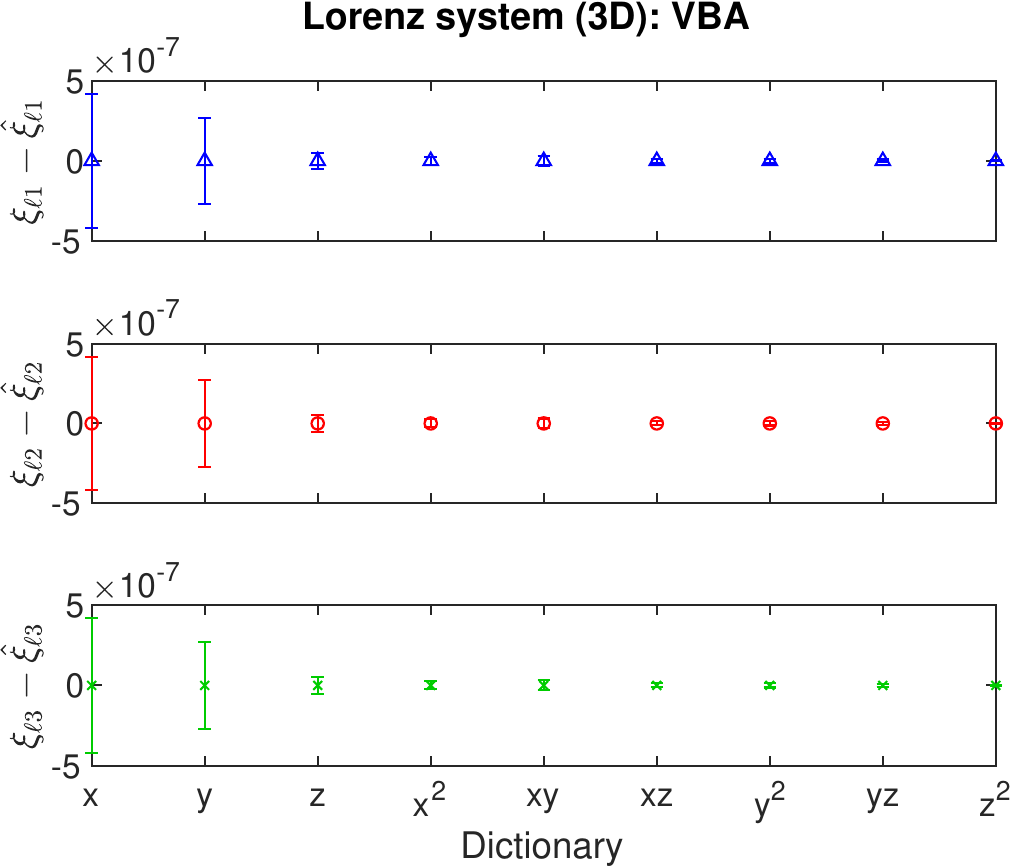} } 
  \put(0,0){\small (c)}
  \put(370,0){\includegraphics[height=60mm]{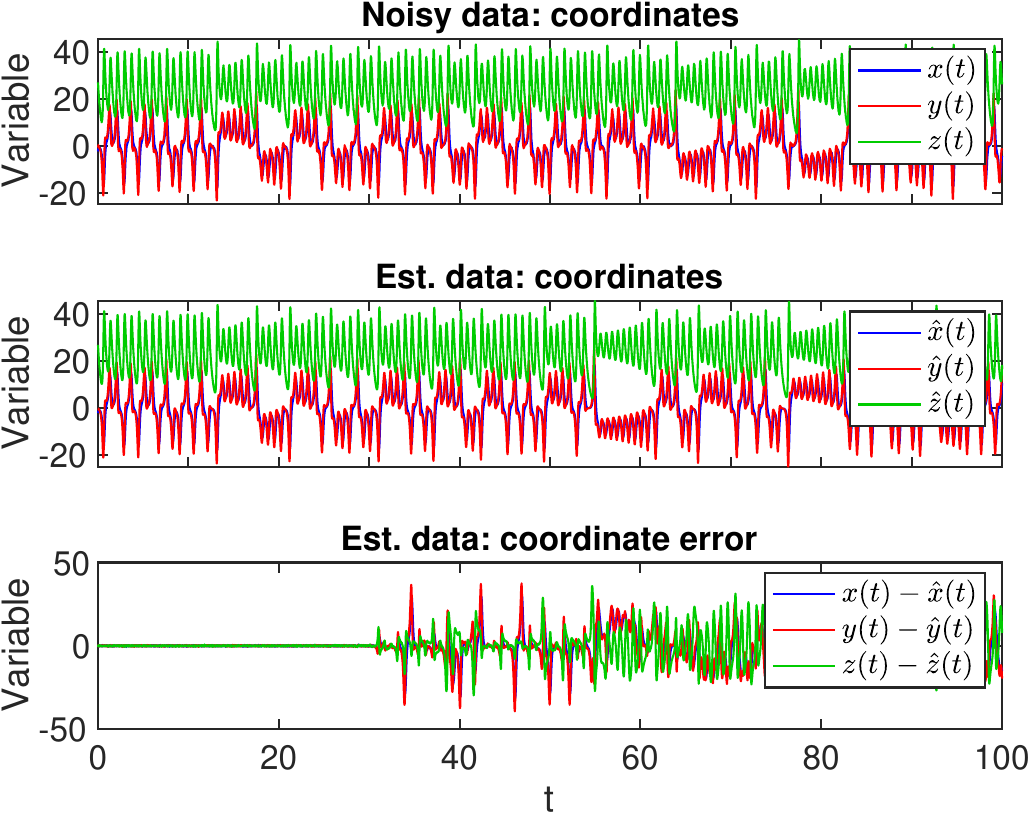} } 
  \put(370,0){\small (d)}
 \end{picture}
\end{center}
\caption{Lorenz {system with added Gaussian noise}: second-order polynomial regularization using VBA, $T=100$, $t_{step}=0.01$, $\varepsilon=0.2$, showing the iteration sequence with decreasing $\mathsf{E}_{\eizero}$, including 
(a) 2-norm residual, regularization and objective functions \eqref{eq:modJ}, showing the optimal iteration ($k=6$); 
(b) Gaussian norms for the prior, likelihood, posterior and evidence, showing the optimal iteration ($k=6$); 
(c) optimal error in predicted coefficients $\matparamc_{ij}-\hat{\matparamc}_{ij}$, with error bars from the posterior covariance \eqref{eq:posterior_estimators}; and
(d) original and optimal predicted data and their differences.}
\label{fig:SI_Lorenz_Gauss_VBA}
\end{figure*}

\begin{figure*}[h]
\begin{center}
\setlength{\unitlength}{0.6pt}
 \begin{picture}(800,300)
  \put(0,0){\includegraphics[height=65mm]{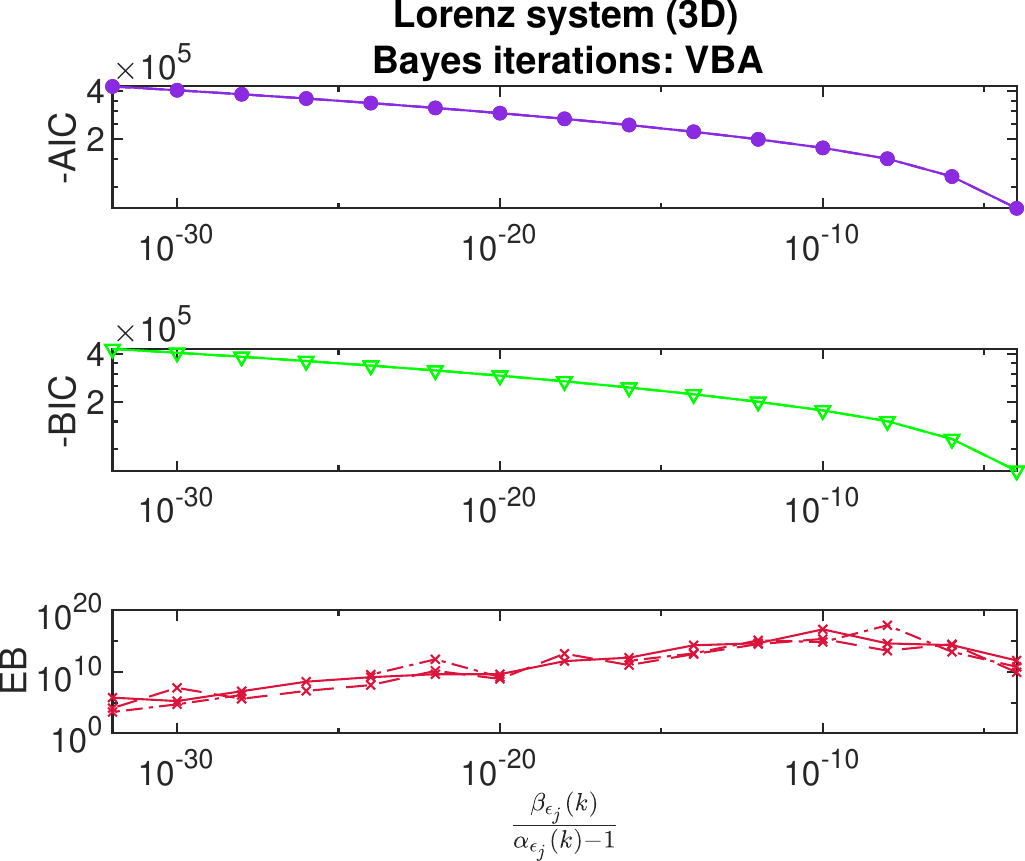} }
  \put(0,0){\small (a)}
  \put(400,0){\includegraphics[height=65mm]{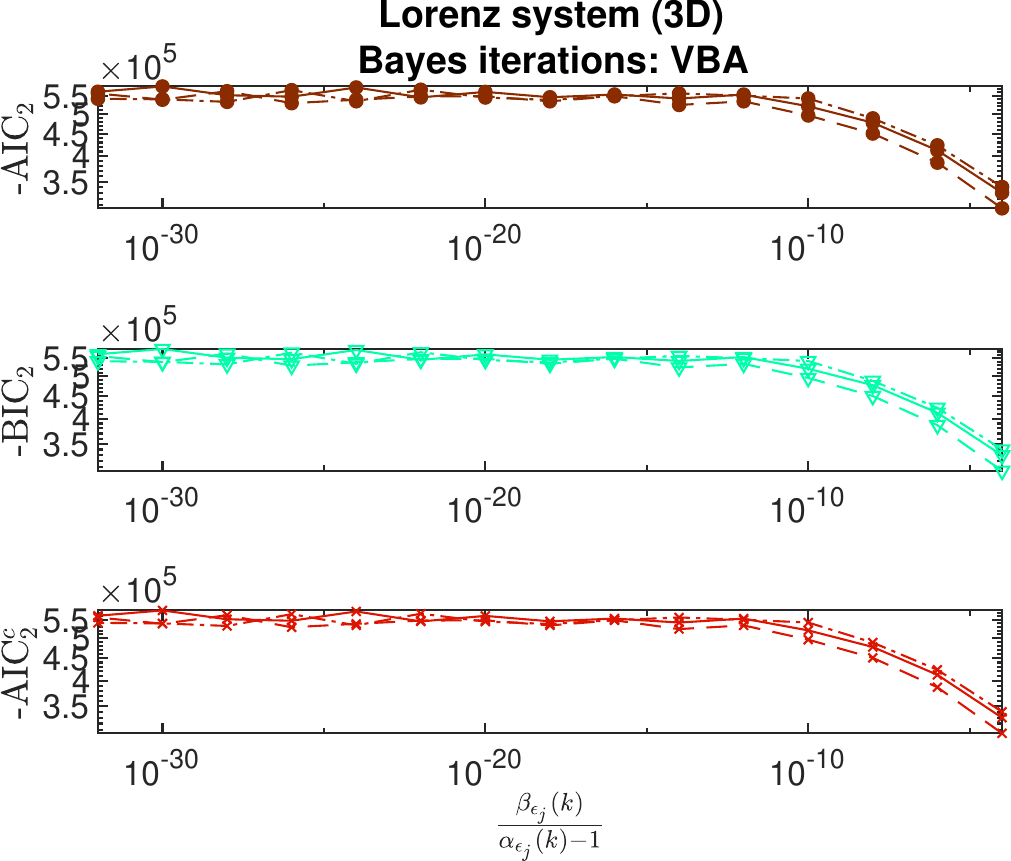} }
  \put(400,0){\small (b)}
 \end{picture}
\end{center}
\caption{Lorenz {system with added Gaussian noise}: second-order polynomial regularization using VBA, $T=100$, $t_{step}=0.01$, $\varepsilon=0.2$, showing the alternative metrics by iteration sequence: (a) AIC, BIC and EB \eqref{eq:AIC_BIC}-\eqref{eq:EB}, and (b) 2-norm approximations to AIC, BIC and AIC$_c$.}
\label{fig:SI_Lorenz_Gauss_VBA_metrics}
\end{figure*}

\begin{figure*}[h]
\begin{center}
\setlength{\unitlength}{0.6pt}
 \begin{picture}(800,580)
  \put(0,300){\includegraphics[height=60mm]{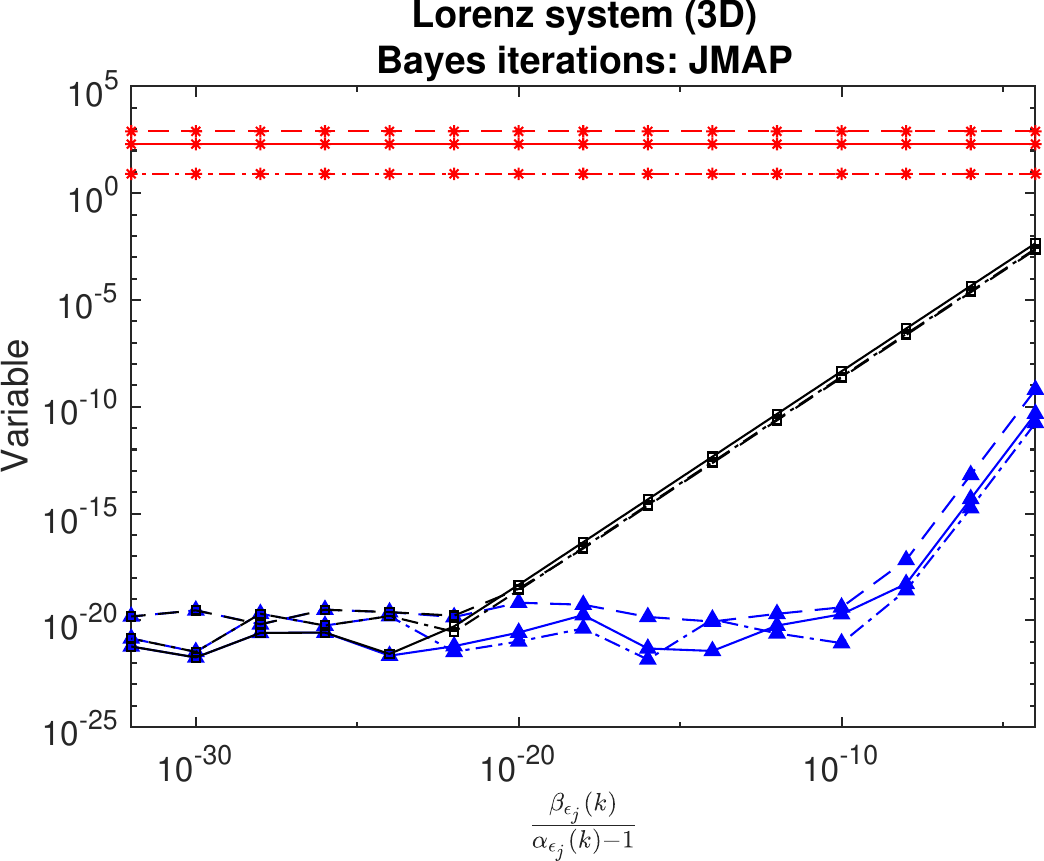} }
  \put(55,400){\includegraphics[height=25mm]{figs_suppl/figleg/Legend_2norms} }
  \put(264,362){\includegraphics[height=6mm]{figs_suppl/figleg/circle} }
  \put(0,300){\small (a)}
  \put(370,300){\includegraphics[height=60mm]{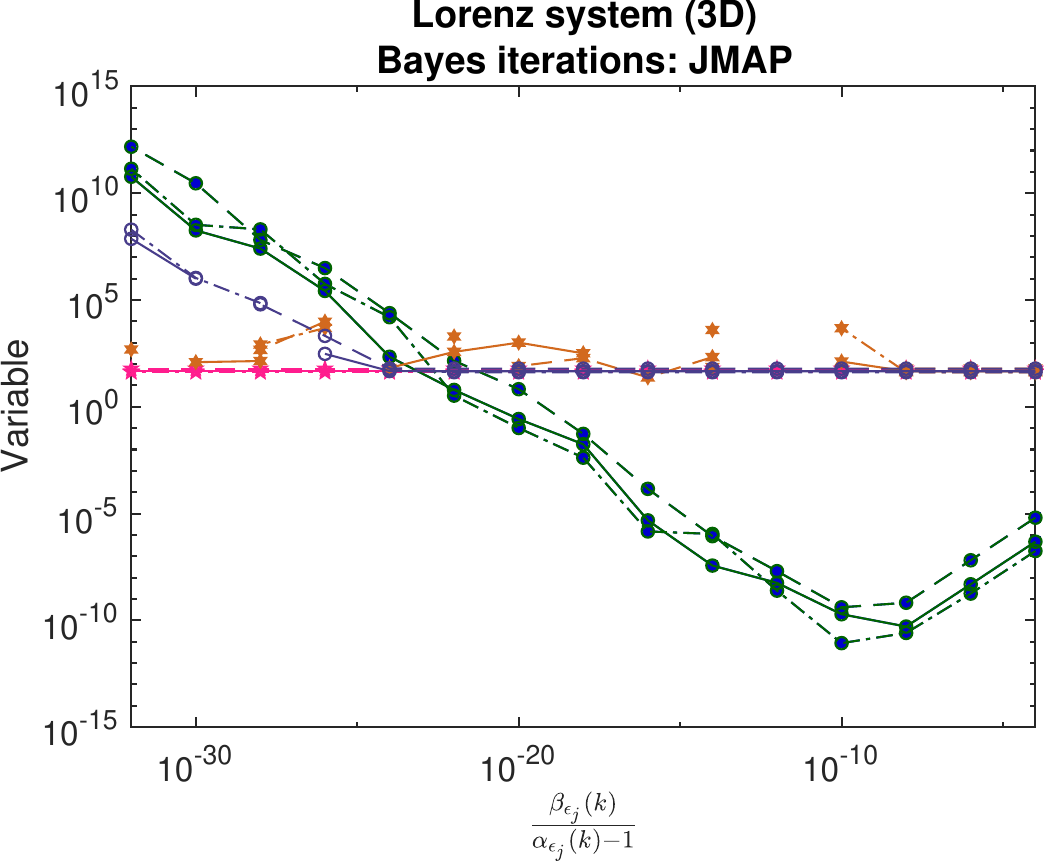} }
  \put(720,362){\includegraphics[height=41mm]{figs_suppl/figleg/Legend_Gnorms1} }
  \put(635,362){\includegraphics[height=6mm]{figs_suppl/figleg/circle} }
  \put(370,300){\small (b)}
  \put(0,0){\includegraphics[height=60mm]{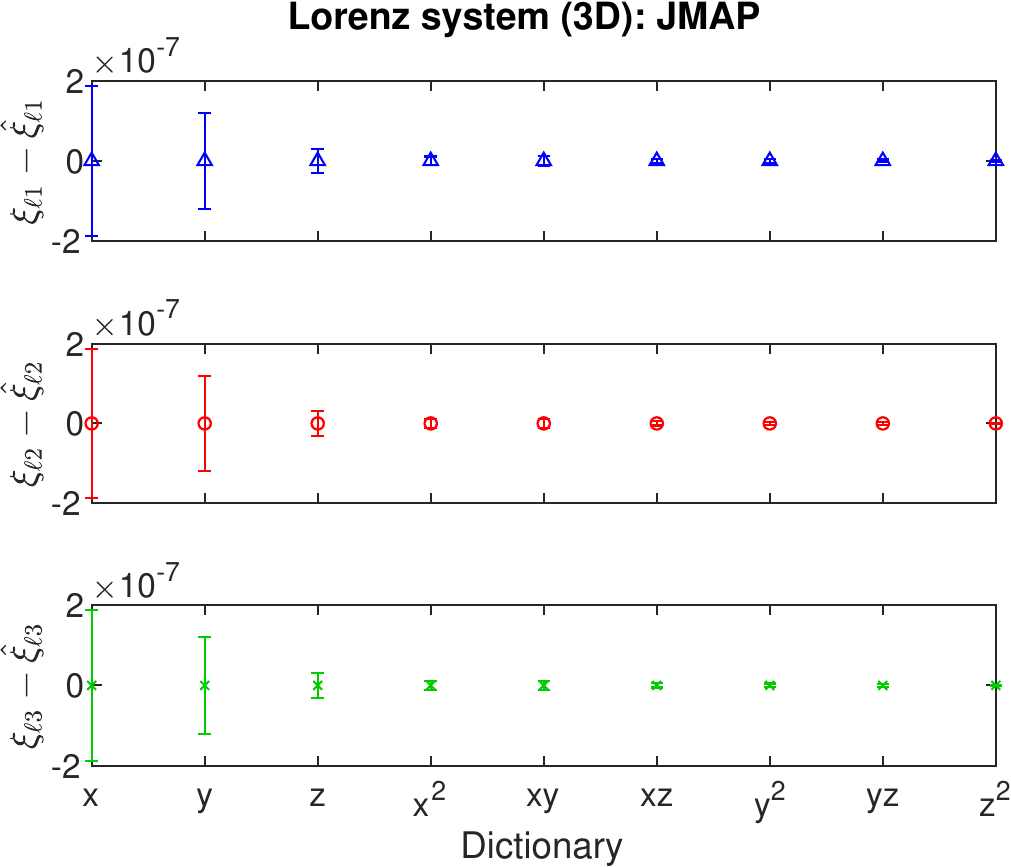} } 
  \put(0,0){\small (c)}
  \put(370,0){\includegraphics[height=60mm]{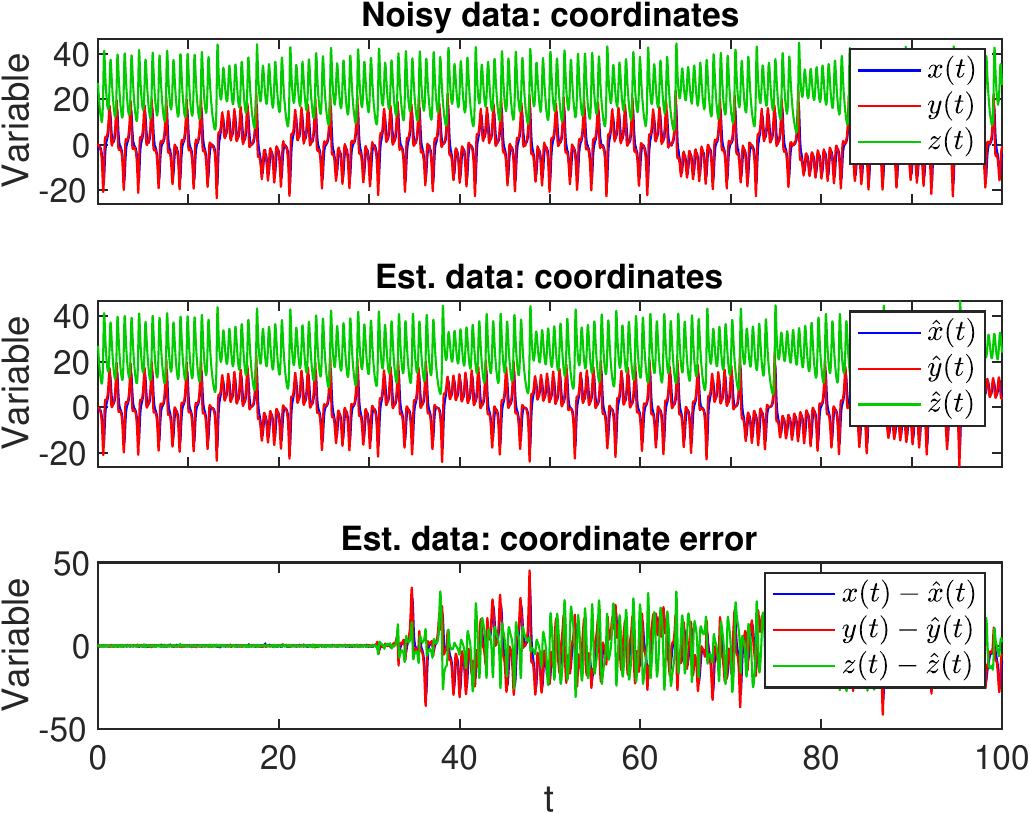} } 
  \put(370,0){\small (d)}
 \end{picture}
\end{center}
\caption{{Lorenz system with added Laplace noise: second-order polynomial regularization using JMAP, $T=100$, $t_{step}=0.01$, $\varepsilon=0.2$, showing the iteration sequence with decreasing $\mathsf{E}_{\eizero}$, including 
(a) 2-norm residual, regularization and objective functions \eqref{eq:modJ}, showing the optimal iteration ($k=5$); 
(b) Gaussian norms for the prior, likelihood, posterior and evidence, showing the optimal iteration ($k=5$); 
(c) optimal error in predicted coefficients $\matparamc_{ij}-\hat{\matparamc}_{ij}$, with error bars from the posterior covariance \eqref{eq:posterior_estimators}; and
(d) original and optimal predicted data and their differences.}}
\label{fig:SI_Lorenz_Laplace_JMAP}
\end{figure*}

\begin{figure*}[h]
\begin{center}
\setlength{\unitlength}{0.6pt}
 \begin{picture}(800,580)
  \put(0,300){\includegraphics[height=60mm]{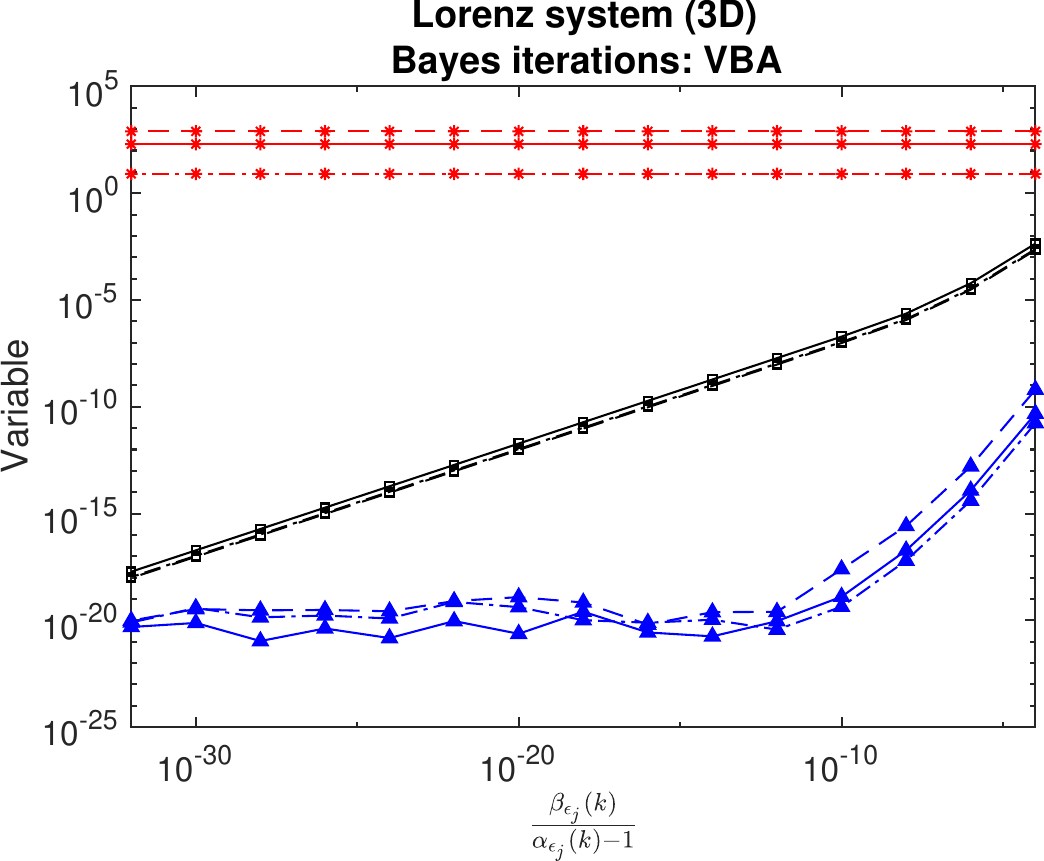} }
  \put(55,400){\includegraphics[height=25mm]{figs_suppl/figleg/Legend_2norms} }
  \put(243,366){\includegraphics[height=6mm]{figs_suppl/figleg/circle} }
  \put(0,300){\small (a)}
  \put(370,300){\includegraphics[height=60mm]{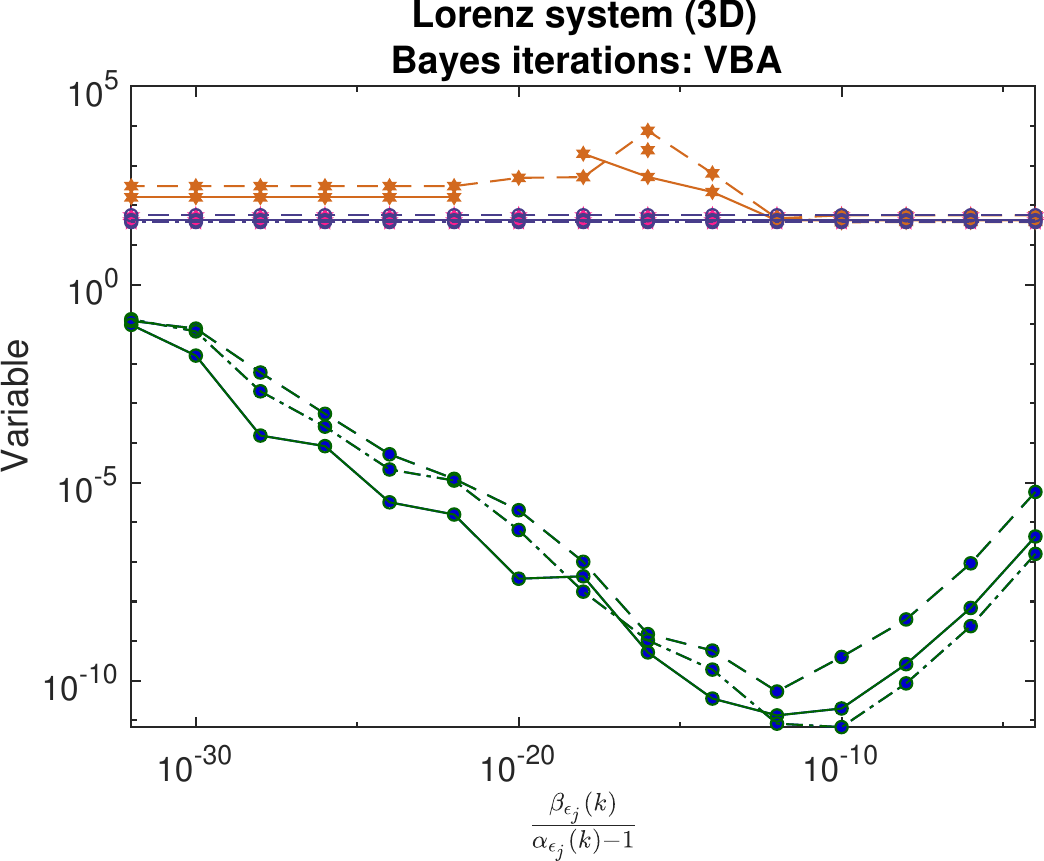} }
  \put(720,362){\includegraphics[height=41mm]{figs_suppl/figleg/Legend_Gnorms1} }
  \put(613,338){\includegraphics[height=6mm]{figs_suppl/figleg/circle} }
  \put(370,300){\small (b)}
  \put(0,0){\includegraphics[height=60mm]{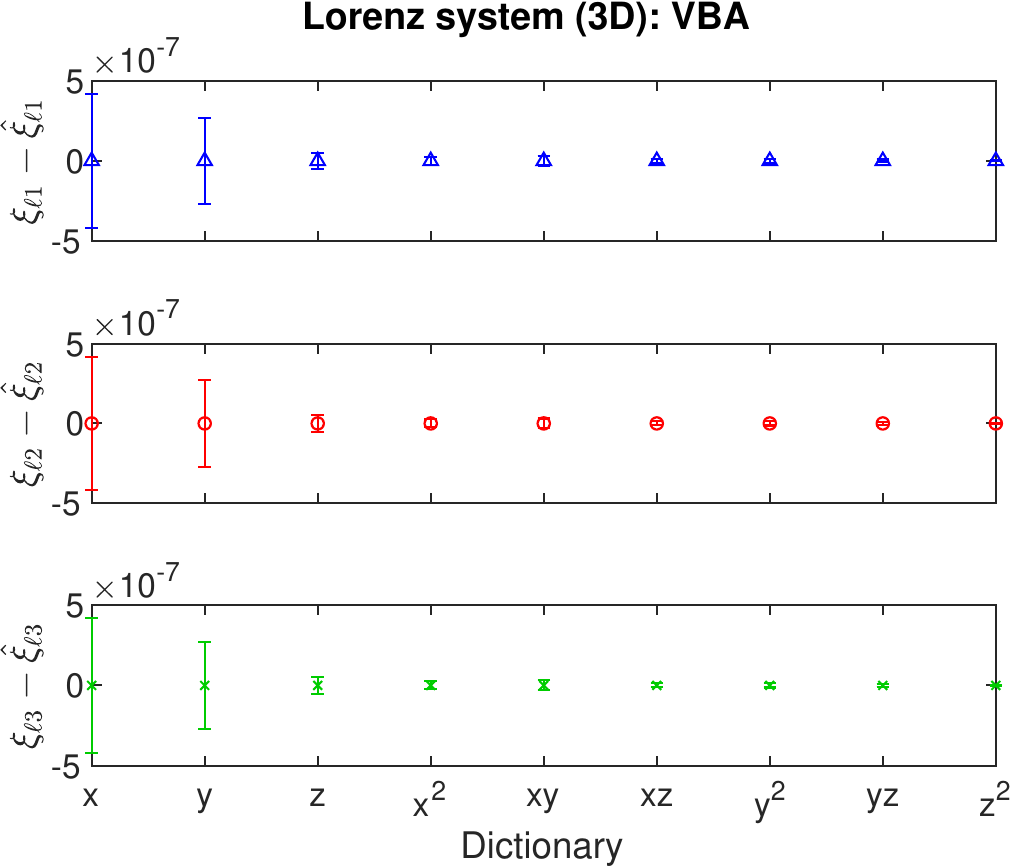} } 
  \put(0,0){\small (c)}
  \put(370,0){\includegraphics[height=60mm]{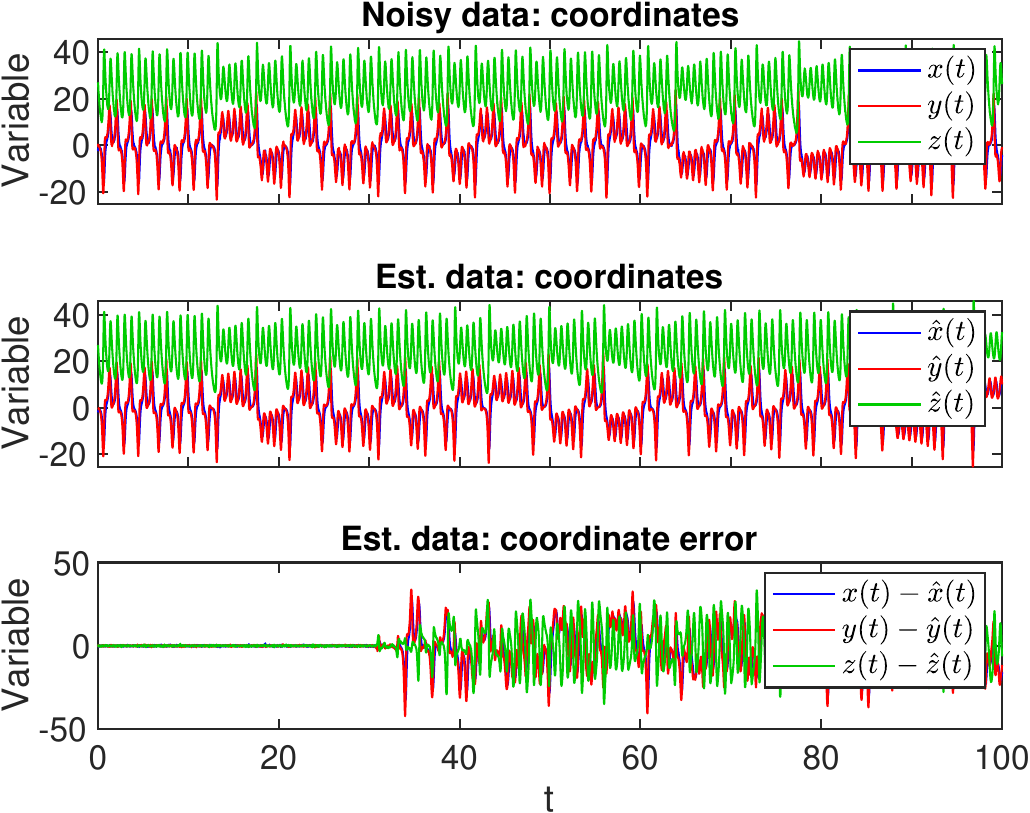} } 
  \put(370,0){\small (d)}
 \end{picture}
\end{center}
\caption{{Lorenz system with added Laplace noise: second-order polynomial regularization using VBA, $T=100$, $t_{step}=0.01$, $\varepsilon=0.2$, showing the iteration sequence with decreasing $\mathsf{E}_{\eizero}$, including 
(a) 2-norm residual, regularization and objective functions \eqref{eq:modJ}, showing the optimal iteration ($k=6$); 
(b) Gaussian norms for the prior, likelihood, posterior and evidence, showing the optimal iteration ($k=6$); 
(c) optimal error in predicted coefficients $\matparamc_{ij}-\hat{\matparamc}_{ij}$, with error bars from the posterior covariance \eqref{eq:posterior_estimators}; and
(d) original and optimal predicted data and their differences.}}
\label{fig:SI_Lorenz_Laplace_VBA}
\end{figure*}

\begin{figure*}[h]
\begin{center}
\setlength{\unitlength}{0.55pt}
 \begin{picture}(800,940)
  \put(20,640){\includegraphics[height=60mm]{figs_suppl/figLt100_e0.2/Fig233} } 
  \put(20,640){\small (a)}
  \put(410,640){\includegraphics[height=60mm]{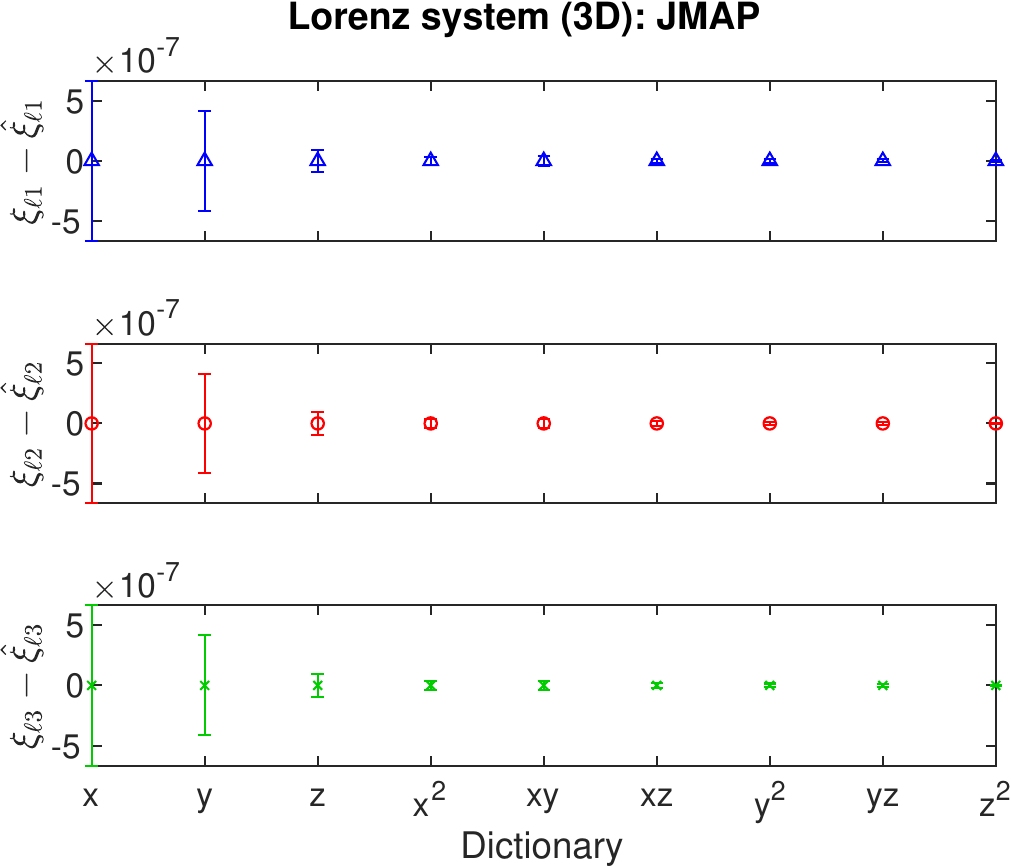} } 
  \put(410,640){\small (b)}
  \put(20,320){\includegraphics[height=60mm]{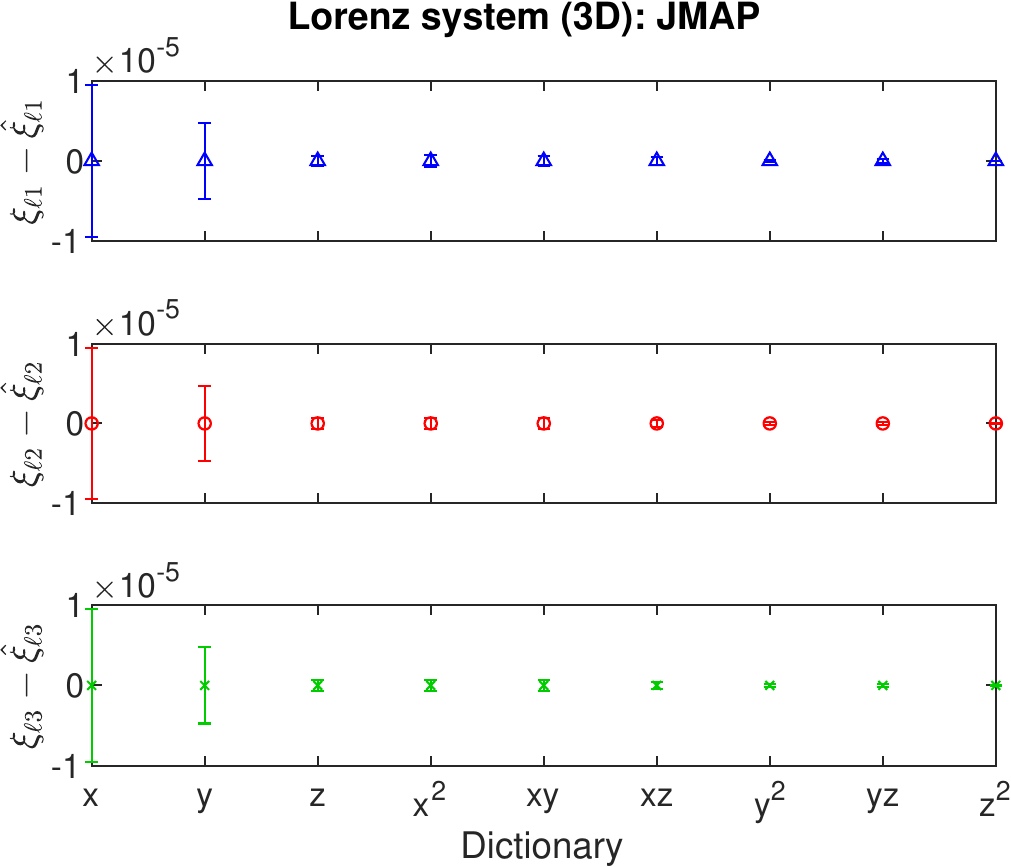} } 
  \put(20,320){\small (c)}
  \put(410,320){\includegraphics[height=60mm]{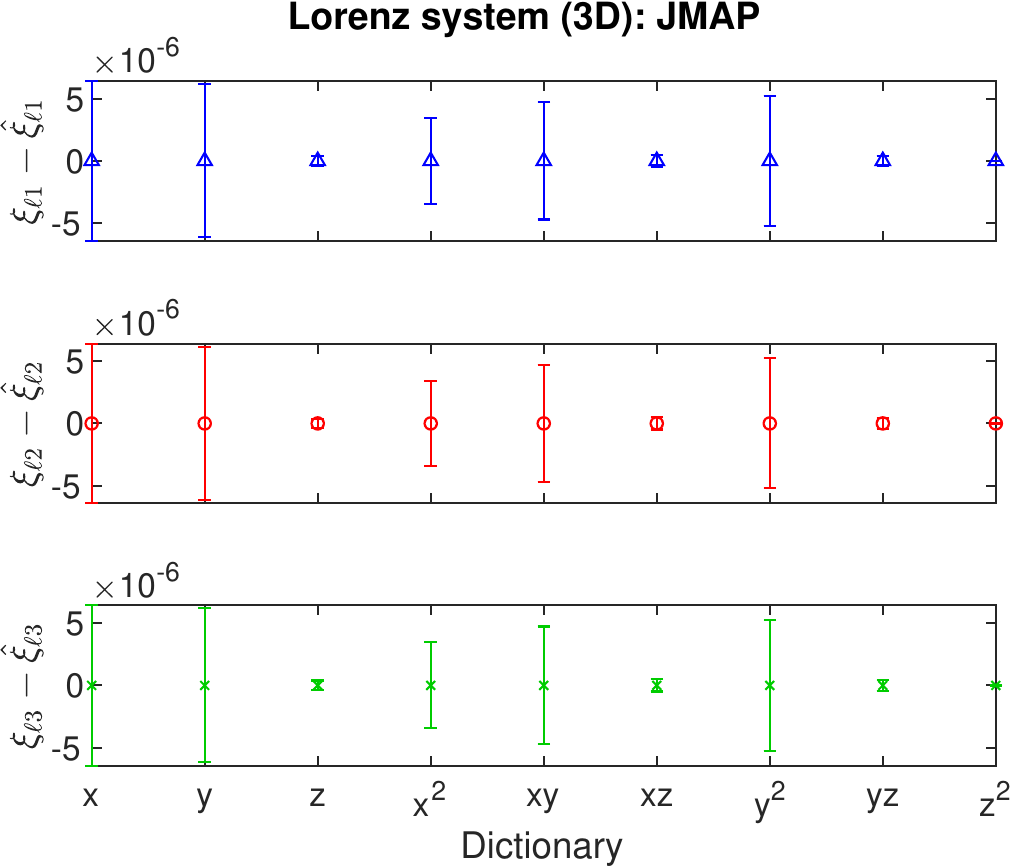} } 
  \put(410,320){\small (d)}
  \put(0,0){\includegraphics[height=60mm]{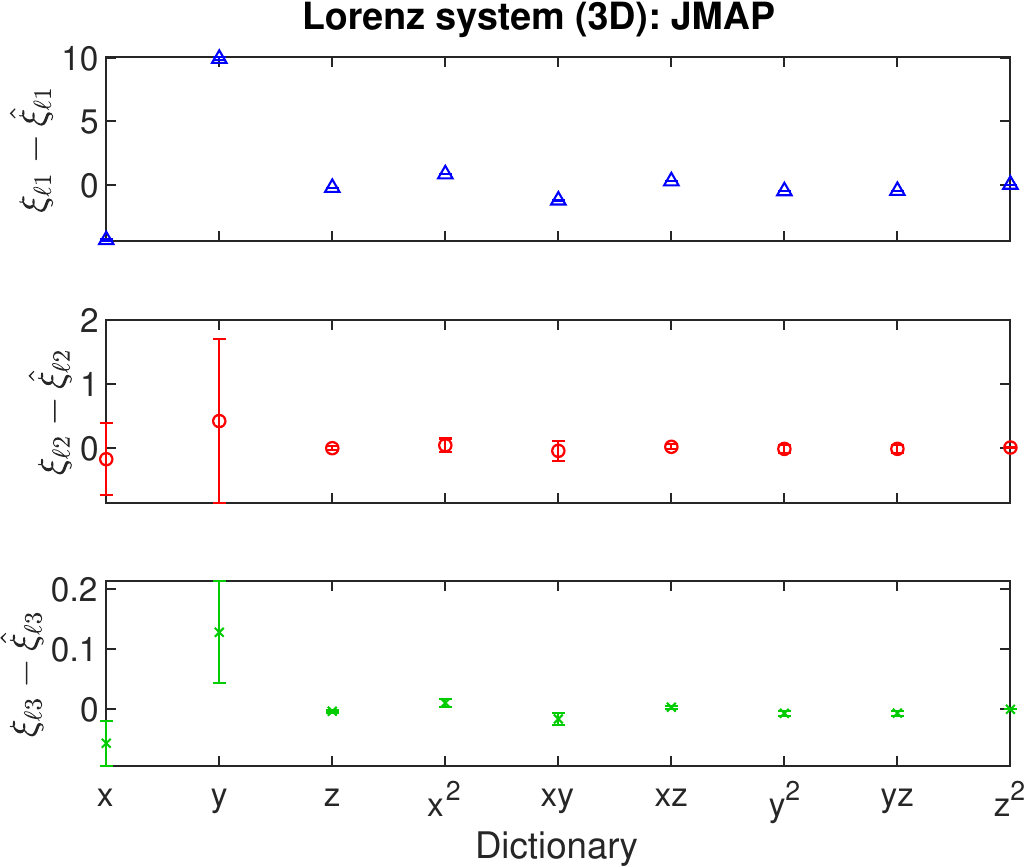} } 
  \put(0,0){\small (e)}
  \put(410,0){\includegraphics[height=60mm]{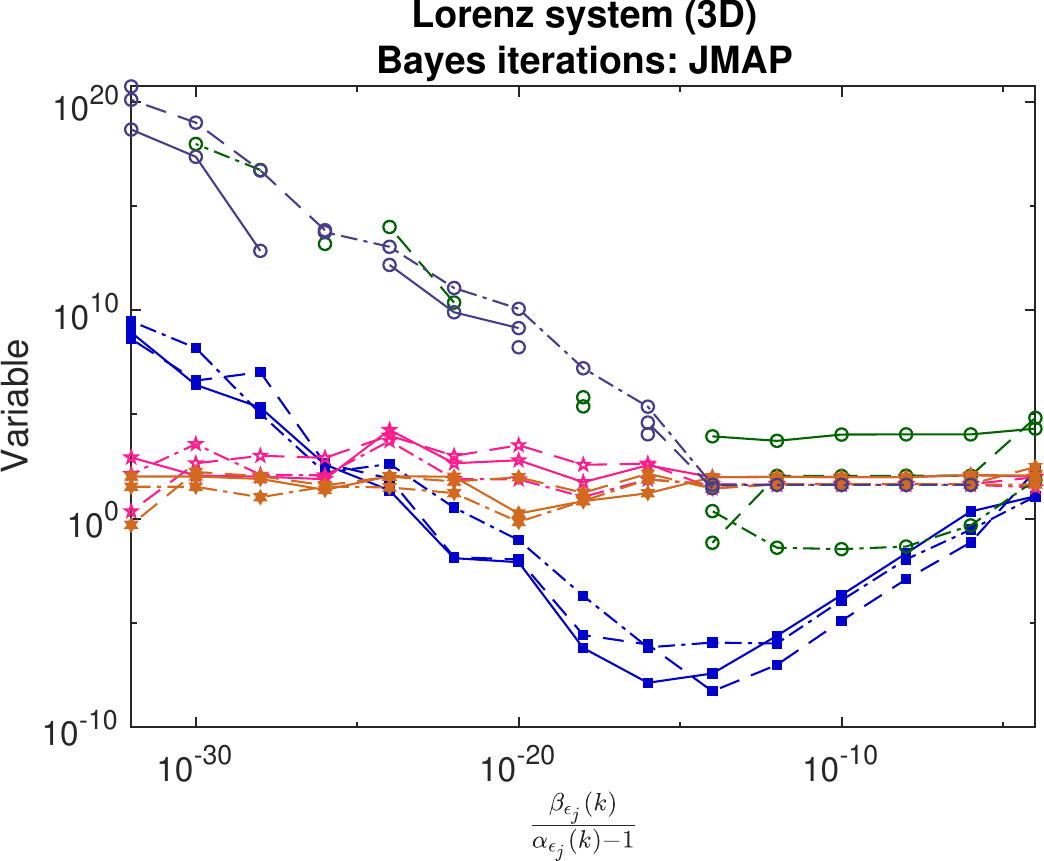} }
  \put(790,68){\includegraphics[height=41mm]{figs_suppl/figleg/Legend_Gnorms1} }
  \put(660,160){\includegraphics[height=6mm]{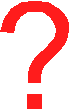} }
  \put(651,55){\includegraphics[height=6mm]{figs_suppl/figleg/circle} }
  \put(410,0){\small (f)}
  \end{picture}
\end{center}
\caption{Lorenz system {with added Gaussian noise}: effect of time-series length on JMAP regularization for $t_{step}=0.01$, $\varepsilon=0.2$: 
(a) $T=100$; 
(b) $T=10$; 
(c) $T=1$; 
(d) $T=0.1$; and 
(e) $T=0.08$; 
(f) Gaussian norms for the prior, likelihood, posterior and evidence for $T=0.08$, $t_{step}=0.01$, $\varepsilon=0.2$, showing breakdown of the JMAP algorithm for this small number of points ($m=8$).}
\label{fig:SI_Lorenz_Gauss_JMAP_time}
\end{figure*}

\begin{figure*}[h]
\begin{center}
\setlength{\unitlength}{0.55pt}
 \begin{picture}(850,960)
  \put(0,640){\includegraphics[height=60mm]{figs_suppl/figLt100_e0.2/Fig380} }
  \put(380,707){\includegraphics[height=41mm]{figs_suppl/figleg/Legend_Gnorms1} }
  \put(289,707){\includegraphics[height=6mm]{figs_suppl/figleg/circle} }
  \put(0,640){\small (a)}
  \put(500,640){\includegraphics[height=60mm]{figs_suppl/figLt100_e0.2/Fig233} } 
  \put(500,640){\small (b)}
  \put(0,320){\includegraphics[height=60mm]{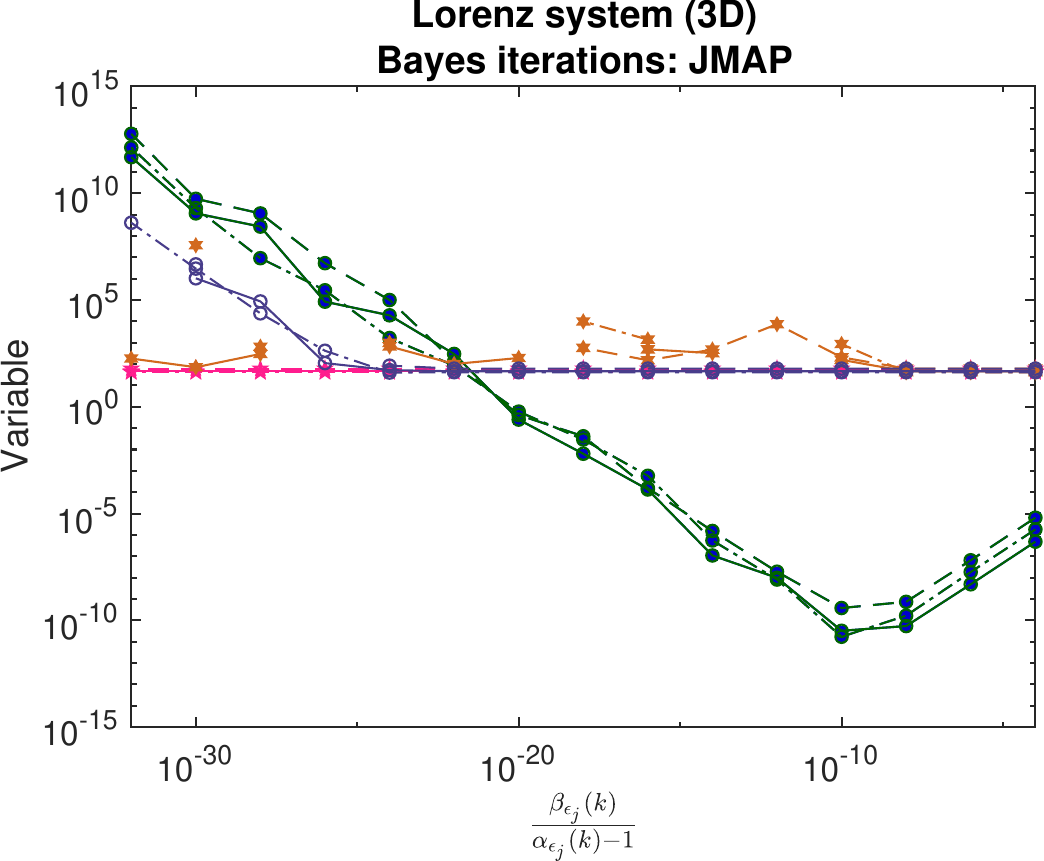} }
  \put(288,390){\includegraphics[height=6mm]{figs_suppl/figleg/circle} }
  \put(0,320){\small (c)}
  \put(500,320){\includegraphics[height=60mm]{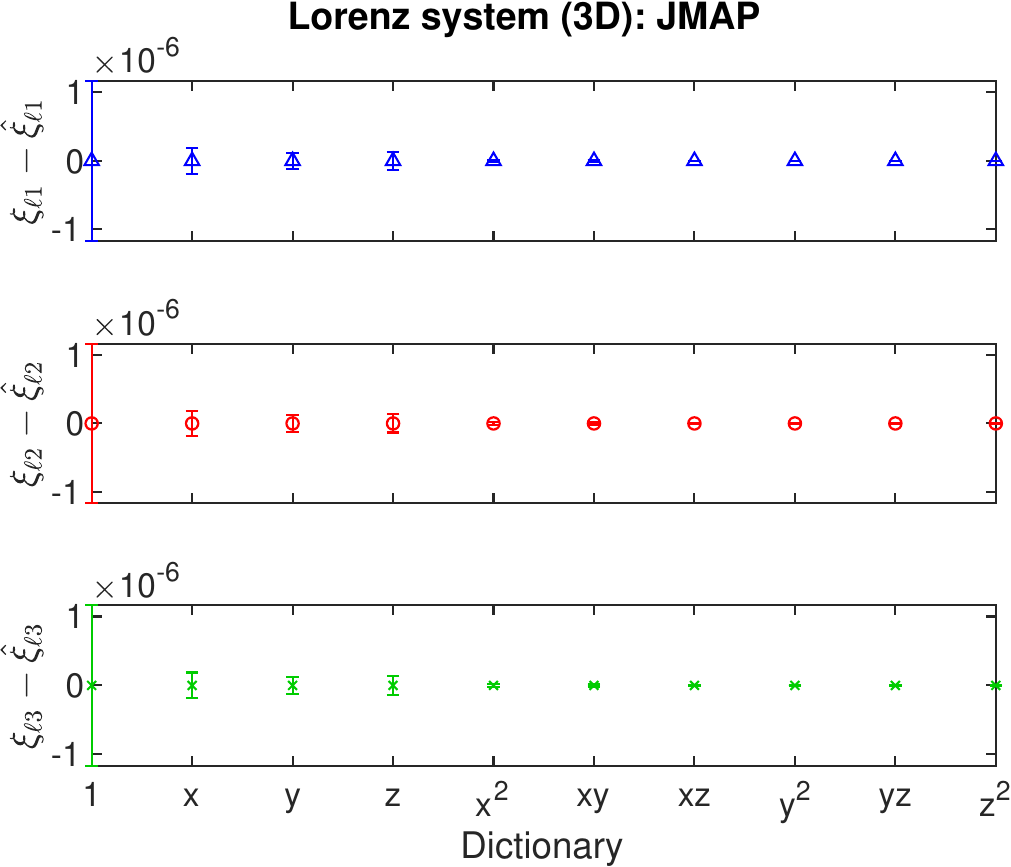} } 
  \put(500,320){\small (d)}
  \put(0,0){\includegraphics[height=60mm]{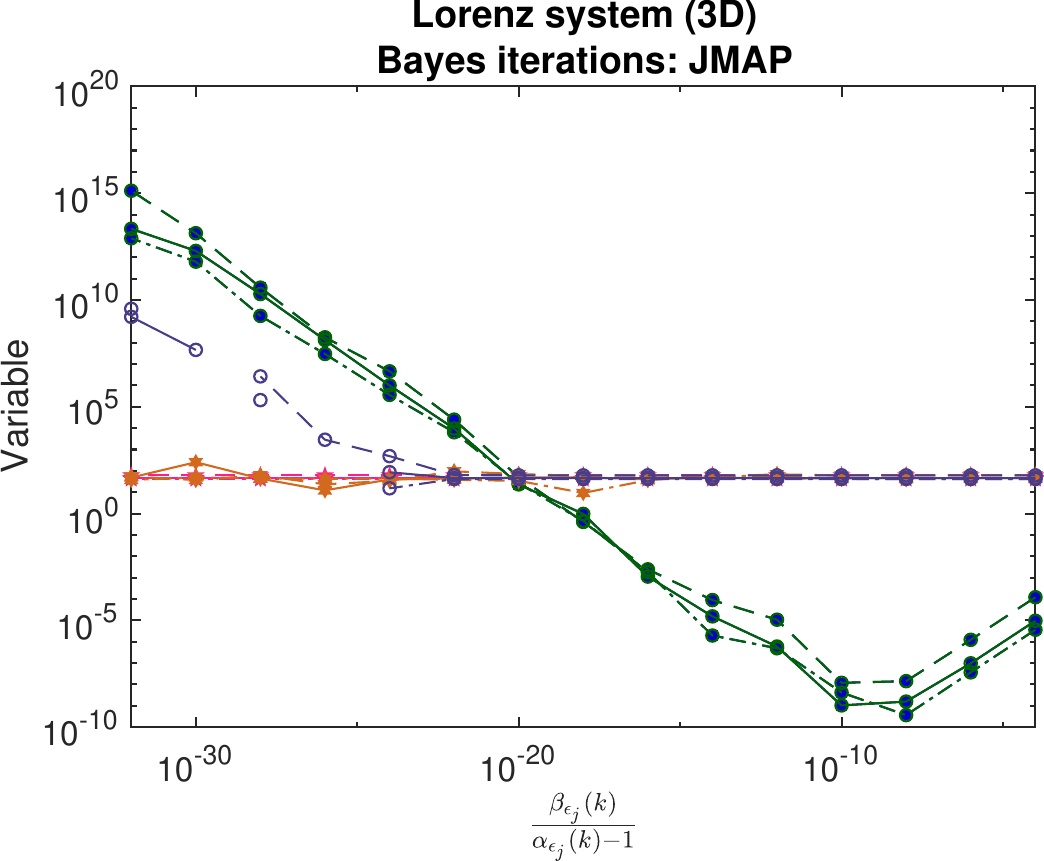} }
  \put(311,43){\includegraphics[height=6mm]{figs_suppl/figleg/circle} }
  \put(0,0){\small (e)}
  \put(500,0){\includegraphics[height=60mm]{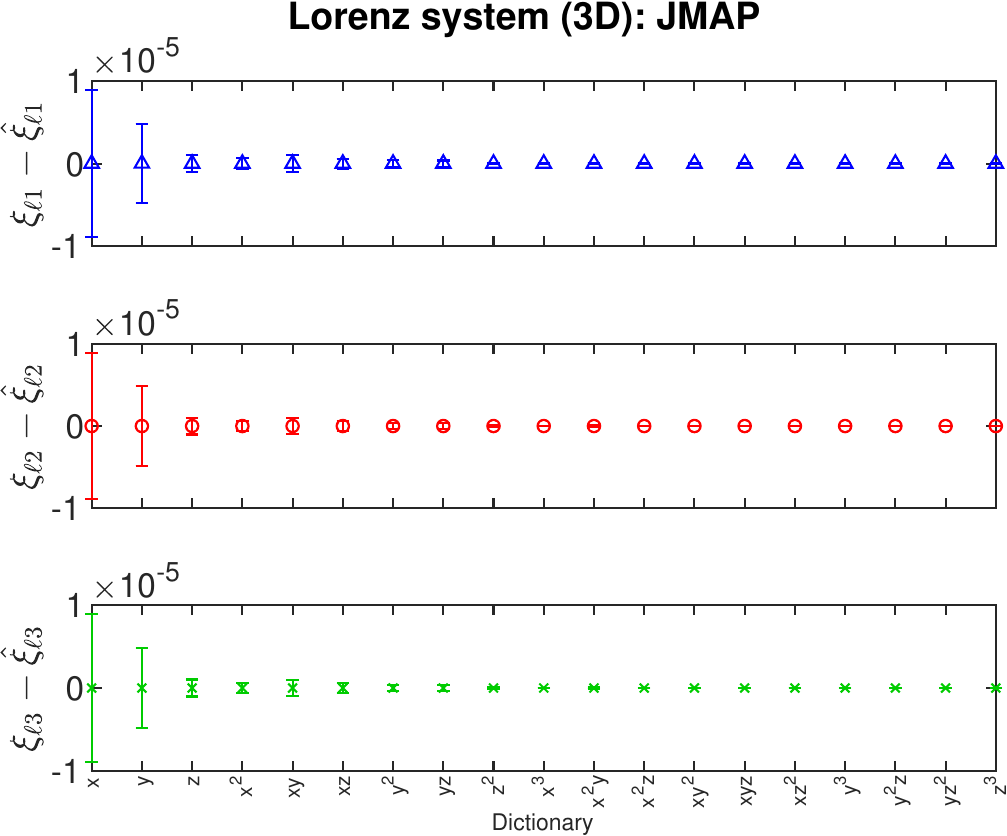} } 
  \put(500,0){\small (f)}
 \end{picture}
\end{center}
\caption{Lorenz system {with added Gaussian noise}, effect of different alphabets on JMAP regularization for $T=100$, $t_{step}=0.01$, $\varepsilon=0.2$ with (a)-(b) polynomials up to second order, 
(c)-(d) polynomials up to second order with a column of 1s, and 
(e)-(f) polynomials up to third order. 
These show 
(a),(c),(e) Gaussian norms for the prior, likelihood, posterior and evidence, with the optimal iteration; and
(b),(d),(f) optimal error in predicted coefficients $\matparamc_{\ell j}-\hat{\matparamc}_{\ell j}$, with error bars calculated from the posterior covariance \eqref{eq:posterior_estimators}.}
\label{fig:SI_Lorenz_Gauss_JMAP_poly}
\end{figure*}

\begin{figure*}[h]
\begin{center}
\setlength{\unitlength}{0.6pt}
 \begin{picture}(800,580)
  \put(0,300){\includegraphics[height=60mm]{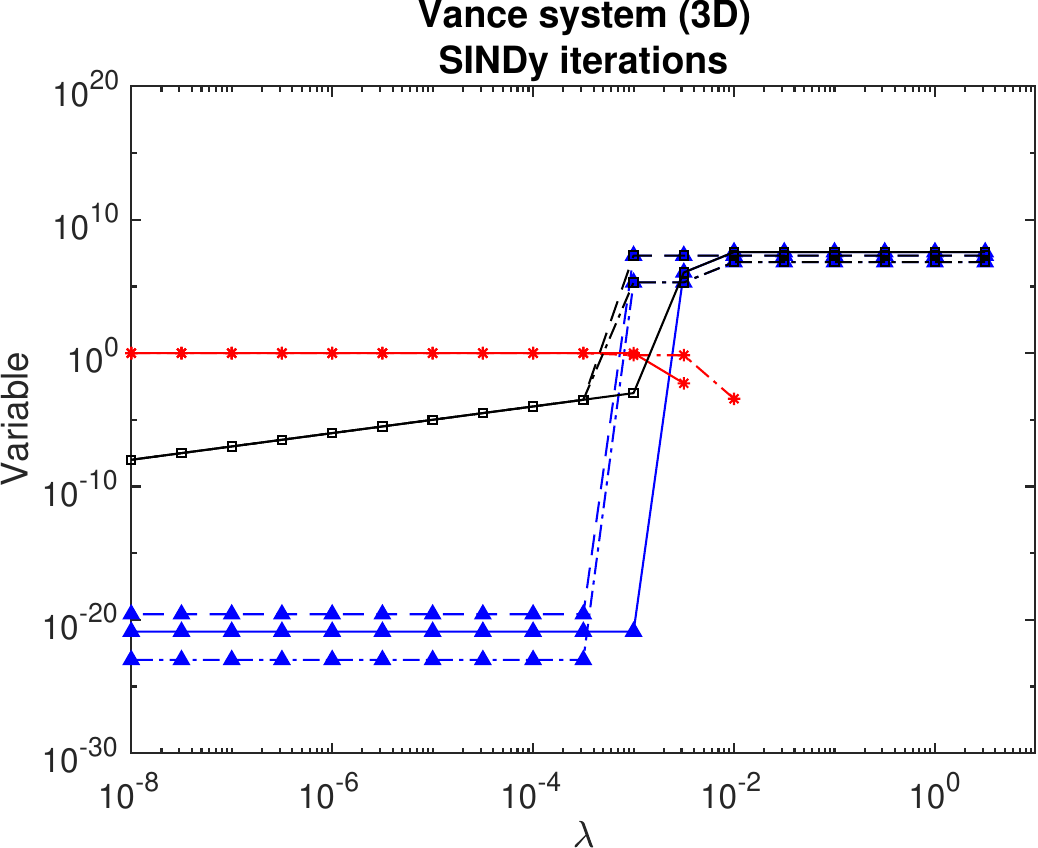} }
  \put(55,505){\includegraphics[height=9mm]{figs_suppl/figleg/Legend_2norms2} }
  \put(181,356){\includegraphics[height=6mm]{figs_suppl/figleg/circle} }
  \put(0,300){\small (a)}
  \put(370,300){\includegraphics[height=60mm]{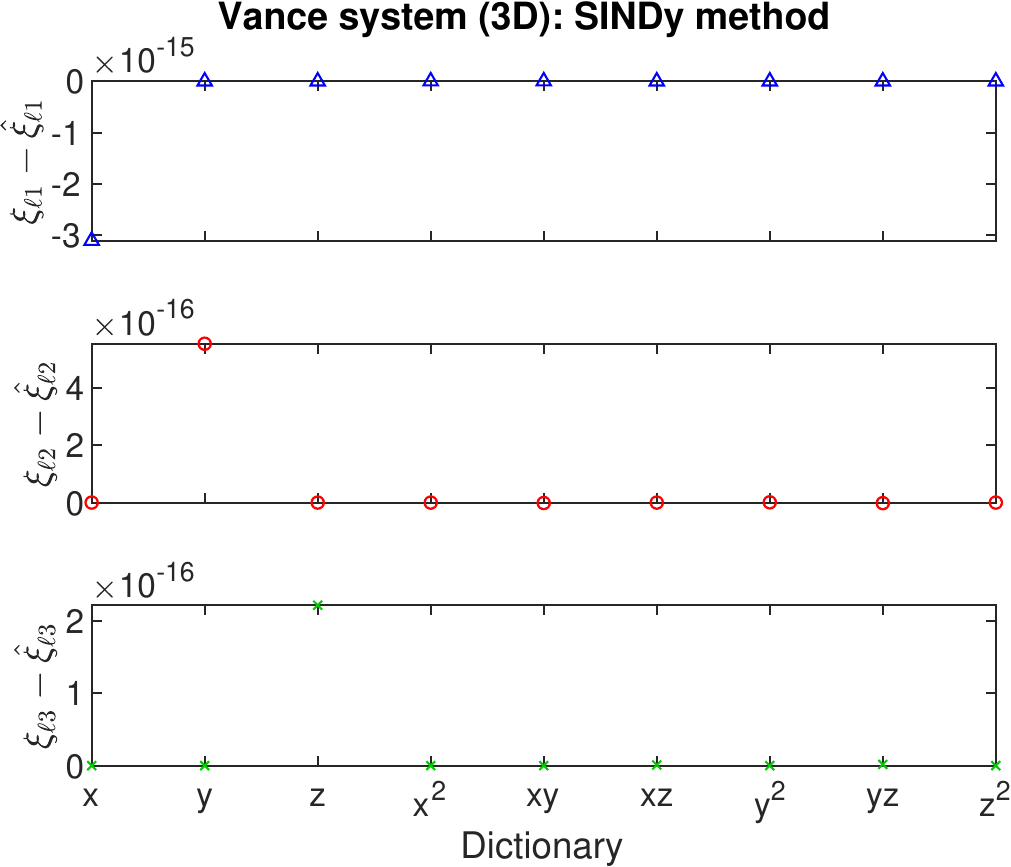} }
  \put(370,300){\small (b)}
  \put(0,0){\includegraphics[height=60mm]{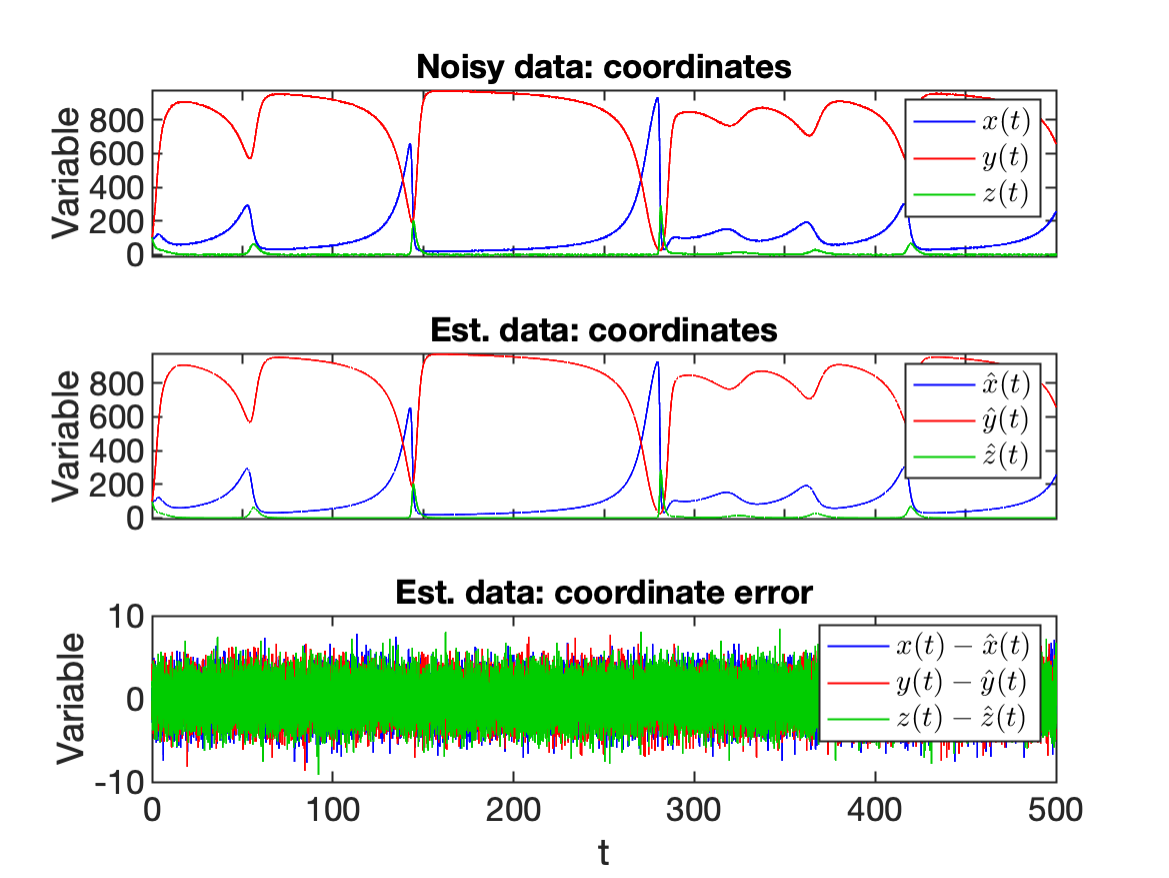} }
  \put(0,0){\small (c)}
 \end{picture}
\end{center}
\caption{Vance {system with added Gaussian noise}: second-order polynomial regularization using SINDy, $T=500$, $t_{step}=0.02$, $\varepsilon=2.0$: 
(a) plots of 2-norm residual, regularization and objective functions \eqref{eq:modJ} 
with decreasing $\lambda$, showing the optimal iteration ($k=10)$; 
(b) optimal error in predicted coefficients $\matparamc_{ij}-\hat{\matparamc}_{ij}$; and
(c) original and optimal predicted data and their differences.}
\label{fig:SI_Vance_SINDy}
\end{figure*}

\begin{figure*}[h]
\begin{center}
\setlength{\unitlength}{0.6pt}
 \begin{picture}(800,580)
  \put(0,300){\includegraphics[height=60mm]{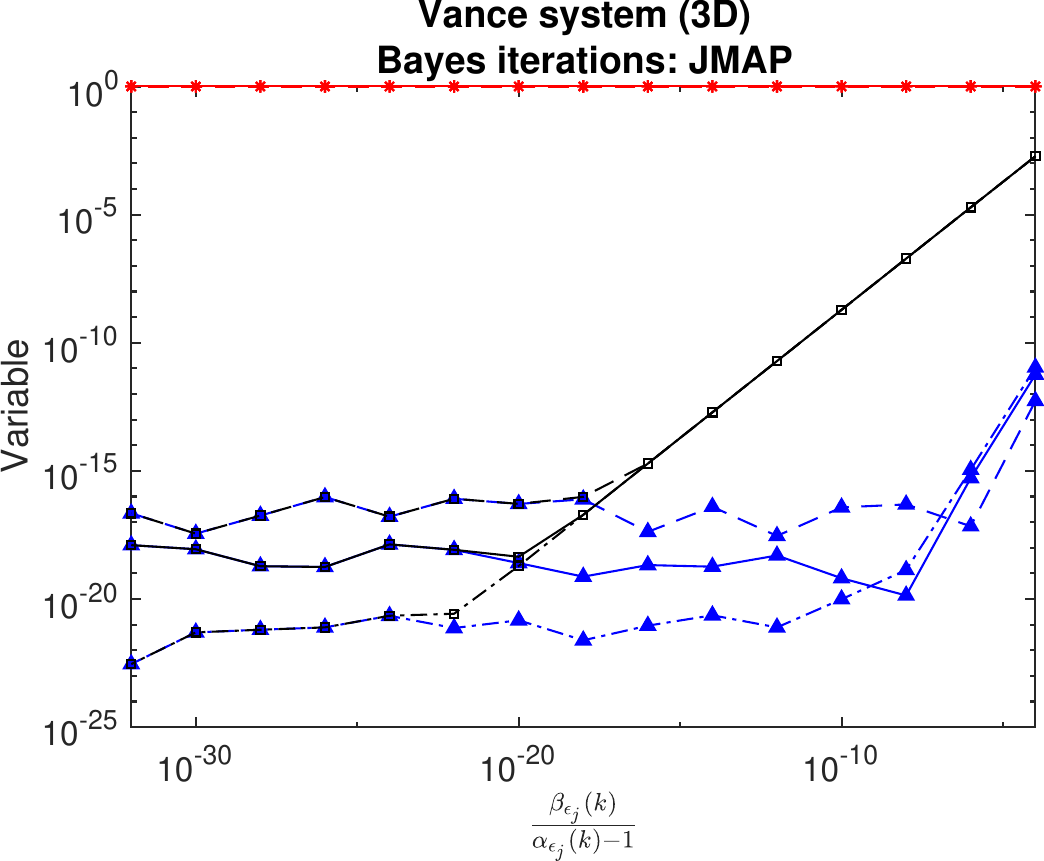} }
  \put(55,430){\includegraphics[height=25mm]{figs_suppl/figleg/Legend_2norms} }
  \put(285,378){\includegraphics[height=11mm, width=6mm]{figs_suppl/figleg/circle} }
  \put(0,300){\small (a)}
  \put(370,300){\includegraphics[height=60mm]{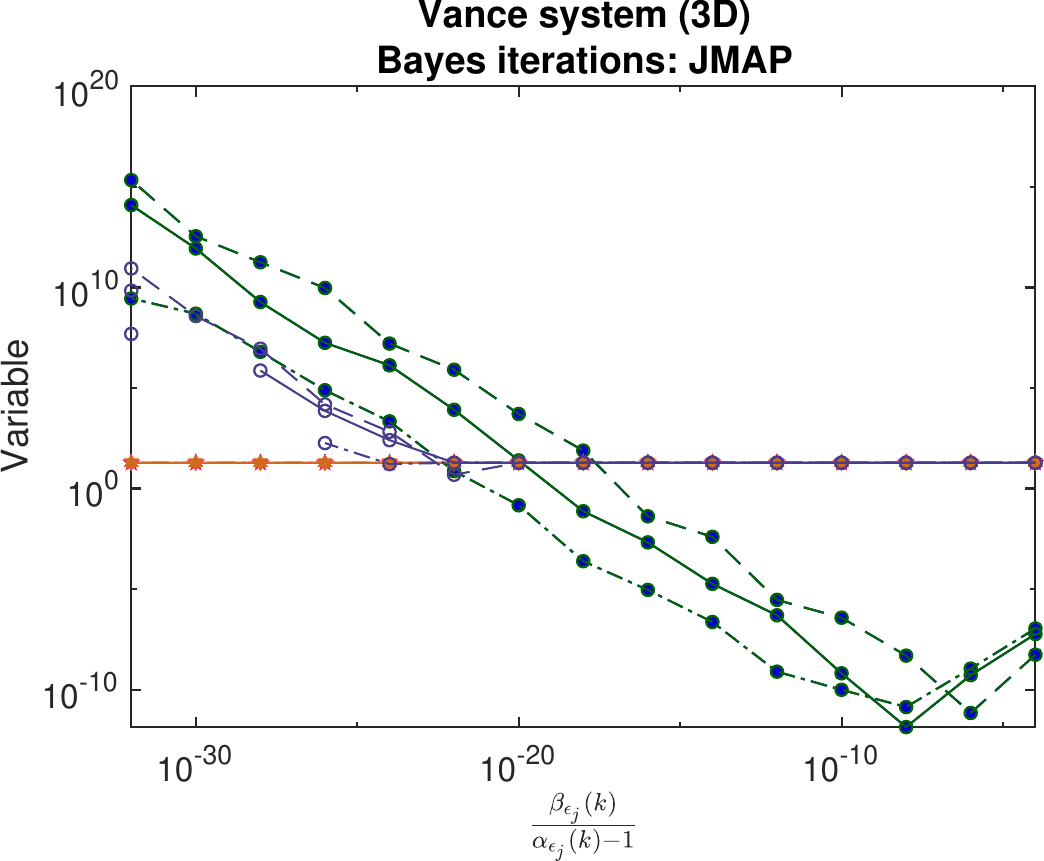} }
  \put(720,362){\includegraphics[height=41mm]{figs_suppl/figleg/Legend_Gnorms1} }
  \put(655,336){\includegraphics[height=9mm, width=6mm]{figs_suppl/figleg/circle} }
  \put(370,300){\small (b)}
  \put(0,0){\includegraphics[height=60mm]{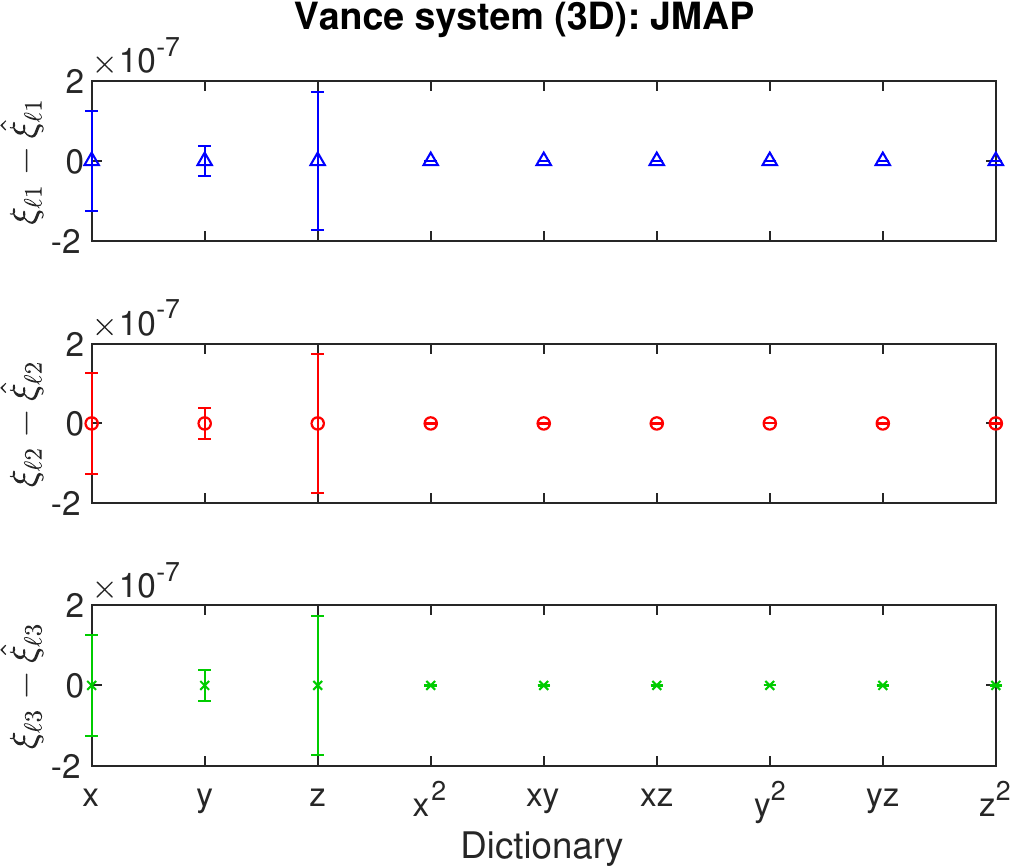} }  
  \put(0,0){\small (c)}
  \put(370,0){\includegraphics[height=60mm]{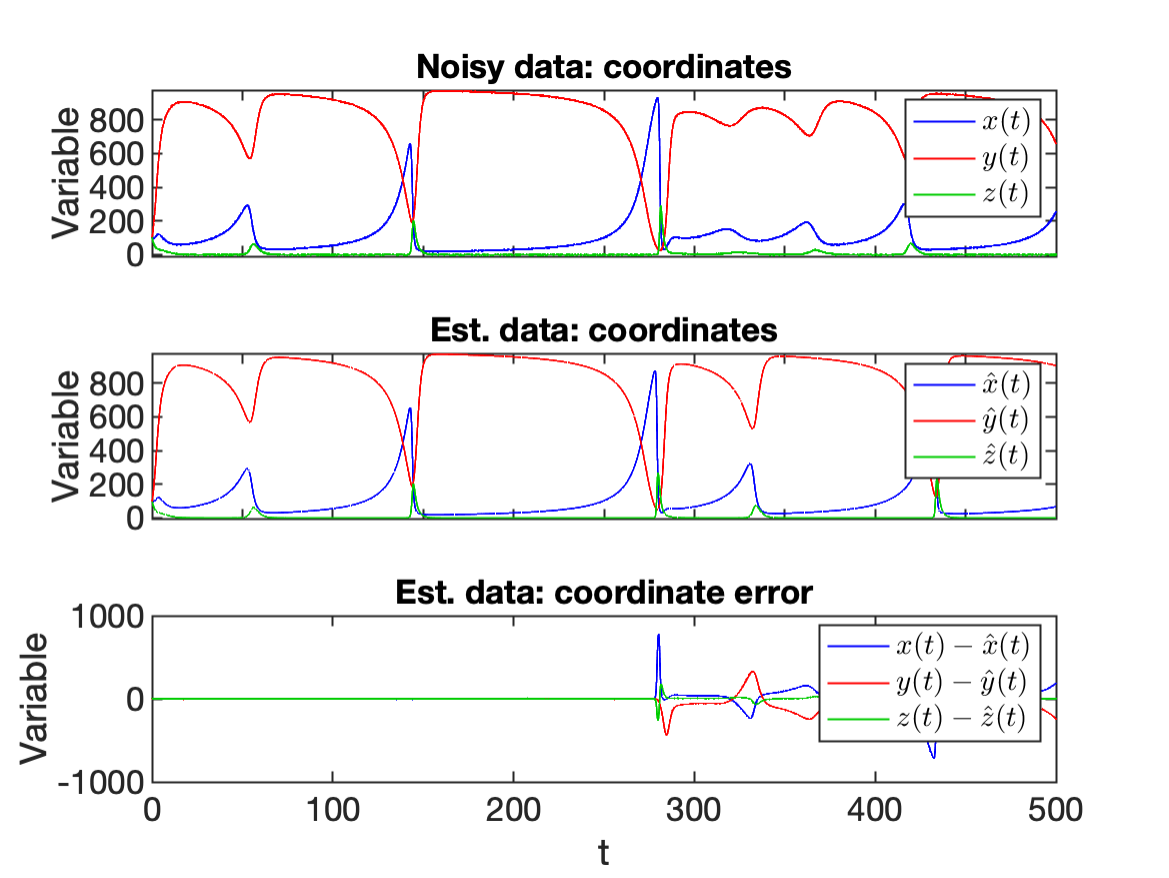} } 
  \put(370,0){\small (d)}
 \end{picture}
\end{center}
\caption{Vance {system with added Gaussian noise}: second-order polynomial regularization using JMAP, $T=500$, $t_{step}=0.02$, $\varepsilon=2.0$, showing the iteration sequence with decreasing $\mathsf{E}_{\eizero}$, including 
(a) 2-norm residual, regularization and objective functions \eqref{eq:modJ}, showing the optimal iteration ($k=4$); 
(b) Gaussian norms for the prior, likelihood, posterior and evidence, showing the optimal iteration ($k=4$); 
(c) optimal error in predicted coefficients $\matparamc_{ij}-\hat{\matparamc}_{ij}$, with error bars from the posterior covariance \eqref{eq:posterior_estimators}; and
(d) original and optimal predicted data and their differences.}
\label{fig:SI_Vance_JMAP}
\end{figure*}

\begin{figure*}[h]
\begin{center}
\setlength{\unitlength}{0.6pt}
 \begin{picture}(800,300)
  \put(0,0){\includegraphics[height=65mm]{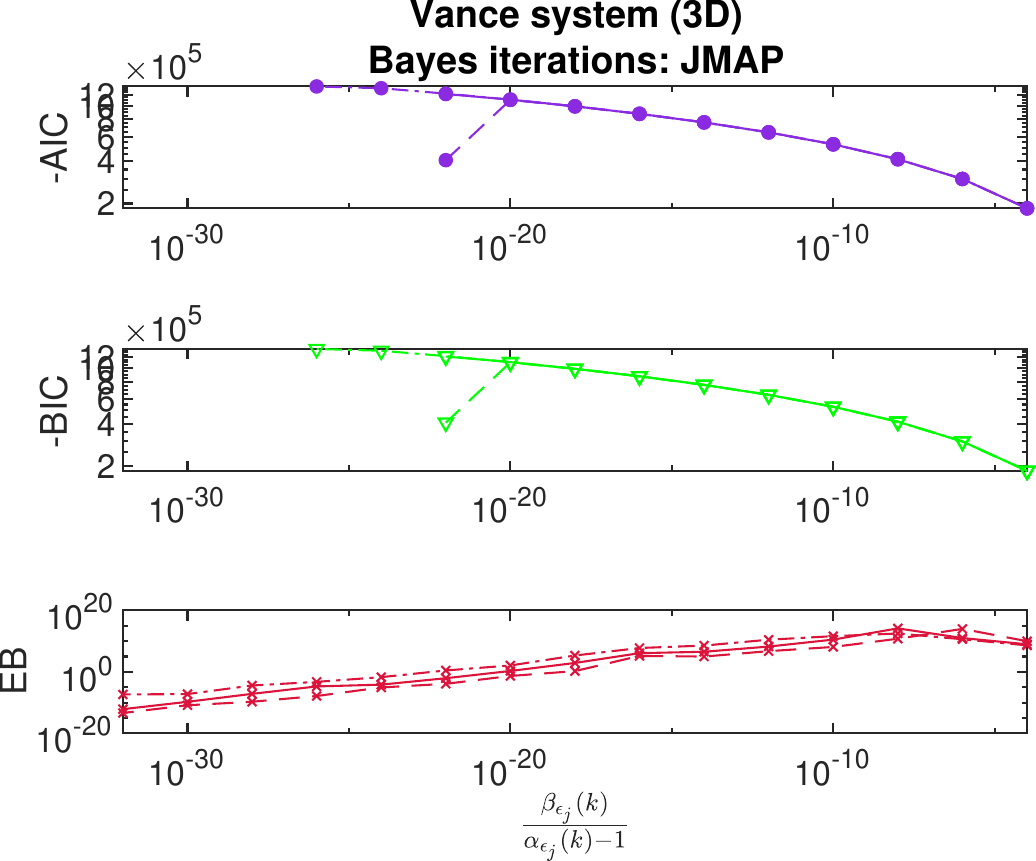} }
  \put(0,0){\small (a)}
  \put(400,0){\includegraphics[height=65mm]{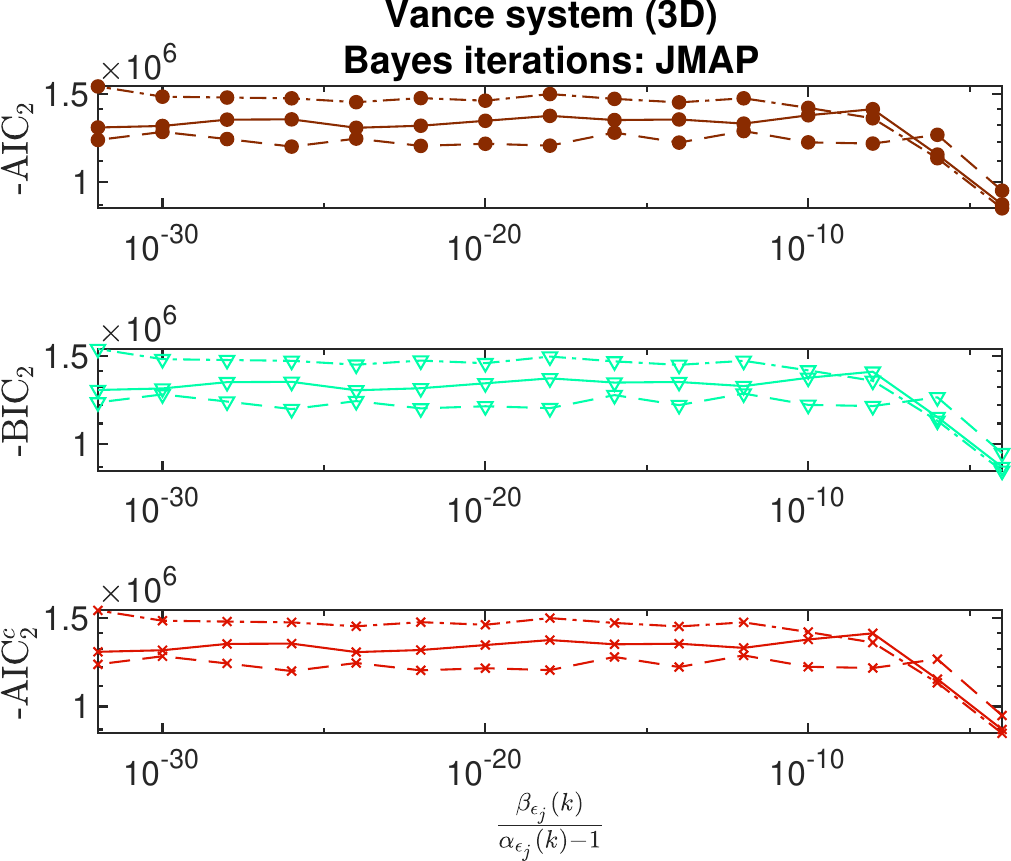} }
  \put(400,0){\small (b)}
 \end{picture}
\end{center}
\caption{Vance {system with added Gaussian noise}: second-order polynomial regularization using JMAP, $T=500$, $t_{step}=0.02$, $\varepsilon=2.0$, showing the alternative metrics by iteration sequence: (a) AIC, BIC and EB \eqref{eq:AIC_BIC}-\eqref{eq:EB}, and (b) 2-norm approximations to AIC, BIC and AIC$_c$.}
\label{fig:SI_Vance_sys_JMAP_metrics}
\end{figure*}

\begin{figure*}[h]
\begin{center}
\setlength{\unitlength}{0.6pt}
 \begin{picture}(800,580)
  \put(0,300){\includegraphics[height=60mm]{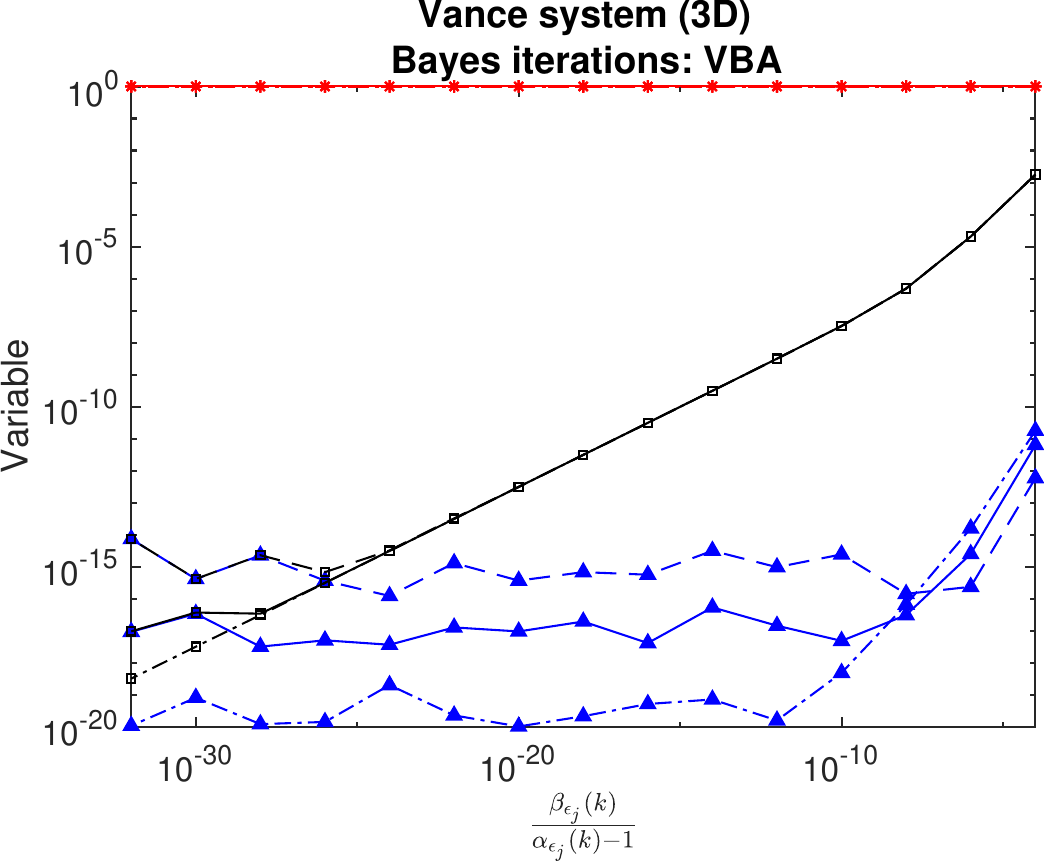} }
  \put(55,430){\includegraphics[height=25mm]{figs_suppl/figleg/Legend_2norms} }
  \put(285,371){\includegraphics[width=6mm]{figs_suppl/figleg/circle} }
  \put(0,300){\small (a)}
  \put(370,300){\includegraphics[height=60mm]{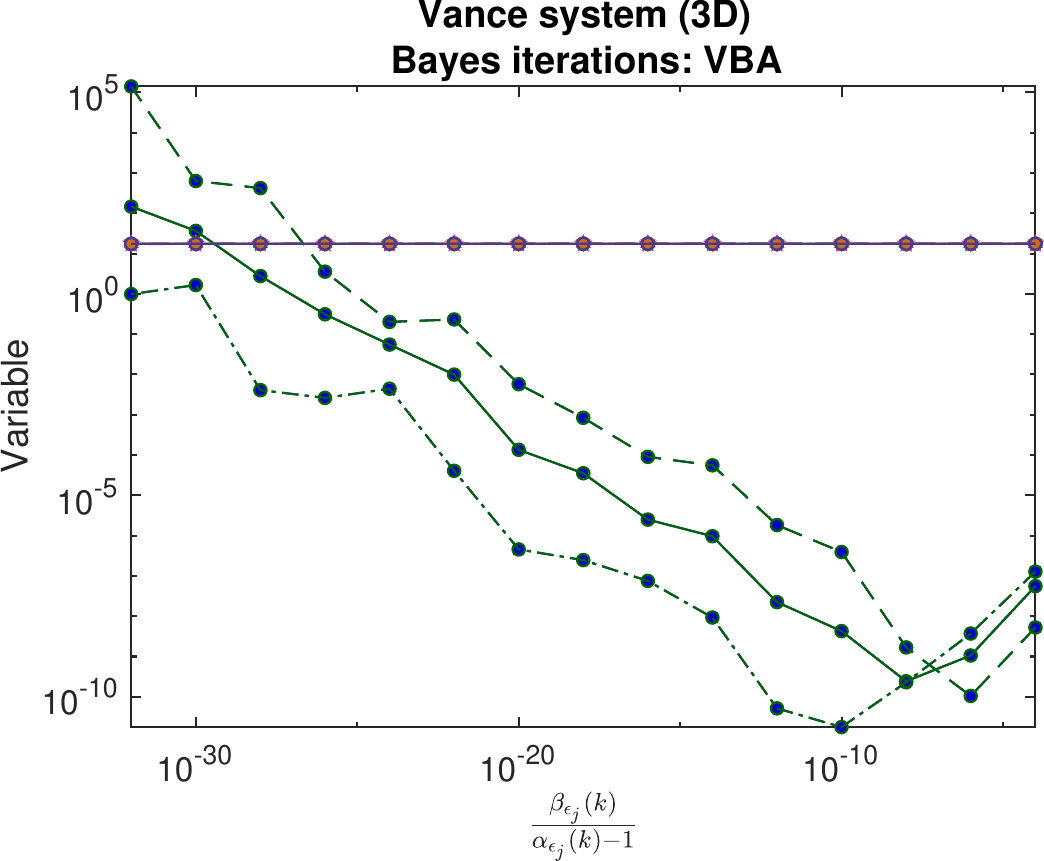} }
  \put(720,362){\includegraphics[height=41mm]{figs_suppl/figleg/Legend_Gnorms1} }
  \put(655,348){\includegraphics[width=6mm]{figs_suppl/figleg/circle} }
  \put(370,300){\small (b)}
  \put(0,0){\includegraphics[height=60mm]{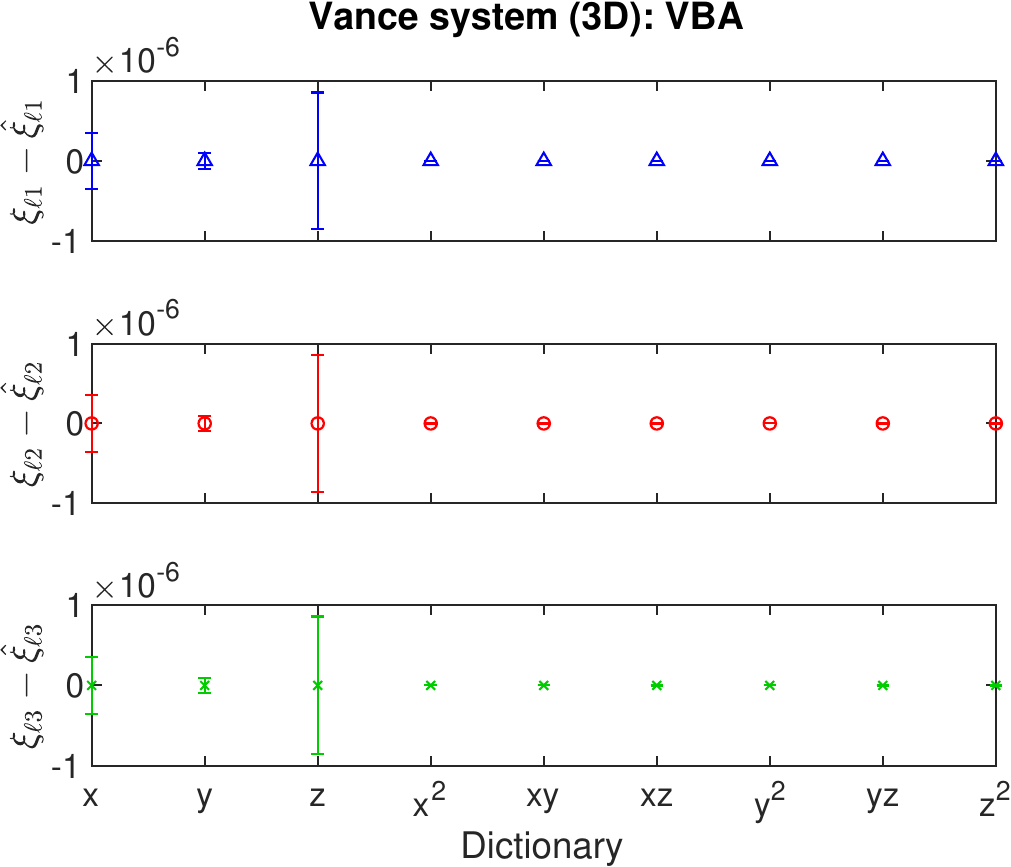} } 
  \put(0,0){\small (c)}
  \put(370,0){\includegraphics[height=60mm]{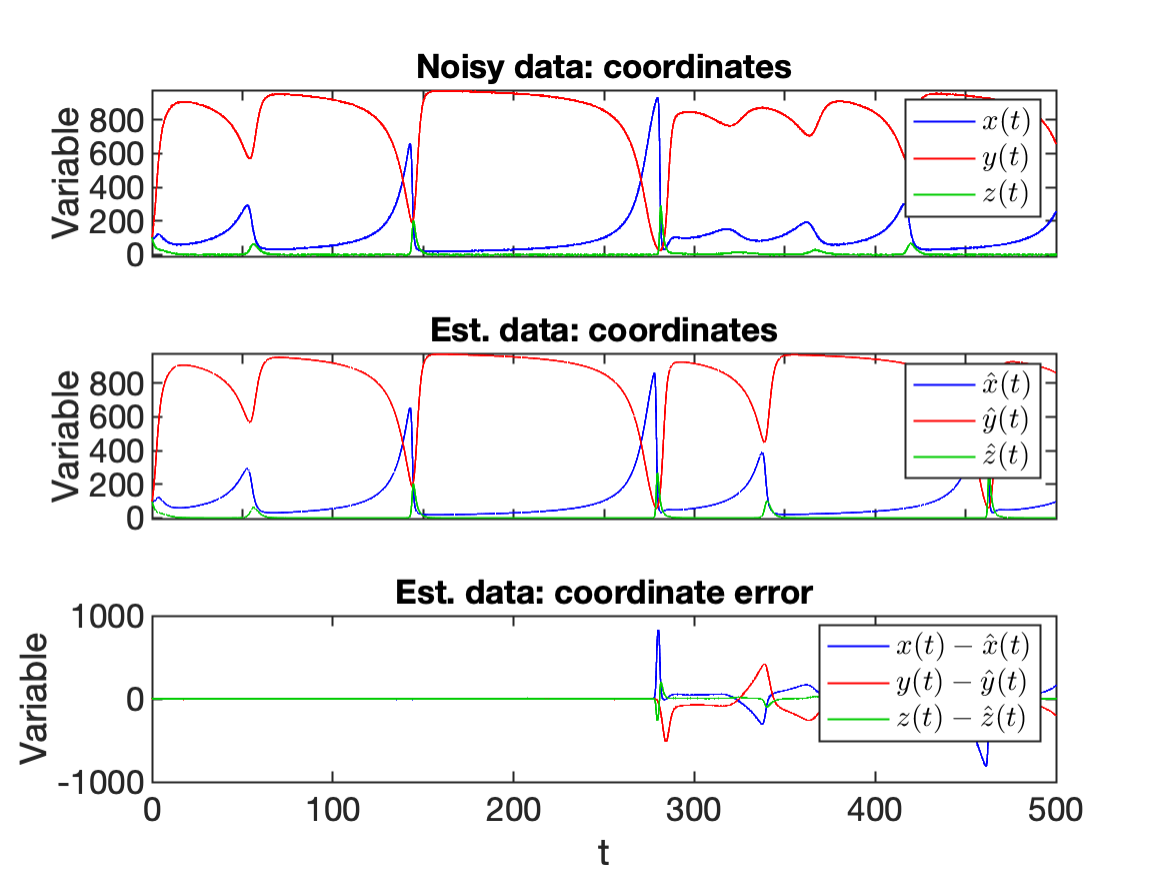} } 
  \put(370,0){\small (d)}
 \end{picture}
\end{center}
\caption{Vance {system with added Gaussian noise}: second-order polynomial regularization using VBA, $T=500$, $t_{step}=0.02$, $\varepsilon=2.0$, showing the iteration sequence with decreasing $\mathsf{E}_{\eizero}$, including 
(a) 2-norm residual, regularization and objective functions \eqref{eq:modJ}, showing the optimal iteration ($k=4$); 
(b) Gaussian norms for the prior, likelihood, posterior and evidence, showing the optimal iteration ($k=4$); 
(c) optimal error in predicted coefficients $\matparamc_{ij}-\hat{\matparamc}_{ij}$, with error bars from the posterior covariance \eqref{eq:posterior_estimators}; and
(d) original and optimal predicted data and their differences.}
\label{fig:SI_Vance_VBA}
\end{figure*}

\begin{figure*}[h]
\begin{center}
\setlength{\unitlength}{0.6pt}
 \begin{picture}(800,300)
  \put(0,0){\includegraphics[height=65mm]{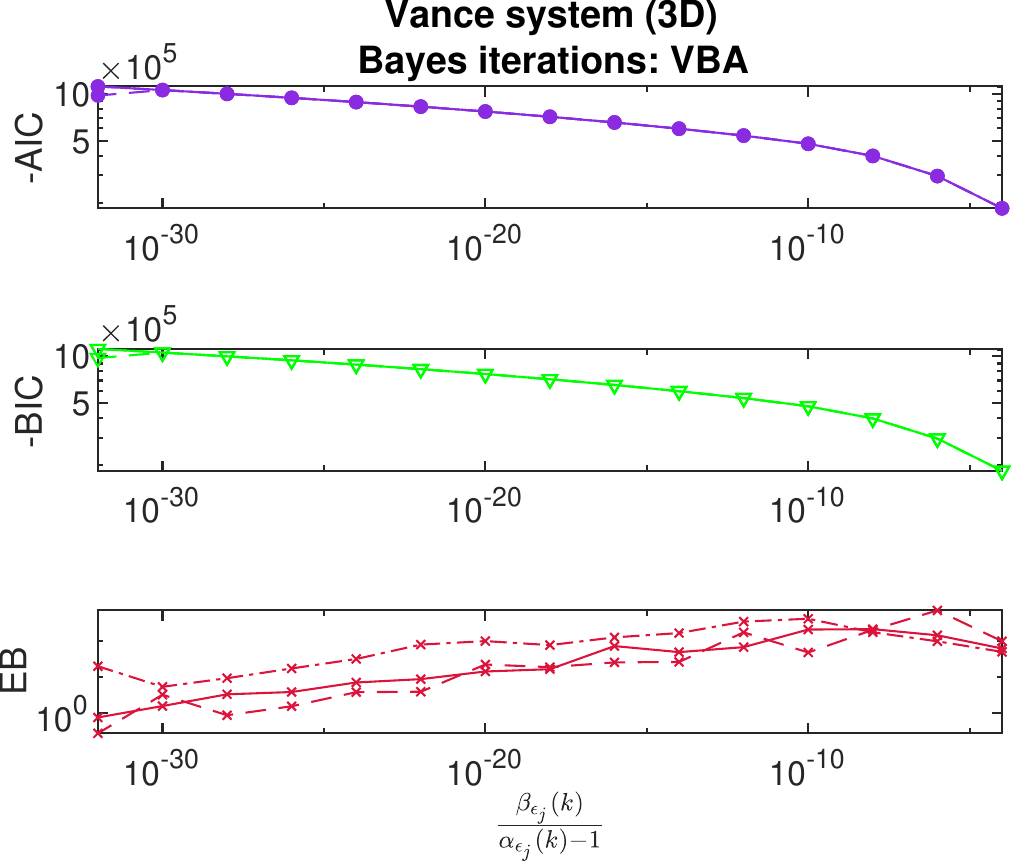} }
  \put(0,0){\small (a)}
  \put(400,0){\includegraphics[height=65mm]{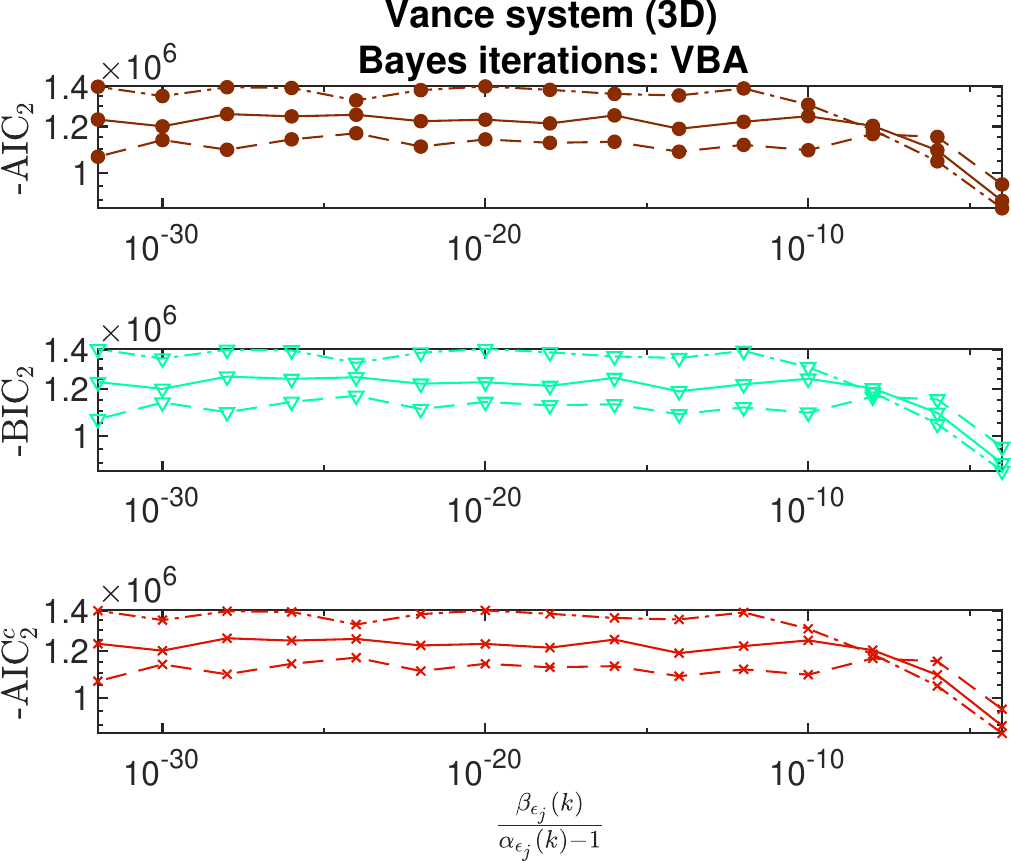} }
  \put(400,0){\small (b)}
 \end{picture}
\end{center}
\caption{Vance {system with added Gaussian noise}: second-order polynomial regularization using VBA, $T=500$, $t_{step}=0.02$, $\varepsilon=2.0$, showing the alternative metrics by iteration sequence: (a) AIC, BIC and EB \eqref{eq:AIC_BIC}-\eqref{eq:EB}, and (b) 2-norm approximations to AIC, BIC and AIC$_c$.}
\label{fig:SI_Vance_sys_VBA_metrics}
\end{figure*}

\begin{figure*}[h]
\begin{center}
\setlength{\unitlength}{0.6pt}
 \begin{picture}(800,580)
  \put(0,300){\includegraphics[height=60mm]{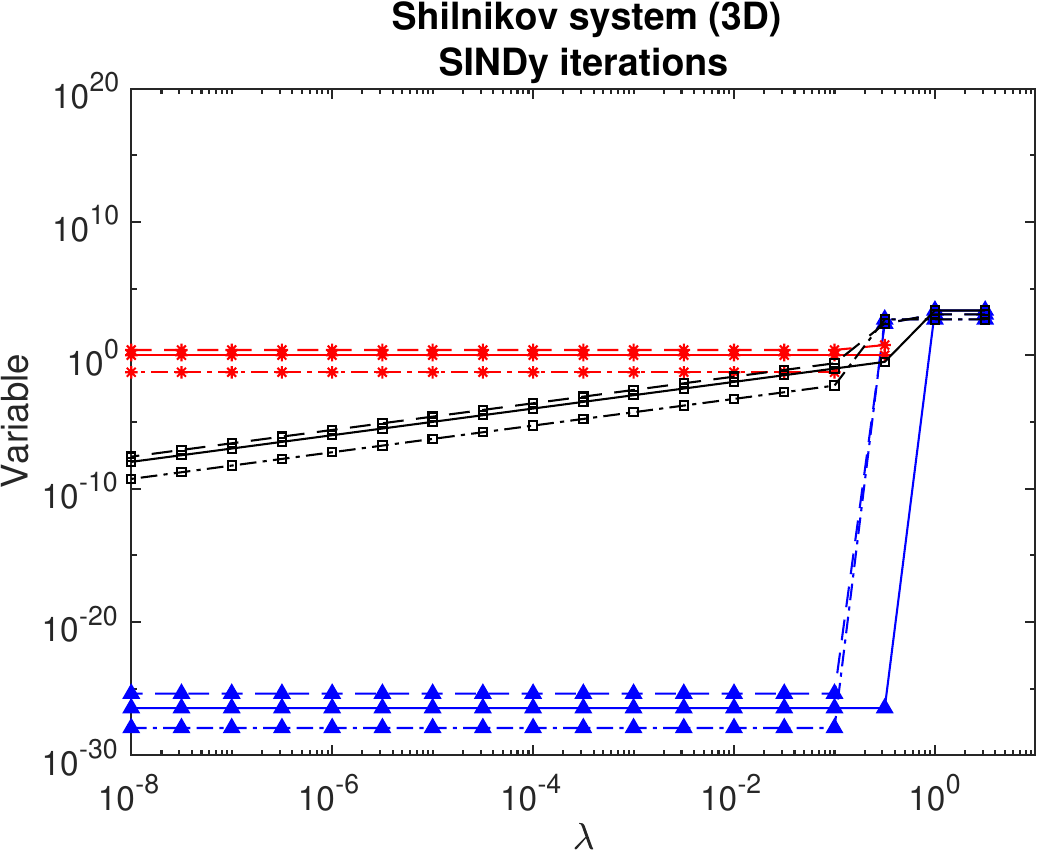} }
  \put(55,505){\includegraphics[height=9mm]{figs_suppl/figleg/Legend_2norms2} }
  \put(264,332){\includegraphics[height=6mm]{figs_suppl/figleg/circle} }
  \put(0,300){\small (a)}
  \put(370,300){\includegraphics[height=60mm]{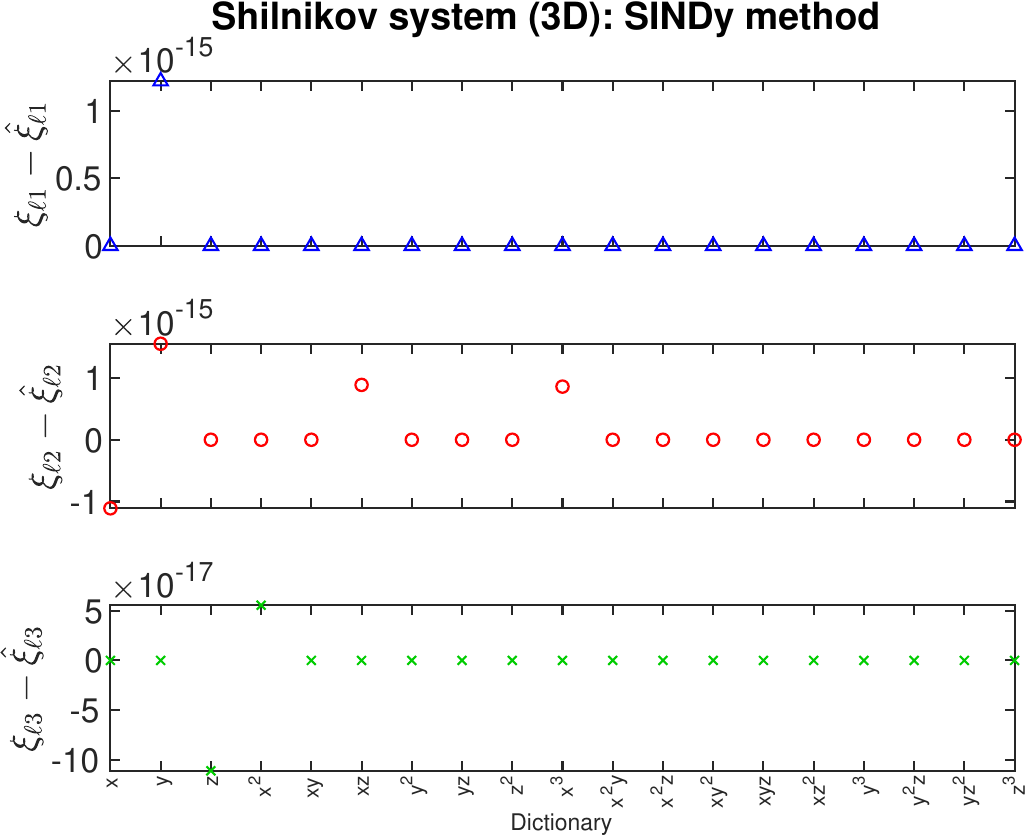} }
  \put(370,300){\small (b)}
  \put(0,0){\includegraphics[height=60mm]{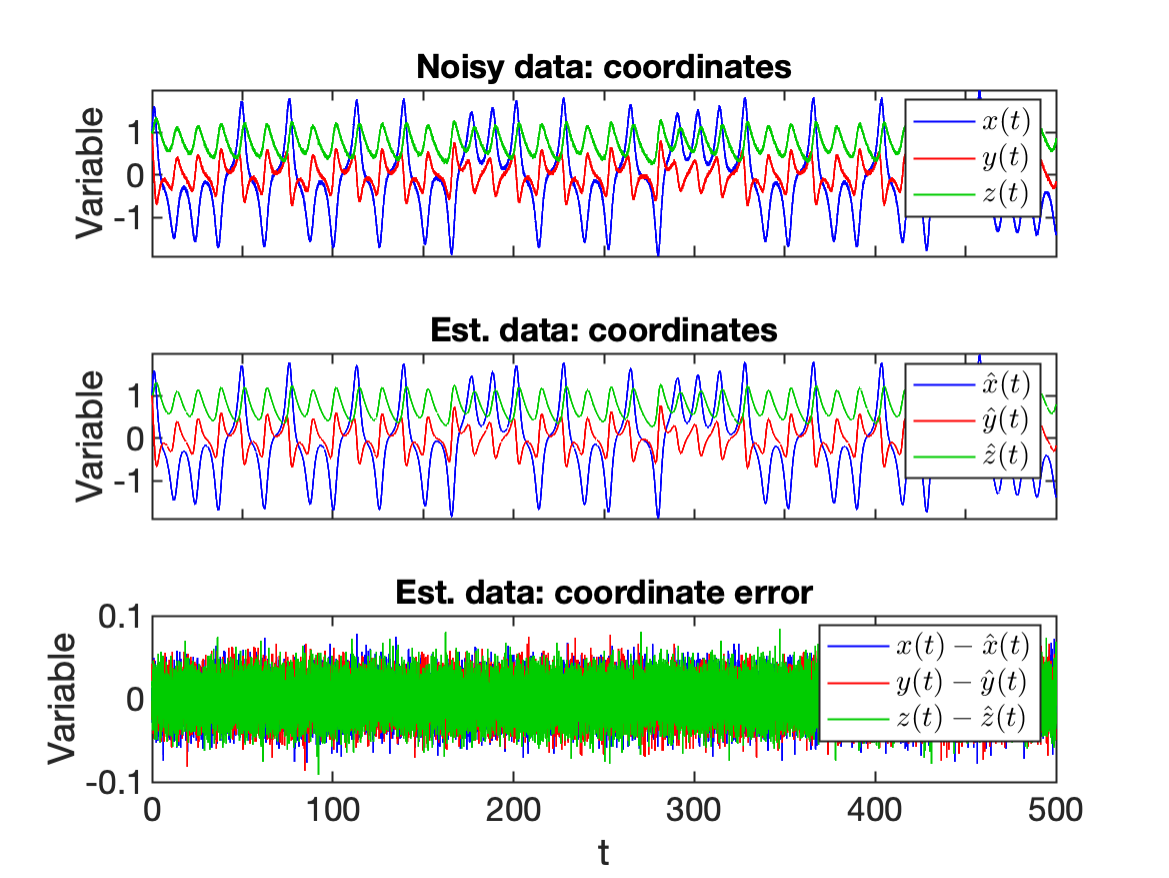} }
  \put(0,0){\small (c)}
 \end{picture}
\end{center}
\caption{Shilnikov {system with added Gaussian noise}: third-order polynomial regularization using SINDy, $T=500$, $t_{step}=0.02$, $\varepsilon=0.02$:
(a) plots of 2-norm residual, regularization and objective functions \eqref{eq:modJ} 
with decreasing $\lambda$, showing the optimal iteration ($k=5)$; 
(b) optimal error in predicted coefficients $\matparamc_{ij}-\hat{\matparamc}_{ij}$; and
(c) original and optimal predicted data and their differences.}
\label{fig:SI_Shilnikov_SINDy}
\end{figure*}

\begin{figure*}[h]
\begin{center}
\setlength{\unitlength}{0.6pt}
 \begin{picture}(800,580)
  \put(0,300){\includegraphics[height=60mm]{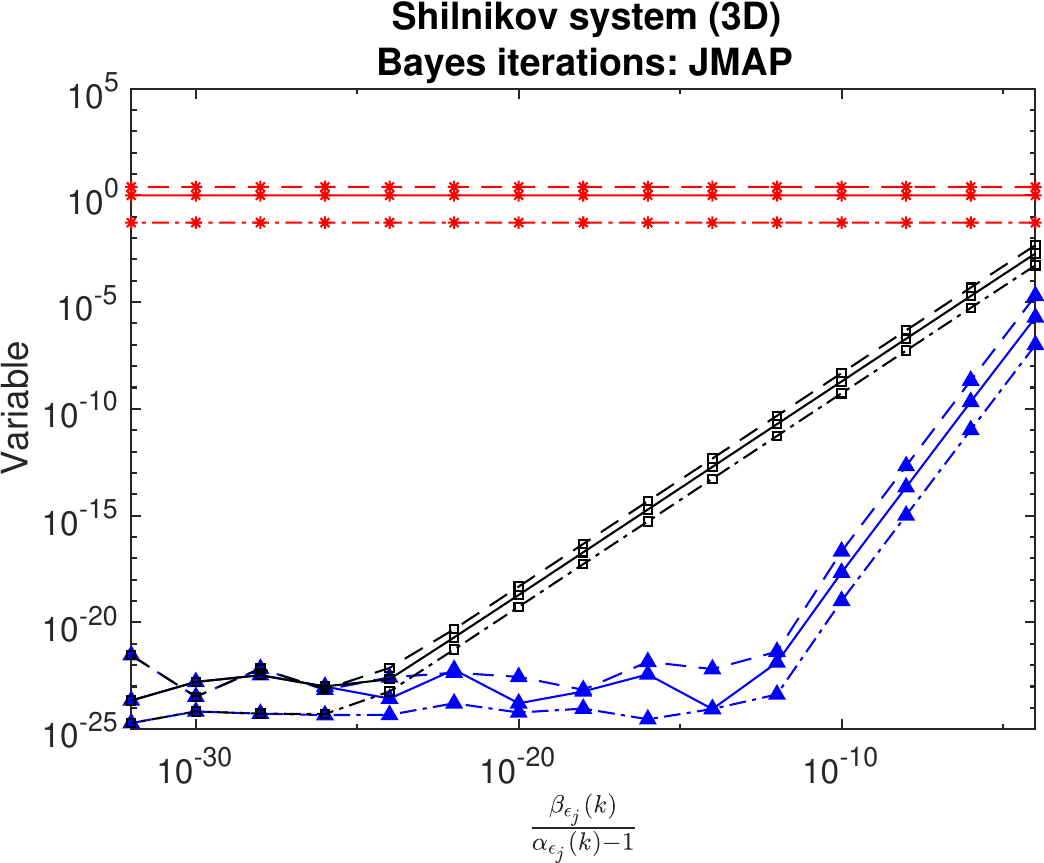} }
  \put(55,380){\includegraphics[height=25mm]{figs_suppl/figleg/Legend_2norms} }
  \put(243,349){\includegraphics[height=6mm]{figs_suppl/figleg/circle} }
  \put(0,300){\small (a)}
  \put(370,300){\includegraphics[height=60mm]{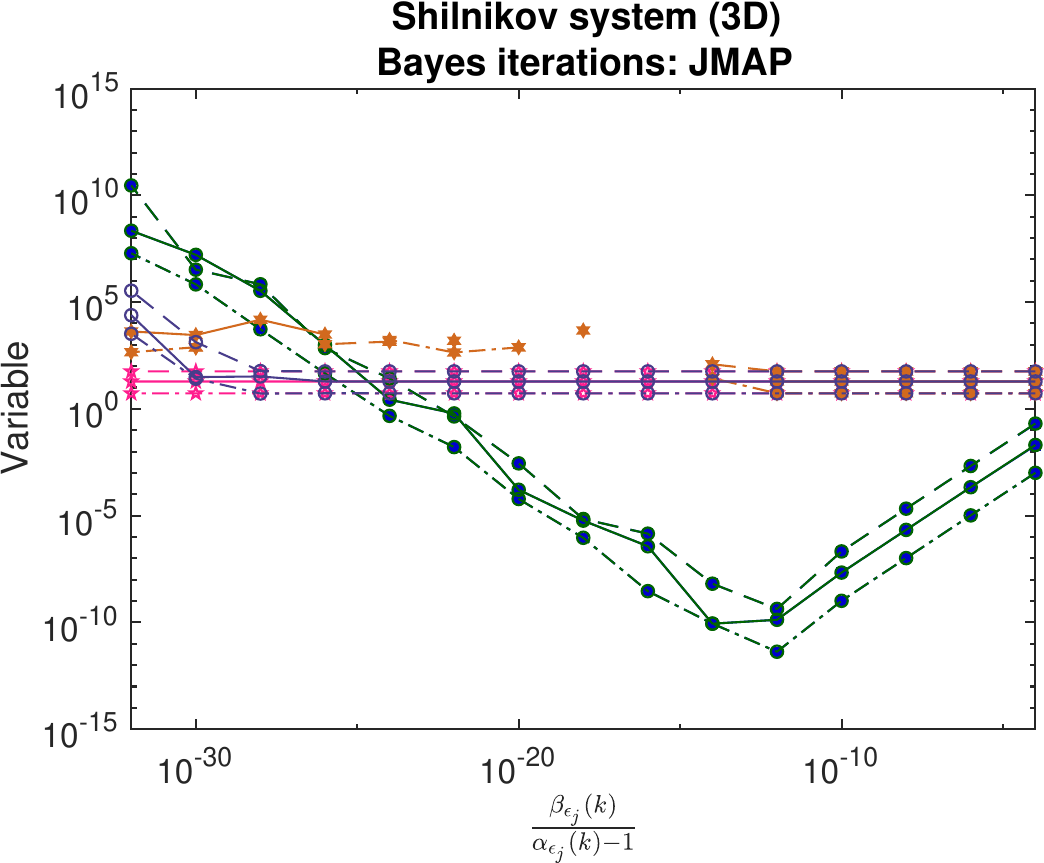} }
  \put(720,362){\includegraphics[height=41mm]{figs_suppl/figleg/Legend_Gnorms1} }
  \put(612,362){\includegraphics[height=6mm]{figs_suppl/figleg/circle} }
  \put(370,300){\small (b)}
  \put(0,0){\includegraphics[height=60mm]{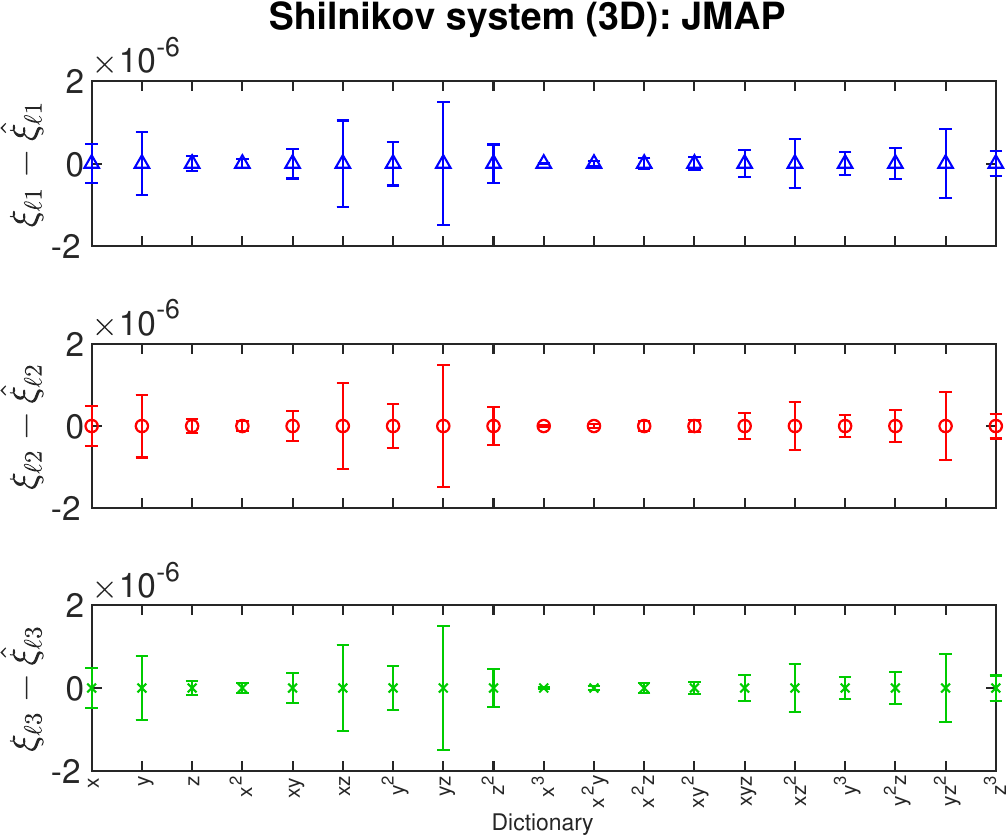} }  
  \put(0,0){\small (c)}
  \put(370,0){\includegraphics[height=60mm]{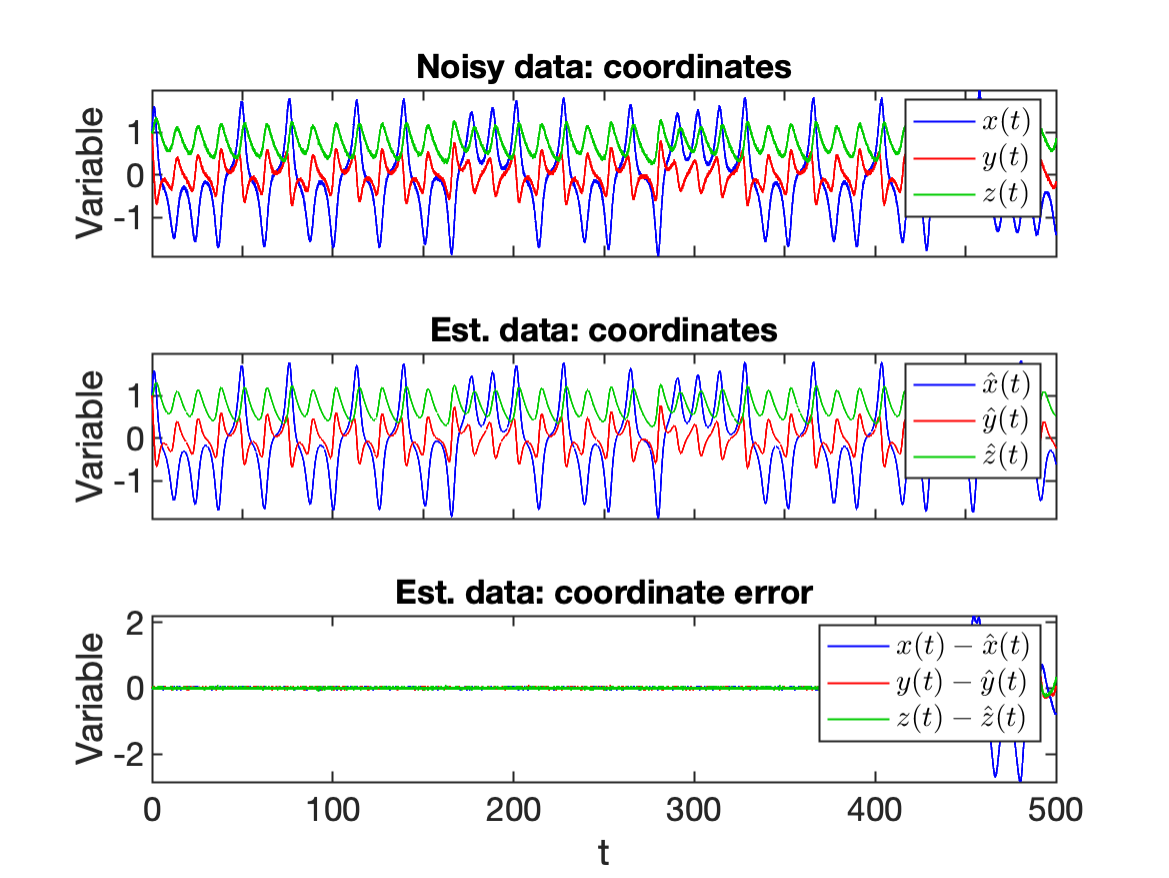} } 
  \put(370,0){\small (d)}
 \end{picture}
\end{center}
\caption{Shilnikov {system with added Gaussian noise}: third-order polynomial regularization using JMAP, $T=500$, $t_{step}=0.02$, $\varepsilon=0.02$, showing the iteration sequence with decreasing $\mathsf{E}_{\eizero}$, including 
(a) 2-norm residual, regularization and objective functions \eqref{eq:modJ}, showing the optimal iteration ($k=6$); 
(b) Gaussian norms for the prior, likelihood, posterior and evidence, showing the optimal iteration ($k=6$); 
(c) optimal error in predicted coefficients $\matparamc_{ij}-\hat{\matparamc}_{ij}$, with error bars from the posterior covariance \eqref{eq:posterior_estimators}; and
(d) original and optimal predicted data and their differences.}
\label{fig:SI_Shilnikov_JMAP}
\end{figure*}

\begin{figure*}[h]
\begin{center}
\setlength{\unitlength}{0.6pt}
 \begin{picture}(800,300)
  \put(0,0){\includegraphics[height=65mm]{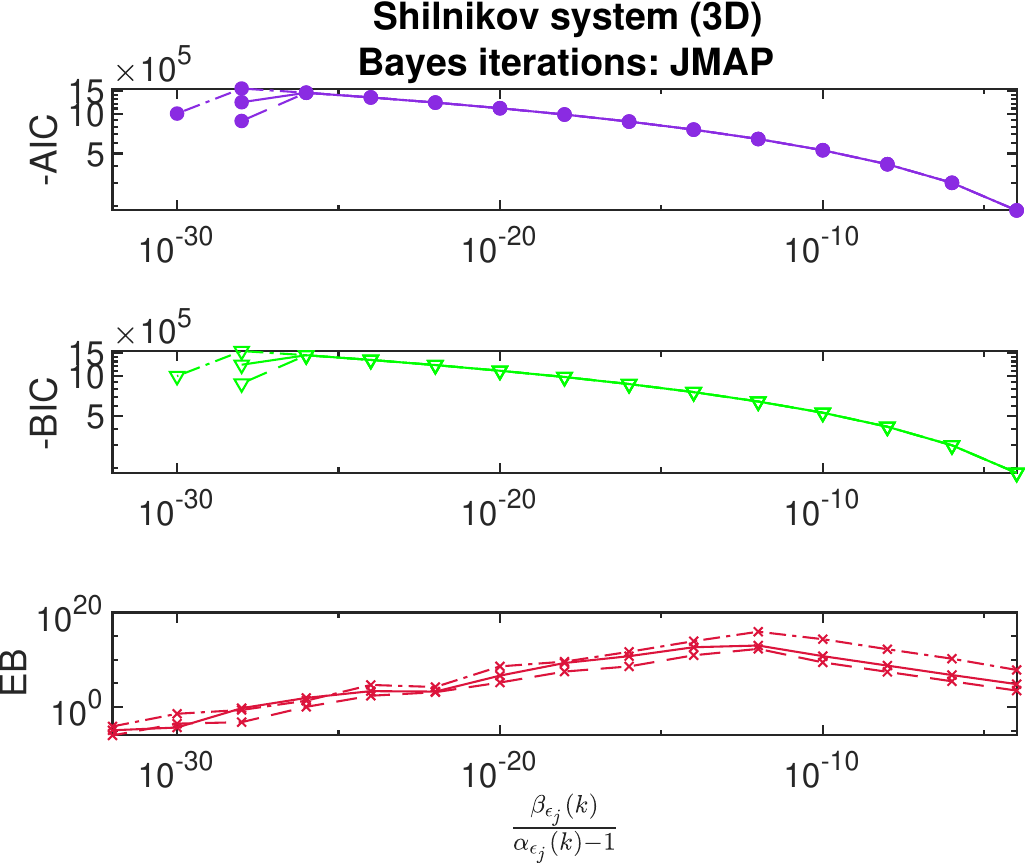} }
  \put(0,0){\small (a)}
  \put(400,0){\includegraphics[height=65mm]{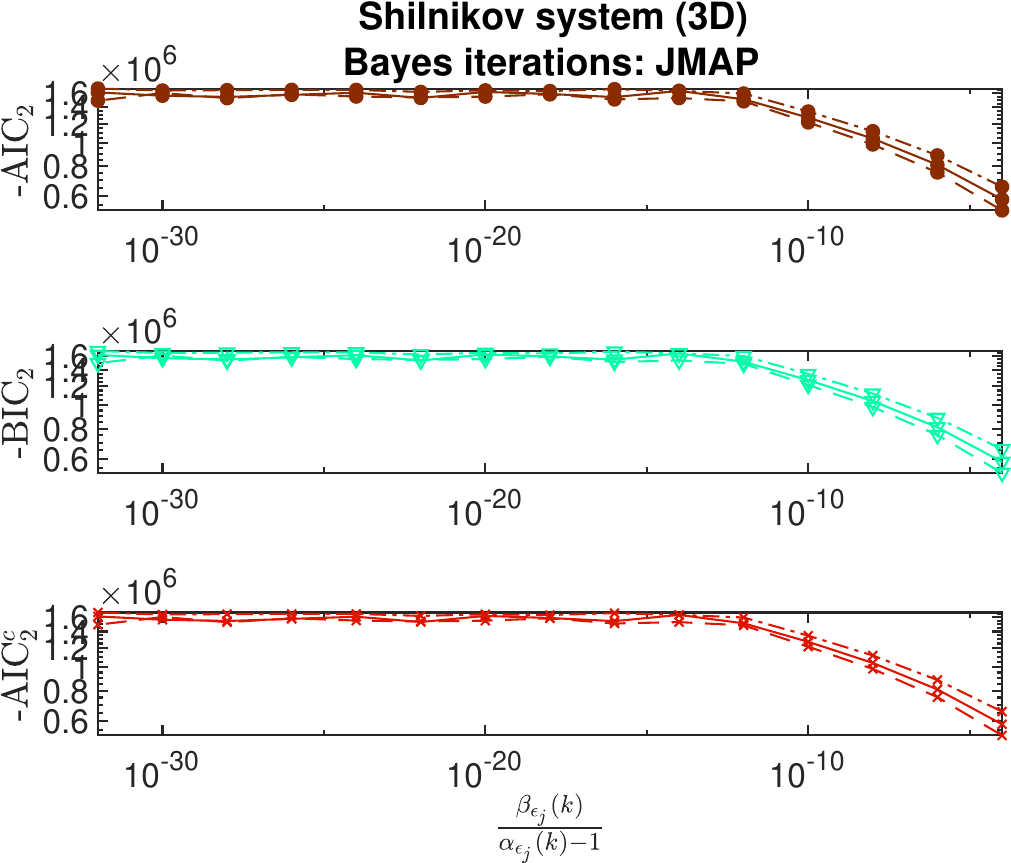} }
  \put(400,0){\small (b)}
 \end{picture}
\end{center}
\caption{Shilnikov {system with added Gaussian noise}: third-order polynomial regularization using JMAP, $T=500$, $t_{step}=0.02$, $\varepsilon=0.02$, showing the alternative metrics by iteration sequence: (a) AIC, BIC and EB \eqref{eq:AIC_BIC}-\eqref{eq:EB}, and (b) 2-norm approximations to AIC, BIC and AIC$_c$.}
\label{fig:SI_Shilnikov_sys_JMAP_metrics}
\end{figure*}

\begin{figure*}[h]
\begin{center}
\setlength{\unitlength}{0.6pt}
 \begin{picture}(800,580)
  \put(0,300){\includegraphics[height=60mm]{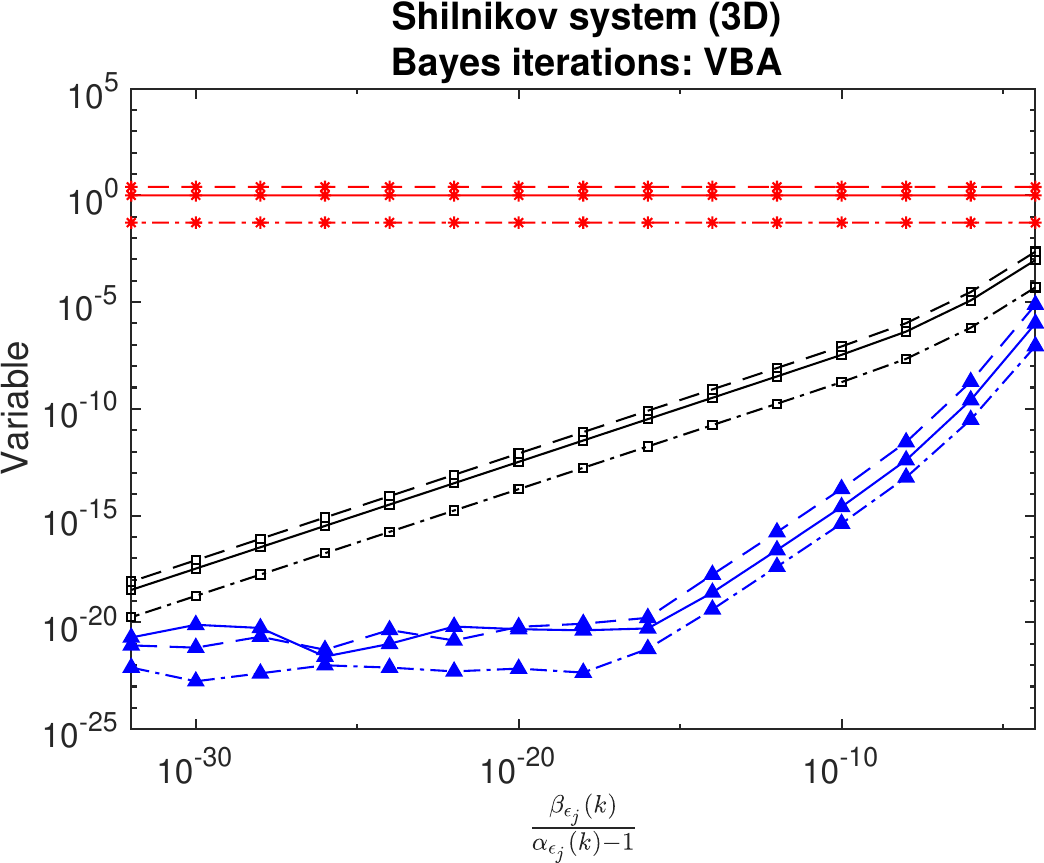} }
  \put(55,385){\includegraphics[height=25mm]{figs_suppl/figleg/Legend_2norms} }
  \put(200,359){\includegraphics[height=6mm]{figs_suppl/figleg/circle} }
  \put(0,300){\small (a)}
  \put(370,300){\includegraphics[height=60mm]{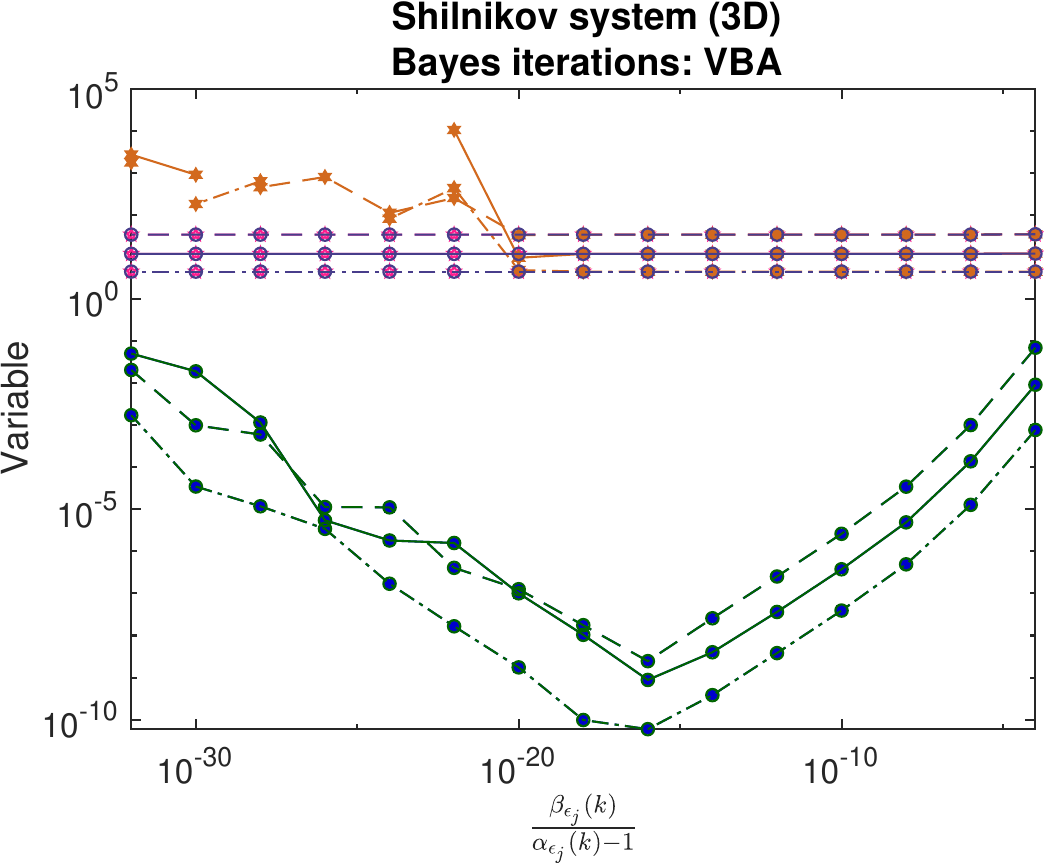} }
  \put(720,362){\includegraphics[height=41mm]{figs_suppl/figleg/Legend_Gnorms1} }
  \put(569,338){\includegraphics[width=6mm, height=8mm]{figs_suppl/figleg/circle} }
  \put(370,300){\small (b)}
  \put(0,0){\includegraphics[height=60mm]{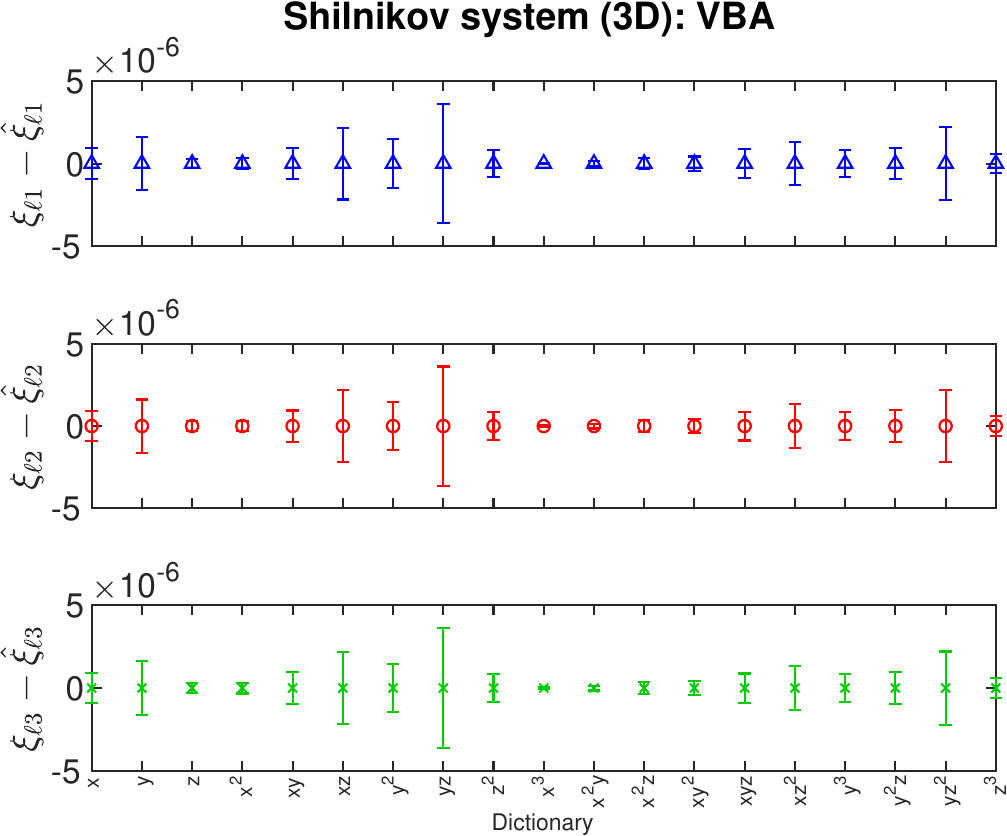} } 
  \put(0,0){\small (c)}
  \put(370,0){\includegraphics[height=60mm]{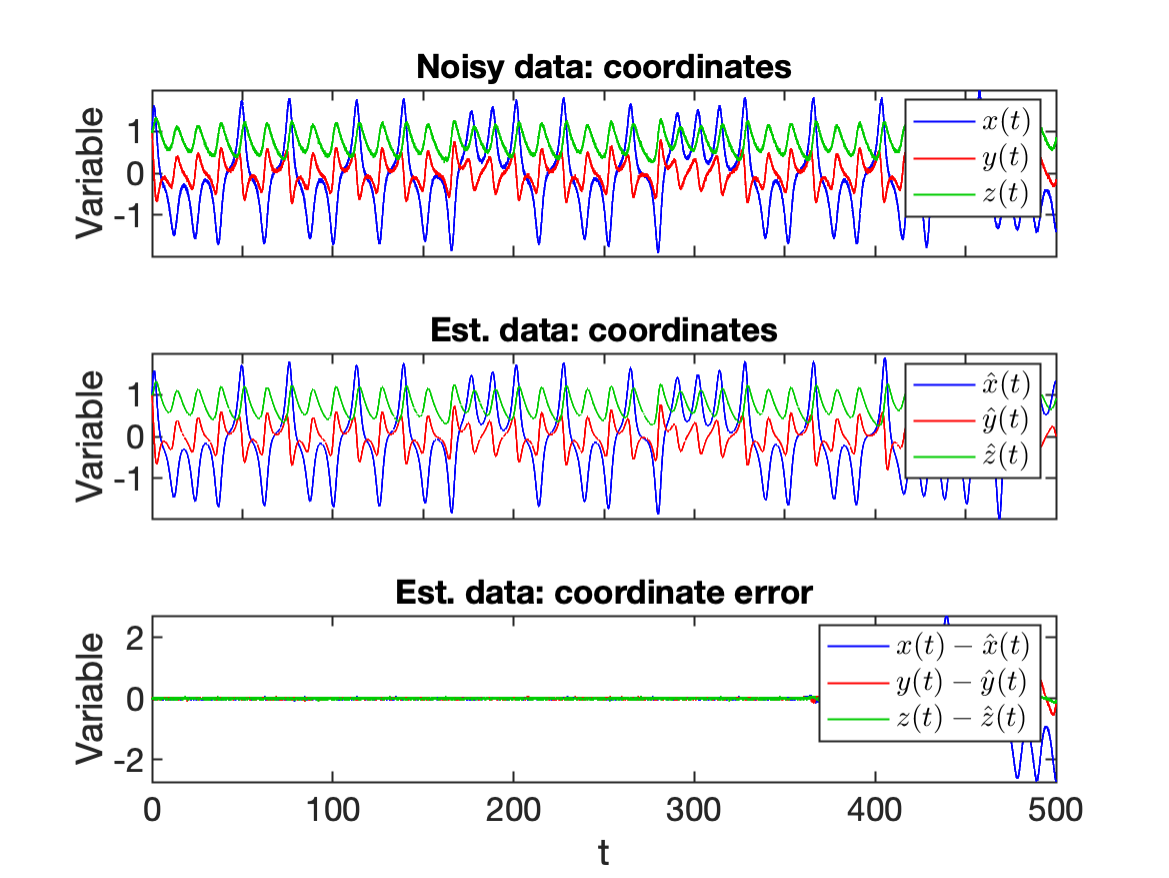} } 
  \put(370,0){\small (d)}
 \end{picture}
\end{center}
\caption{Shilnikov {system with added Gaussian noise}: third-order polynomial regularization using VBA, $T=500$, $t_{step}=0.02$, $\varepsilon=0.02$, showing the iteration sequence with decreasing $\mathsf{E}_{\eizero}$, including 
(a) 2-norm residual, regularization and objective functions \eqref{eq:modJ}, showing the optimal iteration ($k=8$); 
(b) Gaussian norms for the prior, likelihood, posterior and evidence, showing the optimal iteration ($k=8$); 
(c) optimal error in predicted coefficients $\matparamc_{ij}-\hat{\matparamc}_{ij}$, with error bars from the posterior covariance \eqref{eq:posterior_estimators}; and
(d) original and optimal predicted data and their differences.}
\label{fig:SI_Shilnikov_VBA}
\end{figure*}

\begin{figure*}[h]
\begin{center}
\setlength{\unitlength}{0.6pt}
 \begin{picture}(800,300)
  \put(0,0){\includegraphics[height=65mm]{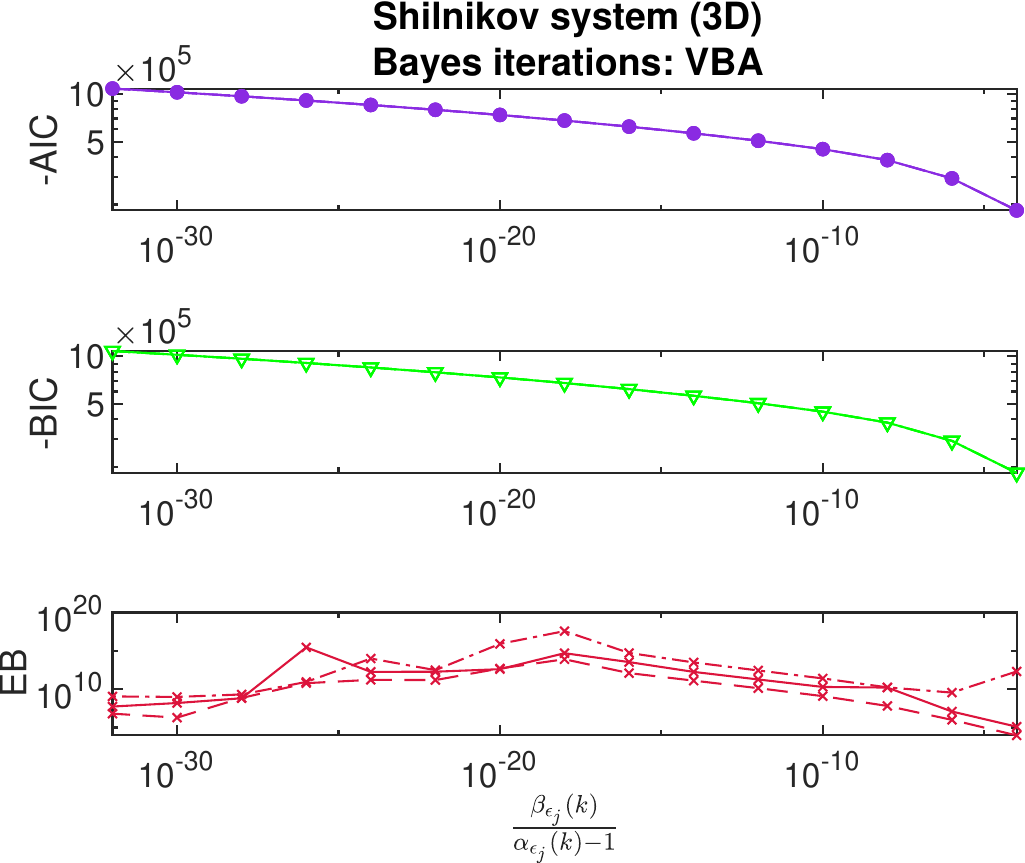} }
  \put(0,0){\small (a)}
  \put(400,0){\includegraphics[height=65mm]{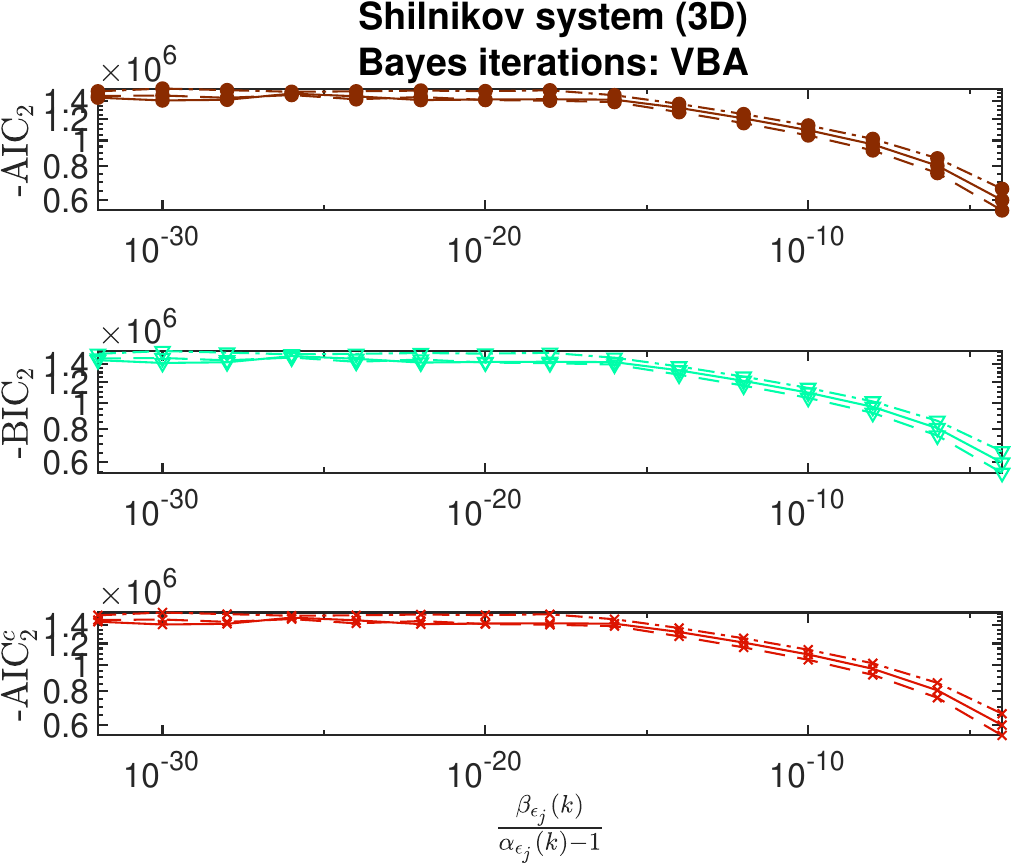} }
  \put(400,0){\small (b)}
 \end{picture}
\end{center}
\caption{Shilnikov {system with added Gaussian noise}: third-order polynomial regularization using VBA, $T=500$, $t_{step}=0.02$, $\varepsilon=0.02$, showing the alternative metrics by iteration sequence: (a) AIC, BIC and EB \eqref{eq:AIC_BIC}-\eqref{eq:EB}, and (b) 2-norm approximations to AIC, BIC and AIC$_c$.}
\label{fig:SI_Shilnikov_sys_VBA_metrics}
\end{figure*}

The supplementary information, given in Figures \ref{fig:SI_noise}-\ref{fig:SI_Shilnikov_sys_VBA_metrics}, provides a more complete set of figures for the case studies examined in the manuscript.



\end{document}